***OSIRIS-REx Returned a Pristine Sample of Asteroid Bennu: Takeaways from the Mission's Contamination Control and Knowledge Program***

submitted to *Space Science Reviews*

Jason P. Dworkin[1,2*] ORCID: 0000-0001-7779-7765
Scott A. Sandford[3] ORCID: 0000-0002-6034-9816
Nicole G. Lunning[4] ORCID: 0000-0002-9845-5104
Kevin Righter[4,5] ORCID: 0000-0002-6075-7908
Charles C. Lorentson[1]
Daniel P. Glavin[1] ORCID: 0000-0001-7779-7765
Kathie L. Thomas-Keprta[4,6] ORCID: 0000-0001-9402-3606
Loan Le[4,7] ORCID: 0000-0002-6760-2431
Catherine M. Corrigan[8] ORCID: 0000-0002-0151-1457
Timothy J. McCoy ORCID: 0000-0002-4573-3553
Edward B. Bierhaus[9] ORCID: 0000-0001-5890-9821
José C. Aponte[1] ORCID: 0000-0002-0131-1981
Angel Mojarro[1,2] ORCID: 0000-0003-4547-4747
Hannah L. McLain[1,2] ORCID: 0000-0002-4306-4608
Eric T. Parker[1] ORCID: 0000-0001-7825-5941
Denise K. Buckner[1,10] ORCID: 0000-0002-5160-5857
Frédéric Seguin[1,2] ORCID: 0009-0003-3306-4350
Radford L. Perry[1]
Heather V. Graham[1] ORCID: 0000-0002-1986-1885
Havishk Tripathi[1,11] ORCID: 0000-0003-0371-3527
Pierre Haenecour[12] ORCID: 0000-0001-5671-5388
Piers Koefoed[13] ORCID: 0000-0002-4036-6088
Kun Wang[13] ORCID: 0000-0003-0691-7058
Cody Schultz[14] ORCID: 0000-0002-9525-9191
Takahiro Hiroi[14] ORCID: 0000-0003-2866-9091
Ralph E. Milliken[14] ORCID: 0000-0003-3240-4918
Simon J. Clemett[4,7] ORCID: 0000-0002-5187-1653
Bumsoo Kim[4,7] ORCID: 0000-0002-0533-1330
Christopher J. Snead[4]
Eve L. Berger[4] ORCID: 0000-0002-9902-439X
Michael Liss[15,16]
Jessica J. Barnes[12] ORCID: 0000-0002-7151-1014
Sandra Freund[9] ORCID: 0000-0002-2417-6575
Zoe A. Wilson[9]
Ryan D. Dubisher[9]
Natasha V. Almeida[17] ORCID: 0000-0003-4871-8225
Kimberly K. Allums[4,6]
Erika H. Blumenfeld[4,18] ORCID: 0000-0001-7183-0472
Joseph E. Aebersold[4]
Rachel C. Funk[4,19]

Carla P. Gonzales[4,20]
Julia L. Plummer[4,20]
Maritza Montoya[4,20]
Daniel J. Rohmer[4,6]
Erika V. Rivera[4,21] ORCID: 0000-0002-9253-0337
Jannatul Ferdous[4,20]
Gabriel C. Garcia[4,20]
Melissa Rodriguez[4,20] ORCID: 0000-0002-6567-0801
José L. Ponce[4]
Ronald Bastien[4]
Wayland D. Connelly[4,22] ORCID: 0009-0005-6168-3858
Jemma Davidson[4] ORCID: 0000-0002-3725-2960
Marghaleray Amini[23] ORCID: 0000-0003-1582-4881
Vivian Lai[23] ORCID: 0000-0003-3758-2561
Dominique Weis[23] ORCID: 0000-0002-6638-5543
Josh Wimpenny[24] ORCID: 0000-0002-7637-0978
Jan Render[24] ORCID: 0000-0002-6534-9129
Gregory A. Brennecka[24] ORCID: 0000-0002-0852-5595
Yoshihiro Furukawa[25] ORCID: 0000-0001-7241-530X
Toshiki Koga[26] ORCID: 0000-0001-5728-0590
Yoshinori Takano[26] ORCID: 0000-0003-1151-144X
Hiroshi Naraoka[27,28] ORCID: 0000-0002-2373-8759
Yasuhiro Oba[29] ORCID: 0000-0002-6852-3604
Kenneth J. Domanik[12] ORCID: 0009-0002-0184-1816
Yao-Jen Chang[12] ORCID: 0000-0002-5515-7254
Zoe Zeszut[12] ORCID: 0000-0003-3886-3365
Thomas J. Zega[12] ORCID: 0000-0002-9549-022X
George D. Cody[30] ORCID: 0000-0003-4176-3648
Conel M. O'D. Alexander[30] ORCID: 0000-0002-8558-1427
Ian A. Franchi[31] ORCID: 0000-0003-4151-0480
Monica M. Grady[31] ORCID: 0000-0001-8402-3717
Danielle N. Simkus[1,2,32] ORCID: 0000-0002-8272-623X
Jamie E. Elsila[1] ORCID: 0000-0002-0008-2590
Philippe Schmitt-Kopplin[15,16,33] ORCID: 0000-0003-0824-2664
Allison A. Baczynski[34] ORCID: 0000-0001-7490-6620
Katherine H. Freeman[34] ORCID: 0000-0002-3350-7671
Amy E. Hofmann[35] ORCID: 0000-0001-6869-5118
Aaron S. Burton[4,36] ORCID: 0000-0002-1194-4336
Lindsay P. Keller[4] ORCID: 0000-0003-1560-2939
Sara S. Russell[17] ORCID: 0000-0001-5531-7847
Anjani T. Polit[12] ORCID: 0000-0003-4071-7834
Francis M. McCubbin[4] ORCID: 0000-0002-2101-4431
Harold C. Connolly Jr.[12,37,38] ORCID: 0000-0002-1803-5098
Dante S. Lauretta[12] ORCID: 0000-0002-2597-5950

[*]Corresponding author’s email address: jason.p.dworkin@nasa.gov

[1] NASA Goddard Space Flight Center, Greenbelt, Maryland, USA
[2] Center for Space Science and Technology, University of Maryland Baltimore County, Baltimore, Maryland, USA
[3] NASA Ames Research Center, Moffett Field, California, USA
[4] NASA Johnson Space Center, Houston, Texas, USA
[5] Department of Earth and Environmental Science, University of Rochester, Rochester, New York, USA
[6] Barrios/JETS II Contract, NASA Johnson Space Center, Houston, Texas, USA
[7] JETS II Contract, NASA Johnson Space Center, Houston, Texas, USA
[8] Smithsonian Institution National Museum of Natural History, Washington, DC, USA
[9] Lockheed Martin Space, Littleton, Colorado, USA
[10] NASA Postdoctoral Program, Oak Ridge Associated Universities, Oak Ridge, Tennessee, USA
[11] Buseck Center for Meteorite Studies, Arizona State University, Tempe, AZ, USA
[12] Lunar and Planetary Laboratory, University of Arizona, Tucson, Arizona, USA
[13] McDonnell Center for the Space Sciences, Department of Earth, Environmental, & Planetary Sciences, Washington University, St. Louis, Missouri, USA
[14] Department of Earth, Environmental, and Planetary Sciences, Brown University, Providence, Rhode Island, USA
[15] Technical University Munich, Freising, Germany
[16] Research Unit Analytical Biogeochemistry, Helmholtz Munich, Neuherberg, Germany
[17] Natural History Museum, London, UK
[18] LZ Technology, JETS Contract, NASA Johnson Space Center, Houston, Texas, USA
[19] GeoControl Systems, Inc./JETS II Contract, NASA Johnson Space Center, Houston, Texas, USA
[20] Amentum/JETS II Contract, NASA Johnson Space Center, Houston, Texas, USA
[21] Mclaurin Aerospace/JETS II Contract, NASA Johnson Space Center, Houston, Texas, USA
[22] MB Solutions, Inc/JETS II Contract, NASA Johnson Space Center, Houston, Texas, USA
[23] PCIGR, Department of Earth, Ocean and Atmospheric Sciences, The University of British Columbia, Vancouver, British Columbia, Canada
[24] Lawrence Livermore National Laboratory, Livermore, CA, USA
[25] Department of Earth Science, Tohoku University, Sendai, Japan
[26] Biogeochemistry Research Center, Japan Agency for Marine-Earth Science and Technology (JAMSTEC), Natsushima, Yokosuka, Japan
[27] Department of Earth and Planetary Sciences, Kyushu University, Fukuoka, Fukuoka, Japan
[28] Institute of Space and Astronautical Science, Japan Aerospace and Exploration Agency (JAXA), Sagamihara, Kanagawa, Japan
[29] Institute of Low Temperature Science, Hokkaido University, Sapporo, Hokkaido, Japan
[30] Earth and Planets Laboratory, Carnegie Institution for Science, Washington, DC, USA
[31] School of Physical Sciences, Open University, Milton Keyes, UK
[32] Queen's University Kingston, Ontario, Canada
[33] Center for Astrochemical Studies, Max Planck Institute for Extraterrestrial Physics, Garching, Germany
[34] Department of Geosciences, Pennsylvania State University, University Park, PA, USA

[35] Jet Propulsion Laboratory, California Institute of Technology, Pasadena, CA, USA
[36] NASA Headquarters, Washington, DC, USA
[37] Department of Geology, Rowan University, Glassboro, New Jersey, USA
[38] Department of Earth and Planetary Sciences, American Museum of Natural History, New York, New York, USA

**Abstract**

NASA's OSIRIS-REx mission had the objective of delivering a pristine sample from asteroid (101955) Bennu to Earth for scientific analysis—where "pristine" signifies the absence of foreign materials that could affect sample measurements. OSIRIS-REx returned 121.6 g of regolith in September 2023; this study documents the systematic investigation of suspected contaminants encountered during the mission's sample analysis phase. Most suspected contaminants did not originate from the spacecraft or sample curation. Some were introduced during laboratory analyses, reinforcing the importance of procedural blanks and a strategic approach to sharing samples across laboratories with different analytical targets. Several suspected contaminants, such as phosphate and sodium fluoride particles, were ultimately identified as indigenous to Bennu, highlighting the critical role of contamination knowledge in preventing the dismissal of valuable scientific data. We find that the returned sample meets the definition of pristine, except for an isolated 1.24% (by mass) that escaped the sample container and was thereby contaminated with spacecraft particulates. These findings demonstrate the effectiveness of systematically applied contamination science and engineering practices and provide lessons and approaches to help maximize the scientific integrity of future planetary sample return missions.

**Keywords**

OSIRIS-REx, Bennu, Asteroid, Sample Return, Contamination, Sample Analysis

# 1 Introduction

The Origins, Spectral Interpretation, Resource Identification, and Security–Regolith Explorer (OSIRIS-REx) mission is the third New Frontiers–class mission of the National Aeronautics and Space Administration (NASA). It was designed to characterize the ~500-m-diameter carbonaceous near-Earth asteroid (101955) Bennu and collect a regolith sample for analysis on Earth (Lauretta et al. 2017; Lauretta et al. 2021; Lauretta & Connolly et al. 2024). The mission was selected by NASA in May 2011, launched in September 2016, began mapping Bennu in late 2018, and collected a sample of surface rocks and dust in the spacecraft's Touch-and-Go Sample Acquisition Mechanism (TAGSAM) on 20 October 2020. A total of 121.6 g of bulk sample collected in the TAGSAM sampler head (plus fine particles embedded within 24 contact pads) and witness plates for recording potential sample contamination exposures were returned to Earth in the Sample Return Capsule (SRC) on 24 September 2023 (Lauretta & Connolly et al. 2024). The SRC landed at the Utah Test and Training Range (UTTR) and was transported to the OSIRIS-REx curation facility (Righter et al. 2023) at NASA's Johnson Space Center (JSC) in Houston, Texas, under nitrogen gas ($GN_2$) purge.

The value of the Bennu sample comes not only from retrieving it from a known and characterized asteroid, but also from keeping it pristine (defined below) since collection. Ensuring the pristine nature of the sample is achieved by meeting two Level 1 mission requirements (Lauretta et al. 2017):

1. Return ≥60 g of pristine bulk sample from [Bennu]. No more than 25% of the total returned mass will be used by the mission team to meet its science objectives. "Pristine" is defined to mean that no foreign material introduced into the sample hampers the scientific analysis of the sample.
2. Document the contamination of the sample acquired from collection, transport, curation, and distribution.

The methods used to verify that the spacecraft and ground support equipment could meet these contamination control requirements are described in Dworkin et al. (2018). Briefly, all hardware likely to come into contact with the sample was not permitted to have surface contamination of more than 534 ng/cm$^2$ total carbon, nor more than 180 ng/cm$^2$ of amino acids or hydrazine. There were three principal hardware components: (i) The TAGSAM head, which collected and contained the sample. (ii) The sample canister, a chamber inside the SRC that housed the TAGSAM head after sampling to keep it isolated from the rest of the SRC. The sample canister consisted of a lid and avionics deck connected by a hinge and held closed by latches. (iii) The launch container, which housed the TAGSAM head once it was affixed to the spacecraft, through launch, and until just prior to arrival at Bennu.

The carbon and amino acid contamination control requirements for the spacecraft were met before launch. The surfaces of the TAGSAM head, sample canister, and launch container respectively had 281, 503, and 134 ng/cm$^2$ total carbon and 0.96, 13.1, and 2.32 ng/cm$^2$ of amino acids (Dworkin et al. 2018). The hydrazine requirement was expected to be met based on pre-launch models of its release during the sampling maneuver (Dworkin et al. 2018); these models were updated after sample collection with actual spacecraft thruster burn times to demonstrate that the requirement was indeed met, with ≤62 ng/cm$^2$ of hydrazine that could have come into contact with the TAGSAM head (Glavin & Dworkin et al. 2025).

Although the contamination control requirements were necessary to help ensure the pristinity of the sample collected from Bennu, they are insufficient to determine whether OSIRIS-REx returned a pristine sample. We must also evaluate the findings from the analysis of the sample on Earth with knowledge of the chemistry of potential contamination to ensure that no foreign material impacted the findings. This contamination knowledge relies on an understanding of the materials which could have been introduced into the sample at any point from collection by the spacecraft through analysis in laboratories.

# 2 Contamination Knowledge Strategy in Preparation for Sample Analysis

## *2.1 Materials Archiving and Characterization*

Samples of potentially contaminating material from the spacecraft, ground support equipment, SRC recovery, and sample curation were archived or characterized in support of contamination knowledge. A watchlist of known and expected contaminants was maintained (**Table 1**). In addition, contamination knowledge witness plates were deployed in the spacecraft assembly, test, and launch operations (ATLO) environment and analyzed monthly or at each major operation. Witness plates were also deployed in the curation facility (**Supplement 1**) and flown on the TAGSAM head and in the sample canister (Section 3). A total of 464 items of flight hardware and ground support equipment plus 343 witness plates to date were archived at JSC for this work and for future contamination knowledge investigations (**Supplement 2**).

**Table 1** Known and expected sample contaminants based on materials used for spacecraft construction, operations, sample recovery, and curation. Samples of most of these materials can be found in the materials archive (**Supplement 2**; JSC 2024a).

| Brief Description | Source |
|---|---|
| **Assembly** | |
| Isopropanol | Cleaning |
| **Sample Collection** | |
| 6061 Aluminum | TAGSAM body |
| 304L Stainless steel | TAGSAM perf plate |
| Mylar | TAGSAM flap |
| Steel and titanium | TAGSAM pyrotechnic device (pyrovalve) |
| 304 Stainless steel | TAGSAM convoluted tube |
| 316L Stainless steel | TAGSAM gas feed lines and convoluted tube |
| Elgiloy (Fe-Co-Cr-Ni-Mo alloy) | TAGSAM contact pads |
| Aluminum, sapphire | TAGSAM and canister witness plates |
| **Stow, Return, Recovery** | |
| Braycote 601EF | Avionics deck fasteners |
| Henkel Loctite EA9394 | Avionics deck and canister air filter |
| Cytec FM 300-2U epoxy | Avionics deck |
| Polypropylene fibers with Ti particles | Canister air filter |
| Scotch-Weld 2216 and Cab-O-Sil M-5 fumed silica | SRC fastener staking compound |
| Aluminized Kapton, glass fiber batting, and Kapton and Aramid stitching with trace acrylic adhesive | SRC batting |
| Isopropanol | Cleaning of recovery equipment |
| **Curation** | |
| $GN_2$, $\delta^{15}N_{AIR} = -0.88‰$ [a] | Curation gas |
| Glass, PTFE, 6061 and 6063 aluminum, 316 and 316L stainless steel, Hypalon | Curation glovebox |
| 316L and 304 stainless steel, 6061 aluminum, Viton, glass, aluminum foil | Curation storage and shipping containers |

[a] McCubbin et al. 2016

In the year leading up to sample return, the OSIRIS-REx science team prepared by conducting a Sample Analysis Readiness Test (SART). The SART was also an opportunity to characterize some of the expected contaminants to streamline the potential identification of them in the sample. The TAGSAM mylar flap, TAGSAM convoluted tube, and Henkel Loctite EA9394 were anticipated as sources of particulate contamination (**Table 1**). Coordinated analyses were performed on material from the collection of archived spacecraft materials (JSC 2024) and are presented in **Supplement 3** as resources for future analysts.

### *2.2 Curation Facilities*

A temporary cleanroom was constructed in U.S. Army Dugway Proving Ground building 1012 within the Avery Complex of the UTTR, where the SRC was disassembled in preparation for transport to JSC (Lauretta & Connolly et al. 2024). For permanent curation, a custom ISO 5 cleanroom was constructed at JSC (Righter et al. 2023), including bespoke $GN_2$ gloveboxes for opening and disassembly of the sample canister and TAGSAM head, sample retrieval and manipulation, and sample and witness plate storage. The Bennu sample will continue to be curated and allocated for scientific analysis at this facility for decades to come.

Starting one month before sample return, gas samples from $GN_2$ purge systems for the gloveboxes in both the UTTR and JSC cleanrooms were regularly collected and analyzed using the same procedure as for characterizing ambient gases prior to launch (Dworkin et al. 2018) (**Supplement 1**).

The JSC cleanroom was monitored monthly starting one year before sample return and the UTTR cleanroom starting one month before sample return. Monitoring was performed by analysis of witness plates (**Supplement 1**) and gas samples (**Supplement 4**). Gas samples were collected in evacuated 480 mL valved metal canisters and promptly analyzed by the JSC Toxicology and Environmental Chemistry group via gas chromatography mass spectrometry against 94 target compounds (**Supplement 4**; Sandford et al. 2026).

Like the spacecraft, the UTTR and JSC cleanrooms had materials restrictions (**Supplement 5**). These were derived from the spacecraft contamination control plan and proven techniques previously used by the JSC curation team (Allen et al. 2011; McCubbin et al. 2019). When the science team proposed to add sample analysis devices (Fulford et al. 2024; Macke et al. 2024) to the TAGSAM $GN_2$ glovebox in the JSC cleanroom, the designs had to undergo a contamination review for prohibited materials and testing with witness plates under simulated usage prior to the arrival of the sample. The result of the procedure was summarized in a Contamination Memo (**Supplement 6**) written and approved by the Mission Sample Scientist and Curation Lead prior to integration of the devices.

### *2.3 SRC Recovery*

Preparations for recovering the SRC after landing included careful attention to contamination risks (Sandford et al. 2026). In the event of an SRC breach, landing in standing water, or other environmental ingress, a formal environmental sampling plan was established for contamination knowledge of the SRC landing site (**Supplement 7**). Execution of this plan was rehearsed by all SRC recovery personnel.

The SRC landed at 14:52 UTC on 24 September 2023, after a descent that involved unexpected tumbling and nutation due to the drogue parachute not deploying at the expected time (Lauretta &

Connolly et al. 2024). Environmental samples were collected at the landing site (**Supplement 7**) (Sandford et al. 2026) as the SRC was wrapped in layers of polytetrafluoroethylene (PTFE, e.g., Teflon™) and flown on a helicopter longline to the temporary UTTR cleanroom. Fortunately, the SRC was not breached in any way during landing, so the solid environmental samples were archived without analysis (Sandford et al. 2026) although the gas samples were analyzed (**Supplement 4**). These solid environmental samples are available to analysts, if needed (JSC 2024).

In the UTTR cleanroom, the SRC was partially disassembled to permit safe travel to JSC (**Supplement 8**) and was placed on $GN_2$ purge at 18:52 UTC, exactly 4 hours after touchdown. Disassembly involved removing pyrotechnics and batteries for safety and removing the sample canister from the heatshield and backshell of the SRC (Righter et al. 2023; Sandford et al. 2026). It also included vacuuming and storing loose particles external to the canister (**Fig. 1**). Later, Scanning Electron Microscopy with Energy-Dispersive X-ray Spectroscopy (SEM/EDX) analysis of a split of this loose material (sample ID OREX-220001-1; **Supplement 9**) revealed abundant char and SRC capsule debris, as expected. However, Bennu particles were also identified among this material (**Fig. 2**). It is unknown whether these particles escaped the sample canister during the tumbling and nutation of the SRC or whether they were shed from the TAGSAM head during sample stowing (Lauretta et al. 2022).

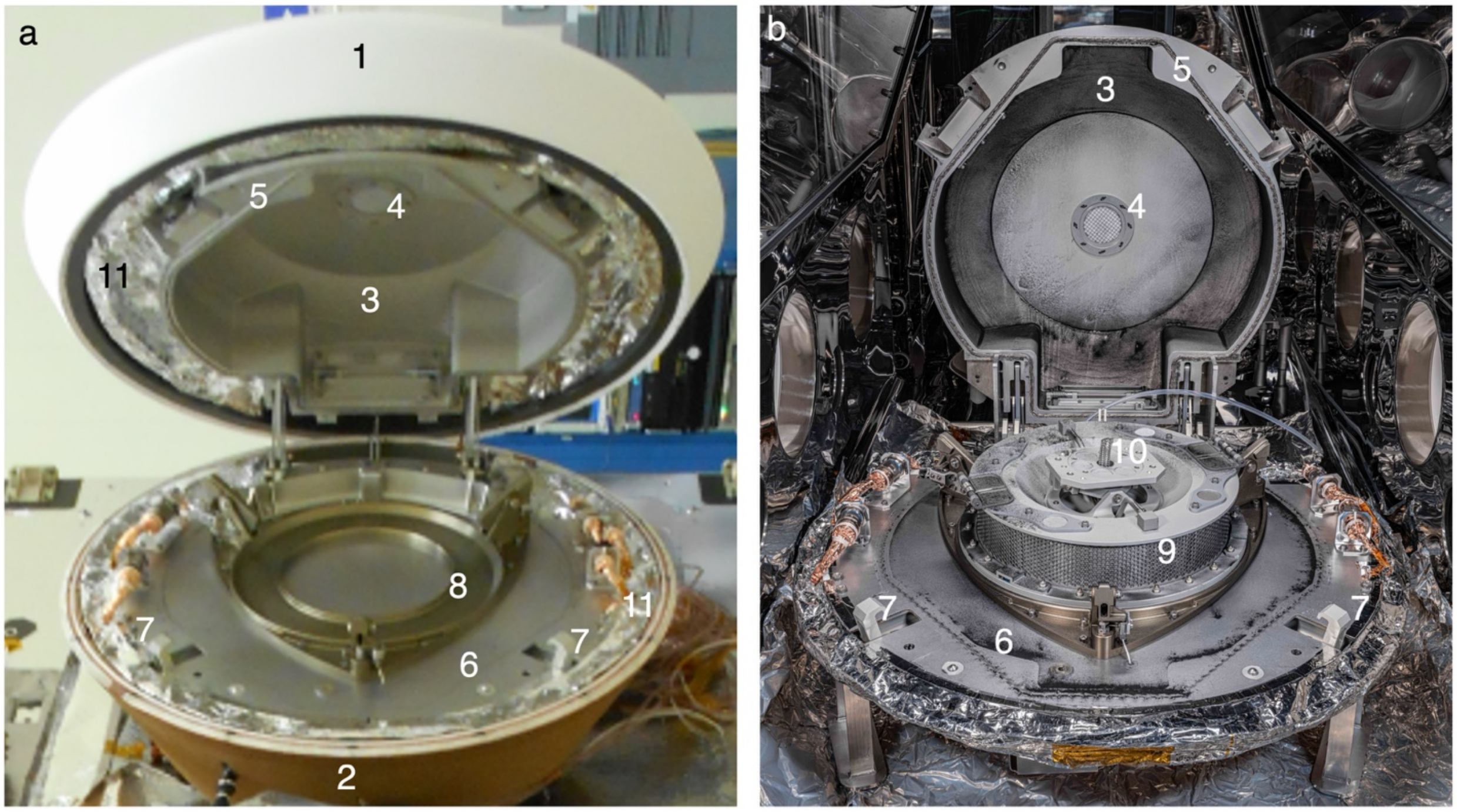

**Fig. 1** SRC and sample canister. **a**. The SRC prior to spacecraft integration during ATLO (after Dworkin et al. 2017). **b**. The sample canister at JSC in a $GN_2$ glovebox (after Lauretta & Connolly et al. 2024). Labels: 1, heatshield; 2, backshell; 3, sample canister lid; 4, canister air filter; 5, sample canister tortuous-path seal; 6, avionics deck; 7, canister latch; 8, capture ring; 9, TAGSAM head (32 cm diameter); 10, cut convoluted tube; and 11, SRC blanket Kapton with batting.

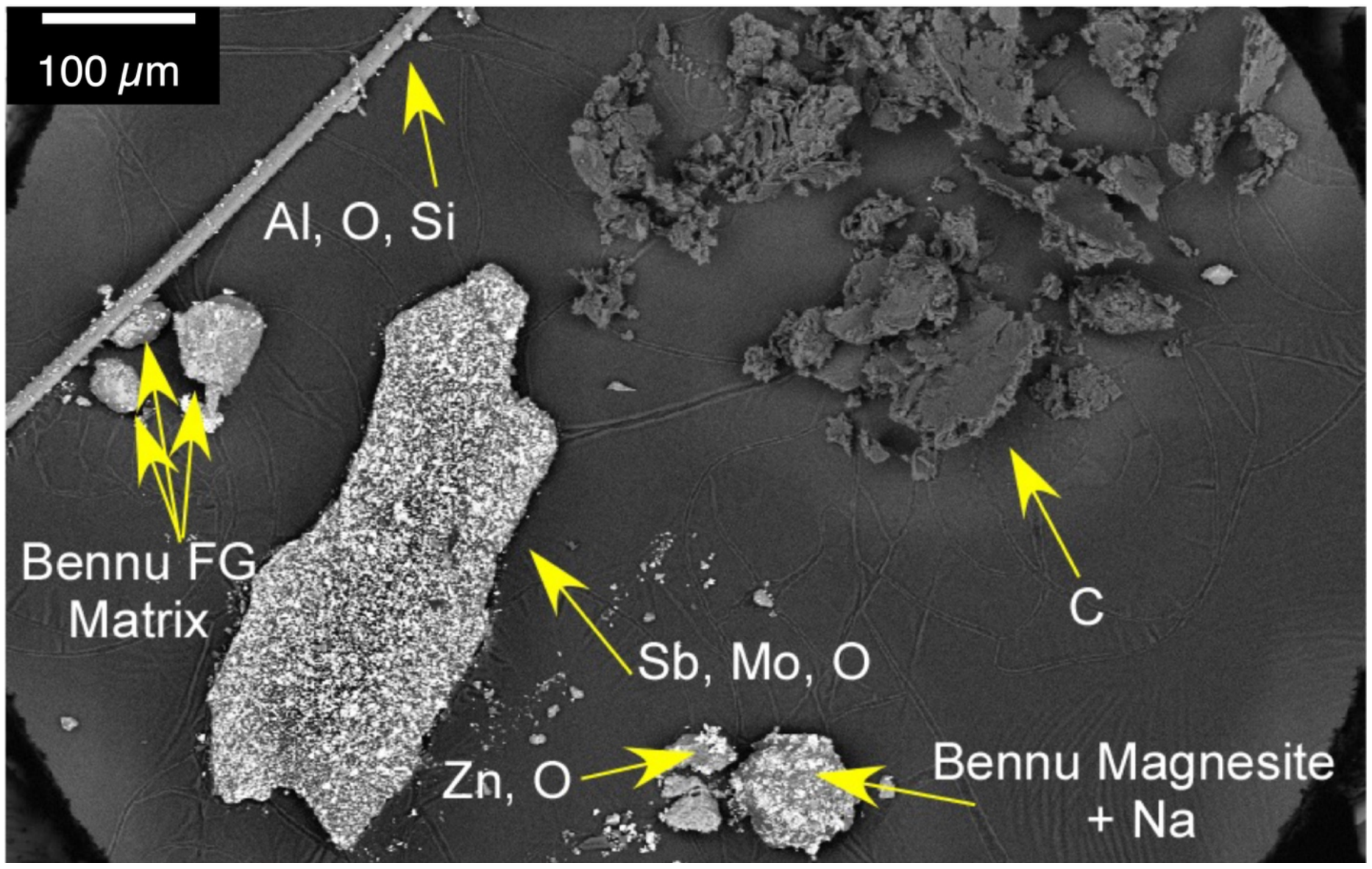

**Fig. 2** Low-angle backscatter electron (LABE) imaging (**Supplement 9**) of sample OREX-220001-1 consisting of particles obtained from inside the SRC but exterior to the sample canister — including fragments of Bennu, SRC char, and metal. FG, fine-grained.

To mitigate possible exposure of the sample to humid air upon arrival in Houston, the canister lid latches were released in the UTTR cleanroom so the PTFE-wrapped canister could be transported directly to the JSC OSIRIS-REx cleanroom, eliminating the need to take the canister off the purge to release the latches in Houston. This meant that the canister lid was held in place by its own weight during transportation but kept in a high $GN_2$ flow. Since the lid was not latched the seal could chatter from vibrations from the airplane, trucks, and carts. Indeed, after transport, concentrations of black dust were observed at the sample canister's tortuous-path seal, and in spots breaching it (**Fig. 3**). The released latches and vibration during travel had provided the means for loose material on the avionics deck to migrate under the tortuous-path seal; the aggressive purge might have also acted as an air bearing and helped open the seal. This scenario is the likely pathway for the SRC debris trapped under the lip of the canister lid that escaped vacuuming at UTTR to have been transported across the seal during the vibrations encountered while being transported to JSC.

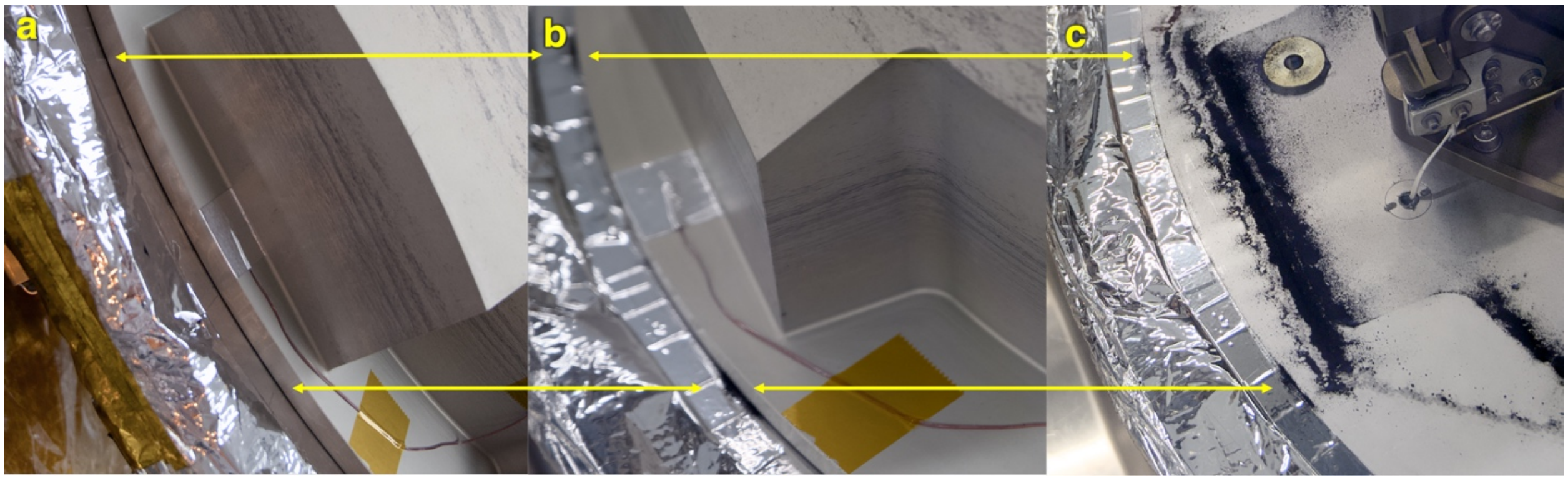

**Fig. 3** Fine black particles of Bennu material leaked out of the tortuous-path seal at the front edge of the canister between departure from UTTR and arrival at JSC, showing a path of escape Bennu material and the likely path for SRC contamination to the avionics deck. Arrows show locations (**a**) with no evidence of black material prior to shipment from UTTR, (**b**) with leaking black dust before sample canister opening at JSC, and (**c**) with an accumulation of Bennu material after canister opening. The fine needle in the metal annulus visible in **c** was the entry point for the $GN_2$ purge. An indication of scale is the orange Kapton tape visible in panels a and b, which is 15x20 mm.

# 3 Flight Witness Plates

## *3.1 Witness Plate Design*

Witness plates were made from high-purity aluminum and sapphire mounted on the interior of the sample canister lid and TAGSAM head. The exposure timing of the witness plates varied, so they provided knowledge of the contamination environment during different phases of the mission (**Fig. 4**).

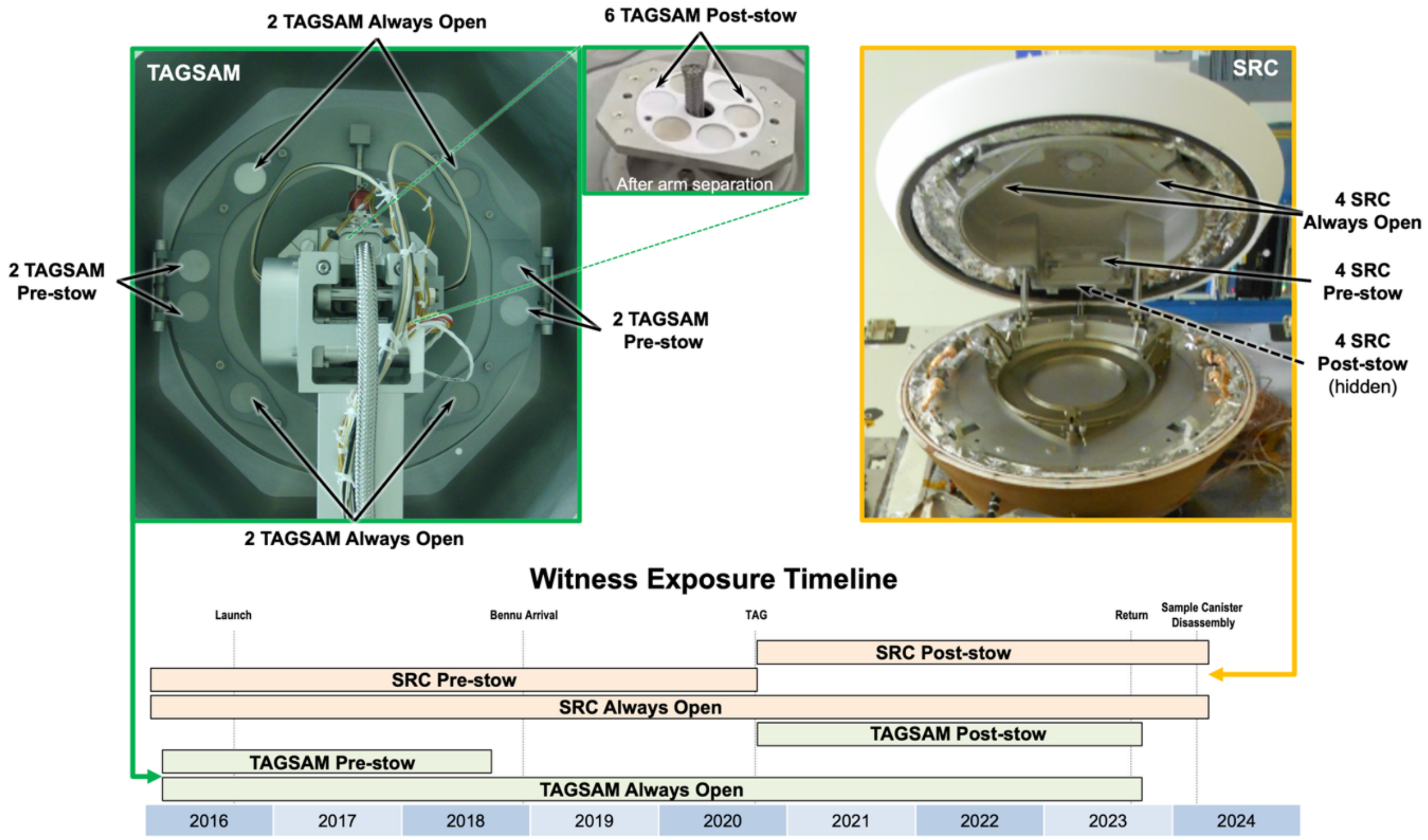

**Fig. 4** Schematic of the timing of witness plate exposure (after Dworkin et al. 2018).

The flight witness plates were deployed in stages for separate identification of terrestrial and spacecraft-related contaminants (Dworkin et al. 2018). Some of these witness plates were covered with a 37 µm steel screen to prevent Bennu particles larger than this mesh size from contacting the plate (**Supplement 10**). Not all of the witness plates had screens because the physical properties of Bennu regolith were essentially unknown, so the effectiveness of a screen at excluding Bennu material but not contamination had not been tested.

Witness plate thickness varied to help prevent confusion between plates after their removal or if they were broken during a potential SRC hard landing — a lesson learned from the Genesis mission (Burnett et al. 2019). Therefore, some witness plates had a 1145-0 aluminum shim behind them to maintain constant thickness within the witness plate holder. The witness plate stack, screen, and/or shim were held in place by mounting brackets and fasteners (Bierhaus et al. 2018). Although the attachment to underlying hardware should have kept the witness plates, shims, and screens stationary for much of the mission, it is possible that during shock (e.g., Earth atmospheric entry) or high-frequency acceleration (e.g., airplane or truck transport), some abrasion could have occurred between witness plate components. A frosted finish was abraded to mark which face of the plate was exposed, slightly increase the surface area of collection, and eliminate specular reflection into the OSIRIS-REx camera system during flight. Details of the witness plates, including locations, exposures, and associated screens and shims, are shown in **Supplement 10**.

### *3.2 Witness Plate Challenges*

A known challenge from mission development was the question of how to subdivide the witness plates for multiple analyses and curation. The plan had been to develop the methods of witness plate subdivision after a successful sample collection at Bennu in 2020. However, the global COVID-19 pandemic delayed and greatly shortened the time allotted.

To test the subdivision approach, flight-like witness plates were precision-cleaned and -cut at JSC in an ISO 6 cleanroom. The curation team used methods developed for the Genesis mission (Burkett et al. 2013) to subdivide the sapphire witness plates, using a diamond scribe to score the back of the witness plates and then break them along the score line. The aluminum plates were cut with stainless steel shears. Half of the cut plates were precision-cleaned again as a control. A pair cleaned then cut (test) and cleaned then cut then cleaned (control) of aluminum and sapphire plates was sent to the science team for double-blind contamination testing.

The curation team observed that cutting the thicker plates, particularly in the case of aluminum, was difficult. We had required the witness plates to have different thicknesses to distinguish the different exposures, but we neglected to specify a maximum thickness to enable subdivision. A foil that could be subdivided by tearing, as was used in the ground witness plates, would have been simpler to cleanly subdivide but would have had the disadvantage of being more complicated to securely mount for spaceflight.

Using the methods and analytes described in Glavin & Dworkin et al. (2025), during SART testing it was demonstrated that these methods of cutting did not add additional organic contamination. However, the addition of steel particles from the tools used to hold and cut the plates — as detected at JSC by SEM/EDX analyses — presents a potential complication for witness plate characterization of inorganic contamination. During OSIRIS-REx mission design, several options, including pre-scored and pre-cut witness plates, had been considered but rejected as presenting too great a risk of generating debris. However, the probability of debris generation was assumed, not tested. Thus, future missions should look for better ways to design witness plates that can be more easily subdivided after return without increasing the risk of them breaking during flight and releasing debris in space. The witness plates should be designed in collaboration between the curators who will need to subdivide the plates, the science team who have specific analytical targets and sensitivities, and the engineers who understand the necessities of flight implementation.

Unfortunately, it appears that some flight witness plates had unanticipated exposures. First, **Fig. 5** shows that a spray of Bennu dust hit the TAGSAM head, likely when the SRC was flipped for disassembly at the UTTR, as suggested by the absence of dust under TAGSAM gas tubing, creating a defined “clean shadow.” If the dust spray occurred during reentry, then it should have been deposited on the canister lid during deceleration or widely distributed during tumbling/nutation. The screens did not prevent the fine, friable Bennu dust from getting onto witness plate surfaces during disassembly. Additionally, many of the witness plate covers failed to keep Bennu dust out during flight (**Fig. 6**).

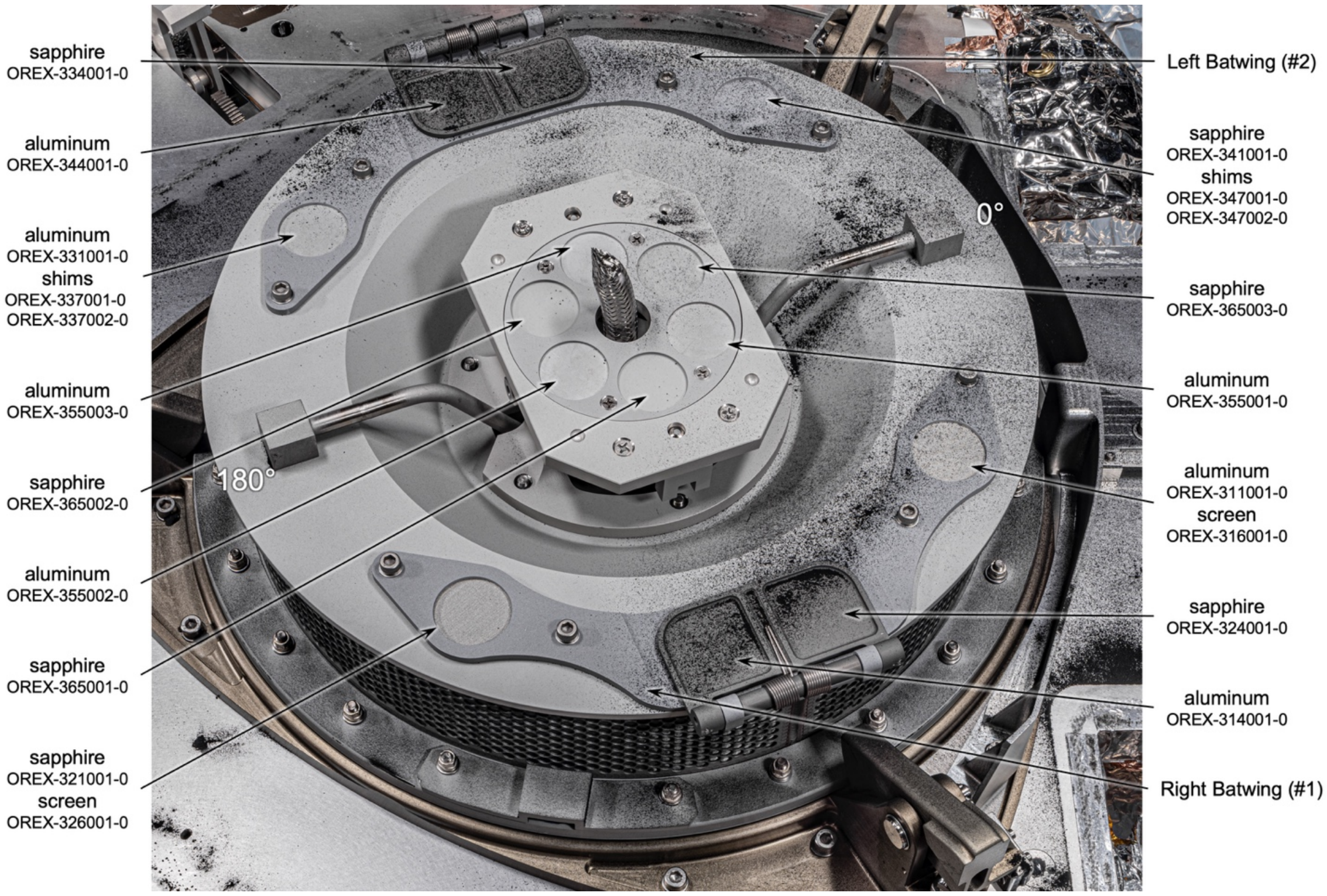

**Fig. 5** Location of witness plates on the TAGSAM head. The spray of dust covers all witness plates. The "clean shadow" below the gas tube at 0° shows the direction of the dust impingement. Witness plates OREX-314001-0, OREX-324001-0, OREX-334001-0, and OREX-344001-0 are beneath covers that closed at Bennu arrival and cannot be seen in the image.

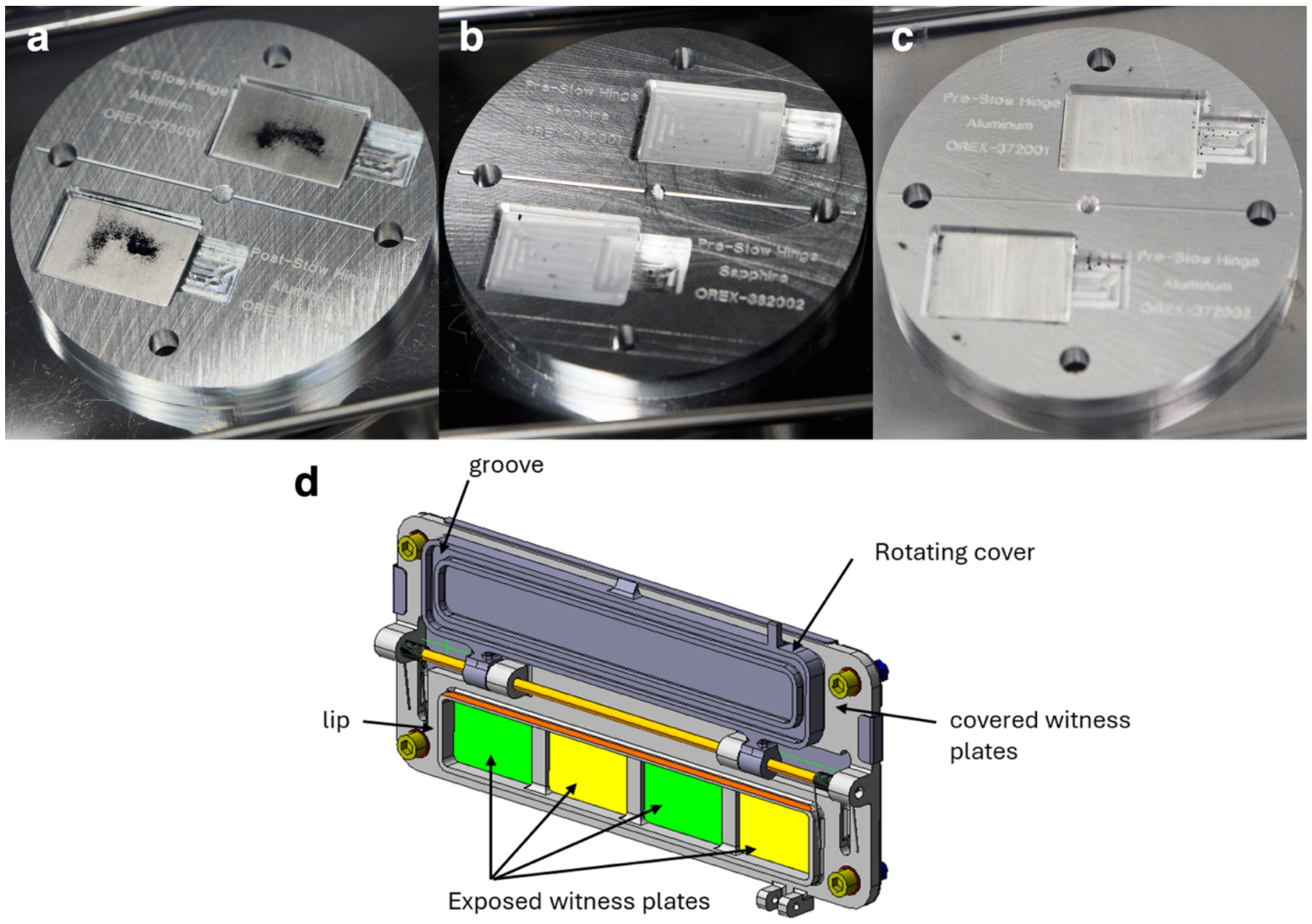

**Fig. 6 a**. Dust, presumably from Bennu, on aluminum witness plates OREX-372001-0 and OREX-372002-0, which were both covered by a 37 µm screen. **b**. Dust, presumably from Bennu, on sapphire witness plates OREX-382001-0 and OREX-382002-0. **c**. Aluminum witness plates OREX-372001-0 and OREX-372002-0, which were closed prior to stowing the filled TAGSAM head via the mechanism illustrated in **d**.

**Fig. 6b** shows two sapphire witness plates in the canister that were covered prior to sample stow. The opening of the SRC in space to prepare for stowing the TAGSAM head triggered the covering of these witness plates via spring actuation, but these were not hermetic seals. Instead, witness plate covers were designed to prevent losing or gaining material via a tortuous-path seal. These witness plates remained covered until disassembly by the curation team in the glovebox after all Bennu sample had been removed and contained. Thus, there is no reasonable path for sample material to reach the witness plates other than via leaks through the covers.

Only the covers of the sample canister witness plates appear to have leaked; the covers on the TAGSAM witness plates did not, though the cover mechanisms were the same (Bierhaus et al. 2018). One reason for the different outcome may be that the closed TAGSAM witness plate covers were horizontal during travel from UTTR to JSC, whereas the closed sample canister witness plate covers were vertical with the hinge pointed down (**Fig. 6d**). It is plausible that vibration chatter permitted Bennu dust to leak in through the seal. It is also possible that during the tumbling/nutation phase of SRC descent, dust happened to leak into the sample canister's covered witnesses but not the TAGSAM ones. Either way, these tortuous-path seals should have been designed to hold fast during the vibrations of return, recovery, and transport. If future missions employ similar covers, the cause of the cover failure should be modeled in more detail.

Sample analysts had planned to analyze the relevant witness plates in parallel with Bennu samples for multiple reasons. First, measuring samples and witness plates at the same time using the same extraction and analytical procedures produces more comparable data. In addition, this allows for the ability to re-analyze Bennu samples in light of witness plate results. Finally, there is an economy of scale in the analyses where the addition of an extra sample in the laboratory schedule is faster and cheaper than analyzing the sample and witnesses at different times. The science team had assumed, but did not require, that the witness plates would be allocated promptly. However, supply chain problems delayed the acquisition and installation of the glovebox dedicated for witness plate allocation, and Bennu sample requests were filled in the meantime to manage curation resources and time. Because the sample allocations were prioritized, the glovebox for processing the witness plates was delayed by almost two years.

Thus, witness plate allocation took place towards the end of the mission's sample analysis phase, and only one witness plate, from the TAGSAM head, covered since Bennu arrival, and without obvious Bennu dust (OREX-324001-0), was allocated in time for analysis by a subset of the science team. A sample canister witness plate shim (OREX-377025-0) was also allocated to be used as a comparison. This shim was exposed to the ATLO environment but not directly to the space environment.

### *3.3 Witness Plate Analysis Results*

The methods of subdivision, extraction, and analysis of witness plates and the corresponding controls, with extended data tables, are given in **Supplement 11**.

TAGSAM sapphire witness plate OREX-324001-0 was opened on 18 April 2016 (before launch) and closed on 18 October 2018 when the TAGSAM head was unstowed from its launch container as the spacecraft approached Bennu, for a total exposure of 914 days to the ATLO, launch, and outbound cruise environments. This witness plate appears free of Bennu dust (**Fig. 7a**). Aluminum shim OREX-377025-0 (**Fig. 7b**) was located underneath sample canister witness plate OREX-372002-0 (**Fig. 7c**), which was exposed from 18 April 2016 to 27 October 2020, when the SRC opened for stowage of the sample, for a total of 1,686 days. This flight shim had no line-of-sight exposure outside its witness plate stack since ATLO, so it was essentially exposed only to the interior of the witness plate stack for 2,914 days — from 18 April 2016 to 4 April 2024, when the canister witness plates were disassembled.

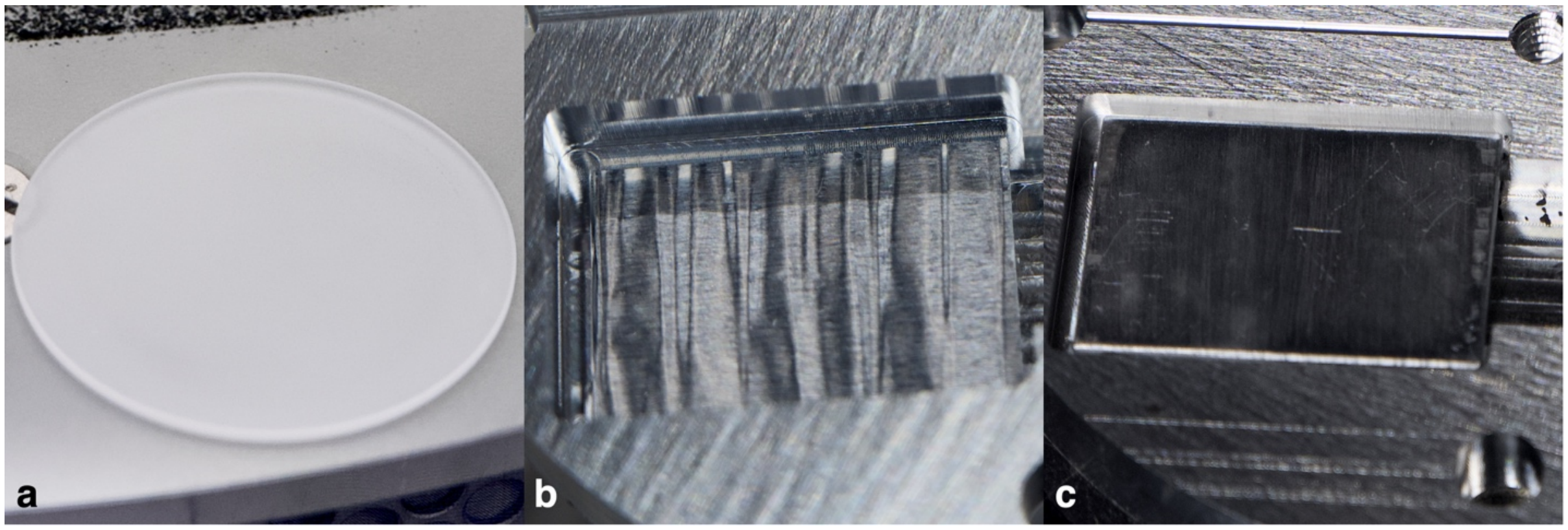

**Fig. 7 a**. TAGSAM sapphire witness plate OREX-324001-0. **b**. Sample canister aluminum shim OREX-377025-0. **c**. Canister aluminum witness plate OREX-372002-0, which was above shim OREX-377025-0, not analyzed in this study.

Liquid chromatography–mass spectrometry (LC-MS) of a 6M HCl vapor-hydrolyzed hot water extract of TAGSAM sapphire witness plate OREX-324001-0 (subsample OREX-324001-100) detected trace methylamine, but no other amines. It is not impossible that the methylamine was derived from Henkel Loctite EA9394, which has a 3:1 ratio of methylamine to ethylamine in hot water extracts (**Supplement 3**), and that the ethylamine was below detection limits.

Three amino acids (L-aspartic acid, D-serine, L-serine, and L-threonine) were detected at low levels in OREX-324001-100, for a total of 0.48 ± 0.17 ng/cm$^2$. The 6M HCl vapor-hydrolyzed hot water extract of the shim (subsample OREX-337025-1) contained glycine, β-alanine, L-alanine, L-valine, L-leucine, and L-isoleucine for a total of 0.58 ± 0.14 ng/cm$^2$. The control sapphire 0 and control aluminum 1 had <0.04 ng/cm$^2$ and 0.11 ± 0.08 ng/cm$^2$ amino acids, respectively, which were subtracted from the above witness plate and shim values to correct for laboratory background.

Prior to launch, proxy witness plates for the TAGSAM and sample canister were found to have higher concentrations of amino acids (mostly glycine) — 0.96 ng/cm$^2$ and 13.1 ng/cm$^2$, respectively — than the witness plate or shim. The lower abundances of amino acids on the flight witness plate and shim than measured on witness plates during ATLO could be attributed to amino acid evaporation in space, except this explanation is not consistent with the differences in amino acid ratios. So, the reason for the differences in amino acids measured on the witness plate, shim, and ATLO witness plates are unclear. Regardless, the relative amino acid abundances in the aluminum shim and the sapphire witness plate extracts do not resemble the amino acid distributions found in a Bennu aggregate (bulk unsorted) sample (Glavin & Dworkin et al. 2025).

Gas chromatography–mass spectrometry (GC-MS) for carboxylic acids in the hot water extract of OREX-324001-100 showed traces of isobutyric acid ($1.7 \times 10^{-6} \pm 1.7 \times 10^{-6}$ nmol/cm$^2$) and succinic acid ($1.5 \times 10^{-5} \pm 1.0 \times 10^{-6}$ nmol/cm$^2$) at concentrations greater than those measured for the control levels. Two-dimensional high-resolution GC-MS (GC×GC-HRMS) performed on dichloromethane (DCM) and methanol (MeOH) (4:1 v/v) extracts of the sapphire subsample OREX-324001-101 and the aluminum shim subsample OREX-377025-3 showed no compounds detected above the low levels measured in their respective controls (sapphire 1 and aluminum 3).

However, pyrolysis GC-MS analysis of TAGSAM sapphire witness plate subsample OREX-324001-103 showed the presence of isopropanol, naphthalene, 1- and 2-methylnaphthalene, and a sequence of alkene pairs that were not detected by GC×GC-HRMS. Traces of benzene, toluene, $C_2$ and $C_3$ alkylbenzenes, and thiophene were detected, but these species were also present in the control. Pyrolysis GC-MS abundances between the sample and control were the same within error, indicating that these are probably laboratory background. Most compounds were identified by their retention time and characteristic multiple reaction monitoring (MRM) transitions. However, the isopropanol could only be identified by a single ion (*m/z* 59) and retention time, so the identification is less confident.

The aluminum shim showed 28 ± 16 ng/cm$^2$ of non-volatile residue (NVR), within error of the control aluminum, 91 ± 52 ng/cm$^2$. Likewise, the TAGSAM witness plate showed 45 ± 25 ng/cm$^2$ of NVR, within error of the control sapphire, 65 ± 37 ng/cm$^2$. The flight TAGSAM head was measured by a proxy witness plate prior to launch to have 220 ng/cm$^2$ NVR, which was higher than the NVR measured in the sample canister (180 ng/cm$^2$) and in the launch container (100 ng/cm$^2$) (Dworkin et al. 2018). This suggests that either the NVR burden was overestimated in ATLO or that the NVR on the witness plate inside the launch container decreased by evaporation during outbound cruise.

### *3.4 Witness Plate Lessons*

A known problem with witness plates is that analyte quantitation is a function of surface area while sample quantitation is a function of sample mass. This means that a direct subtraction of witness plate contaminants from sample contaminants is not possible. This limitation of witness plate analysis was compounded by the science team's inability to measure more than one witness plate type or exposure time and the unavailability of the witness plates for analysis in parallel with the sample so that the instruments were operating under identical parameters, and contaminants in the witness plate could be targeted in sample analysis.

When the archiving of flight and ground support hardware began in 2014, there was no permanent contamination knowledge archive storage location and no laboratory where contamination-related work could be done. Without proper facilities, the development of witness plate handling procedures and subdivision plans could have been hampered. A subsequent upgrade of the cleanrooms in Building 31 at JSC included an advanced curation cleanroom space that is used for a wide variety of development and testing curation projects (e.g., McCubbin et al. 2019) to support contamination control and knowledge.

Witness plate subdivision technique development would have benefited from starting earlier in the mission than the SART to enable iterative testing. In addition, there was limited personnel support for curation activities until two years before sample arrival, when numerous other requirements were competing for the curation team's capacity.

While the witness plates had different surface finishes to preserve knowledge of which surface was exposed, there was not an equivalent distinguishing characteristic for the screen position and orientation. During testing as part of the SART, it was realized that such documentation would have been desirable. Unfortunately, this was not possible since the flown witness plates and screens lacked fiducials, and the circular plates could rotate during flight or while the stack was removed by hand.

There was only one flight-like sapphire witness plate in the materials archive. Without a reliable method of subdivision, this unique control was viewed as too valuable to be consumed by the science team. However, it could have been important in discriminating between any contamination that resulted from the witness plate processing on the ground versus witness plate contamination that occurred during flight. Instead, spare aluminum shim and sapphire witness plates were ashed at 500 °C in air at NASA's Goddard Space Flight Center and used as a processing control for any contamination of the flight witness plates that occurred during extraction and analysis.

## 4 Suspected Contaminants

### *4.1 Suspected Contaminant Reporting Procedures*

If a suspected contaminant was observed during Bennu sample analysis, its existence was presented to the Contamination Science Lead, who maintained a log of all suspected contaminants. During preliminary examination of the first samples retrieved (the "quick-look" analyses; Lauretta & Connolly et al. 2024), any suspected contaminants were reported by the analysts directly to the Contamination Science Lead; after the initial rush of new findings, suspected contaminant reporting followed the sequence shown in **Fig. 8**.

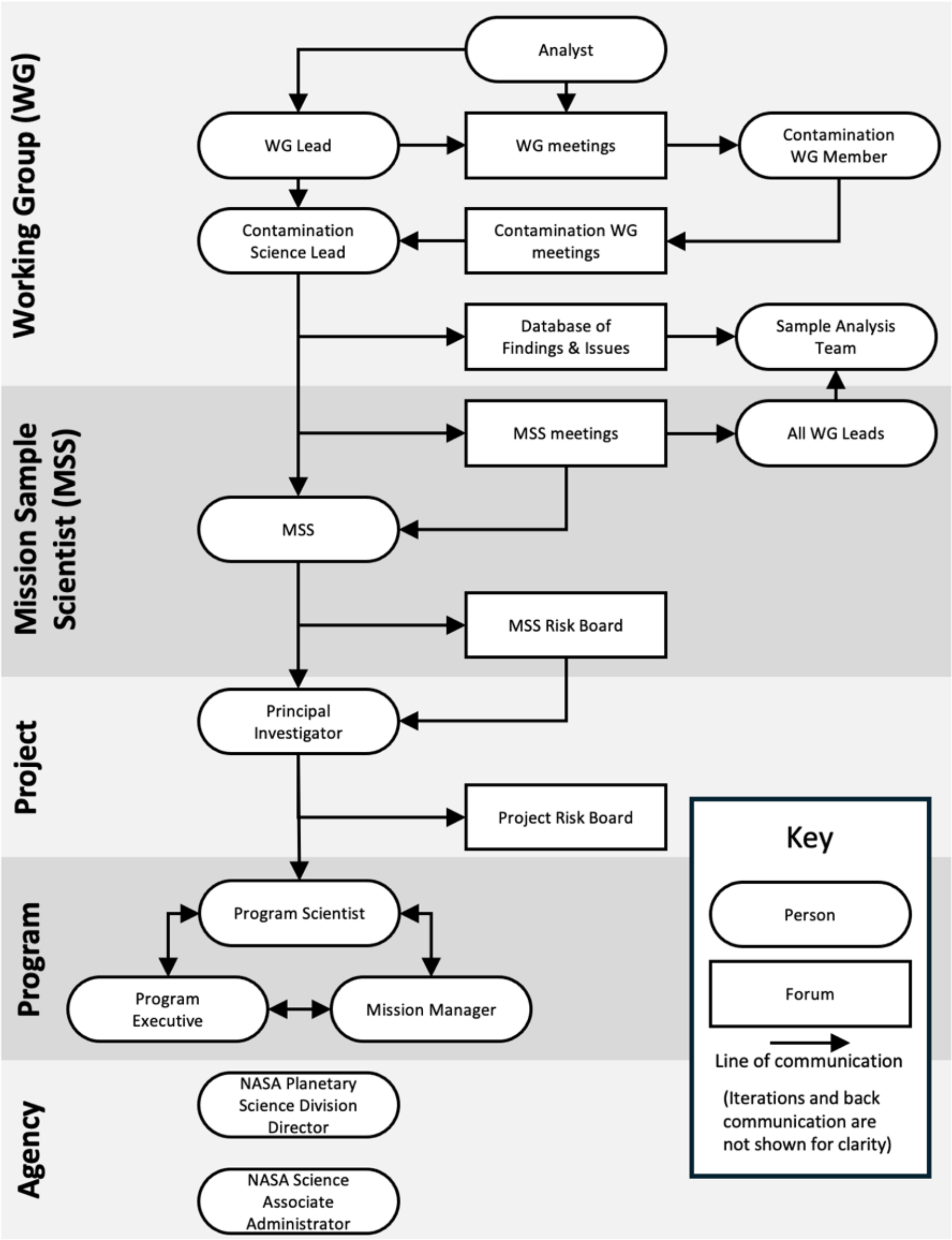

**Fig. 8** The reporting pathway for a suspected contaminant. No suspected contaminants required Project, Program, or Agency involvement.

Collaboration between the Mission Sample Scientist and Contamination Science Lead led to the identification of suspected contaminants and their potential science impacts. No suspected contaminants required elevation to a formal risk board. Each suspected contaminant was assigned a fixed number associated with hypotheses, data, and a fishbone/Ishikawa diagram to assess the likely source. This information and all supporting data were maintained in a database accessible to all team members. Furthermore, monthly summaries of the evolving list of potential contaminants were provided to the Principal Investigator (PI).

During Bennu sample analysis, the science team reported a total of 31 observations of suspected contaminants in the bulk sample (**Table 2**); no suspected contaminants were observed on the contact pads. Potential contamination sources were divided into the following categories for analysis via an Ishikawa diagram: canister or TAGSAM head materials, SRC materials, spacecraft materials, materials associated with curation processes, materials associated with laboratory processes, and Bennu material misidentified as a contaminant. In addition, it was noted if the suspected contaminant only appeared in samples recovered from the avionics deck or in those retrieved from the TAGSAM head, and if the suspected contaminant was observed with multiple techniques or in multiple laboratories.

**Table 2** Potential contaminants examined and their deduced origins.

| # | Brief Description | Origin |
|---|---|---|
| | **Isolated from the Sample** | |
| 1 | Glossy particles | SRC |
| 2 | Toluene and other volatiles | SRC |
| | **Laboratory Artifacts** | |
| 3 | Metals on planchette | Laboratory |
| 4 | Late peaks in pyrolysis | Laboratory |
| 5 | High amino acid abundance | Laboratory |
| 6 | Metallic flake | Laboratory |
| 7 | Gold particles | Laboratory |
| 8 | Phthalates | Laboratory |
| 9 | High Mo and W | Laboratory |
| 10 | Quartz crystal and organic-rich kaolin particle | Laboratory |
| 11 | Biofilaments | Laboratory |
| 12 | Stainless steel particle | Curation |
| | **Indigenous to Bennu** | |
| 13 | Magnesium sodium phosphate particles | Bennu |
| 14 | NaF particles | Bennu |
| 15 | Red-orange particle | Bennu |
| 16 | High $NH_3$ | Bennu |
| 17 | Low $^{15}N/^{14}N$ translucent material | Bennu |
| | **Avionics Deck** | |
| 18 | Al-rich hotspots in particles | SRC |
| 19 | Al-rich particles | SRC |
| 20 | Zn-rich particles | SRC |
| 21 | Ni-rich particles | SRC |
| 22 | Si, Ti, Ag, In, Sn, Ir, Pb particles | SRC |
| 23 | High B, Zr, Sb, Ba, La, Hf, Th, U | SRC |
| 24 | Various SEM anomalies | SRC |
| 25 | Fibers | SRC |
| 26 | Single blue particle | SRC |
| 27 | Angular white particles | SRC |
| | **Retaining Ring** | |
| 28 | Metallic particles | Spacecraft (TAGSAM pyrovalve) |
| | **Laboratory Artifact or Indigenous to Bennu** | |
| 29 | L-Valine enantiomeric excess | Bennu or Laboratory |
| | **Possible Trace** | |
| 30 | Isopropanol | Spacecraft |
| 31 | Polyolefins | Ubiquitous |

### *4.2 Contaminants Isolated from the Sample*

Two suspected contaminants were confirmed not to have had access to the sample. The first was glossy particles (**Table 2 #1**) that were found to be Scotch-Weld 2216 thickened with untreated Cab-O-Sil M-5 fumed silica staking compound used to hold various spacecraft fasteners. This material was only found outside of the sample canister (**Fig. 1**). Most was dislodged at the UTTR during preliminary SRC disassembly; remaining staking compound that shed from the exterior of the canister in the canister glovebox was isolated by the curation team (OREX-210000-0, OREX-590025-0, OREX-590035-0, and OREX-590076-0). There was no visible evidence of any missing or damaged compound on fasteners, and no evidence of this contaminant in the analyzed samples.

The second (**Table 2 #2**) were hydrocarbons, primarily toluene (Aponte et al. 2026; Clemett et al. submitted). Though these species were generated during SRC re-entry, they were isolated from the sample by being trapped by the sample canister air filter (Sandford et al. 2026). The prime evidence confirming the isolation was that the toluene in the air filter exhibited a terrestrial δD (–73.1 to –120.4‰) (Sandford et al. 2026), whereas the toluene extracted from Bennu aggregate OREX-800107-128, measured for this work with the identical method (δD = +148 ± 44‰ and $\delta^{13}C$ = +181 ± 34‰), were in the range of other Bennu organics (Glavin & Dworkin et al. 2025).

As discussed in Section 3, toluene (as well as benzene, thiophene, and alkylbenzenes) was detected on TAGSAM sapphire witness plate OREX-324001-0 via pyrolysis GC-MS at concentrations only a few times greater than that on the control sapphire, though not detected by GC×GC-HRMS (**Supplement 11**). The presence of these species on the witness plate may therefore not be meaningful given the semi-quantitative nature of pyrolysis GC-MS and the arbitrary size of the sliver of sapphire witness plate and control sapphire analyzed. Notably, toluene and the other observed species are also likely to be pyrolysis products of laboratory polymers, such as polystyrene, due to their similarity to the family of oils generated by the pyrolysis of plastics (Gonzalez-Aguilar et al. 2023). Furthermore, lack of detection of these species in the witness plate extract by GC×GC-HRMS supports the hypothesis that the witness plate toluene observed by pyrolysis GC-MS is a laboratory artifact.

Isotopically light acetone with a $\delta^{13}C$ value of −20.3 ± 2.2‰ (**Table 2 #29**) was detected in a 2.6 g sample of homogenized aggregate OREX-800107-127 (Baczynski & Mcintosh et al. 2026). The acetone was more $^{13}C$-depleted than that of the Murchison carbonaceous chondrite, $\delta^{13}C$ = +8.9 ± 1.4 (Baczynski & Mcintosh et al. 2026), raising the possibility that the acetone could be a terrestrial contaminant. Supporting this scenario, acetone was also detected in the environmental gas samples (**Table 3**. The acetone might have derived from the oxidation of isopropanol, which was also detected in the environmental gas samples (**Table 3**), used in the cleaning of the spacecraft or ground support equipment. This oxidation reaction is very unfavorable at the modest temperatures during ATLO and experienced by the spacecraft (Kulkarni and Wachs 2002), so no more than a small fraction of isopropanol could be converted to acetone.

However, 2-butanone in the same Bennu sample also had a light $\delta^{13}C$ value ($\delta^{13}C$ = −19.3 ± 1.5, versus −3.4 ± 1.2 in Murchison), but this compound does not have a similarly plausible contamination source, suggesting that the acetone in the sample is indigenous. No acetone was detectable in analyses of sapphire witness plate subsample OREX-324001-103, suggesting that there was no significant acetone contamination imparted during ATLO. Although no witness plates exposed during SRC atmospheric reentry were available, the effectiveness of the canister air filter was demonstrated for toluene as described above. Even if the air filter had partially failed, the amount of acetone contamination should be proportional to the amount of isopropanol contamination. Isopropanol is the most abundant organic compound detected in the environmental gas samples (**Table 3, Supplement 4**), and only a trace of isopropanol was tentatively detected in the sample (discussed in **section 4.8**, below). Thus, there cannot be a significant amount of acetone contamination.

**Table 3** Gas sample analysis (parts per billion by volume unless otherwise specified) at different mission phases (Sandford et al. 2026) compared with Bennu aggregate samples OREX-800107-183 (Baczynski & Mcintosh et al. 2026), OREX-800107-128 (Aponte et al. 2026), OREX-800128-104, and TAGSAM glovebox exhaust trap OREX-590040-0 (**Supplement 9**). The latter two samples were analyzed in this study and are shown in *italics*.

| Environment | Location | Isopropanol 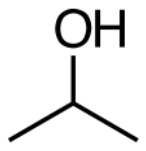 | Acetone 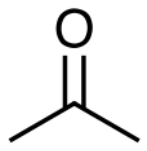 | Benzene 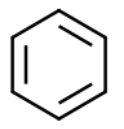 | Toluene 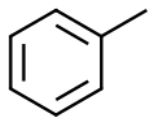 |
|---|---|---|---|---|---|
| Bennu samples | OREX-800107-0 subsamples [a] | [b] | 36.02 | 13.5 | 114.5 |
| | *OREX-800128-104* [c] | *Tentative* | *No* | [d] | [d] |
| | *Curation Glovebox Exhaust* [c] | [d] | *No* | *No* | *Yes* |
| Pre-launch | TAGSAM witness plate [c] | Yes | No | Yes | Yes |
| | LMS purge | 25 | 19 | <8 | <7 |
| | LMS highbay air | 45 | 22 | <8 | <7 |
| | PHSF air | 320 | 15 | <8 | <7 |
| | ULA Fairing air | 3,500 | 16 | <8 | <7 |
| Sampling | TAGSAM pyrovalve ground test | 3,300 | 63 | <8 | <7 |
| Before landing | UTTR purge before return | 130 | 6.1 | <6.0 | <3.0 |
| | UTTR cleanroom before return | 13,000 | 46 | <6.0 | <6.0 |
| | UTTR upwind of landing site | 2,100 | 22 | <6.0 | <6.0 |
| | UTTR helicopter rotorwash | <10 | 6.21 | <6.0 | <6.0 |
| Trapped | Canister Air Filter [c] | Yes | Yes | Yes | Yes |
| After landing | Heat shield, UTTR landing site | <10 | 13 | <6.0 | <6.0 |
| | Backshell, UTTR landing site | <10 | 14 | <6.0 | <6.0 |
| | Vent 1, UTTR landing site | <10 | 18 | <6.0 | <6.0 |
| | Vent 2, UTTR landing site | <10 | 15 | <6.0 | <6.0 |
| SRC heatsoak | Vent 1 before purge, UTTR cleanroom | 15,000 | 38 | <6.0 | 22 |
| | Vent 2 before purge, UTTR cleanroom | 16,000 | 35 | <6.0 | 6.9 |
| SRC outgassing | Sample canister filter exterior under purge, UTTR cleanroom | 10,000 | 67 | <6.0 | <6.0 |

[a] Concentration in nmol/g
[b] Not investigated
[c] Values are detected (Yes), not detected (No), or tentatively detected (Tentative) by pyrolysis GC-MS.
[d] Abundance may not be significantly higher than blank.

### *4.3 Laboratory-Introduced Artifacts*

Eight of the reported suspected contaminants were determined to be contamination imparted in the laboratories of analysts, after the samples were released from curation. Such laboratory-introduced contaminants are referred to as artifacts. Examples of artifacts resulting from expedited processing of the initial quick-look sample from the avionics deck (Lauretta & Connolly et al. 2024) include metal flakes on a hastily cleaned SEM planchet (**Table 2 #3**), organics escaped from ill-fitting gloves or garments (**Table 2 #4,5**), and a flake of aluminum foil (**Table 2 #6**). Subsequent sample analyses demonstrated that none of these artifacts affected the results.

Other artifacts came from standard laboratory procedures that only impacted downstream analyses, such as gold flakes from plated forceps (**Table 2 #7**) and common laboratory background (**Table 2 #8–10**). Phthalates are ubiquitous (Tunney et al. 2023; Zainol et al. 2026); though they were detected in Bennu sample extracts (**Table 2 #8**), the ratios of the specific phthalates in the same homogenized sample varied by laboratory, indicating that they were an artifact from the laboratory environment, rather than from the sample, spacecraft, or curation.

The main source of laboratory-introduced contamination occurred when a laboratory performing one type of analysis introduced an artifact or byproduct inconsequential to that analysis. This type of contamination was discovered later by another laboratory performing a different type of analysis on the same sample that is sensitive to the introduced contamination. Determining the source(s) of such potential artifacts was relatively simple for techniques that routinely employ a control that has been exposed to the same environments as the sample. For example, the detection of anomalously high abundances of the elements La, Ce, and W in Bennu homogenized aggregate subsample OREX-800107-101 (Wang et al. submitted) was explained by the analysis of blank fused silica sample which had been prepared and shipped in parallel. These elements were likely derived from the muffle furnace used in the preparation of the glassware involved in the homogenization of the 6.425 g aggregate parent sample OREX-800107-0, where the prevention of organic—not inorganic—contamination prevention was the priority (Ryan et al. 2026).

In other cases, cross-laboratory comparisons of the same sample and procedural blanks can be used to determine instrument artifacts, as in the case of W and Mo in Bennu aggregate OREX-803015-0 (Wang et al. submitted) (**Table 2 #9**). However, particularly in cases where the method was non-quantitative, not every sample in this analysis chain had a corresponding procedural blank. It is advisable for researchers to analyze and process blank samples in parallel with precious sample return material and to pass along those blanks with samples that are undergoing sequential analyses.

Another example was the detection of a single quartz crystal and an organic-rich kaolin particle in OREX-803124-0 (**Table 2 #10**). This sample was allocated for bulk optical and infrared spectroscopic analyses. As such, the sample-handling environment did not need to be rigorously controlled, and only laboratory coats and hair nets were employed. During the first of five transfers between institutions before the suspected contaminants were discovered in this sample, researchers noted that the container was difficult to open and when forced fine Bennu material was deposited on lid and around the Viton seal. It was material from the lid that showed the anomalous particles in analyses. The spectra of the organic-rich kaolin particle were compared to the only known source of kaolin on the spacecraft, staking compound (**Table 2 #1**). The staking compound is a mixture of Scotch-Weld 2216 with untreated Cab-O-Sil M-5 which contains kaolin and organics but also abundant amorphous silica. However, there was no evidence in the particle's spectra of the strong characteristic amorphous silica infrared bands in the kaolin particle spectra. For the quartz particle, records were searched to determine whether there had been any anomalies with the quartz halogen lamps used during spacecraft testing, but there were none. Other possible spacecraft sources of quartz had no viable pathway to the SRC. Neither type of particles was observed in any other sample. The most parsimonious explanation for two chemically unrelated particles found in only this sample was that they could have been shed into the laboratory, and consequently the sample, during the struggle to open the container. The particles were separated and did not contribute to subsequent analyses.

Carbon- and nitrogen-rich hydrated asteroid (Genge et al. 2025) and meteorite (Steele et al. 1999) samples can become colonized by fungal hyphae and other biology if permitted. There was evidence of fungal DNA in witness plates exposed during ATLO (Regberg et al. 2020). This detection was unsurprising because there were no bioburden requirements on the spacecraft, and the mission amino acid contamination requirement was a low 160 ng/cm$^2$. Resilient species exist in cleanrooms, even those with bioburden requirements (Schultz et al. 2025). Though unlikely in the dry $GN_2$ curation environment, it is possible that biological activity could occur over time. Fortunately, 4.9% of the sample has been curated at –80 °C since shortly after sample return to prevent volatile loss and biological activity. Furthermore, the JSC curation facilities regularly undergo biological contamination monitoring (McCubbin et al. 2019. Curation cleaning practices are also optimized for organic, inorganic, and biological cleanliness, which reduces the risk of biological contamination of the samples that are curated at room temperature.

Biofilaments (**Table 2 #11)**, likely fungal hyphae, were observed on a polished surface of Bennu stone subsample OREX-800055-1 (**Fig. 9**). Subsamples of OREX-800055-0 did not show evidence of anomalous organic compounds indicative of biology (Mojarro et al. 2026). Biofilaments were not found on any other samples. The origin of the biofilaments are the comparatively dirty sample sectioning and polishing procedures, which are not typically performed under organically or biologically clean conditions and do not include the preparation of blank samples (e.g., Harrington et al. 2025).

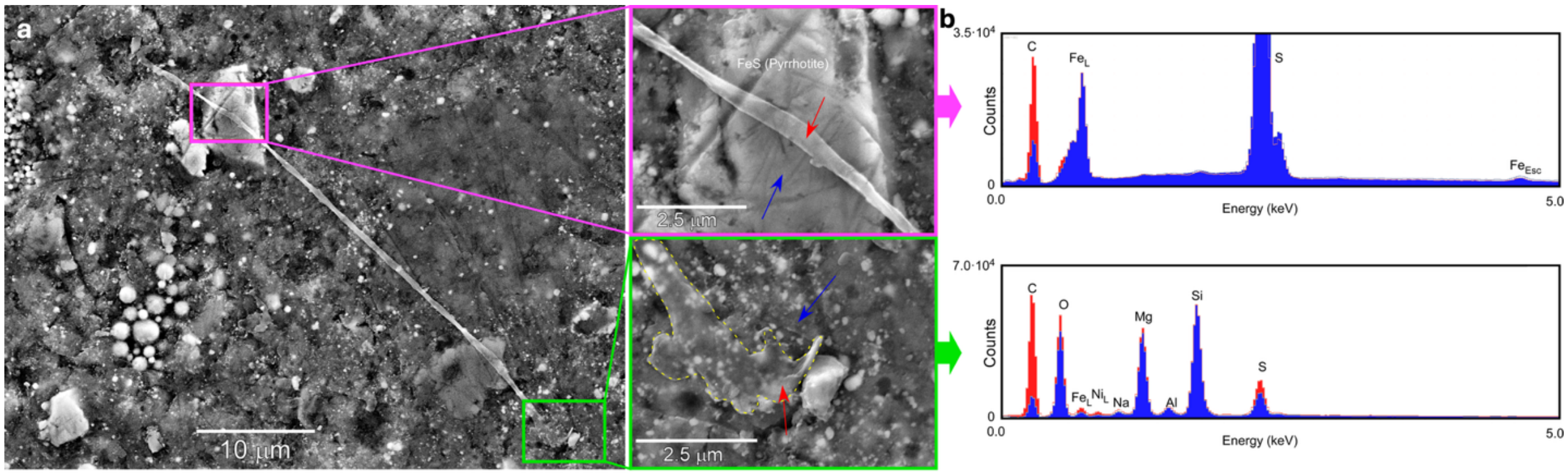

**Fig. 9 a.** SEM/LABE image (**Supplement 9**) of a polished surface of OREX-800055-1, showing the location of a biogenic, filamentous feature ~ 40 µm in length. Regions highlighted by the magenta and green boxes are magnified in the LABE views at right. Top, magenta: A magnified image of the upper region of the filament overlying a pyrrhotite grain. Bottom, green: A magnified image of the lower region of the filament overlying the phyllosilicate matrix. Red and blue arrows designate spot locations of EDX spectra. **b.** Overlapping EDX spectra of the filament (red) and background (blue), showing the filament is comparatively enriched in C.

An anomalous ~0.75 mm particle (OREX-590073-0) (**Table 2 #12**) was detected by the curation team while performing X-ray computed tomography (XCT) under curation-pristine conditions (Eckley et al. 2026). It consisted of 316 stainless steel with a small amount of aluminum metal and fluoropolymer (consistent with PTFE) and Bennu dust (**Fig. 10**). This particle was detected in stone OREX-800150-6, which had been split by the curation team from OREX-800150-0 at the request of the science team. The particle was removed and was of no impact to the remainder of the sample. This particle was not present in the initial XCT characterization of the parent particle, OREX-800150-0, so OREX-590073-0 must have been introduced during the storage, manipulation, or splitting of OREX-800150-0.

The most likely source of OREX-590073-0 is the curation glovebox, as it is furnished with 316 stainless steel hinges, and both PTFE and aluminum foil are used extensively in the gloveboxes (**Supplement 5**). The Eagle Stainless containers used for shipping samples from the curation facility to analysts under $GN_2$ were made of 316 stainless steel, aluminum foil dunnage, and a fluoropolymer gasket and some analysts reported difficulty opening the sealed containers, which could have led to damage. However, these containers were returned and precision-cleaned before entering the cleanroom or $GN_2$ glovebox, which makes them an unlikely origin of particle OREX-800150-0.

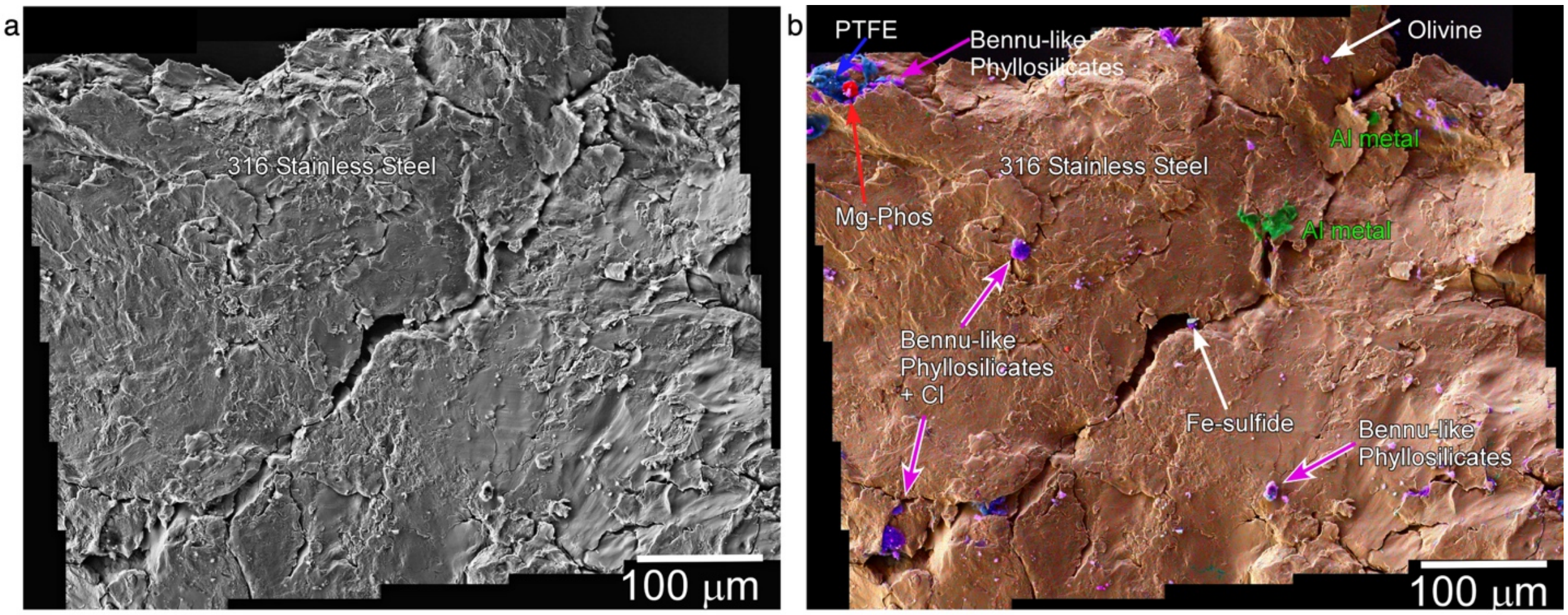

**Fig. 10 a.** SEM/LABE image (**Supplement 9**) view of a piece of stainless steel OREX-590073-0. A fracture is visible in the particle center. Surface particles are present, though contrast constraints limit our ability to resolve them. **b.** Overlapping colorized EDX element maps superimposed on the SEM/LABE image. The stainless steel surface (orange) is dotted with silicates (magenta), phosphate (red), fluorocarbon polymer consistent with PTFE $(C_2F_4)_n$ (blue), Al metal (green), and Fe-sulfides (cyan).

Since the discovery of OREX-590073-0, the curation team has adopted a policy of more aggressively cleaning debris from inside the glovebox between sample exposures. It is worth noting that OREX-590073-0 was detected by the curation team during the execution of their standard operating procedures for cataloging and distributing samples, and it was able to be isolated before allocation and subsequent analysis. Furthermore, this is the only anomaly detected since Bennu sample curation began, suggesting that the impact of the slow degradation of curation hardware was successfully mitigated.

It should be noted that none of the laboratory derived artifacts impacted the sample analysis results or conclusions.

### *4.4 Indigenous to Bennu*

Five of the potential contaminants on the list were ultimately determined to in fact be indigenous Bennu material. That these species were initially assumed to be contaminants underscores the value of sample return, where the opportunities for contamination are carefully controlled, documented, and characterized. Had these species been observed in a meteorite, they would have been assumed to be from the terrestrial environment, and important science would have been lost.

The first such material encountered by the team was white particles, initially identified as a magnesium phosphate, in the quick-look sample from the avionics deck (**Table 2 #13**). All plausible contamination sources in the manufacture and implementation of the SRC, TAGSAM, and spacecraft were carefully examined in cooperation with the engineering team at LMS. Subsequent testing by the science team ultimately demonstrated that the magnesium sodium phosphate material was from Bennu (Lauretta & Connolly et al. 2024). LMS continued their contract support for OSIRIS-REx until one month after sample return; a lesson learned is that missions should provide a financial instrument to permit contamination science personnel to occasionally consult with flight system engineers throughout the sample analysis phase (see Section 5).

Similar examples include villiaumite (NaF) grains in OREX-803174-0 (**Table 2 #14**), subsequently determined to be a product of late-stage aqueous alteration on Bennu's parent body (McCoy & Russell et al. 2025), and a red-orange particle (**Table 2 #15**) in sample OREX-803124-0 (JSC 2024b), shown by infrared spectroscopy to be Fe,Mn-bearing magnesite (Hiroi et al. 2025; Hamilton et al. 2026).

Elevated ammonia ($NH_3$; **Table 2 #16**), a potential contaminant associated with hydrazine thruster exposure, was found in a hot water extract of Bennu aggregate sample OREX-803001-0. However, this $NH_3$ was later shown to be enriched in $^{15}N$ and isotopically above the typical terrestrial range (Glavin & Dworkin et al. 2025; Baczynski & Mcintosh et al. 2026). Prior to isotopic analysis, the possibility existed that this $NH_3$ was the result of exposure to flight hydrazine propellant, which produces abundant $NH_3$ when used. Only by having collected and analyzed flight hydrazine propellant to know its $^{15}N$ abundance as part of the mission's contamination control plan (Dworkin et al. 2018) was it possible to compare $NH_3$ observed in a sample and deduce its outer solar system origin.

Lastly, what looked suspiciously like bits of translucent plastic with low $^{15}N/^{14}N$ (**Table 2 #17**), was observed in some Bennu samples (Sandford & Gainsforth et al. 2025). However, these materials were shown to be isotopically, elementally, and structurally inconsistent with the polymers in the SRC. All SRC lubricants were fluorinated, unlike the polymeric substance observed in the sample. The multilayered structures of this substance are unlike the distinctive microscopic structures of the epoxy adhesive used in the SRC, Henkel Loctite EA9394 (**Supplement 3**). Furthermore, there was no evidence of polymeric contamination in the pyrolysis GC-MS analysis of Bennu samples (e.g., Mojarro et al. 2025). Thus, the polymeric substance was determined to be indigenous and proposed to have formed by cryochemistry very early in the history of Bennu's parent body (Sandford & Gainsforth et al. 2025).

### *4.5 Avionics Deck Contaminants*

A small amount of Bennu sample (1.503 g, or 1.24% of the total 121.6 g of collected sample) was recovered from the avionics deck and other areas outside of the TAGSAM head, some of which was used in quick-look analyses (Lauretta & Connolly et al. 2024) before the TAGSAM head was opened. Of the 31 total suspected contaminants investigated, 10 appeared only in avionics deck samples (**Table 2, #18–27**). Establishing the source of these suspected contaminants was given a lower priority because they did not represent a concern for the majority of the sample to be studied.

To search for the source of unusual metallic particles observed in the avionics deck material (**Table 2 #18–24**), material from exterior to the sample canister that had been vacuumed at the UTTR was examined by SEM/EDX (**Supplement 9**). This exterior sample was found to contain an assortment of metal-rich particles consistent with the metal anomalies found in quick-look samples from the avionics deck (e.g., **Fig. 2**), indicating SRC components ingested into the sample canister as the likely origin.

The avionics deck samples OREX-500002-0 and OREX-501082-0 were observed to contain fibers (**Table 2 #25**). These fibers were individually removed by the analysts from the samples prior to the quick-look analyses (Lauretta & Connolly et al. 2024; Glavin & Dworkin et al. 2025). Pyrolysis GC-MS determined that the fibers did not contain organic compounds. LMS was helpful in identifying their possible sources in the SRC via Ishikawa diagram. The fibers were compared with materials archive samples of candidate contaminants—namely, filtrete fibers from the canister air filter (materials archive sample OR-MA-0019,72), SRC blanket Kapton (materials archive sample OR-MA-0168), damaged SRC blanket Kapton (Genesis mission materials archive sample E50067), and SRC blanket fiberglass batting (materials archive sample OR-MA-0172). SEM/EDX (**Supplement 9**) indicated that the SRC blanket fiberglass batting was the source of the fibers. The specification sheet for this batting indicated that it was borosilicate, but SEM/EDX analysis of the fibers and the archived batting sample demonstrated that this is not the case — offering a reminder that manufacturer-provided information may be incorrect.

A single blue particle was observed in a photograph of the avionics deck (**Fig. 11a, Table 2 #26**). The blue color suggests it may have come from SRC wire insulation, a Ti-6Al-4V alloy washer from the TAGSAM head (materials archive sample OR-MA-0269), or Solithane 113 epoxy (materials archive sample OR-MS-0134) used in the OSIRIS-REx Laser Altimeter (OLA) instrument. These materials can be distinguished by their compositions. However, the blue particle was not separated from but rather mixed with the bulk avionics deck sample, and it was not found in any allocations to the science team. However, if the blue particle is found later, it could be readily separated, and its origin could be identified, if desired, due to the distinct difference in composition of these three potential sources.

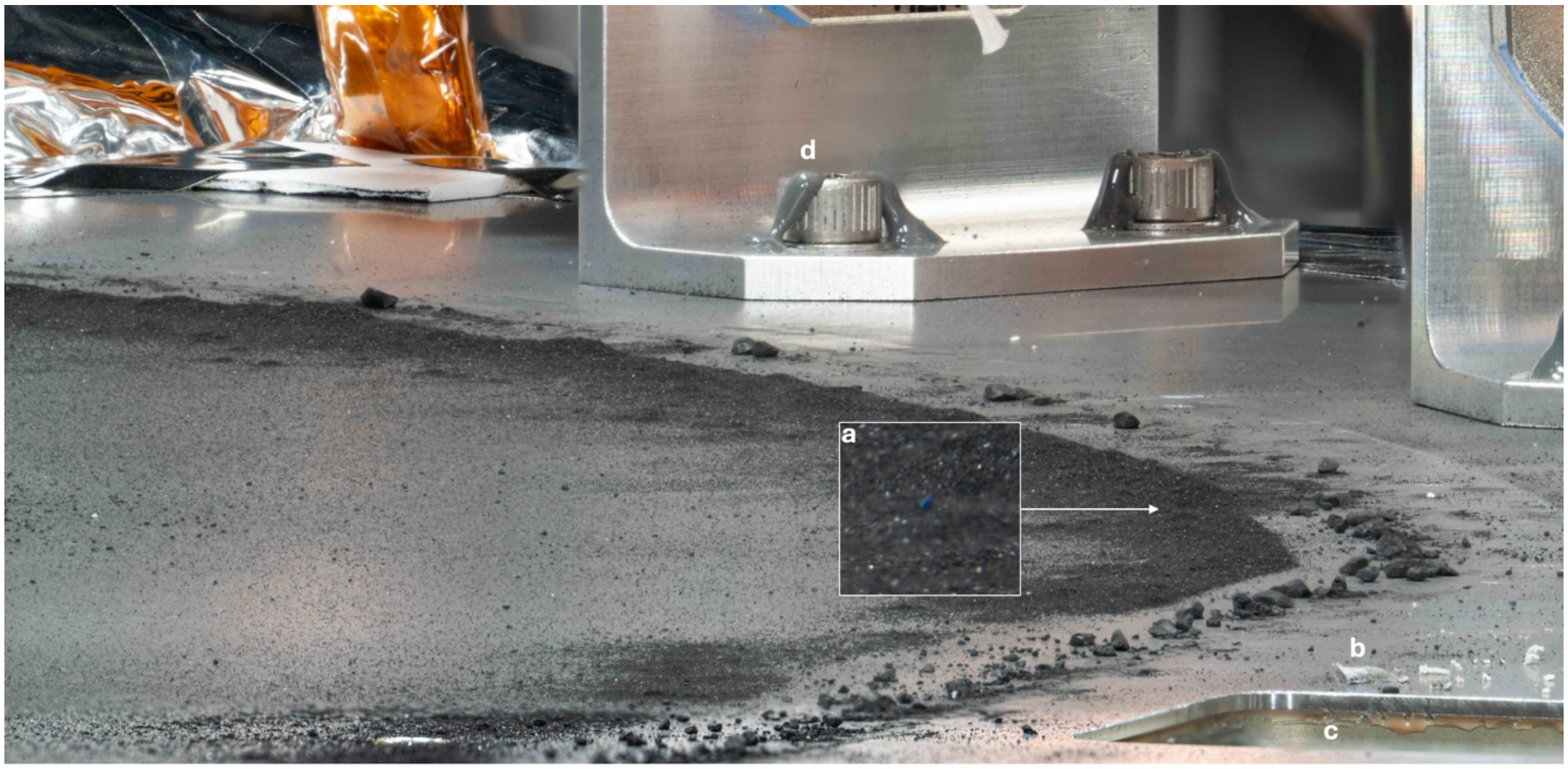

**Fig. 11** Some of the potential contaminants observed on the avionics deck. **a.** Blue particle, **Table 2 #26**. **b.** Angular white particles, **Table 2 #27**. **c.** Avionics deck adhesive Cytec FM 300-2U epoxy. **d.** Staking compound Scotch-Weld 2216 and Cab-O-Sil M-5 fumed silica **Table 2 #1**. The latter two were not found in any samples but, because of their presence on the avionics deck, were compared against unknown materials as part of Ishikawa diagram analyses.

Several white angular fragments were observed outside the canister seal but within line of sight to sample on the avionics deck, making mixing a possibility (**Fig. 11**, **Table 2 #27**). Their source has yet to be determined, as they were not found in the UTTR vacuumed sample OREX-220001-

1. Their composition is also unknown, in part because they have only been identified in the imaging of the sample canister and not during sample analysis.

Thus, samples recovered from the avionics deck should be viewed as contaminated with material from the spacecraft. Importantly, this very small fraction of the total Bennu sample was isolated from the bulk material in the TAGSAM head. There has been no evidence of any of these contaminants beyond the avionics deck.

Note that in addition to the 121.6 g of Bennu material in the official collection, there is an additional 1.155 g of adhering dust that the curation team removed, with difficulty, from the avionics deck, canister lid, and canister hinges in the spring of 2024. This extra sample was not included in the official total collected material since it could only be removable by scraping the flight hardware stainless steel tools and is likely contaminated with metal particles (the recovered material shows light specks mixed with the black Bennu dust). The majority of the material (1.112 g) is in samples OREX-510002-0 to OREX-510005-0 with the remainder in OREX-500013-0 to OREX-500015-0, OREX-590055-0 to OREX-590060-0, OREX-590062-0, OREX-590065-0, and OREX-590067-0 to OREX-590070-0. This material, as well as the additional material that still adheres to other flight hardware items at the time of publication, likely also contains contaminants observed on the avionics deck.

### *4.6 Retaining Ring Contaminant*

When TAGSAM collected the sample, pressurized $GN_2$ (with 10% He) was released by the firing of a NASA standard initiator (NSI) pyrotechnic device (Bierhaus et al. 2018) (**Fig. 12a**). Several metallic-appearing flakes were observed in the gas pathway when the retaining ring holding the contact pads were removed (**Table 2 #28**). These particles were collected (OREX-580001-0) and isolated from the bulk sample. **Fig. 12b** shows one metallic particle prior to its removal from the retaining ring. A known risk was stainless steel fragments from the pyrovalve of the TAGSAM gas bottle (Dworkin et al. 2018); melted and splintered fragments of the mechanism could have been carried with the gas flow. The shapes of these metal particles resemble those of particles captured in a contamination test conducted in 2017 (**Fig. 12c,d**), though at different scales (Dworkin et al. 2018). The design of the test selected for small particles, as the large particles punctured the PTFE filter used for collection.

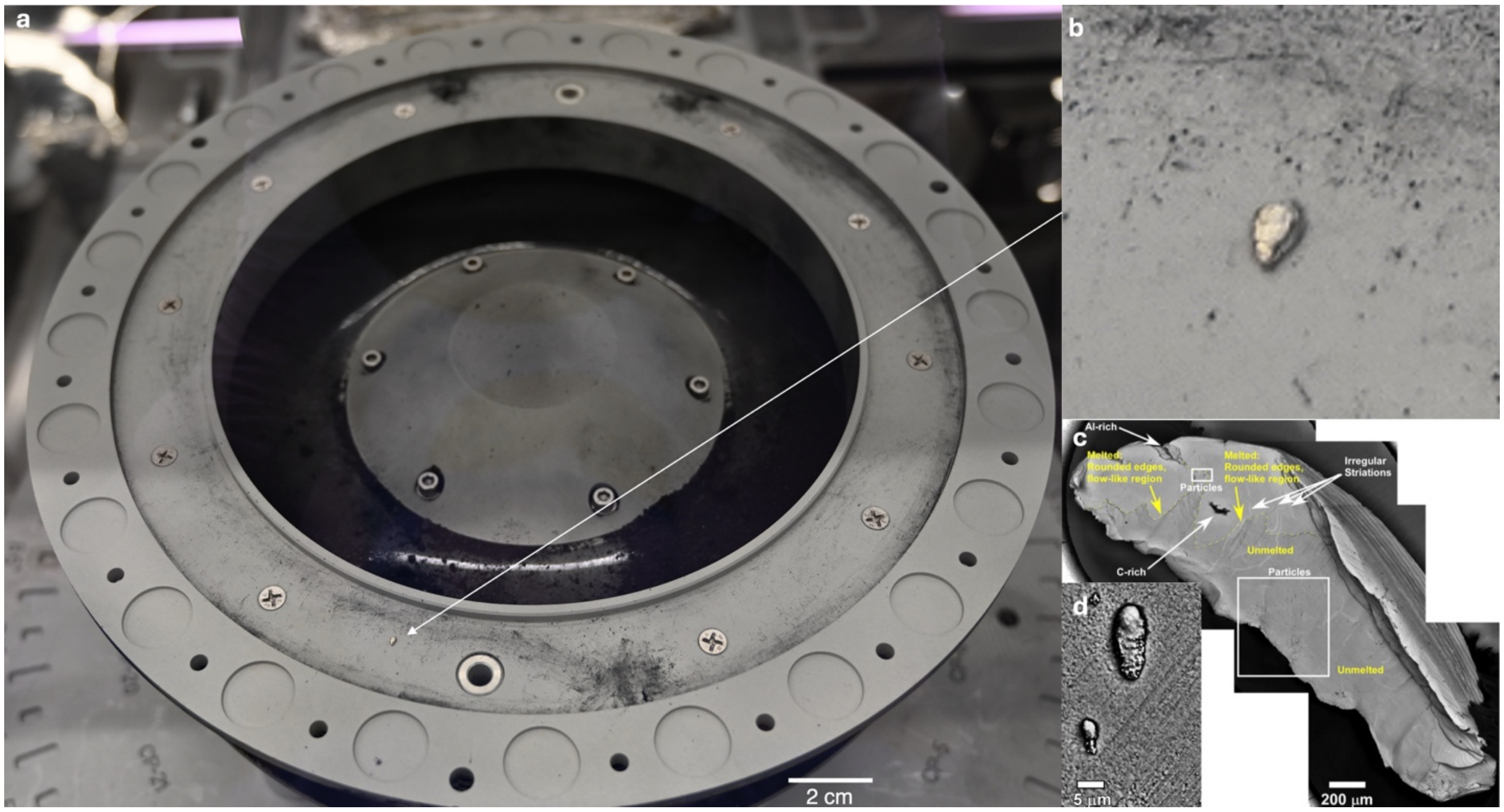

**Fig. 12 a**. The cavity beneath the TAGSAM retaining ring captured metallic particles. **b.** One such particle. **c.** SEM/LABE image of a metal fragment from a pyrovalve contamination test indicating regions of melt, abrasion, and adhering melted particles. **d.** two melted particles.

### *4.7 Laboratory Artifact or Indigenous to Bennu?*

A possible laboratory artifact is an apparent L-valine excess (**Table 2 #29**) in acid hydrolyzed extracts of Bennu (Glavin & Dworkin et al. 2025). In this case, the sample had been handled to minimize organic contamination, and a negative control was used, as well as a positive control (Murchison meteorite) and standards to verify that the analyses were operating as expected. Though all samples and controls were processed in an ISO-5 flowbench, all glassware was heated in air at 500 °C overnight, and reagents were of the highest available purity, some level of amino acid contamination is unavoidable in highly sensitive analyses even under the most careful preparation. The low abundances of valine in the Bennu samples analyzed, together with the low level of L-valine contamination, presented a problem in the sample analysis statistics, which is why the apparent enantiomeric excess of L-valine was not claimed as significant (Glavin & Dworkin et al. 2025). The variability in contamination from sample to sample makes it difficult to rely on blank-subtraction for quantitation since the background L-valine is close to the low L-valine abundance in the sample.

The L-valine excess is unlikely to be due to a biological or industrial contaminant; we are not aware of any peptides or industrial polymers used in manufacturing the OSIRIS-REx spacecraft containing L-valine or its precursors. Furthermore, any peptide contaminants with L-valine could not also contain excesses of other chiral amino acids to be consistent with the otherwise racemic (D=L) results. Moreover, trace L-excesses of valine were detected in both hydrolyzed and unhydrolyzed extracts of OREX-800107-127, whereas a preliminary search for oligopeptides did not show detectable amino acid dimers or larger peptides (Mojarro et al. 2025; Parker et al. in preparation), indicating an absence of biological contamination. Although nucleobases (Glavin &

Dworkin et al. 2025; Oba et al. 2026) and sugars (Furukawa et al. 2026) are present in Bennu samples, there was no evidence of a biological origin of these compounds, nor is there any evidence of nucleosides or nucleotides with current techniques.

As discussed previously, analysis of TAGSAM witness plate OREX-324001-0 detected the amino acids L-aspartic acid, L-serine, D-serine, and L-threonine (**Supplement 11**). That these L-excesses were below the detection limit for the report of racemic amino acids in Bennu material (Glavin & Dworkin et al. 2025) demonstrates that this contamination had no measurable impact on the sample. The only amino acid reported in Bennu samples that was not racemic was valine, which is below the detection limit in the witness plate. The D- and L-valine abundances reported by Glavin & Dworkin et al. (2025) in a hydrolyzed Bennu extract were 0.16 ± 0.01 and 0.32 ± 0.06 nmol/g, respectively. These abundances were so low that the small but ubiquitous laboratory background of L amino acids, not spacecraft contamination, may have contributed to this excess of L-valine.

Thus, we have established that the enantiomeric excess of L-valine (**Table 2 #29**) is not attributable to spacecraft contamination. It may reflect typical laboratory background levels; it is also possible that there is an indigenous valine enantiomeric excess in Bennu. A similar excess was observed in samples from Ryugu (Naraoka et al. 2023). As with Bennu, the total valine abundance was low. Note, however, that the instrument that detected the excess in Ryugu samples used L-valine in the stationary phase (Furusho et al. 2024), whereas a method that did not employ L-valine did not detect any valine (Parker et al. 2023). Though suspected, neither a laboratory nor biological source of the valine enantiomeric excess could be verified.

In preliminary analyses of sub-millimeter mineral grains that were extracted from angular (OREX-800151-102) and hummocky (OREX-800099-103) type Bennu samples (Lauretta & Connolly et al. 2024), the angular sample shows no D- and L-valine detected above the background level, whereas the hummocky sample shows an excess of ~20% L-valine (Kim et al. 2026). The lack of either enantiomer of valine in the angular sample shows that this laboratory has sufficiently low L-valine background to be reliable. It further suggests that the enantiomeric excess of L-valine (in hummocky material) may be indigenous to Bennu. The only report of an excess of L-valine among carbonaceous chondrites is a modest 2.2% in the Murchison meteorite (Pizzarello and Cronin 2000). If the large L-valine excess detected in the preliminary analysis of hummocky Bennu material (Kim et al. 2026) is confirmed, it has important implications for the origin of symmetry breaking (Glavin et al. 2020 and references therein) and should be examined in detail.

### *4.8 Possible Trace Sample Contaminants*

Methylamine, D-serine, L-serine, L-threonine, L-aspartic acid, isobutyric acid, and succinic acid were detected in sapphire witness plate subsample OREX-324001-100 (**Supplement 11**). The ratios of these low-abundance compounds are distinct from those observed in Bennu samples (**Fig. 13**). If the entire interior surface area (1916 $cm^2$) of the TAGSAM head (Dworkin et al. 2018) contained that amount of contamination, and delivered it all to the bulk sample, it would result in only a small amount of sample contamination. The worst case is D-serine, which by this logic could be responsible for up to ~10% of the observed D-serine in the Bennu sample analyzed by Glavin & Dworkin et al. (2025), OREX-803001-0. However, the L-serine extracted from the sapphire witness plate is more abundant than D-serine by an order of magnitude, yet L-serine was below detection limits in the hydrolyzed hot water extract of Bennu aggregate OREX-803001-0, making this rationale unlikely. Similarly, the aspartic acid in Bennu aggregates OREX-803001-0 (Glavin

& Dworkin et al. 2025) and OREX-800107-183 (Baczynski & Mcintosh et al. 2026) were racemic, suggesting that any contribution of L-aspartic acid contamination was insignificant. Threonine was detected but not quantified in OREX-803001-0 (Glavin & Dworkin et al. 2025) and subsamples of OREX-800023-0, OREX-800055-0, OREX-800088-0, and OREX-800107-0 (Mojarro et al. 2025). However, it is chemically very similar to serine, differing by only a methylene group; so, the rationale for serine should apply to threonine.

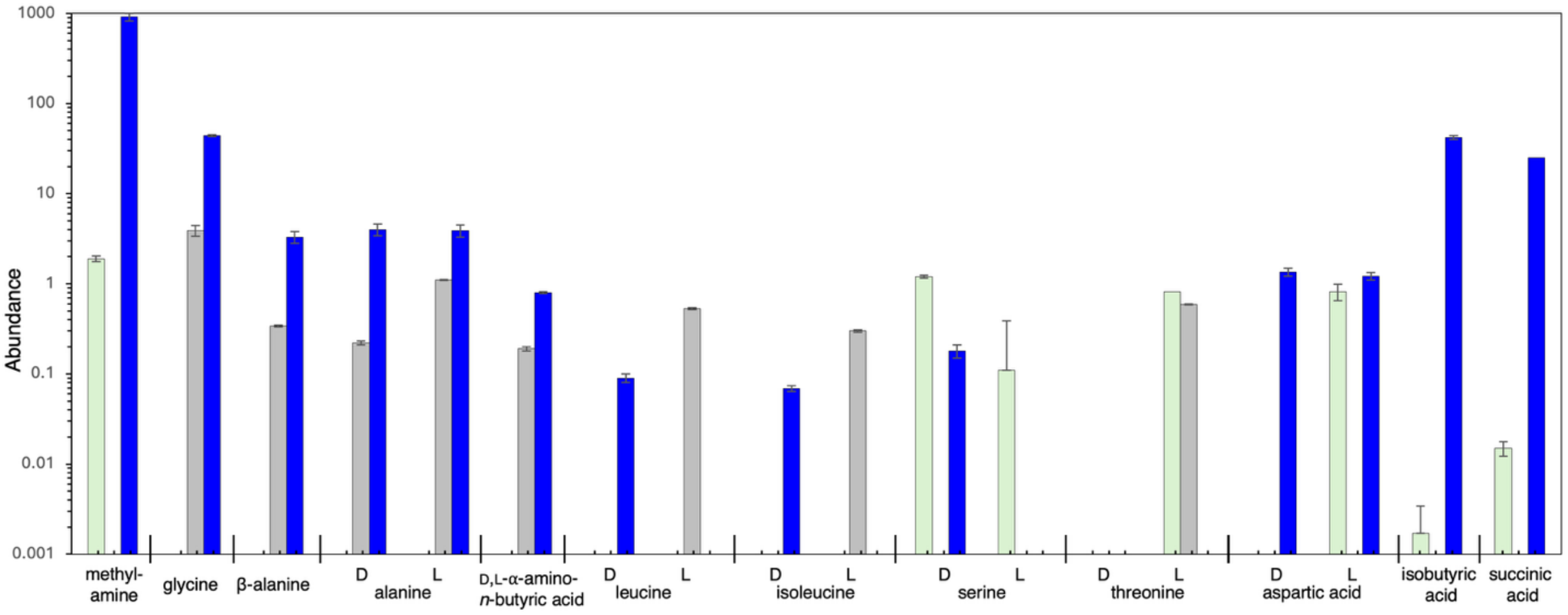

**Fig. 13** Blank-corrected abundances of amines, amino acids, and carboxylic acids detected in the hot water extracts of sapphire witness plate OREX-324001-100 (green), aluminum shim OREX-377025-1 (gray), and Bennu aggregate samples OREX-803001-0 (blue) (Glavin & Dworkin et al. 2025) and OREX-800107-111 (Buckner et al. 2026) (**Supplement 11**). Amino acids were acid hydrolyzed prior to analysis. The witness plate and shim abundances are in units of pmol/cm$^2$; those of the Bennu aggregates are in units of nmol/g.

The only achiral species detected in sapphire witness plate OREX-803001-0 are methylamine, isobutyric acid, and succinic acid, preventing the use of chirality as a check of potential contamination. Using the same assumptions of the delivery of the full TAGSAM surface area to the sample would produce a maximum contamination of ~3 × $10^{-3}$% methylamine, ~6 × $10^{-5}$% isobutyric acid, and ~9 × $10^{-4}$% succinic acid.

Isopropanol, an anticipated potential contaminant (**Table 1**), was found in the sample canister air filter (Sandford et al. 2026). Analysis of air and $GN_2$ samples from spacecraft construction, integration, and SRC recovery (**Table 3**) also showed isopropanol. The highest concentration of isopropanol in the air samples was detected near the air vents of the SRC in the UTTR cleanroom, but not at the landing site at UTTR. As the heat shield on the SRC dissipates heat, the rest of the SRC warms. The delayed heating (heatsoak) could be responsible for the release of isopropanol. Curiously, although isopropanol is an ablation product, there was no detectable isopropanol at the SRC when measured shortly after landing (**Table 3**).

Pyrolysis GC-MS of Bennu aggregate OREX-800128-104 found a small peak tentatively identified as isopropanol based on retention time and a single ion (**Fig. 14a**), not an MRM transition. Isopropanol was also observed in the TAGSAM glovebox exhaust at a slightly higher abundance than in the quiescent blank collected prior to sample arrival (**Supplement 9**). Thus, isopropanol is a plausible but unconfirmed sample contaminant, given the tentative peak identification and low abundances (**Table 2 #30)**.

The pyrolysis GC-MS of sapphire witness plate subsample OREX-324001-103 showed small aromatic compounds (styrene, biphenyl, and naphthalenes) at abundances much lower than in Bennu aggregate OREX-800128-104 **Fig. 14b–e**), thus are not significant contaminants. This analysis also showed low levels of alkenes paired with the corresponding alkanes, though only trace levels of $C_6$–$C_{16}$ alkenes were found in Bennu aggregate OREX-800128-104 (**Table 2 #31**, **Fig. 14f**).

The low levels of $C_6$–$C_{16}$ alkenes (**Fig. 14f**) detected on witness plate subsample OREX-324001-103 had not been noticed in the Bennu samples until, prompted by the witness plate results, closer investigation was performed via pyrolysis GC-MS on Bennu aggregate OREX-800128-104 and by GC×GC-HRMS on Bennu aggregate OREX-800107-128 (**Supplement 11**). These species were only detectable in Bennu samples by pyrolysis GC-MS.

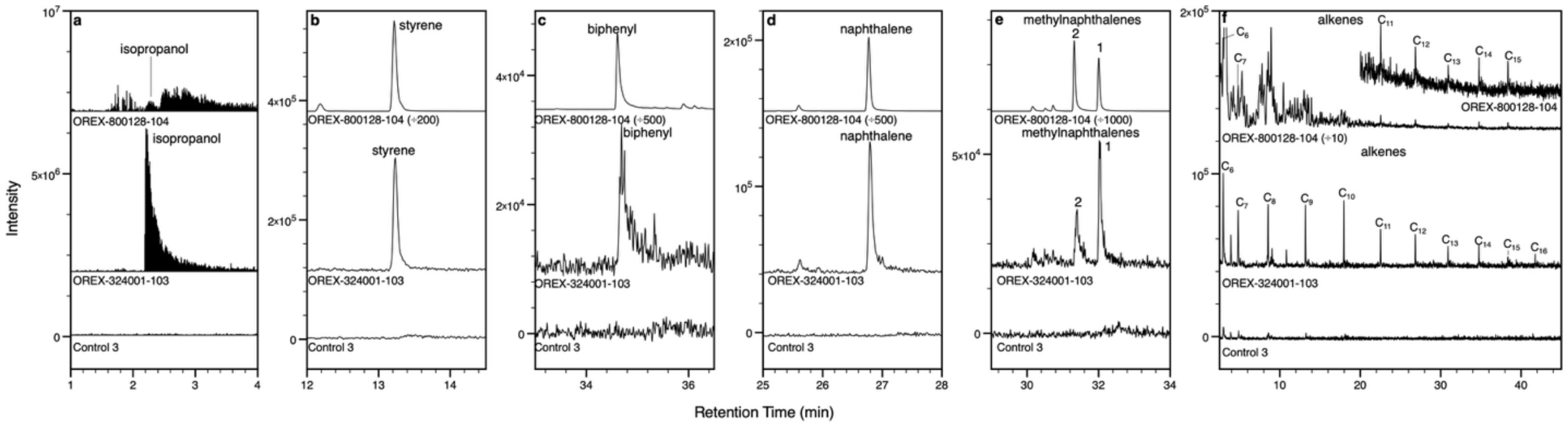

**Fig. 14** Pyrolysis GC-MS of compounds not detected in sapphire control 3 but seen in sapphire witness plate subsample OREX-324001-103 and Bennu aggerate sample OREX-800128-104. All peaks are identified based on their retention time compared with standards. **a.** Single ion chromatogram of *m/z* 59 consistent with isopropanol. **b.** Chromatogram of the MRM 104.1 → 78.1 transition characteristic of styrene. The Bennu aggregate chromatogram has been scaled down by a factor of 500 for clarity. **c.** Chromatogram of the MRM 154.1 → 153.1 transition characteristic of biphenyl. The Bennu aggregate chromatogram has been scaled down by a factor of 500 for clarity. **d.** Chromatogram of the MRM 128.1 → 127.1 transition characteristic of naphthalene. The Bennu aggregate chromatogram has been scaled down by a factor of 500 for clarity. **e.** Chromatogram of the MRM 142.1 → 141.1 transition characteristic of methylnaphthalenes. The methylnaphthalene isomers are indicated by number, and the Bennu aggregate chromatogram has been scaled down by a factor of 1,000 for clarity. **f.** Chromatogram of the MRM 55.1 → 29.1 transition characteristic of alkanes. The Bennu aggregate chromatogram has been scaled down by a factor of 10 for clarity due to the large peaks co-eluting with alkenes. These coelutions prevent the identification of $C_8$–$C_{10}$ alkenes. The unscaled chromatogram is shown after 20 minutes.

It is notable that there does not appear to be a pattern in the abundances of these alkenes (**Fig. 15**), other than the gradual decrease in approximate abundance in both the pyrolyzed witness plate and the pyrolyzed Bennu aggregate with increasing molecular weight. The molecular distribution of alkenes observed is inconsistent with the chain length observed in biology (e.g., Zhu et al. 2018).

However, alkane/alkene pairs are a diagnostic pyrolysis fingerprint of polyethylene and other polyolefins (Soják et al. 2007), though the pyrolysis GC-MS method in this study was not optimized to distinguish the isomers. Polyethylene is a ubiquitous laboratory and packaging contaminant used in spacecraft construction (Dworkin et al. 2018) and curation (**Supplement 5**). The absence of alkane or alkenes detectable by GC×GC-HRMS of solvent extracts of the witness plate supports the hypothesis that these contaminants are pyrolysis products of polymers and that there was no meaningful contamination contribution to the analysis of alkanes in Bennu samples (Aponte et al. 2026).

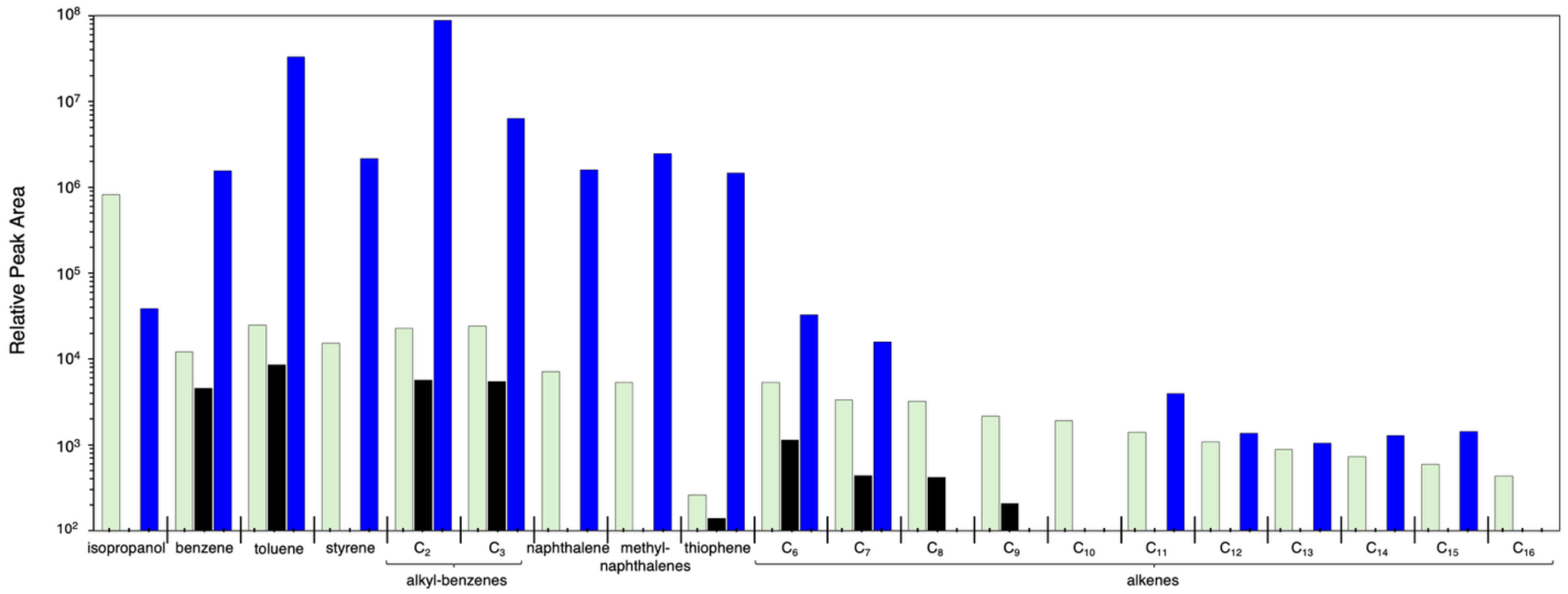

**Fig. 15** Peak area in arbitrary units of compounds detected in the sapphire witness plate OREX-324001-103 (green), sapphire control 3 (black), and Bennu aggregate sample OREX-800128-104 (blue) by semi-quantitative pyrolysis GC-MS (**Supplement 11**).

The identical argument can be made for naphthalene and methylnaphthalenes, which were seen only in the pyrolysis GC-MS of the witness plate but are abundant in both the pyrolysis and solvent extraction of Bennu aggregates. The ratio of the trace abundance of the 1- and 2-methylnaphthalene isomers observed in the witness plate is reversed from those observed in the Bennu samples by both pyrolysis GC-MS (Mojarro et al. 2025) and GC×GC-HRMS (Aponte et al. 2026), providing evidence that the witness plate methylnaphthalenes did not significantly contaminate the sample. Thus, future researchers interested in alkenes should avoid pyrolysis.

## 5 Discussion

The basis for the mission's definition of pristine — no foreign material introduced in an amount that hampers the ability to analyze the chemistry and mineralogy of the sample — is the National Research Council (NRC) recommendation to achieve ±30% precision and accuracy on measurements (NRC 2007) (Dworkin et al. 2018). The analyses presented here show that all contamination identified within the sample returned from Bennu is either readily separable, isolated to the 1.24% of mass that escaped from the TAGSAM head onto the avionics deck, and/or below the limit of quantitation in Bennu material. Thus, the vast majority of the returned sample — 98.76%, or just over 120 g — meets the definition of pristine. Moreover, ~120 g is more than twice the mass of pristine material that the mission was required to return (Lauretta et al. 2017).

The strategies for contamination knowledge applied by OSIRIS-REx are summarized in **Table 4**. It is heartening that lessons from NASA's Stardust comet sample return mission (Sandford et al. 2010) were largely learned and applied for the OSIRIS-REx mission. Of the 10 lessons from Stardust (Sandford et al. 2010), only the timely distribution of the witness plates and the implementation of witness plate subdivision were lacking from OSIRIS-REx. However, the efficient distribution of materials archive samples and flight hardware, which was a problem for Stardust, was a strength of OSIRIS-REx thanks to planning and cooperation.

**Table 4** Strategies implemented by OSIRIS-REx of relevance to future missions.

| | Strategy Implemented | Application to Future Missions |
|---|---|---|
| 1 | Testable and justified definition of contamination early in mission development, applied throughout the mission lifecycle. | Establish a clear, mission-specific definitions of "pristine" based on science objectives and available resources. |
| 2 | Understanding and buy-in across the entire team (e.g., managers, scientists, engineers, technicians) and stakeholders of the importance of CC and CK to meeting mission objectives. | Integrate contamination awareness into all project phases and relevant subsystems. Develop a culture invested in the importance of contamination control and knowledge for mission success. |
| 3 | Elevating CC and CK to a Level 1 mission requirement, with appropriate descopes. | Prevent the erosion of contamination requirements without a project-level review. |
| 5 | Designated individual (e.g., Project Scientist, Deputy PI) with responsibility and authority to implement the Level 1 requirement, with contacts and support in management, engineering, curation, and science. | Assign oversight of CC and CK implementation to one person integrated in relevant areas across the project. |
| 5 | Including contamination control and knowledge as part of the project and subsystem lifecycle reviews. | Add contamination to project reviews and risk boards. |
| 6 | Archiving ample flight materials for comparison with sample suspected contaminants. | Construct a comprehensive materials archive during the development of flight hardware for contamination knowledge, with enough material to be allocated for multiple studies. |
| 7 | Having ample flight witness plates and relevant controls. | Design redundant witness plates to permit allocation for multiple studies and archiving. |
| 8 | Close partnering between contamination science and contamination engineering leads. | Maintain ongoing collaboration between scientists and engineers throughout the mission lifecycle. |
| 9 | Understanding that foreign materials which are readily separable and do not impact sample analysis are not contaminants. | Recognize when a foreign substance is not a contamination risk. |
| 10 | Analyzing the sample and relevant controls for contamination as part of the science necessary for understanding not only sample integrity, but also the diversity of indigenous materials. | Communicate that a sufficiently rigorous CK plan can lead to new discoveries that would otherwise have been discounted as contamination. |

CC = contamination control, CK = contamination knowledge

In addition, OSIRIS-REx sample analysis has imparted further lessons which can be applied to future sample return missions, summarized in **Table 5**. While there is always room for improvement, and it is impossible to remove all sources of organic contamination (Chan et al. 2020), it is important to stress that the work of the OSIRIS-REx team in control and knowledge of contamination from conceptualization through curation was effective and worthy of emulation by other projects. Some of the lessons we identify would have required additional resources and personnel or knowledge of Bennu that did not exist at the time of execution.

**Table 5** Lessons learned from OSIRIS-REx with recommendations for future missions.

| | Observation | OSIRIS-REx Impact | Application to Future Missions |
|---|---|---|---|
| 1 | Witness plates were not available for analysis by the science team at the same time as the Bennu samples. | There was no economy of scale with simultaneous witness plate and sample analysis. There was no ability for witness plate findings to impact sample analysis strategy or curation process. | Curation staffing and infrastructure should be sufficient to distribute witness plate and other contamination knowledge material requests at a priority commensurate with sample return material. |
| 2 | Most witness plates contained visible Bennu sample fines on their surfaces. | Most witness plates could not be analyzed for contamination without the ability to exclude contributions from the Bennu sample. | If possible, design witness plate barrier to better permit contamination to pass while excluding sample. This requires knowledge of sample physical properties during mission design, which may not exist (as was the case for Bennu). |
| 3 | Canister witness plate covers failed. | These witness plates could not be used as intended. | Design witness plate covers to seal throughout their entire lifecycle, including after return. |
| 4 | Witness plates could not be easily subdivided without imparting contamination. | Particulate analysis of witness plates was not possible with the resources available. Subdividing thick witness plates was very challenging. | Reserve funding sufficient for the curation and science teams to permit timely development and testing of witness plate subdivision. Specify requirements on witness plate thickness and fiducials early in design. |
| 5 | Most contaminants were found in the avionics deck samples. | The first (quick-look) samples to be analyzed were contaminated with spacecraft materials. Though they ultimately did not impact the findings, they were a distraction. | Understand that the first available sample may be the most contaminated and apply resources and expectations accordingly. |
| 6 | Avionics deck contaminants were isolated and did not impact subsequent sample analysis. | The goal of identifying all contaminants and their sources on the avionics sample was relaxed. | The origin of every contaminant without significant impact to the sample does not need to be determined. |
| 7 | Some suspected contaminants were laboratory artifacts. | Suspected contaminants were more confidently identified in laboratories that utilized procedural blanks. | Include procedural blanks for all analyses. These processing blanks should follow the sample during laboratory transfers. Have sufficient sample for parallel analyses. |
| 8 | LMS support was only available during quick-look analysis. | Analyses of materials and designs were less efficient than if flight system engineers had been involved. | Reserve funding to permit occasional consultation by the relevant engineers during sample analysis. |
| 9 | The contamination analysis budget was assumed to be a component of sample analysis. | The witness plate was allocated near the end of sample analysis. Contamination analyses of the witness plate and SRC debris occurred after the completion of sample analysis. | Provide contingency sample science budget for contamination analyses and complex suspected contaminants. |
| 10 | No spacecraft contaminants were identified other than those from the SRC and TAGSAM. | Much of the archived spacecraft hardware did not need to be used. | Contamination control of the spacecraft was sufficient. |

## 6 Conclusion

The OSIRIS-REx mission achieved its objective of returning at least 60 g of pristine regolith from asteroid Bennu, where pristine is defined as free from foreign materials that impact scientific measurements. In fact, the mission returned more than 120 g of pristine regolith, enabling significant discoveries by the science team in the immediate term and providing material for generations of scientists to come.

This achievement was made possible by prioritizing contamination control and contamination knowledge in mission design and implementation. It required cooperation and effective communication across the project including management, science, engineering, and technicians. By clearly deriving and implementing the definition of pristine relevant for the target, the analytical objectives, and the resources available (Dworkin et al. 2018), the path to success was charted and successfully navigated.

Major contributors to OSIRIS-REx successfully returning a pristine sample were a mission Contamination Science Lead advocate for the importance of contamination control and knowledge within the management structure, and the willingness of numerous stakeholders to support contamination control and knowledge efforts. For future sample return missions, this success is most likely to be repeated by management commitment to a strong and active contamination science program with support from all levels of the mission team, from mission development through sample curation.

The diligent work of the curation team kept the sample pristine while characterizing, cataloging, and distributing samples to the science team. The science team effectively reported observations and unusual findings to the Contamination Science Lead, who tracked and investigated potential contaminants in collaboration with the Mission Sample Scientist. Hypotheses for the origin of potential contaminants were posed and tested. In some cases, species suspected to be contamination were found to be indigenous to Bennu. Without a mission that protected the sample from uncontrolled exposure to the terrestrial environment, and paid careful attention to contamination control and knowledge, these discoveries would have been discounted.

It is also significant to remember that the act of analyzing the sample can introduce contamination. As such, careful subdivision of sample is necessary to prevent different analyses from interfering with each other. Even with careful planning, there can be unintended consequences of otherwise good decisions or trades. A prime example is the release of the sample canister latches at the UTTR, which likely prevented humid air from imparting moisture contamination to the entire sample but also permitted particulate debris mixing with the sample on the avionics deck during transport from the UTTR to JSC.

Planetary science follows a path. “Methodically we work — flyby, orbit, land, rove, return samples. We’ve done that with the moon, and we’re well on our way of doing that with Mars. There’ll be other objects in the future we’ll do the same process with” (Green 2017). This roadmap for solar system exploration puts sample return at the pinnacle of exploration. NASA’s Artemis crewed missions to the Moon and JAXA’s MMX robotic mission to Mars’ moon Phobos are poised to return samples. China National Space Administration’s Tianwen-2 sample return mission to asteroid (469219) Kamoʻoalewa is in progress. Missions to return samples from Mars, Ceres, and a comet are being conceptualized. It is hoped that the observations presented here not only give confidence to future analysts of Bennu samples, but also provide strategies and lessons for future mission implementation to ensure the pristinity and maximize the scientific value of returned samples.

## 7 Statements and Declarations

The authors declare no non-financial interests that are directly or indirectly related to this work.

## 8 Acknowledgements

This material is based upon work supported by the National Aeronautics and Space Administration (NASA) under award NNH09ZDA007O and contracts NNG12FD66C and NNM10AA11C issued through the New Frontiers Program. NASA's Participating Scientist Program award no. 80NSSC22K1689 supported K.W. and P.K.; award no. 80NSSC22K1690 supported K.H.F. and A.A.B; and 21-ORSAPS21_2-0009 supported A.E.H. The Jet Propulsion Laboratory is operated by the California Institute of Technology under contract with NASA (contract no. 80NM0018D0004). Part of this work was performed under the auspices of the U.S. Department of Energy by Lawrence Livermore National Laboratory under contract DE-AC52-07NA27344 with release number LLNL-JRNL-LLNL-JRNL-2020170. PCIGR was funded by Canadian Space Agency grant 22EXPORUBC. This research is partly supported by the Japan Society for the Promotion of Science under KAKENHI grant no. 25H00677 for Y.O., 21KK0062 for Y.T. and T.K., and 20H00202 for H.N. Funding for P.S.-K. and M.L. was from German Research Foundation—project-ID 364653263—TRR 235 (CRC 235) and project-ID 521256690—TRR 392/1 2024 (CRC 392/1 B2). We also appreciate the careful review of the manuscript by C.W.V. Wolner and K.E. Dworkin and helpful comments by C. Herd and an anonymous reviewer.

We wish to thank the hundreds of people and their families who labored and sacrificed to make OSIRIS-REx a reality. We honor the memory of Michael J. Drake, the first Principal Investigator of the OSIRIS-REx mission.

## 9 Data availability

The instrument data supporting the experimental results in this study are available at https://astromat.org via the DOIs given in **Supplement 12**.

### *Supplemental Information for*

# *OSIRIS-REx Returned a Pristine Sample of Asteroid Bennu: Takeaways from the Mission's Contamination Control and Knowledge Program*

Authors and affiliations

Jason P. Dworkin[1,2*]
Scott A. Sandford[3]
Nicole G. Lunning[4]
Kevin Righter[4,5]
Charles C. Lorentson[1]
Daniel P. Glavin[1]
Kathie L. Thomas-Keprta[4,6]
Loan Le[4,7]
Catherine M. Corrigan[8]
Timothy J. McCoy
Edward B. Bierhaus[9]
José C. Aponte[1]
Angel Mojarro[1,2]
Hannah L. McLain[1,2]
Eric T. Parker[1]
Denise K. Buckner[1,10]
Frédéric Seguin[1,2]
Radford L. Perry[1]
Heather V. Graham[1]
Havishk Tripathi[1,11]
Pierre Haenecour[12]
Piers Koefoed[13]
Kun Wang[13]
Cody Schultz[14]
Takahiro Hiroi[14]
Ralph E. Milliken[14]
Simon J. Clemett[4,7]
Bumsoo Kim[4,7]
Christopher J. Snead[4]
Eve L. Berger[4]
Michael Liss[15,16]
Jessica J. Barnes[12]
Sandra Freund[9]
Zoe A. Wilson[9]
Ryan D. Dubisher[9]
Natasha V. Almeida[17]
Kimberly K. Allums[4,6]
Erika H. Blumenfeld[4,18]
Joseph E. Aebersold[4]
Rachel C. Funk[4,19]

Carla P. Gonzales[4,20]
Julia L. Plummer[4,20]
Maritza Montoya[4,20]
Daniel J. Rohmer[4,6]
Erika V. Rivera[4,21]
Jannatul Ferdous[4,20]
Gabriel C. Garcia[4,20]
Melissa Rodriguez[4,20]
José L. Ponce[4]
Ronald Bastien[4]
Wayland D. Connelly[4,22]
Jemma Davidson[4]
Marghaleray Amini[23]
Vivian Lai[23]
Dominique Weis[23]
Josh Wimpenny[24]
Jan Render[24]
Gregory A. Brennecka[24]
Yoshihiro Furukawa[25]
Toshiki Koga[26]
Yoshinori Takano[26]
Hiroshi Naraoka[27,28]
Yasuhiro Oba[29]
Kenneth J. Domanik[12]
Yao-Jen Chang[12]
Zoe Zeszut[12]
Thomas J. Zega[12]
George D. Cody[30]
Conel M. O'D. Alexander[30]
Ian A. Franchi[31]
Monica M. Grady[31]
Danielle N. Simkus[1,2,32]
Jamie E. Elsila[1]
Philippe Schmitt-Kopplin[15,16,33]
Allison A. Baczynski[34]
Katherine H. Freeman[34]
Amy E. Hofmann[35]
Aaron S. Burton[4,36]
Lindsay P. Keller[4]
Sara S. Russell[17]
Anjani T. Polit[12]
Francis M. McCubbin[4]
Harold C. Connolly Jr.[12,37,38]
Dante S. Lauretta[12]

[*]Corresponding author's email address: jason.p.dworkin@nasa.gov

[1] NASA Goddard Space Flight Center, Greenbelt, Maryland, USA
[2] Center for Space Science and Technology, University of Maryland Baltimore County, Baltimore, Maryland, USA
[3] NASA Ames Research Center, Moffett Field, California, USA
[4] NASA Johnson Space Center, Houston, Texas, USA
[5] Department of Earth and Environmental Science, University of Rochester, Rochester, New York, USA
[6] Barrios/JETS II Contract, NASA Johnson Space Center, Houston, Texas, USA
[7] JETS II Contract, NASA Johnson Space Center, Houston, Texas, USA
[8] Smithsonian Institution National Museum of Natural History, Washington, DC, USA
[9] Lockheed Martin Space, Littleton, Colorado, USA
[10] NASA Postdoctoral Program, Oak Ridge Associated Universities, Oak Ridge, Tennessee, USA
[11] Buseck Center for Meteorite Studies, Arizona State University, Tempe, AZ, USA
[12] Lunar and Planetary Laboratory, University of Arizona, Tucson, Arizona, USA
[13] McDonnell Center for the Space Sciences, Department of Earth, Environmental, & Planetary Sciences, Washington University, St. Louis, Missouri, USA
[14] Department of Earth, Environmental, and Planetary Sciences, Brown University, Providence, Rhode Island, USA
[15] Technical University Munich, Freising, Germany
[16] Research Unit Analytical Biogeochemistry, Helmholtz Munich, Neuherberg, Germany
[17] Natural History Museum, London, UK
[18] LZ Technology, JETS Contract, NASA Johnson Space Center, Houston, Texas, USA
[19] GeoControl Systems, Inc./JETS II Contract, NASA Johnson Space Center, Houston, Texas, USA
[20] Amentum/JETS II Contract, NASA Johnson Space Center, Houston, Texas, USA
[21] Mclaurin Aerospace/JETS II Contract, NASA Johnson Space Center, Houston, Texas, USA
[22] MB Solutions, Inc/JETS II Contract, NASA Johnson Space Center, Houston, Texas, USA
[23] PCIGR, Department of Earth, Ocean and Atmospheric Sciences, The University of British Columbia, Vancouver, British Columbia, Canada
[24] Lawrence Livermore National Laboratory, Livermore, CA, USA
[25] Department of Earth Science, Tohoku University, Sendai, Japan
[26] Biogeochemistry Research Center, Japan Agency for Marine-Earth Science and Technology (JAMSTEC), Natsushima, Yokosuka, Japan
[27] Department of Earth and Planetary Sciences, Kyushu University, Fukuoka, Fukuoka, Japan
[28] Institute of Space and Astronautical Science, Japan Aerospace and Exploration Agency (JAXA), Sagamihara, Kanagawa, Japan
[29] Institute of Low Temperature Science, Hokkaido University, Sapporo, Hokkaido, Japan
[30] Earth and Planets Laboratory, Carnegie Institution for Science, Washington, DC, USA
[31] School of Physical Sciences, Open University, Milton Keyes, UK
[32] Queen’s University Kingston, Ontario, Canada
[33] Center for Astrochemical Studies, Max Planck Institute for Extraterrestrial Physics, Garching, Germany
[34] Department of Geosciences, Pennsylvania State University, University Park, PA, USA

[35] Jet Propulsion Laboratory, California Institute of Technology, Pasadena, CA, USA
[36] NASA Headquarters, Washington, DC, USA
[37] Department of Geology, Rowan University, Glassboro, New Jersey, USA
[38] Department of Earth and Planetary Sciences, American Museum of Natural History, New York, New York, USA

# Table of Contents

## **Supplement 1:** Curation environment $GN_2$, amino acid, particle, and NVR results

### OSIRIS-REx Cleanroom Monitoring Report

#### *Monitoring of the OSIRIS-REx Temporary Cleanroom at the Utah Testing and Training Range*

The temporary cleanroom (ISO 7) was installed in June 2023 in building 1012 at the Utah Testing and Training Range (UTTR), which was the same building used to house temporary cleanrooms for the Genesis and Stardust Missions. The temporary cleanroom for OSIRIS-REx recovery operations differed from those of Genesis and Stardust in that the OSIRIS-REx temporary cleanroom had a diamond-aluminum-raised floor (6061 aluminum rather than vinyl) and a small antechamber for gowning on one side. The antechamber was added by the curation team after installation of the cleanroom by the vendor (LASCO), because a readiness review in March 2023 recommended the addition of a gowning antechamber. The procurement process for the temporary cleanroom was already finalized in March 2023, and the scope of work could not reasonably be changed. The astromaterials curation team had this cleanroom antechamber in storage, which was previously used to support frozen processing for the Apollo Next Generation Sample Analysis (ANGSA) program. The ability to support the readiness review recommendation was made possible by previous (non-OSIRIS-REx) curation work to coordinate and support storage of this equipment, which is frequently challenging within floor space limitations.

The temporary cleanroom was installed in June 2023 so it would be in place for the first UTTR-based recovery rehearsal in July 2023. This rehearsal included activities that went up to the temporary cleanroom double doors (no antechamber), but as a contamination control measure, this rehearsal did not include activities inside the cleanroom. The SRC disassembly activities were rehearsed separately in a cleanroom space at Lockheed Martin Space (LMS) with both LMS and curation team members participating. However, other activities were taking place within building 1012 until mid-August 2023, which included opening external high bay doors for vehicles to access the same room that housed the temporary cleanroom. It was recognized in July 2023 that opening the high bay doors of 1012 represented a potential contamination vector for the temporary cleanroom. The curation team was back at UTTR at the end of August 2023 to support another recovery rehearsal, at which time particle counts were collected while the high bay doors were open to evaluate the impact on the cleanroom. The August 2023 particle counts with the doors open spiked above the ISO 7 limit (average = 36,572 particle count sized 0.5 µm ($\#/m^3$)); this measurement supported work to socialize across the recovery team and media support teams the importance of keeping the high bay doors and other exterior doors closed to protect the cleanroom and to protect the Bennu sample. In addition, a team of curation cleaning technicians came to UTTR the week of 12 September 2023 to conduct a rigorous deep cleaning of the temporary cleanroom prior to SRC landing on Sunday 24 September 2023.

#### *Temporary Cleanroom Monitoring Prior to SRC Landing*

The particle counts for the temporary cleanroom were consistently within ISO 7 levels prior to SRC landing, with the exception of the particle counts collected with the high bay doors open on 31 August 2023 (Fig. S1-1). The particle counts were collected when the curation team was onsite at UTTR for rehearsals in July and August 2023, as well as during cleaning and recovery preparations between 12 September-24 September 2024.

As a complement to the organic compounds mass $>C_7$ ($ng/cm^2$) values determined from Balazs pre-baked silicon wafers deployed for 24 hours in the cleanroom, we also deployed air samplers from Balazs™ NanoAnalysis (hereafter Balazs) using a sampling pump set at a flow rate of 100 mL/minute deployed for 6 hours. Both the silicon wafers and air samples were analyzed by Balazs, respectively yielding measurements in ng/cm$^2$ and ng/L. The organic compounds with mass >$C_7$ (ng/cm$^2$) in the temporary cleanroom were slightly higher than those reported for the JSC OSIRIS-REx ISO 5 cleanroom (Righter et al. 2023) but remained below or within the range determined for the Genesis ISO 4 cleanroom between 2000-2011 (Fig. S1-2). However, the mass >$C_7$ (ng/L) determined via air sampling were more than an order of magnitude greater than those determined for the JSC OSIRIS-REx ISO 5 between 2021-2026 (Fig. S1-3), which will be discussed below in the section on that cleanroom.

***Particle Counts on Landing Day***

After the regular particle counts were measured at multiple locations in the temporary UTTR cleanroom, the particle counter was set up to run back-to-back measurements starting at **7:21 am Utah local time** on the morning of 24 September 2023 (Fig. S1-5). This particle counter remained in one corner of the cleanroom and its display was visible through one of the cleanroom windows by the curation team. Each particle count measurement took two minutes and there was 15 seconds between each measurement. During development of the SRC recovery procedure, the recovery leaders from GSFC, Lockheed Martin, the SAT, and curation discussed the appropriate response if the particle counts exceeded ISO 7 levels. The possibility of pausing SRC disassembly until particle counts dropped below the ISO 7 was considered, but ultimately, there was a consensus that establishing the nitrogen purge on the sample canister and continuing disassembly to get the sample canister into a curation glovebox at JSC as quickly as safely possible needed to be a higher priority for sample protection.

All times referenced in this section are Utah local time (U.S. Mountain Daylight Time).

At **10:58 am**, the cleanroom double doors (side without the antechamber for gowning) were opened to bring the SRC into the cleanroom. The cleanroom recovered to within ISO 7 particle counts by 11:02 am.

The majority of the particle counts between **11:45 am-12:14 pm** are above the ISO 7 limit. At this time, vent covers were being removed from the SRC back shell. These covers needed to be dug out of the back shell. Debris with a range of particle sizes including fine-sized debris shed during this process onto the SRC, the stand for the SRC, and the cleanroom floor. The curation team archived the vent covers. In addition, debris from the back shell vent covers was isolated using the "dirty" cleanroom vacuum cleaner (used for materials that were not potential Bennu samples). At this point, the back shell was still on the heatshield so there was tortuous path between these particles and the Bennu sample inside the sample canister.

There are two particle counts above the ISO 7 limit between **12:44-12:46 pm**. At this point in disassembly of the SRC, items were being removed from the avionics deck to support establishing the nitrogen purge of the sample canister. These activities required manipulation and removal of some of the surrounding Multi-Layer Insulation (MLI) blankets.

The particle counts were within ISO 7 values at **12:51 pm** when the nitrogen purge was established on the sample canister.

There are three particle counts above ISO 7 levels during the interval that the SRC battery was removed, which involved removing MLI blankets and cutting wires (**13:11-13:15**). There are two additional time intervals (**14:40-14:42 and 15:00**) with spikes in particle counts above ISO 7 levels, which occurred during subsequent removal of items from the avionics deck and manipulation of MLI blankets. At the above points, the sample canister was still inside the heat shield.

The particle counts were within ISO 7 levels when the sample canister on nitrogen purge was fully enclosed in its shipping container at **16:56**.

There are three particle counts above the ISO 7 limit between **17:13-17:18.** During this time the MLI blankets within the headshield were manipulated to containerize debris, which included using the "clean" cleanroom vacuum to collect potential Bennu sample from inside of the heat shield, the filter of which was officially named OREX-220001-0. After the debris was collected the MLI blankets were removed at **17:18**.

There are three particle counts above the ISO 7 limit between **17:33-17:52.** At this time, the team in the cleanroom were containerizing the heat shield for transport, which may have released char particles. The final particle counts of the day above the ISO 7 limit are likely associated with beginning packing the cleanroom and preparing for transport early the next day.

***OSIRIS-REx JSC Cleanroom Monitoring***

The ISO 5 cleanroom constructed for the OSIRIS-REx samples at NASA JSC has been previously described in detail by Righter et al. (2023). The OSIRIS-REx ISO 5 cleanroom in which the pristine Bennu samples are handled in nitrogen gloveboxes and stored in nitrogen isolating desiccators has maintained the initial baseline reported in Righter et al. (2023). This lab was certified as ISO 5 upon completion in November 2021. Particle counts were collected at six locations within this lab on a weekly cadence (Fig. S1-4) starting on 19 May 2022. The first three months of this data set were previously reported in Righter et al. (2023). The particle counts have consistently been below the particle limit for ISO 5 cleanrooms (<3520 particles sized 0.5 µm per m$^3$), with one exception measured on 1 June 2023 in which the average of the six measured locations in the cleanroom was slightly above the limit at 3650 particles sized 0.5 µm per m$^3$. This 1 June 2023 particle count was almost four months before the Bennu sample arrived at JSC and it was during an intense push to have the gloveboxes installed and functional, which included curation personnel working long days and weekends in the cleanroom. This intense period of work in the cleanroom was needed to support a clean hardware disassembly rehearsal the following week (mission milestone Curation Rehearsal #3), which included project managers from NASA GSFC/NASA HQ and mission sample science leadership traveling to JSC to participate in and/or observe the rehearsal.

The levels of organic compounds with mass >$C_7$ (ng/cm$^2$) have remained close to the initial measurement for the ISO 5 OSIRIS-REx Cleanroom (Fig. S1-3) reported by Righter et al. (2023). The organic compounds with mass >$C_7$ (ng/cm$^2$) have remained below or at the lower end of the range determined for the Genesis ISO 4 cleanrooms between 2000-2011 (Fig. S1-3).

As a complement to the organic compounds mass >$C_7$ (ng/cm$^2$) values determined from Balazs pre-baked silicon wafers deployed for 24 hours in the cleanroom, we also deployed Balazs air samplers using a sampling pump set at a flow rate of 100 mL/minute deployed for 6 hours. Both the silicon wafers and air samples were analyzed by Balazs, respectively yielding measurement sin ng/cm$^2$ and ng/L. The silicon wafers and air samples deployed in the cleanroom were measured along with shipping blanks. The organic compounds mass >$C_7$ (ng/L) determined from the air samples were consistently low, but there is not a complementary Genesis data set. The organic compounds masses >$C_7$ (ng/L) from the air samples do not consistently trend with those determined from the silicon wafers (Fig. S1-3). The initial measurement in this data set was collected during commissioning of the lab, which was six months after completion of the epoxy flooring in March 2021. This data set from OSIRIS-REx may be useful for future curation cleanroom construction planning.

Amino acids were monitored in the ISO 5 OSIRIS-REx cleanroom using baked out high-purity aluminum witness foils, which were analyzed using methods comparable to those used for Assembly Testing and Launch Operations (ATLO) witness foils described in Dworkin et al. (2018). For the monitoring period between 27 October 2022 and 10 August 2023, the witness foils were changed roughly every month with the exact deployment dates listed in Table S1-3. The total amino acids (ng/cm$^2$) were consistently well below the mission threshold of <200 ng/cm$^2$ (Fig. S1-6). The total amino acid values were slightly elevated over the previous values for the period between 17 May 2023 and 10 August 2023, which was a period of intense outfitting, rehearsals, in situ glovebox cleaning, and equipment testing in the cleanroom prior to sample arrival. Witness foils that could be analyzed for amino acids have been deployed during preliminary examination and continued curation operations, albeit for longer deployments as the curation team balanced schedule pressures.

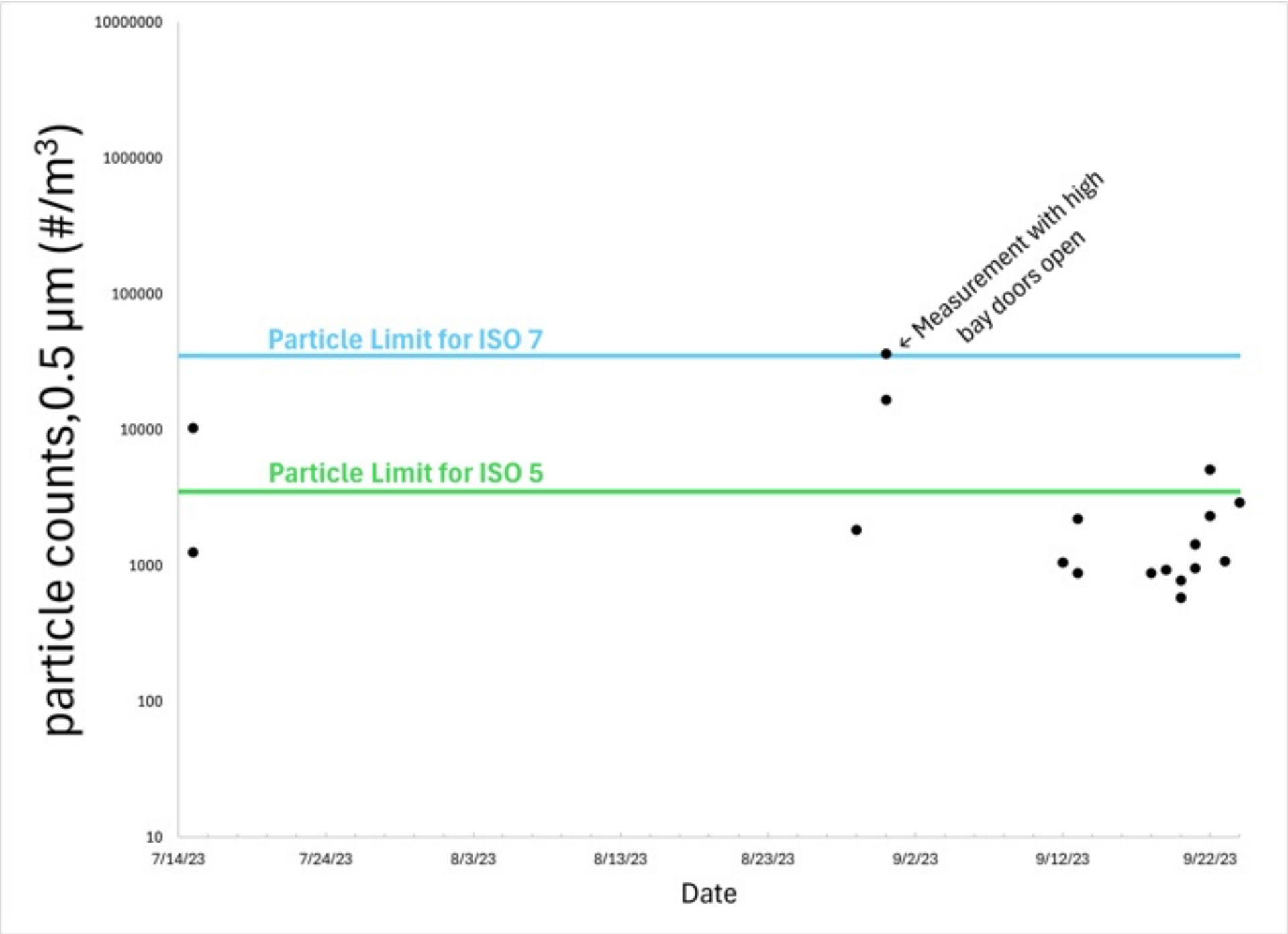

**Fig. S1-1** Particle counts in the temporary cleanroom at UTTR prior to SRC landing. Particle counts were collected when the curation team was onsite at UTTR for rehearsals (July and August 2023) and for pre-landing cleaning and preparations (September 2023).

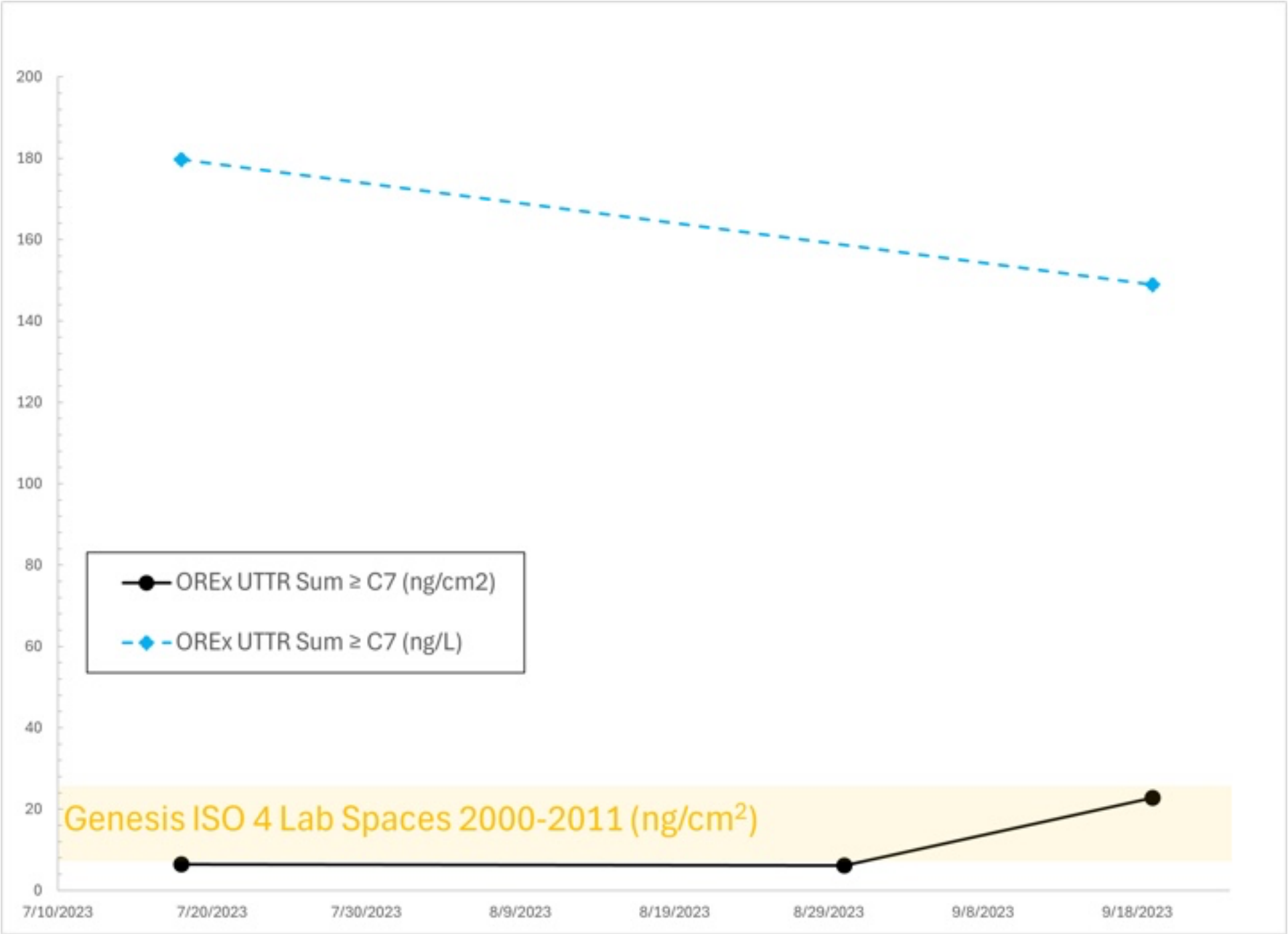

**Fig. S1-2** UTTR temporary cleanroom organic monitoring of sum $\geq C_7$ via Balazs silicon wafer (ng/cm$^2$) and Balazs air sampler (ng/L). The yellow field covers the range of values sum $\geq C_7$ (ng/cm$^2$) determined for the Genesis ISO 4 curation lab spaces between the years 2000 and 2011 reported by Righter et al. (2023). The data in this figure are listed in Table S1-1.

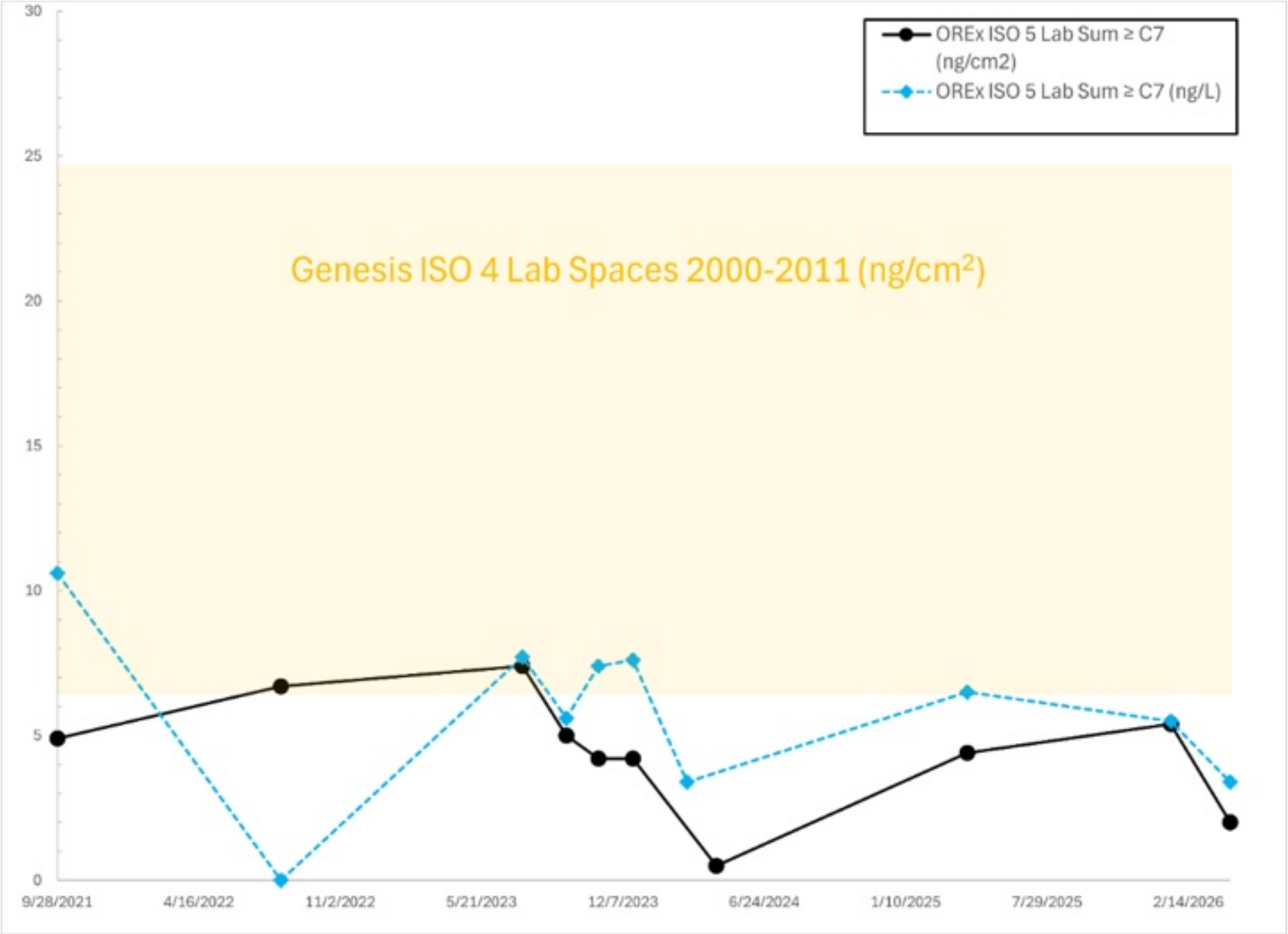

**Fig. S1-3** OSIRIS-REx cleanroom organic monitoring of sum $\geq C_7$ via Balazs silicon wafer (ng/cm$^2$) and Balazs air sampler (ng/L). The yellow field covers the range of values sum $\geq C_7$ (ng/cm$^2$) determined for the Genesis ISO 4 curation lab spaces between the years 2000 and 2011 reported by Righter et al. (2023). The data in this figure are listed in Table S1-2.

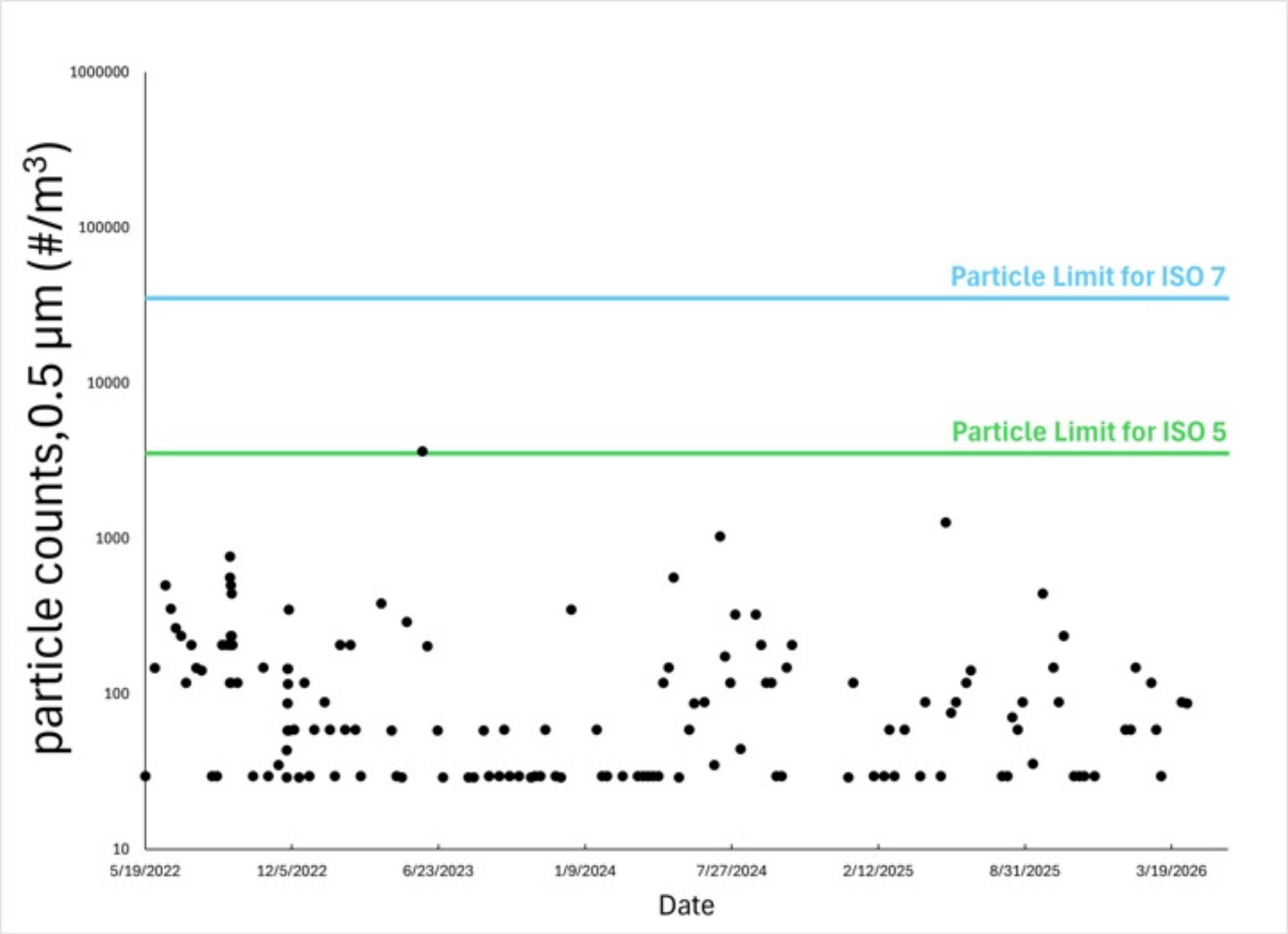

**Fig. S1-4** OSIRIS-REx ISO 5 cleanroom at JSC particle counts from 2021-2026. These particle counts (black circles) include data points from 2022 that were previously reported by Righter et al. (2023). The particle count value slightly above the ISO 5 limit is from 1 June 2023 during a period of intense outfitting in the lab to prepare for mission milestone Curation Rehearsal #3, which happened the following week.

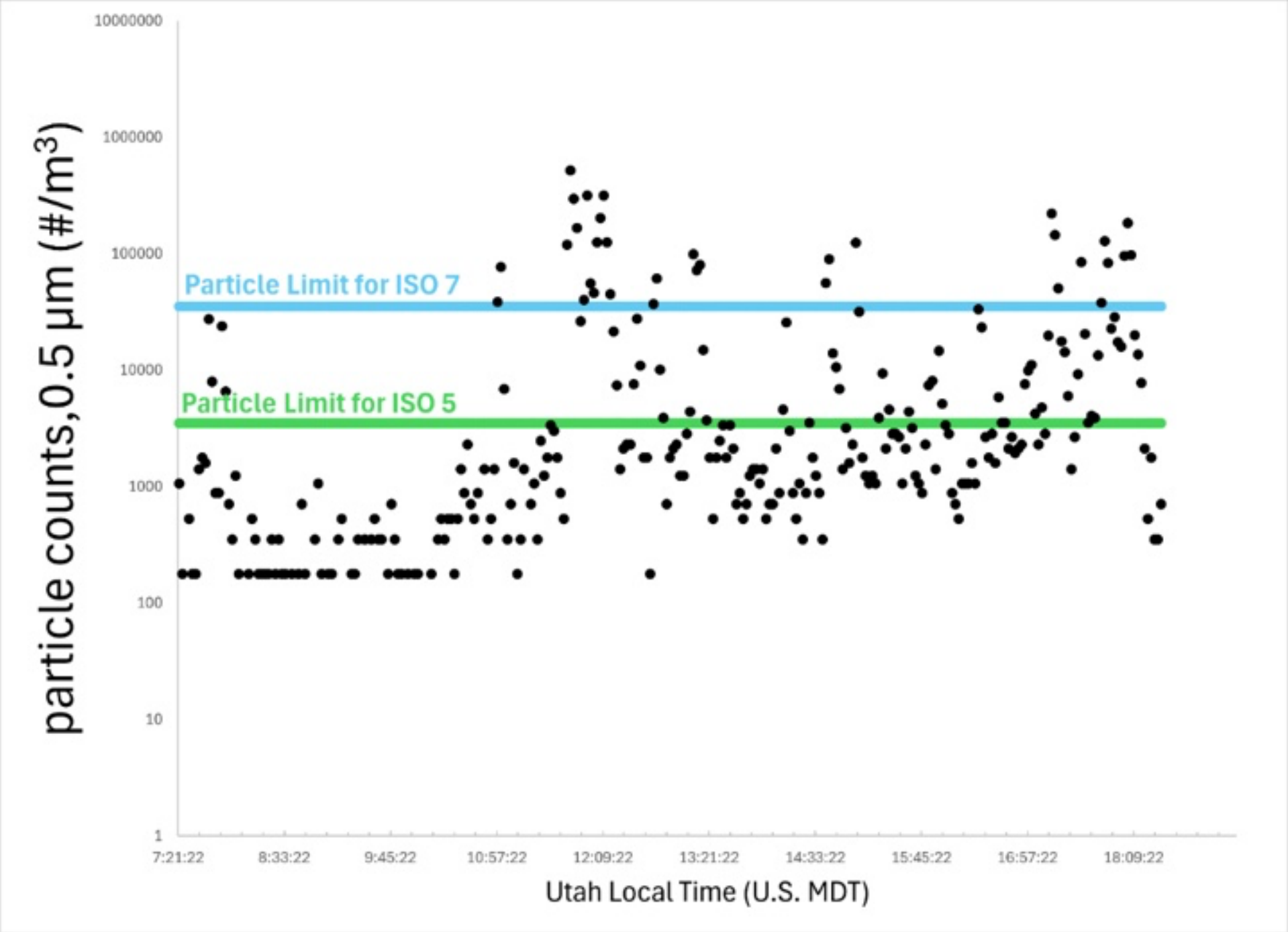

**Fig. S1-5** UTTR temporary cleanroom (ISO 7) back-to-back particle counts on 24 September 2023 during SRC disassembly activities.

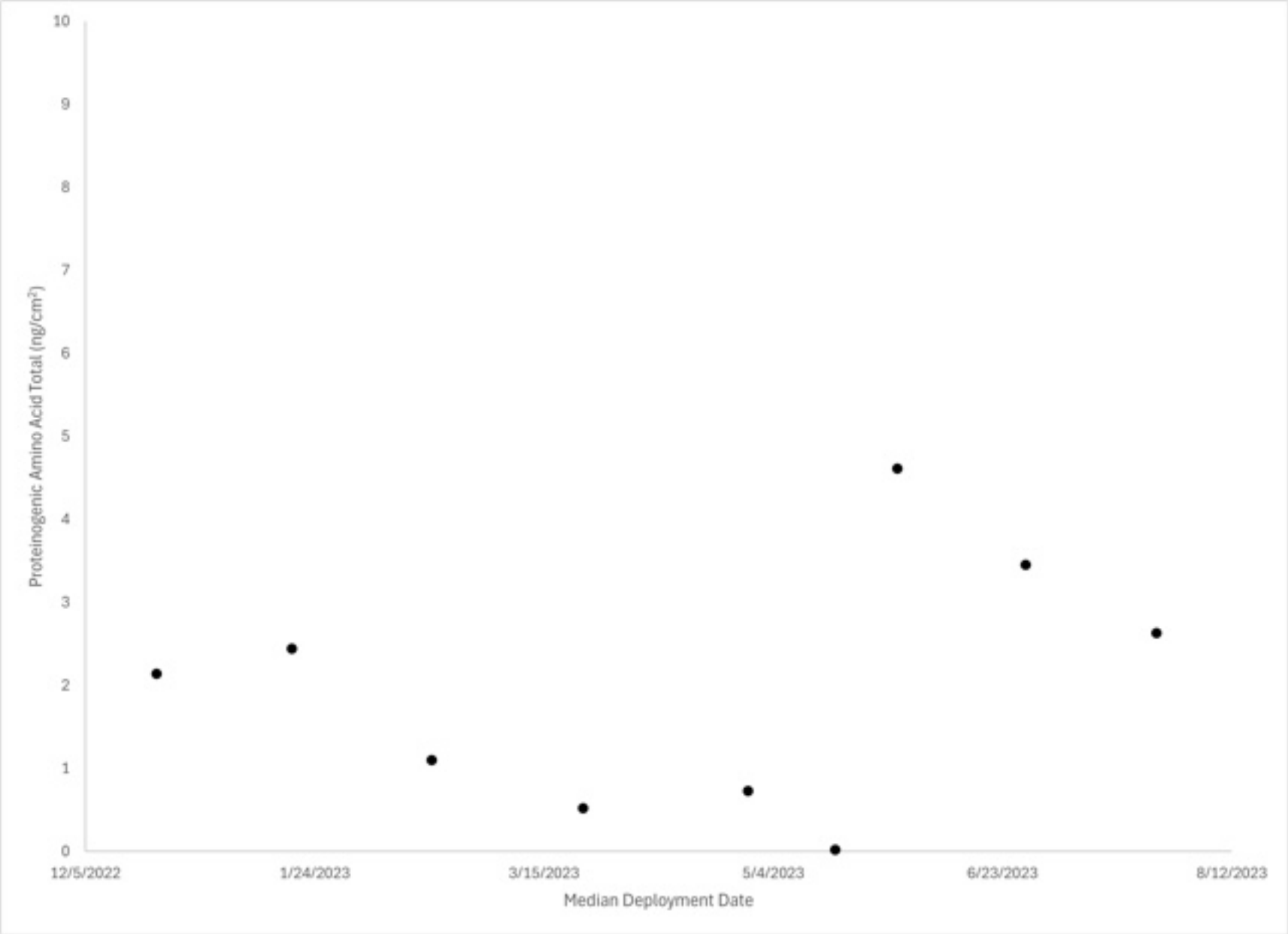

**Fig. S1-6** Total proteinogenic amino acids measured on aluminum witness foils deployed in the ISO 5 OSIRIS-REx cleanroom. The witness plates in the above were deployed between 28-40 days and are represented by the median date within their deployment period. All values are well below the mission limit of >200 ng/cm$^2$. The data set is included in Table S1-3.

**Table S1-1** UTTR temporary cleanroom organic monitoring of sum $\geq C_7$ via Balazs silicon wafer (ng/cm$^2$) and Balazs air sampler (ng/L).

| | OSIRIS-REx UTTR Sum $\geq C_7$ (ng/cm$^2$) | OSIRIS-REx UTTR Sum $\geq C_7$ (ng/L) |
|---|---|---|
| 7/18/2023 | 6.4 | 179.7 |
| 8/30/2023 | 6.1 | n/a |
| 9/19/2023 | 22.8 | 148.9 |

**Table S1-2** JSC OSIRIS-REx ISO cleanroom organic monitoring of sum $\geq C_7$ via Balazs silicon wafer (ng/cm$^2$) and Balazs air sampler (ng/L). These two types of monitoring were typically started at the same time, with the exceptions of 7 March 2024 and 17 April 2024 because the curation team's bandwidth was heavily oversubscribed and the kits were ordered separately.

| Start Date | OSIRIS-REx ISO 5 Lab Sum $\geq C_7$ (ng/cm$^2$) | OSIRIS-REx ISO 5 Lab Sum $\geq C_7$ (ng/L) |
|---|---|---|
| 9/28/2021 | 4.9 | 10.6 |
| 8/10/2022 | 6.7 | 0 |
| 7/18/2023 | 7.4 | 7.7 |
| 9/18/2023 | 5 | 5.6 |
| 11/2/2023 | 4.2 | 7.4 |
| 12/21/2023 | 4.2 | 7.6 |
| 3/7/2024 | n/a | 3.4 |
| 4/17/2024 | 0.5 | n/a |
| 4/7/2025 | 4.4 | 6.5 |
| 1/20/2026 | 5.4 | 5.5 |
| 4/14/2026 | 2 | 3.4 |

**Table S1-3** Amino Acid Monitoring in the OSIRIS-REx ISO 5 Cleanroom

| Deployment start | | 7/12/2023 | 6/14/2023 | 5/17/2023 | 4/10/2023 | 3/6/2023 | 2/3/2023 | 1/4/2023 | 12/6/2022 | 10/27/2022 |
|---|---|---|---|---|---|---|---|---|---|---|
| Deployment end | | 8/10/2023 | 7/12/2023 | 6/14/2023 | 5/17/2023 | 4/10/2023 | 3/6/2023 | 2/3/2023 | 1/4/2023 | 12/6/2023 |
| | #days | 29 | 28 | 28 | 37 | 35 | 31 | 30 | 30 | 40 |
| Gly | avg. | 0.16 | 0 | 0 | 0 | 0 | 0 | 0 | 0.06 | 0 |
| | S.D. | | | 0 | 0 | 0 | 0 | 0 | 0.01 | 0 |
| β-Ala | avg. | 0.02 | 0.1 | 0 | 0 | 0 | 0 | 0 | 0.01 | 0 |
| | S.D. | | | 0 | 0 | 0 | 0 | 0 | 0.02 | 0 |
| Ala | avg. | 0.07 | 0.47 | 0.19 | 0.05 | 0.02 | 0.06 | 0.16 | 0.16 | 0 |
| | S.D. | | | 0.02 | 0 | 0 | 0 | 0.02 | 0.01 | 0 |
| γ-ABA | avg. | 0.03 | 0.22 | 0.12 | 0.15 | 0.15 | 0.11 | 0.16 | 0.18 | 0 |
| | S.D. | | | 0.01 | 0 | 0.01 | 0.01 | 0.01 | 0.01 | 0 |
| α-AIB | avg. | 0.03 | 0.3 | 0.1 | 0.12 | 0.04 | 0.13 | 0.17 | 0.1 | 0 |
| | S.D. | | | 0.01 | 0.01 | 0.01 | 0.03 | 0.01 | 0.01 | 0 |
| Ser | avg. | 0.04 | 0.04 | 0.15 | 0.03 | 0 | 0.1 | 0.27 | 0.19 | 0 |
| | S.D. | | | 0.01 | 0.01 | 0 | 0.13 | 0.01 | 0.01 | 0 |
| Pro | avg. | 0.03 | 0.05 | 1.94 | 0.07 | 0.21 | 0.05 | 0.13 | 0.13 | 0 |
| | S.D. | | | 0.09 | 0 | 0.02 | 0.01 | 0 | 0.01 | 0 |
| 5-APA | avg. | 0 | 0 | 0.02 | 0 | 0 | 0 | 0 | 0 | 0 |
| | S.D. | | | 0.02 | 0 | 0 | 0 | 0 | 0 | 0 |
| Iva | avg. | 0 | 0 | 0 | 0 | 0 | 0 | 0 | 0 | 0 |
| | S.D. | | | 0 | 0 | 0 | 0 | 0 | 0 | 0 |
| Val | avg. | 0 | 0 | 0.19 | 0.08 | 0.03 | 0.08 | 0.19 | 0.18 | 0 |
| | S.D. | | | 0.01 | 0.01 | 0.01 | 0 | 0.01 | 0.01 | 0 |
| Thr | avg. | 0.02 | 0.03 | 0.03 | 0.02 | 0 | 0.06 | 0.12 | 0.09 | 0 |
| | S.D. | | | 0 | 0.01 | 0 | 0.01 | 0.02 | 0 | 0 |
| Cys | avg. | 0 | 0 | 0 | 0 | 0 | 0 | 0 | 0 | 0 |
| | S.D. | | | 0 | 0 | 0 | 0 | 0 | 0 | 0 |
| Ile | avg. | 1.81 | 1.63 | 0.19 | 0.06 | 0.02 | 0.08 | 0.14 | 0.16 | 0 |
| | S.D. | | | 0.01 | 0.01 | 0.01 | 0.01 | 0.01 | 0.01 | 0 |
| Leu | avg. | Co-eluted | Co-eluted | 0.28 | 0.1 | 0.05 | 0.11 | 0.24 | 0.24 | 0 |
| | S.D. | w/Ile | w/Ile | 0.02 | 0.01 | 0 | 0.01 | 0.01 | 0.02 | 0 |
| Asn | avg. | 0 | 0 | 0 | 0 | 0 | 0 | 0 | 0 | 0 |
| | S.D. | | | 0 | 0 | 0 | 0 | 0 | 0 | 0 |
| Asp | avg. | 0.29 | 0 | 0.06 | 0 | 0 | 0.01 | 0.17 | 0.1 | 0 |
| | S.D. | | | 0.06 | 0 | 0 | 0.06 | 0.06 | 0.05 | 0 |
| Gln | avg. | 0 | 0 | 0 | 0 | 0 | 0 | 0 | 0 | 0 |
| | S.D. | | | 0 | 0 | 0 | 0 | 0 | 0 | 0 |
| Lys | avg. | 0.03 | 0.04 | 0.38 | 0 | 0 | 0.05 | 0.16 | 0.12 | 0 |
| | S.D. | | | 0.02 | 0 | 0 | 0.02 | 0 | 0.04 | 0 |
| Glu | avg. | 0.09 | 0.43 | 0.69 | 0.05 | 0 | 0.23 | 0.44 | 0.34 | 0 |
| | S.D. | | | 0.07 | 0 | 0 | 0.02 | 0.02 | 0.05 | 0 |
| Met | avg. | 0 | 0 | 0 | 0 | 0 | 0 | 0 | 0 | 0 |
| | S.D. | | | 0 | 0 | 0 | 0 | 0 | 0 | 0 |
| His | avg. | 0 | 0 | 0 | 0 | 0 | 0 | 0 | 0 | 0 |
| | S.D. | | | 0 | 0 | 0 | 0 | 0 | 0 | 0 |
| Phe | avg. | 0.02 | 0.13 | 0.05 | 0.01 | 0 | 0.02 | 0.09 | 0.08 | 0 |
| | S.D. | | | 0 | 0 | 0 | 0 | 0 | 0.01 | 0 |
| Arg | avg. | 0 | 0 | 0.22 | 0 | 0 | 0 | 0 | 0 | 0 |
| | S.D. | | | 0.05 | 0 | 0 | 0 | 0 | 0 | 0 |
| Tyr | avg. | 0 | 0 | 0 | 0 | 0 | 0 | 0 | 0 | 0 |
| | S.D. | | | 0 | 0 | 0 | 0 | 0 | 0 | 0 |
| Trp | avg. | 0 | 0 | 0 | 0 | 0 | 0 | 0 | 0 | 0 |
| | S.D. | | | 0 | 0 | 0 | 0 | 0 | 0 | 0 |
| Total ($ng/cm^2$) | | 2.63 | 3.45 | 4.61 | 0.73 | 0.52 | 1.1 | 2.44 | 2.14 | 0.02 |

**Supplement 2:** Archived flight hardware, ground support equipment, and ground witness plates for contamination knowledge.

**Table S2-1** Materials archived for contamination knowledge prior to launch (402 total; after Dworkin et al. 2018).

| Catalog Number | Major Component | Description | Composition | Sample Size |
|---|---|---|---|---|
| **OR-MA-0001** | OTES Instrument | Rotary actuator internal to OTES instrument | Lubricant; Christo-Lube MCG-117 | ~ 1 g |
| **OR-MA-0002** | Spacecraft | Solar array panel structural assy +y | Honeycomb Composite | 10 $cm^2$ |
| **OR-MA-0003** | Spacecraft | Solar array panel structural assy -y | Honeycomb Composite | 10 $cm^2$ |
| **OR-MA-0004** | SRC | Latch claw SRC | Aluminum | 10 $cm^2$ |
| **OR-MA-0005** | SRC | Latch claw SRC | Aluminum | 10 $cm^2$ |
| **OR-MA-0006** | SRC | Hinge toggle link - SRC | Aluminum | 10 $cm^2$ |
| **OR-MA-0007** | SRC | Hinge drive toggle link - SRC | Aluminum | 10 $cm^2$ |
| **OR-MA-0008** | SRC | Toggle link coupon | Aluminum | 10 $cm^2$ |
| **OR-MA-0009** | SRC | Canister control link - SRC | Aluminum | 10 $cm^2$ |
| **OR-MA-0010** | SRC | Bolt, SRC canister/avionics deck hinge assy | Stainless Steel | ~ 1” |
| **OR-MA-0011** | SRC | Washer - SRC sample canister hinge assy | Stainless steel; CRES A286 | ~ 1 cm wide |
| **OR-MA-0012** | SRC | Washer coupon SRC sample canister hinge assy | Stainless steel; CRES A286 | 0.25" |
| **OR-MA-0013** | SRC | Braycote 601 | Lubricant; Perflourinated Polyether | |
| **OR-MA-0014** | SRC | Braycote 601 | Lubricant; Polyether; Perflourinated | |
| **OR-MA-0015** | SRC | SRC filter shell | Aluminum | ~9.5 round x 2.5 cm high |
| **OR-MA-0016** | SRC | SRC filter shell | Aluminum | ~9.5 round x 2.5 cm high |
| **OR-MA-0017** | SRC | SRC filter unit | Aluminum, filter, epoxy | ~9.5 round x 2.5 cm high |
| **OR-MA-0018** | SRC | SRC filter unit | Aluminum, filter, epoxy | ~9.5 round x 2.5 cm high |
| **OR-MA-0019** | SRC | SRC filter unit | Aluminum, filter, epoxy | ~9.5 round x 2.5 cm high |
| **OR-MA-0020** | SRC | SRC filter unit | Aluminum, filter, epoxy | ~9.5 round x 2.5 cm high |
| **OR-MA-0021** | Support Materials | Swab for possible ArcJet testing at Ames | Swab | 2.5 " long |
| **OR-MA-0022** | OCAMS | Colorless First Contact | Polymer; cleaning solution | 31 mm pellet cup |
| **OR-MA-0023** | OCAMS | Dow Corning DC 93-500 | Epoxy - elastomer adhesive | 31 mm pellet cup |
| **OR-MA-0024** | OCAMS | Nusil SCV1-2599 | Epoxy - silicone adhesive | 31 mm pellet cup (half) |
| **OR-MA-0025** | OCAMS | Braycote 602EF | Lubricant - Braycote 602EF | 3 ml syringe |
| **OR-MA-0026** | OCAMS | Nyebar | Lubricant - Nyebar fluorocarbon "epoxy" | 60 ml bottle |
| **OR-MA-0027** | OCAMS | AEROGLAZE Z307 | Paint; black paint | nine 2" x 2" squares |
| **OR-MA-0028** | OCAMS | Witness plates | Glass witness plates | 4" dia. metal can |
| **OR-MA-0029** | OCAMS | Epoxy used on witnesses | Epoxy - Scotchweld 2216B/A | in Aluminum pan |
| **OR-MA-0030** | OCAMS | Alion white paint | Paint; Z93C55 thermal control coating | nine 2" x 2" squares |
| **OR-MA-0031** | SRC | Braided shielding coupon; sample canister hinge assembly | Stainless steel; 316L; LM-DB10911B-0025 | ~ 10” |
| **OR-MA-0032** | TAGSAM | Outer pogo tube coupon; sampler arm pogo subassembly | Aluminum 6061-T6 | ~ 2 cm long section |
| **OR-MA-0033** | TAGSAM | Wrist motor fitting assembly coupon; wrist drive assembly | Aluminum 6061-T652 | 10 $cm^2$ |
| **OR-MA-0034** | SRC | EMI gasket; sample canister hinge assembly | Beryllium copper | ~ 6" long, 0.25" diam. |
| **OR-MA-0035** | TAGSAM | Driven shaft coupon; wrist drive assembly | Stainless steel; A386-CRES | ~ 3.5 cm long |
| **OR-MA-0036** | SRC | Avionics deck connector bracket coupon | Aluminum | 10 $cm^2$ |
| **OR-MA-0037** | TAGSAM | Driven gear key (-022); wrist drive assembly | Stainless steel; A386-CRES | ~ 3.125 cm long |
| **OR-MA-0038** | TAGSAM | Washer coupon; wrist drive CRES | Stainless steel; CRES-NAS620C4 | No.4, 0.032" thick |
| **OR-MA-0039** | TAGSAM | Set screw coupon; wrist drive assembly | Steel-AN565FC428H8 | ~ 1 cm long |
| **OR-MA-0040** | SRC | Tray assembly (-107); capture ring assembly | Aluminum alloy; 7050-T7451 | 10 $cm^2$ |
| **OR-MA-0041** | SRC | Tang bracket assembly (-104); tang assembly | Aluminum | 10 $cm^2$ |
| **OR-MA-0042** | SRC | Spring mandrel (-109); deployable witness plate assembly SRC | Aluminum | 10 $cm^2$ |
| **OR-MA-0043** | SRC | LH spring (-111); deployable witness plate assembly SRC | Stainless steel | ~ 1" long |
| **OR-MA-0044** | TAGSAM | Wrist end fitting assembly (-504); wrist drive assembly | Aluminum 6061-T651 | 10 $cm^2$ |
| **OR-MA-0045** | SRC | Retainer clip deployed (-107); sample canister top assembly | Stainless steel | 10 $cm^2$ |
| **OR-MA-0046** | TAGSAM | Washer coupon; wrist drive assembly | Stainless steel; CRES - NAS620C4 | 0.209" OD, 0.032" thick |
| **OR-MA-0047** | TAGSAM | Screw coupon; wrist drive assembly | Stainless steel; A286-CRES, NAS1352NO4-8 | ~1.5 cm long |
| **OR-MA-0048** | TAGSAM | Washer coupon; wrist drive assembly | Stainless steel; 300 CRES - NAS620CAL | ~0.5 cm wide |
| **OR-MA-0049** | TAGSAM | Nut coupon; wrist drive assembly | Stainless steel; CRES | ~ 1 cm top diameter |
| **OR-MA-0050** | SRC | Locking nut coupon; SRC hinge assembly | Stainless steel; CRES, MS21043-3 | 0.1900" OD |
| **OR-MA-0051** | TAGSAM | Screw coupon; wrist assembly; CRES-NAS1352N04-4 | Stainless steel; CRES-NAS1352N04-4 | 0.8 cm long |
| **OR-MA-0052** | SRC | SRC washer coupon; SRC hinge assembly | Stainless steel; A286 CRES NAS1149E0363R | 0.438" OD |
| **OR-MA-0053** | TAGSAM | Driven shaft spacer coupon (-014); wrist drive assembly | Stainless steel; A386-CRES | ~ 2.5 cm long, ~ 1 cm diameter |
| **OR-MA-0054** | TAGSAM | Wrist end fitting key coupon (-021); wrist drive assembly | Stainless steel; A386-CRES | ~ 8 cm long |
| **OR-MA-0055** | TAGSAM | Detent pin coupon (-006); wrist drive assembly | Stainless steel; A386-CRES | ~ 2 cm long |
| **OR-MA-0056** | Support Materials | AcrJet test piece S-01 | PICA | flakes |
| **OR-MA-0057** | Support Materials | ArcJet test piece S-02 | PICA | flakes |
| **OR-MA-0058** | Support Materials | ArcJet test piece S-03 | PICA | flakes |
| **OR-MA-0059** | SRC | Washer coupon, SRC top assembly | Stainless steel; 300 CRES - NAS620C8L | ~0.5 cm OD |
| **OR-MA-0060** | SRC | SRC canister coupon (-502), sample canister hinge assembly | Aluminum alloy; 7075-T7351 | ~13 cm x 3.5 cm |
| **OR-MA-0061** | TAGSAM | Driven Gear coupon (-006), wrist drive assembly | Stainless steel; 17-4 PH CRES | ~ 4 cm diameter |
| **OR-MA-0062** | TAGSAM | Microswitch bracket coupon (-105), capture ring assembly | Aluminum | 10 $cm^2$ |
| **OR-MA-0063** | TAGSAM | RH Torsion Spring, capture ring assembly | Stainless steel | ~ 2" long |
| **OR-MA-0064** | SRC | Pin Coupon (-103), deployable witness plate assembly SRC | Stainless steel | ~ 12 cm long |
| **OR-MA-0065** | SRC | Witness Plate Bracket Coupon (-106), sample canister top assembly | Aluminum | 10 $cm^2$ |
| **OR-MA-0066** | SRC | RH Spring coupon (-112), Deployable Witness Plate Assembly SRC | Stainless steel | ~ 1" long |
| **OR-MA-0067** | TAGSAM | SARA Panel Structure Coupon | Honeycomb Composite | 5 cm x 5 cm x 3.5 cm |
| **OR-MA-0068** | SRC | Cap Screw, SRC Top Assembly | Stainless steel; CRES, NAS1352N08-8 | ~1.5 cm long |
| **OR-MA-0069** | TAGSAM | Hard Stop Coupon (-104), Capture Ring Assembly | Stainless steel; A286 | ~12 cm long |
| **OR-MA-0070** | SRC | Screw coupon, SRC Top Assembly | Stainless steel; CRES | ~ 1 cm long |
| **OR-MA-0071** | TAGSAM | Cam Coupon (-102), Capture Ring Assembly | Stainless steel | 10 $cm^2$ |
| **OR-MA-0076** | SRC | SARA/SRC TVAC residue from foil witness plates - Control Witness | NVR; residue | in 3" aluminum dish |
| **OR-MA-0077** | SRC | SARA/SRC TVAC residue from foil witness plates - Front TQCM Witness | NVR; residue | in 3" aluminum dish |
| **OR-MA-0078** | SRC | SARA/SRC TVAC residue from foil witness plates - Back of Chamber Witness | NVR; residue | in 3" aluminum dish |
| **OR-MA-0079** | SRC | SARA/SRC TVAC residue from foil witness plates - Below SRC Witness | NVR; residue | in 3" aluminum dish |
| **OR-MA-0080** | SRC | SARA/SRC TVAC residue from foil witness plates - Side TQCM Witness | NVR; residue | in 3" aluminum dish |
| **OR-MA-0081** | SRC | SARA/SRC TVAC residue from foil witness plates - Scavenger Plate Witness | NVR; residue | in 3" aluminum dish |
| **OR-MA-0082** | SRC | Heater, SRC avionics deck (opposite of canister side) | Acrylic adhesive tape | ~8 " plus cable |
| **OR-MA-0083** | SRC | Avionics Deck Connector bracket coupon (-101) Sample canister top assembly | Aluminum 7050-T7451 | 10 $cm^2$ |
| **OR-MA-0084** | Spacecraft | 3M 471 tape, spacecraft structure | Vinyl; blue vinyl; rubber adhesive | 1.5" x 1" |
| **OR-MA-0085** | SRC | SRC shim coupon (-108) | Aluminum foil | 3 cm x 2 cm |
| **OR-MA-0086** | TAGSAM | TAGSAM sapphire witness plate, launch coupon (-111) | Sapphire | 1" diameter |
| **OR-MA-0087** | TAGSAM | Frangibolt catch can coupon (-007) TAGSAM top assembly | Aluminum 6061-T651 | 10 $cm^2$ |
| **OR-MA-0088** | TAGSAM | Gear cover coupon (-011), wrist drive assembly | Aluminum 6061-T651 | 10 $cm^2$ |
| **OR-MA-0089** | TAGSAM | Nut coupon, capture ring assembly | Stainless steel; A286 CRES, NAS509-4C | ~ 1 cm diameter |
| **OR-MA-0090** | TAGSAM | Cam pin coupon (-103), capture ring assembly | Stainless steel; A286 CRES | ~ 10 cm long |
| **OR-MA-0091** | SRC | Washer coupon, sample canister top assembly | Stainless steel; CRES steel, NAS620C10L | 0.354" OD |
| **OR-MA-0092** | SRC | Roll pin coupon, deployable witness plate assembly | Stainless steel; CRES; MS16562-201 | ~ 1 cm long |
| **OR-MA-0093** | SRC | Stow guide bracket assembly right (-102), sample canister top assembly | Aluminum; 7050-T7451 | 10 $cm^2$ |
| **OR-MA-0094** | TAGSAM | Self-locking nut coupon, capture ring assembly | Stainless steel; CRES, NAS1291C02M | 0.2 cm |
| **OR-MA-0095** | SRC | SRC wire mesh screen coupon (-107) | Stainless steel | 3 cm x 2 cm |
| **OR-MA-0096** | SRC | Tang coupon (-105), Tang assembly | Titanium 6AL-4V | 10 $cm^2$ |
| **OR-MA-0097** | TAGSAM | TAGSAM sapphire coupon (-115) | Sapphire | 1 " diameter |
| **OR-MA-0098** | SRC | Witness plate cap screws coupon, SRC top assembly | Stainless Steel; heat-resistant steel, NAS1352N08-10 | ~ 2 cm long |
| **OR-MA-0099** | SRC | Cap screw coupons, capture ring assembly | Stainless Steel; CRES NAS1352N02-10 | ~2 cm long |
| **OR-MA-0100** | SRC | Cap screw coupons | Stainless Steel; CRES NAS1352N02-10 | ~ 2 cm long |
| **OR-MA-0101** | SRC | SRC screw, capture ring assembly | Stainless Steel; A286 CRES | ~0.4" length |
| **OR-MA-0102** | Support Materials | Brass fastener from GCKP stand | Brass | ~ 1.5" |
| **OR-MA-0103** | Support Materials | Si wafer blanks | Silicon wafer | ~ 1 cm x 1 cm |
| **OR-MA-0104** | Support Materials | Si wafer blanks from cleaning | Silicon wafer | ~ 1 cm x 1 cm |
| **OR-MA-0105** | Support Materials | Si wafers | Silicon wafer | ~ 1 cm x 1 cm |
| **OR-MA-0106** | Support Materials | Tape lift from TAGSAM Qual head | Tape Lift from TAGSAM Qual head | |
| **OR-MA-0107** | TAGSAM | Flight TAGSAM Head TVAC residue | NVR; chamber pre-cert - blank | 2" round Al pan |
| **OR-MA-0108** | TAGSAM | Flight TAGSAM Head TVAC residue | NVR; chamber pre-cert | 2" round Al pan |
| **OR-MA-0109** | TAGSAM | Flight TAGSAM Head TVAC residue | NVR; TAGSAM bakeout - blank | 2" round Al pan |
| **OR-MA-0110** | TAGSAM | Flight TAGSAM Head TVAC residue | NVR; TAGSAM bakeout | 2" round Al pan |
| **OR-MA-0111** | TAGSAM | Washer coupon | Stainless steel; CRES NAS620C8L | 5/16" diam |
| **OR-MA-0112** | SRC | Sapphire coupon, pre-open | Sapphire | 1.2" x 0.8" |
| **OR-MA-0113** | Support Materials | Polyester lacing tape | Polyester | ~ 4" long |
| **OR-MA-0114** | Support Materials | Ultratape 1110 (Orange glove tape) | Polyethylene; tape, acrylic adhesive | 1.5 x 3 " |
| **OR-MA-0115** | TAGSAM | TAGSAM rivet, POGO arm assembly | Stainless Steel; A286 CRES | 0.388" |
| **OR-MA-0116** | Launch Container | Launch container base coupon | Aluminum 6061-T651; A286 stainless steel | 10 $cm^2$ |
| **OR-MA-0117** | SRC | Washer, capture ring assembly | Stainless Steel; CRES; NAS1149EN232R | 0.25" diam. |
| **OR-MA-0118** | SRC | SRC avionics deck connector bracket | Aluminum 7050-T7451 | 10 $cm^2$ |
| **OR-MA-0119** | SRC | SRC aluminum witness plate coupon post-open | Aluminum | 1.2" x 0.8 " |
| **OR-MA-0120** | Support Materials | Kimtech nitrile glove NxT | Nitrile butadiene rubber | ~ 8" long |
| **OR-MA-0121** | Support Materials | TX1009 AlphaWipe | Polyester 100% | 9" x 9" |
| **OR-MA-0122** | Support Materials | Shipping container desiccant DESI-PAK | Tyvek packet; bentonite clay | 8" x 5" bag |
| **OR-MA-0123** | REXIS | TiNi switch washer | Ti Al-4V, 302SS, Ag/Cu wire, Tetzel | 2 7/8" x 5/8" |
| **OR-MA-0124** | REXIS | Frangibolt Ti bolt | Titanium 6Al-4V | 1.33" x 0.183" |
| **OR-MA-0125** | REXIS | Frangibolt Housing Iridite | Iridite [Henkel Alodine 1132 Touch-n-prep pen] | 4 1/8" x 7/8" |
| **OR-MA-0126** | REXIS | Torlon standoff | Polymer - Torlon 5030 | 1.058" x 0.394" |
| **OR-MA-0127** | OLA | Solithane 113 | Epoxy - urethane pre-polymer | 1/2 of 3" pan |
| **OR-MA-0128** | OLA | CV-2646 | Polymer - electrically conductive RTV silicone | 1/2 of 3" pan |
| **OR-MA-0129** | OLA | CV-2646 | Polymer - electrically conductive RTV silicone | 1/2 of 3" pan |
| **OR-MA-0130** | OLA | CV-2646 | Polymer - electrically conductive RTV silicone | 1/2 of 3" pan |
| **OR-MA-0131** | OLA | Hysol EA 934NA | Epoxy adhesive | 1/2 of 3" pan |
| **OR-MA-0132** | OLA | Hysol EA 934NA | Epoxy adhesive | 1/2 of 3" pan |
| **OR-MA-0133** | OLA | Solithane 113 | Epoxy - urethane pre-polymer | 1/2 of 3" pan |
| **OR-MA-0134** | OLA | Solithane 113 | Epoxy - urethane pre-polymer | 1/2 of 3" pan |
| **OR-MA-0135** | OLA | Solithane 113 | Epoxy - urethane pre-polymer | 1/2 of 3" pan |
| **OR-MA-0136** | OLA | Solithane 113 | Epoxy - urethane pre-polymer | 1/2 of 3" pan |
| **OR-MA-0137** | OLA | Solithane 113 | Epoxy - urethane pre-polymer | 1/2 of 3" pan |
| **OR-MA-0138** | OLA | Solithane 113 | Epoxy - urethane pre-polymer | 1/2 of 3" pan |
| **OR-MA-0139** | OLA | Solithane 113 | Epoxy - urethane pre-polymer | 1/2 of 3" pan |
| **OR-MA-0140** | OLA | Solithane 113 | Epoxy - urethane pre-polymer | 1/2 of 3" pan |
| **OR-MA-0141** | OLA | Hysol EA 934NA | Epoxy adhesive | 1/2 of 3" pan |
| **OR-MA-0142** | TAGSAM | TAGSAM shim | Aluminum foil | 3 cm dia. |
| **OR-MA-0143** | SRC | SRC sapphire witness plate coupon, post-open | Sapphire | 1.2" x 0.8" |
| **OR-MA-0144** | SRC | SRC aluminum witness plate coupon, always | Aluminum | 1.2" x 0.8" |
| **OR-MA-0145** | SRC | SRC sapphire witness plate coupon, always | Sapphire | 1.2" x 0.8" |
| **OR-MA-0146** | SRC | SRC screw coupon | Stainless Steel; CRES | 0.053" long |
| **OR-MA-0147** | SRC | SRC Cam coupon | Aluminum; 7050-T7451 | 10 $cm^2$ |
| **OR-MA-0148** | SRC | Flap coupon, SRC deployable witness plate assembly | Aluminum | 10 $cm^2$ |
| **OR-MA-0149** | OVIRS | Braycote 601 EF lubricant | Lubricant; Braycote 601 EF | > 10 g |
| **OR-MA-0150** | OVIRS | Z93C55 conductive white paint | Paint; Z93C55 conductive white silicate paint | two 2" x 2" surfaces |
| **OR-MA-0151** | OVIRS | EPOTEK 353 solar mesh epoxy | Epoxy; EPOTEK 353 | ¼" x 3" syringe |
| **OR-MA-0152** | OVIRS | EPOTEK 353ND epoxy - cured 1/4" x 3" syringe | Epoxy; EPOTEK 353ND epoxy - cured | >10 g |
| **OR-MA-0153** | REXIS | Braycote 601EF lubricant | Lubricant; Braycote 601EF | ~ 3 g |
| **OR-MA-0154** | REXIS | Hysol EA 934 NA | Epoxy - Hysol 934 on REXIS Main Electronics Board | ~ 1.5 cm plug |
| **OR-MA-0155** | Support Materials | LF8900C Ziplock bag | Polyethylene with fire retardant additives - Sb2O3 and ethane | 16" x 16" |
| **OR-MA-0156** | Support Materials | TAGSAM simulant | Tagish Lake simulant from A. Hildebrand | 10 g |
| **OR-MA-0157** | TAGSAM | Washer, Sampler Head U-joint Subassembly | Stainless Steel; CRES NAS620C10L | 0.354" OD |
| **OR-MA-0158** | Spacecraft | Solar Cell Coupon, Solar Array | Smart ZTJ material | 3" diameter |
| **OR-MA-0159** | Spacecraft | Hydrazine Dry Down #1 | NVR; hydrazine dry down | |
| **OR-MA-0160** | Spacecraft | Hydrazine Dry Down #2 | NVR; hydrazine dry down #2 | |
| **OR-MA-0161** | Spacecraft | Hydrazine Dry Down #3 | NVR; Hydrazine Dry Down #3 | |
| **OR-MA-0162** | Spacecraft | Hydrazine Dry Down #4 | NVR; Hydrazine Dry Down #4 | |
| **OR-MA-0163** | Spacecraft | Hydrazine Dry Down blank | NVR; Hydrazine Dry Down blank | |
| **OR-MA-0164** | Spacecraft | Hydrazine Dry Down blank | NVR; Hydrazine Dry Down blank | |
| **OR-MA-0165** | Spacecraft | Hydrazine Dry Down unused vial | NVR; Hydrazine Dry Down unused Teflon vial | |
| **OR-MA-0166** | OCAMS | OCAMS CDA motor | Motor for all 3 OCAMS imagers | 3"x2"x1" |
| **OR-MA-0167** | Support Materials | End cap nylon bag from Lufkin AC | Nylon | 8" x 8" |
| **OR-MA-0168** | SRC | Manniglas MLI SRC | MLI blanket | 6" x 6" + braid |
| **OR-MA-0169** | TAGSAM | Reinforced black Kapton blanket | Kapton, adhesive, Kapton tape | 8" x 8" |
| **OR-MA-0170** | TAGSAM | Diffuse PTFE film | PTFE film, diffuse | 8" x 8" |
| **OR-MA-0171** | SRC | Avionics deck coupon | Honeycomb, aluminum + face sheet | ~ 5" long |
| **OR-MA-0172** | SRC | Batt blanket coupon | Kapton, braid, solder | 6" x 6" |
| **OR-MA-0173** | OLA | Black aluminum foil | Aluminum foil, flat black epoxy | ~ 1 ft2 |
| **OR-MA-0174** | SRC | Thermocouple coupon, SARA installation | Copper- constantan wire; PFA insulation | ~12" long |
| **OR-MA-0175** | Spacecraft | RTV adhesive - solar arrays LM-BF10898A-0100 | Epoxy - adhesive, silicone RTV | 1.5" Al pan |
| **OR-MA-0176** | Support Materials | Braycote TAGSAM gas delivery system | Lubricant; Braycote 601EF | ~ 1 g |
| **OR-MA-0177** | TAGSAM | Flight head build sponge count | Miscellaneous | ~ 20" bag |
| **OR-MA-0178** | Support Materials | Gloves from TAGSAM flight head installation (see note) | Nitrile gloves | ~ 20 " bag |
| **OR-MA-0179** | Support Materials | Swabs from baffle/optical cleaning | Cotton swabs | ~12" bag |
| **OR-MA-0180** | Support Materials | Braycote 601 EF; s/c transport SVIP purge system | Lubricant; Braycote 601 EF | 2 oz tube (empty) |
| **OR-MA-0181** | SRC | Avionics box pressure transducer, insulation; 8530C-15M41-SRC-BK | Miscellaneous; avionics box pressure transducer, trimmed insulation | 12" long pieces |
| **OR-MA-0182** | Support Materials | Contamination event, soldering tool | Glass-filled nylon; soldering tool | ~ 6" long |
| **OR-MA-0183** | SRC | Molybdenum disulfide, MMSN399A100 | Lubricant; Braycote 602EF | ~ 1 g |
| **OR-MA-0184** | Support Materials | Epoxy for fairing doors | Epoxy - CHO-BOND 1030-55 | 1.5 x 1.5 x 1 cm |
| **OR-MA-0185** | Support Materials | Entegris filter | Filter for purge gas on MLP/VIF | ~ 8" x 3" cylinder |
| **OR-MA-0186** | TAGSAM | Convoluted tube - T section | Stainless steel | ~ 12" |
| **OR-MA-0187** | TAGSAM | Convoluted tube - inlet side | Stainless steel | ~ 16" length |
| **OR-MA-0188** | Support Materials | Fluoroclean dry down #1 | NVR; Fluoroclean HE-27152-AEBT | Al pan 2" |
| **OR-MA-0189** | Support Materials | Payload fairing amino acid foil witness plates | Aluminum foil; amino acid witness plates | 12" x 12" |
| **OR-MA-0190** | Support Materials | Payload fairing particle plates filter paper -left | Filter paper | ~2" circles |
| **OR-MA-0191** | Support Materials | Payload fairing particle plates filter paper - right | Filter paper | 2" paper disc |
| **OR-MA-0192** | Support Materials | Fluoroclean dry down #2 | NVR; fluoroclean HE-27152-AEBT | Al pan 2" |
| **OR-MA-0193** | SRC | Epoxy used on sep/spin mechanism repair LM-AE10782A | Epoxy - Scotch Weld EC-2216 | ~ 1 g |
| **OR-MA-0194** | SRC | Epoxy used on SRC avionics deck - connector bracket re-work LM-AE10782A | Epoxy - Scotch Weld EC-2216 | ~ 1 g |
| **OR-MA-0195** | SRC | Epoxy used on avionics deck LM-AE10490A | Epoxy - Henkel Loctite EA9394 | ~ 1 g |
| **OR-MA-0196** | SRC | Epoxy used on avionics deck LM-AE10490A | Epoxy - Henkel Loctite EA9394 | ~ 1 g |
| **OR-MA-0197** | SRC | Epoxy used on avionics deck LM-AE10490A | Epoxy - Henkel Loctite EA9394 | ~ 1 g |
| **OR-MA-0198** | SRC | Epoxy used on avionics deck LM-AE10490A | Epoxy - Henkel Loctite EA9394 | ~ 1 g |
| **OR-MA-0199** | SRC | Epoxy used on avionics deck LM-AE10490A | Epoxy - Henkel Loctite EA9394 | ~ 1 g |
| **OR-MA-0200** | SRC | Epoxy used on avionics deck LM-AE10490A | Epoxy - Henkel Loctite EA9394 | ~ 1 g |
| **OR-MA-0201** | SRC | Epoxy used on avionics deck LM-AE10490A | Epoxy - Henkel Loctite EA9394 | ~ 1 g |
| **OR-MA-0202** | SRC | Epoxy used on avionics deck LM-AE10490A | Epoxy - Henkel Loctite EA9394 | ~ 1 g |
| **OR-MA-0203** | SRC | SRC pushoff fitting re-work; LM-BF10316B | Epoxy; RTV 566 silicone adhesive Momentive | ~ 1 g |
| **OR-MA-0204** | SRC | SRC R&R assembly; LM-BF10316B | Epoxy; RTV 566 silicone adhesive Momentive | ~ 1 g |
| **OR-MA-0205** | SRC | SRC backshell pushoff removed, replaced with 203; LM-BF10316B | Epoxy; RTV 566 silicone adhesive Momentive | ~ 1 g |
| **OR-MA-0206** | SRC | SRC harness install; LM-BF10316B | Epoxy; RTV 566 silicone adhesive Momentive | ~ 1 g |
| **OR-MA-0207** | SRC | SRC backshell parachute seal; LM-BF10316B | Epoxy; RTV 566 silicone adhesive Momentive | ~ 1 g |
| **OR-MA-0208** | SRC | SRC Catch Can install; LM-BF10316B | Epoxy; RTV 566 silicone adhesive Momentive | ~ 1 g |
| **OR-MA-0209** | SRC | SRC backshell inner seal; LM-BF10316B | Epoxy; RTV 566 silicone adhesive Momentive | ~ 1 g |
| **OR-MA-0210** | SRC | SRC backshell outer seal; LM-BF10316B | Epoxy; RTV 566 silicone adhesive Momentive | ~ 1 g |
| **OR-MA-0211** | SRC | SRC backshell vents; LM-BF10316B | Epoxy; RTV 566 silicone adhesive Momentive | ~ 1 g |
| **OR-MA-0212** | SRC | SRC sep/spin thermal test cable; LM-BF10316B | Epoxy; RTV 566 silicone adhesive Momentive | ~ 1 g |
| **OR-MA-0213** | SRC | SRC harness potting; LM-BF10316B | Epoxy; RTV 566 silicone adhesive Momentive | ~ 1 g |
| **OR-MA-0214** | SRC | SRC backshell fitting installation; LM-BF10316B | Epoxy; RTV 566 silicone adhesive Momentive | ~ 1 g |
| **OR-MA-0215** | SRC | SRC backshell seal emulsion; LM-BF10316B | Epoxy; RTV 566 silicone adhesive Momentive | ~ 1 g |
| **OR-MA-0216** | SRC | SRC test seal, providing just in case; LM-BF10316B | Epoxy; RTV 566 silicone adhesive Momentive | ~ 1 g |
| **OR-MA-0217** | SRC | SARA cable fitting to test cable; LM-BF10316B | Epoxy; RTV 566 silicone adhesive Momentive | ~ 1 g |
| **OR-MA-0218** | SRC | Epoxy used on SRC filter assembly LM-AE10490A | Epoxy; Henkel Loctite EA9394 | ~ 1 g |
| **OR-MA-0219** | SRC | Epoxy used on SRC filter assembly LM-AE10490A | Epoxy; Henkel Loctite EA9394 | ~ 1 g |
| **OR-MA-0220** | SRC | Epoxy used on SRC backshell pin plate | Epoxy; Henkel Loctite EA9394 | ~ 1 g |
| **OR-MA-0221** | SRC | Epoxy used on SRC filter assembly LM-AE10490A | Epoxy; Henkel Loctite EA9394 | ~ 1 g |
| **OR-MA-0222** | SRC | Epoxy used on SRC septum bushing LM-AE10490A | Epoxy; Henkel Loctite EA9394 | ~ 1 g |
| **OR-MA-0223** | TAGSAM | Epoxy used on TAGSAM N2 system temp sensors LM-AE10490A | Epoxy; Henkel Loctite EA9394 | ~ 1 g |
| **OR-MA-0224** | SRC | Epoxy used on SRC hinge assembly - spools, shunt insert LM-AE10490A | Epoxy; Henkel Loctite EA9394 | ~ 1 g |
| **OR-MA-0225** | SRC | Epoxy used on SRC filter assembly LM-AE10490A | Epoxy; Henkel Loctite EA9394 | ~ 1 g |
| **OR-MA-0226** | SRC | Epoxy used on SRC stepper motor assembly - heater wire bonding LM-AE10490A | Epoxy; Henkel Loctite EA9394 | ~ 1 g |
| **OR-MA-0227** | SRC | Epoxy used on SRC filter assembly LM-AE10490A | Epoxy; Henkel Loctite EA9394 | ~ 1 g |
| **OR-MA-0228** | SRC | Epoxy used on SRC avionics deck assembly - nut plate install LM-AE10490A | Epoxy; Henkel Loctite EA9394 | ~ 1 g |
| **OR-MA-0229** | SRC | Epoxy used on SRC bathtub fitting (attaches backshell to deck) LM-AE10490A | Epoxy; Henkel Loctite EA9394 | ~ 1 g |
| **OR-MA-0230** | SRC | Epoxy used on avionics deck inserts LM-AE10490A | Epoxy; Henkel Loctite EA9394 | ~ 1 g |
| **OR-MA-0231** | SRC | Epoxy used on Detent insert installation LM-AE10490A | Epoxy; Henkel Loctite EA9394 | ~ 1 g |
| **OR-MA-0232** | SRC | Epoxy used on avionics deck various locations LM-AE10490A | Epoxy; Henkel Loctite EA9394 | ~ 1 g |
| **OR-MA-0233** | SRC | SRC filter unit | Aluminum, filter, epoxy | ~9.5 round x 2.5 cm high |
| **OR-MA-0234** | SRC | Epoxy used on sample canister hinge assembly - new insert installation LM-AE10490A | Epoxy; Henkel Loctite EA9394 | ~ 1 g |
| **OR-MA-0235** | SRC | Epoxy used on sample canister hinge assembly - new insert installation LM-AE10490A | Epoxy; Henkel Loctite EA 9394 | ~ 1 g |
| **OR-MA-0236** | SRC | Epoxy used on SRC filter assembly LM-AE10490A | Epoxy; Henkel Loctite EA9394 | ~ 1 g |
| **OR-MA-0237** | TAGSAM | Epoxy used on TAGSAM N2 gas assembly LM-AE10490A | Epoxy; Henkel Loctite EA9394 | ~ 1 g |
| **OR-MA-0238** | SRC | Epoxy used on SRC lap shear testing LM-AE10490A | Epoxy; Henkel Loctite EA9394 | ~ 1 g |
| **OR-MA-0239** | SRC | Epoxy used in an unknown location | Epoxy; Henkel Loctite EA9394 | ~ 1 g |
| **OR-MA-0240** | SRC | Nusil 2599 LM-AH10314A SRC temp sensor replacement | Epoxy; Nusil 2599 silicone adhesive | ~ 1 g |
| **OR-MA-0241** | SRC | Nusil 2599 LM-AH10314A SRC temp sensor replacement | Epoxy; Nusil 2599 silicone adhesive | ~ 1 g |
| **OR-MA-0242** | SRC | Nusil 2599 LM-AH10314A sep/spin installation - temp sensor | Epoxy; Nusil 2599 silicone adhesive | ~ 1 g |
| **OR-MA-0243** | TAGSAM | Nusil 2599 LM-AH10314A TAGSAM wrist motor thermal assembly | Epoxy; Nusil 2599 silicone adhesive | ~ 1 g |
| **OR-MA-0244** | TAGSAM | Nusil 2599 LM-AH10314A TAGSAM shoulder drive thermal control - temp sensor installation | Epoxy; Nusil 2599 silicone adhesive | ~ 1 g |
| **OR-MA-0245** | SRC | Nusil 2599 LM-AH10314A SRC stepper motor sensor bonding | Epoxy; Nusil 2599 silicone adhesive | ~ 1 g |
| **OR-MA-0246** | TAGSAM | Nusil 2599 LM-AH10314A TAGSAM shoulder drive thermal control - temp sensor install | Epoxy; Nusil 2599 silicone adhesive | ~ 1 g |
| **OR-MA-0247** | TAGSAM | Nusil 2599 LM-AH10314A TAGSAM wrist motor temp sensor install | Epoxy; Nusil 2599 silicone adhesive | ~ 1 g |
| **OR-MA-0248** | SRC | Nusil 2599 LM-AH10314A SRC avionics and battery temp sensor install | Epoxy; Nusil 2599 silicone adhesive | ~ 1 g |
| **OR-MA-0249** | SRC | Nusil 2599 LM-AH10314A sep/spin installation - sep nut thermal install | Epoxy; Nusil 2599 silicone adhesive | ~ 1 g |
| **OR-MA-0250** | Spacecraft | Nusil 2599 LM-AH10314A navcams temp sensor installation | Epoxy; Nusil 2599 silicone adhesive | ~ 1 g |
| **OR-MA-0251** | TAGSAM | Contact pad coupon | Stainless steel Velcro; aluminum | ~ 5/8" dia. |
| **OR-MA-0252** | TAGSAM | Wire mesh screen coupon | Stainless steel | 3 cm |
| **OR-MA-0253** | SRC | NAS1352N08-6; SRC top assembly witness plate screws | Stainless Steel; heat resistant steel; NAS1352N08-6 | ~ 0.375" |
| **OR-MA-0254** | SRC | Razor blade, detent pin cleaning | steel | ~ 1.5" |
| **OR-MA-0255** | SRC | Retainer clip, fixed coupon top assembly witness plate | Stainless steel - CRES | ~ 10 $cm^2$ |
| **OR-MA-0256** | TAGSAM | Braycote 602 EF + Mo disulfide; sine vibe test preparations | Lubricant; Braycote 602 EF (Castrol) + MoS2 (Dow) | ~ 1 g |
| **OR-MA-0257** | Support Materials | Witness plate stand frame | Aluminum 6061-T6 | ~ 12" long |
| **OR-MA-0258** | SRC | Red primer, PR-1200RTV; LM-CJ-10894A-0013 | Silane; used for SRC battery and avionics temp sensors | ~ 1 g |
| **OR-MA-0259** | TAGSAM | Gloves used by Jim Harris for witness plate fit check | Nitrile, butadiene rubber; NxT | 1 large size glove |
| **OR-MA-0260** | TAGSAM | Screw coupon, sampler head subassembly | Stainless steel - CRES; NAS1152E7 | ~ 1.5 cm long |
| **OR-MA-0261** | TAGSAM | Spring coupon, sampler head subassembly | Stainless steel | 0.5" diam. |
| **OR-MA-0262** | SRC | SRC dust cover caplugs | Polyethylene polymer conductive | ~ 1" diam. |
| **OR-MA-0263** | Support Materials | Nitrile glove used in SSL | Nitrile glove blue | 1 glove (M) |
| **OR-MA-0264** | TAGSAM | Washer coupon, sampler head subassembly | Stainless steel - CRES; NAS620C8L | 5/16" diam. |
| **OR-MA-0265** | TAGSAM | Washer coupon, sampler head subassembly | Stainless steel - CRES; NAS620C8 | 0.304" diam. |
| **OR-MA-0266** | TAGSAM | Sampler Head diaphragm coupon | Mylar | 10 $cm^2$ |
| **OR-MA-0267** | TAGSAM | Cap screw coupon, sampler head subassembly | Stainless steel - CRES; NAS1352N08-10 | ~ 0.75" diam. |
| **OR-MA-0268** | TAGSAM | Screw coupon, sampler head subassembly | Stainless steel - CRES; NAS1102E08-10 | 1.5 cm long |
| **OR-MA-0269** | TAGSAM | Washer coupon, TAGSAM, W3-50-TI | Titanium 6AL-4VTI | 0.5" dia. |
| **OR-MA-0270** | TAGSAM | Screw coupon, sampler head subassembly NAS1152E8 | Stainless steel - CRES; NAS1152E8 | 2 cm long |
| **OR-MA-0271** | TAGSAM | Nut coupon, sampler head subassembly | Steel; silver plated; hexagonal nut | 0.75 cm OD |
| **OR-MA-0272** | TAGSAM | Restraint coupon, sampler head sub-assembly | Titanium | 6 cm long |
| **OR-MA-0273** | SRC | Al witness plate, pre-open coupon | Aluminum | 1" x 0.75" |
| **OR-MA-0274** | Support Materials | ESD wrist strap used by ATLO personnel | Metallic, painted | 0.75" long |
| **OR-MA-0275** | SRC | Washer coupon, SRC tang assembly | Stainless steel - CRES; NAS620C2 | 0.149" OD |
| **OR-MA-0276** | Support Materials | Particle fallout plate PFOP unused | Polystyrene; HVWP filter | 47 mm diam. filter |
| **OR-MA-0277** | TAGSAM | Washer coupon, U-joint sub-assembly | Stainless steel - CRES; NAS1149EN632R | 0.375" OD |
| **OR-MA-0278** | Support Materials | Connector used for wire splicing | Glass filled nylon | 1.25" long |
| **OR-MA-0279** | TAGSAM | Screw coupon, sampler head top assembly | Stainless Steel; heat resistant; NAS1351N08-12 | 2.25 cm long |
| **OR-MA-0280** | TAGSAM | Spring spacer coupon, witness plate assembly | Aluminum alloy 6061-T651 | ~ 0.86 g |
| **OR-MA-0281** | SRC | Silicone adhesive hardness coupon; microchip installation | Epoxy - adhesive, silicone RTV-142 | ~ 1 g |
| **OR-MA-0282** | SRC | LH torsion spring coupon, capture ring assembly | Stainless steel | ~ 2 " long |
| **OR-MA-0283** | SRC | LH Spring coupon, SRC top assembly | Stainless steel | ~ 2" long |
| **OR-MA-0284** | SRC | RH Spring coupon, SRC top assembly | Stainless steel | ~ 2" long |
| **OR-MA-0285** | TAGSAM | Detent pin coupon (-006) DUPLICATE; wrist drive assembly | Stainless steel; A386-CRES | ~ 2 cm long |
| **OR-MA-0286** | SRC | Washer coupon DUPLICATE, SRC top assembly | Stainless steel; 300 CRES - NAS620C8L | ~0.5 cm OD |
| **OR-MA-0287** | TAGSAM | Washer coupon, DUPLICATE, TAGSAM, W3-50-TI | Titanium 6AL-4VTI | 0.5" dia. |
| **OR-MA-0288** | TAGSAM | Silicone adhesive LM-AH10314A; Qual shoulder drive assembly | Epoxy; Nusil silicone adhesive | ~ 1 g |
| **OR-MA-0289** | TAGSAM | Nusil silicone adhesive; LM-AH10314A; SCV1-2599; qual N2 gas system temp sensor install | Epoxy; Silicone adhesive | ~ 1 g |
| **OR-MA-0290** | TAGSAM | Nusil silicone adhesive; LM-AH10314A; SCV1-2599; qual wrist drive assembly - non-flight | Epoxy; Silicone adhesive | ~ 1 g |
| **OR-MA-0291** | TAGSAM | Nusil silicone adhesive; LM-AH10314A; SCV1-2599; qual shoulder drive assembly - non-flight | Epoxy; Silicone adhesive | ~ 1 g |
| **OR-MA-0292** | SRC | Backshell coupon | SLA-561V | 3" x 3" |
| **OR-MA-0293** | Launch Container | Epoxy adhesive, sealant for Amphenol connector | Epoxy - UAH 6190 | > 1 g |
| **OR-MA-0294** | Launch Container | Blanket repair tape, payload fairing | Extruded PTFE; silicone adhesive | ~ 2" long |
| **OR-MA-0295** | Spacecraft | Acs thruster assembly nfp30708100 | Valves - Stainless steel 300 series, Thruster Chamber - Inconel, Catalyst - iridium coated aluminum | 8" x 1.75"; Flight mass: 0.36 |
| **OR-MA-0296** | SRC | Tombstone coupon, SRC top assembly | Aluminum alloy 7050-T7451 | 10 $cm^2$ |
| **OR-MA-0297** | TAGSAM | Sampler Head Mass Simulator (aluminum hat) | Aluminum, clear anodized | 3.2 kg |
| **OR-MA-0298** | TAGSAM | Stepper motor, SRC and TAGSAM | Heater, temperature sensor, motor | flight mass 0.39 kg |
| **OR-MA-0299** | SRC | Pica coupon | PICA | unknown, flight mass 9.94 kg |
| **OR-MA-0300** | Support Materials | Unknown foil put in Jamie's cabinet | Unknown | 6" x 6" |
| **OR-MA-0301** | SRC | Washer coupon, capture ring assembly | stainless steel CRES; NAS620C6 | 0.267" o.d. |
| **OR-MA-0302** | TAGSAM | Baseplate coupon, sampler head subassembly | Aluminum 6061-T651 | 10 $cm^2$ |
| **OR-MA-0303** | TAGSAM | Convoluted tube, feedline assembly | Stainless steel | 1 tube |
| **OR-MA-0304** | SRC | Stow guide bracket sample assembly, left | Aluminum / Stainless steel | 10 $cm^2$ |
| **OR-MA-0305** | TAGSAM | Sampler head air tube | Aluminum alloy; 6061-T651 | ~ 2.25" long |
| **OR-MA-0306** | TAGSAM | Sampler head surface plate coupon | Aluminum alloy; 6061-T7451 | 10 $cm^2$ |
| **OR-MA-0307** | TAGSAM | Bolt restraint coupon, sampler head subassembly | Aluminum alloy 6061-T651 | 10 $cm^2$ |
| **OR-MA-0308** | TAGSAM | Fan support plate coupon, sampler head subassembly | Aluminum alloy 6061-T651 | 10 $cm^2$ |
| **OR-MA-0309** | TAGSAM | Standoff coupon, sampler head subassembly | Aluminum 6061-T651 and CRES MS21209C0815 | 5.5 cm long |
| **OR-MA-0310** | TAGSAM | Cover plate coupon, sampler head subassembly | Aluminum alloy 6061-T651 | 10 $cm^2$ |
| **OR-MA-0311** | TAGSAM | Screw coupon, sampler head subassembly | Stainless steel CRES NAS1102E08-6 | 1 cm long |
| **OR-MA-0312** | TAGSAM | Contact pad spool coupon | Stainless steel CRES 304L | 0.75" o.d. |
| **OR-MA-0313** | TAGSAM | Screen coupon, air filter assembly | Stainless steel CRES 304L | 10 $cm^2$ |
| **OR-MA-0314** | TAGSAM | Sapphire witness plate coupon, post-stow | Sapphire | 1" diam. |
| **OR-MA-0315** | TAGSAM | Self-locking nut coupon, U-joint subassembly | Silver plated CRES | ~0.33" o.d. |
| **OR-MA-0316** | SRC | Witness plate bracket coupon | Aluminum alloy 7050-T7451 | 10 $cm^2$ |
| **OR-MA-0317** | TAGSAM | Torsion spring RH coupon, witness plate assembly | Stainless steel 17-7 | ~ 3 cm long |
| **OR-MA-0318** | SRC | Connector bracket enable coupon, SRC top assembly | Aluminum alloy 7050-T7451 | 10 $cm^2$ |
| **OR-MA-0319** | TAGSAM | Witness plate retaining pin coupon | Stainless steel CRES A286 | ~0.1 g |
| **OR-MA-0320** | TAGSAM | U-joint yoke, sampler head subassembly | Aluminum alloy 6061-T651 | 10 $cm^2$ |
| **OR-MA-0321** | SRC | Cap screw coupon, SRC top assembly | Stainless Steel; heat resistant steel, NAS1351N3-8 | 1.75 cm long |
| **OR-MA-0322** | SRC | Septum bushing coupon | Aluminum alloy 6061-T651 | 10 $cm^2$ |
| **OR-MA-0323** | TAGSAM | Perforated plate coupon, air filter assembly | Stainless steel CRES 304L | ~ 1 in2 |
| **OR-MA-0324** | TAGSAM | Witness plate baseplate | Aluminum alloy 6061-T651 | 10 $cm^2$ |
| **OR-MA-0325** | TAGSAM | Screw coupon, sampler head top assembly | Stainless steel - CRES NAS1102E04-5 | ~0.75 cm |
| **OR-MA-0326** | TAGSAM | Shaft coupon, witness plate assembly | Stainless steel CRES A286 | 4.01 g |
| **OR-MA-0327** | TAGSAM | U-joint fan support fitting assembly coupon | Aluminum alloy 6061-T651 | 10 $cm^2$ |
| **OR-MA-0328** | SRC | Septum retainer coupon, SRC top assembly | Stainless Steel; heat resistant steel, NAS1351N6-14 | 1" long |
| **OR-MA-0329** | SRC | Bolt coupon, SRC hinge assembly | Stainless steel CRES | #10 x 0.4375" |
| **OR-MA-0330** | TAGSAM | Airframe bearing, inner POGO assembly | Steel | 0.0375" OD |
| **OR-MA-0331** | Support Materials | ESD packaging, non-flight consumable | Dunshield ESD, used in LC firing to catch the lid - ATLO test | ~ 3" x 3" |
| **OR-MA-0332** | SRC | Screw coupon, SRC top assembly | Stainless steel CRES | ~0.5" long |
| **OR-MA-0333** | SRC | Bolt coupon, SRC hinge assembly | n/a | 0.688" long |
| **OR-MA-0334** | SRC | Latch fitting, avionics deck assembly | Aluminum alloy 7075-T7351 | 10 $cm^2$ |
| **OR-MA-0335** | TAGSAM | Witness coupon retainer coupon, U-joint subassembly | Aluminum alloy 6061-T651 | 10 $cm^2$ |
| **OR-MA-0336** | TAGSAM | Screw coupon, U-joint subassembly | n/a | ~ 1" long |
| **OR-MA-0337** | TAGSAM | Feedline Tee coupon, sampler head subassembly | Stainless steel CRES 316L | 1.5" long |
| **OR-MA-0338** | TAGSAM | Feedline elbow, sampler head subassembly | Stainless steel CRES 316L | 1.25" long |
| **OR-MA-0339** | Launch Container | Launch container cover coupon, cover assembly | Aluminum alloy 6061-T651 | 10 $cm^2$ |
| **OR-MA-0340** | TAGSAM | Contact pad Velcro loop pad coupon | Stainless steel Velcro | 0.75" OD |
| **OR-MA-0341** | TAGSAM | Wrist joint restraints, sampler head subassembly | Titanium | ~ 2.5" long |
| **OR-MA-0342** | TAGSAM | U-joint arm fitting, sampler head subassembly | Aluminum alloy 6061-T651 | 10 $cm^2$ |
| **OR-MA-0343** | TAGSAM | Bolt head capture cleat coupon, sampler head subassembly | Aluminum alloy 6061-T651 | 10 $cm^2$ |
| **OR-MA-0344** | TAGSAM | Witness plate shaft coupon, witness plate assembly | Stainless steel CRES A286 | 4.01 g |
| **OR-MA-0345** | SRC | Cotter pin coupon, capture ring assembly | Stainless steel CRES; MS24665-151 | 0.5" long |
| **OR-MA-0346** | TAGSAM | Torsion spring LH coupon, sampler head witness plate assembly | Stainless steel CRES | ~1.25" long |
| **OR-MA-0347** | TAGSAM | Witness plate post-stow coupon, witness plate assembly | Aluminum 99.999% | ~1: diam. |
| **OR-MA-0348** | TAGSAM | U-joint arm fitting coupon, U-joint subassembly | Stainless steel CRES A286 Cadmium plated MS51830CAL102L | 0.134" diam. |
| **OR-MA-0349** | SRC | Washer coupon, SRC top assembly | Stainless steel CRES; NAS1149CN832R | 0.875" OD |
| **OR-MA-0350** | TAGSAM | Screw coupon, sample head U-joint subassembly | Stainless steel CRES; NAS1151E6 | 0.375" OD |
| **OR-MA-0351** | TAGSAM | Nut coupon, sampler head U-joint subassembly | Stainless steel CRES A286; MS21043-06N | 0/168" OD |
| **OR-MA-0352** | TAGSAM | Feedline tube coupon, sampler head subassembly | Stainless steel CRES 316L | ~0.122 kg |
| **OR-MA-0353** | TAGSAM | Witness plate launch coupon, witness plate assembly | Aluminum 99.999% | ~ 1" diam. |
| **OR-MA-0354** | TAGSAM | Witness plate cover plate coupon | Aluminum alloy 6061-T651 | 10 $cm^2$ |
| **OR-MA-0355** | SRC | Cap screw coupon, SRC top assembly | Stainless steel; nas1352e06-6 | 0.75" long |
| **OR-MA-0356** | SRC | Washer coupon, capture ring assembly | Stainless steel CRES; NAS1149C0416R | 0.500" OD |
| **OR-MA-0357** | TAGSAM | Witness plate always coupon, witness plate assembly | Aluminum 99.999% | ~1" diam |
| **OR-MA-0358** | SRC | Cap screw coupon, Tang assembly | Stainless steel CRES | ~0.25" long |
| **OR-MA-0359** | TAGSAM | U-joint fan support fitting, U-joint subassembly | Stainless steel CRES 303 cadmium plated MS51830CA104L | 0.134" diam |
| **OR-MA-0360** | TAGSAM | Sampler arm wrist shim, Pogo assembly | Laminated aluminum | 6 cm long |
| **OR-MA-0361** | SRC | Washer coupon, sample canister top assembly | Stainless steel CRES; NAS620C6L | 0.267" OD |
| **OR-MA-0362** | Launch Container | Launch container barrel coupon | Aluminum alloy 6061-T651 | 10 $cm^2$ |
| **OR-MA-0363** | TAGSAM | Cap screw coupon, TAGSAM top assembly | Stainless steel CRES; NAS1352N08H6 | ~0.25" long |
| **OR-MA-0364** | Spacecraft | Top deck blanket (+Z) spacecraft | Kapton blanketing, Kapton tape | 4" x 3" |
| **OR-MA-0365** | Support Materials | Nylon tape used to package TAGSAM TVAC NVR samples | Nylon tape, rubber adhesive | 2" wide |
| **OR-MA-0366** | Support Materials | Swabbed particulate from flight TAGSAM head | Unknown, collected with polyethylene swab | ~100-200 microns |
| **OR-MA-0367** | Support Materials | Pink polyethylene in cleanroom room 202 | Polyethylene, anti-static | maybe 2 ft2 |
| **OR-MA-0368** | Spacecraft | Spacecraft USB drive | Metallic | ~ 1.5" long |
| **OR-MA-0369** | Support Materials | TAGCAMS nylon camera covers | Nylon | ~2" diam (2) |
| **OR-MA-0370** | Support Materials | Faraday Cap - cable cutter mate | n/a | ~0.5" long |
| **OR-MA-0371** | Support Materials | Crane Control panel residue; Cape | Unknown | n/a |
| **OR-MA-0372** | TAGSAM | Severed frangibolt fasteners TAGSAM | Stainless steel | ~2.25" long |
| **OR-MA-0373** | Support Materials | TAGSAM TVAC Si wafers | Silicon wafers, polyethylene cases | 3" diam (5) |
| **OR-MA-0374** | Spacecraft | Black Kapton MLI blanketing | Kapton, tapes, films, staples, etc. | 8" x 8" |
| **OR-MA-0375** | Support Materials | Tube cutter, qual TAGSAM head "decapitation" | n/a | ~1.5" long |
| **OR-MA-0376** | Support Materials | Epoxy used to stake tube cutter | Epoxy, Glass | ~ 1 g |
| **OR-MA-0377** | SRC | Temperature sensor assembly coupon | Wire, tape, thread, etc. | ~ 2 ft |
| **OR-MA-0378** | Support Materials | Swabs from detent pin cleaning on SRC | Polyethylene swabs | ~2" swabs |
| **OR-MA-0379** | Launch Container | PLA transducer cap - nylon | Cap material unknown, nylon bagging | ~0.5" long |
| **OR-MA-0380** | TAGSAM | Qual TAGSAM head insert | Ultem, stainless steel fastener | ~3" long |
| **OR-MA-0381** | Support Materials | insert pull throughs, SRC canister | Silicone | ~2" long |
| **OR-MA-0382** | Support Materials | TAGSAM test blanket | Fiberglass re-enforced Teflon | ~ 2 ft long |
| **OR-MA-0383** | TAGSAM | Wire coupon, TAGSAM ordnance installation | Stainless steel CRES 302 | 0.20" diam |
| **OR-MA-0384** | SRC | Temperature strip coupon, SRC thermal installation | Kapton | 0.56" diam |
| **OR-MA-0385** | SRC | Temperature strip coupon, SRC thermal installation | Kapton | 0.38" x 1.5" |
| **OR-MA-0386** | SRC | Temperature strip coupon, SRC thermal installation | Kapton | 0.38" x 0.8" |
| **OR-MA-0387** | Spacecraft | Diode bonding adhesive coupon, solar array | Epoxy - RTV silicone LM-BF10898F | > 1 g |
| **OR-MA-0388** | Spacecraft | Diode bonding adhesive coupon, solar array | Epoxy - RTV silicone LM-BF10898B | > 1 g |
| **OR-MA-0389** | Spacecraft | Wire coupon - red, blue - solar array | Silver coated copper, Kapton insulation | > 1 ft |
| **OR-MA-0390** | Spacecraft | Wire coupon - green - solar array | Silver coated copper, Kapton insulation | > 1 ft |
| **OR-MA-0391** | Support Materials | Ultratape 1153 clean tent tape coupon, VIF | Polyethylene, rubber adhesive | ~ 2" |
| **OR-MA-0392** | Support Materials | Ultratape 1154 | Polyethylene, synthetic rubber adhesive | ~2" |
| **OR-MA-0393** | Support Materials | ULA fairing adhesive | Epoxy - RTV silicone CV3-1142 | >1g |
| **OR-MA-0394** | SRC | Conformal coating coupon, sep/spin mechanism assembly | Polyurethane (Arathane 5750) | >1 g |
| **OR-MA-0395** | TAGSAM | Qual TAGSAM head witness plates (3) | Sapphire | ~1" diam. |
| **OR-MA-0396** | TAGSAM | Qual TAGSAM head witness plate | Sapphire | ~1" diam. |
| **OR-MA-0397** | TAGSAM | Qual TAGSAM head witness plate + shim | Sapphire | ~1" diam. |
| **OR-MA-0398** | TAGSAM | Qual TAGSAM head witness plate | Sapphire | ~1" diam. |
| **OR-MA-0399** | TAGSAM | Qual TAGSAM head witness plate | Sapphire | ~1" diam. |
| **OR-MA-0400** | TAGSAM | Qual TAGSAM head witness plates (3) | Sapphire | ~1" diam. |
| **OR-MA-0401** | Support Materials | Contamination Control samples Box 1 Al foils | Aluminum foil | ~ 1 cm x 2 cm |
| **OR-MA-0402** | Support Materials | Contamination Control samples Box 2 Al foils | Aluminum foil | ~ 1 cm x 2 cm |
| **OR-MA-0403** | Support Materials | Contamination Control samples Box 3 Al foils | Aluminum foil | ~ 1 cm x 2 cm |
| **OR-MA-0404** | Support Materials | Contamination Control samples Box 4 Al foils | Aluminum foil | ~ 1 cm x 2 cm |
| **OR-MA-0405** | Support Materials | Contamination Control samples Box 5 Al foils | Aluminum foil | ~ 1 cm x 2 cm |
| **OR-MA-0406** | Support Materials | Contamination control knowledge samples curation cabinet | Aluminum foil | ~ 1 cm x 2 cm |

**Table S2-2** Materials archived for contamination knowledge after launch (62 total).

| Catalog Number | Major Component | Description | Composition | Sample Size |
|---|---|---|---|---|
| **OR-MA-0407** | TAGSAM | 3D printed TAGSAM head provided to JSC by Lockheed Martin Denver for curation activities, such as sizing of components, disassembly, disassembly panning, etc. | Resin | ~15" diameter x 4" high |
| **OR-MA-0408** | Construction | Aluminum strut from cleanroom walls and ceiling | Aluminum | 2 pieces: 2" x 2" x 2" and 2" x 2" x 5" |
| **OR-MA-0409** | Construction | Fireproofing material (spray) | | |
| **OR-MA-0410** | Construction | Chemlink adhesive/sealant | | |
| **OR-MA-0411** | Construction | Fire caulk | | |
| **OR-MA-0412** | Construction | Epoxy paint used on cleanroom walls | Epoxy paint on stainless steel plate | 6 plates |
| **OR-MA-0413** | Construction | Water pipe insulation | | |
| **OR-MA-0414** | Construction | Duct insulation (non-cleanroom) | SOFTR® Duct Wrap FRK (Owens Corning) | |
| **OR-MA-0415** | Construction | Wall insulation (non-cleanroom) | fiberglass insulation | |
| **OR-MA-0416** | Construction | Fire rated glass for Windows A and B plus Doors 202, 212, 216, and 220 | Fire rated glass 3/8" Schott PYRAN Platinum L Fire Rated (laminated) | |
| **OR-MA-0417** | Construction | Conference room glass | 1/2" clear tempered glass | |
| **OR-MA-0418** | Construction | Glass used for Window C and Doors 213B and 215 | 3/8" laminated clear tempered glass: 3/16 clear glass" + 0.030 PVB clear + 3/16" clear glass | |
| **OR-MA-0419** | Construction | Glass used for Window D and Doors 106, 111, 108, 109, 107, 110, 217,213, 213A, and 214 | 1/4" laminated clear tempered glass: 1/8" + 0.030 PVB clear + 1/8" clear glass | |
| **OR-MA-0420** | Construction | Arizona Polymer Flooring epoxy floor 750 | Flooring samples - three pieces 12"x 12" (2) and smaller 4" x 4" | |
| **OR-MA-0421** | Construction | Fire sprinkler head Model TY9281; TY-FRB Stainless Steel quick response pendent 8.0 K, ¾ in NPT | Fire sprinkler head; stainless steel, Teflon, Be-Ni gasket | |
| **OR-MA-0422** | Construction | Cleanroom duct insulation | Cleanroom duct insulation | 6" x 6" |
| **OR-MA-0423** | Construction | Lasco wall panel mock-up | Lasco wall panel mock-up aluminum | ~9" x ~8" with two struts and a panel |
| **OR-MA-0424** | Construction | Lasco wall panels | Lasco wall panels 5 pieces | 6" x 6" |
| **OR-MA-0425** | Construction | Stonchem 441 LV liner | Stonchem 441 LV liner | 4" x 4" |
| **OR-MA-0426** | Construction | Wall shims | Wall shims to level base of wall panel system | |
| **OR-MA-0427** | Construction | Lasco wall struts | Lasco wall strut - aluminum (4 pieces) | 8" x 2" x 2" |
| **OR-MA-0428** | Construction | Lasco Ceiling strut | Lasco Ceiling strut - aluminum (4 pieces) | 6" x 2" x 2" |
| **OR-MA-0429** | Construction | Teflon gasket for cleanroom ducts | Teflon gasket | 18" |
| **OR-MA-0430** | Construction | SpeedCove | Coving material for wall/floor interface | 5" x 5" |
| **OR-MA-0431** | Construction | SpeedCove adhesive | Adhesive used behind speedcove - 6 pieces | 1" x 1" |
| **OR-MA-0432** | Construction | Texas Access Auto sliding door coupon | Aluminum door coupon | 3" x 4" x 5/32" |
| **OR-MA-0433** | Construction | Spare contact pad | | |
| **OR-MA-0434** | Construction | Spare contact pad | | |
| **OR-MA-0435** | Construction | Spare contact pad | | |
| **OR-MA-0436** | Construction | Arizona Polymer Flooring Epoxy 420 | | |
| **OR-MA-0437** | Support Materials | Gas analyses from JSC Toxicology group, 2016-2023. | This record is an archive of all gas analyses done for the mission by the JSC Toxicology group | |
| **OR-MA-0438** | Support Materials | Clean Apex Torq bit - same as used for SRC disassembly in UTTR cleanroom (reference procedure for relevant steps); cleaned to NVR=0.2 micron | Tool steel | 1" |
| **OR-MA-0439** | Support Materials | Apex Torq bit (not clean) - used for SRC disassembly in UTTR cleanroom (reference procedure for relevant steps). | Tool steel | 1" |
| **OR-MA-0440** | Support Materials | Pycnometer witness foil #1 container X2180 Goddard organic | Aluminum foil | |
| **OR-MA-0441** | Support Materials | Pycnometer witness foil #1 blank container X2181 Goddard organics | Aluminum foil | 0.9" x 0.9" |
| **OR-MA-0442** | Support Materials | Pycnometer witness foil #2 container X2182 sem | Aluminum foil | 0.9" x 0.9" |
| **OR-MA-0443** | Support Materials | Pycnometer witness foil #2 blank container X2183 sem | Aluminum foil | 0.9" x 0.9" |
| **OR-MA-0444** | Support Materials | Pycnometer witness foil %3 container X2186 JSC organics | Aluminum foil | 0.9" x 0.9" |
| **OR-MA-0445** | Support Materials | Pycnometer witness foil #3 container blank X2187 JSC organics | Aluminum foil | 0.9" x 0.9" |
| **OR-MA-0446** | Support Materials | Pycnometer witness foil #4 container X2185 archive | Aluminum foil | 0.9" x 0.9" |
| **OR-MA-0447** | Support Materials | Pycnometer witness foil #4 blank container X2184 | Aluminum foil | 0.9" x 0.9" |
| **OR-MA-0448** | Support Materials | Balazs testing results | Summaries and test result reports from Balazs organic and inorganic wafers and air sampling. | |
| **OR-MA-0449** | Support Materials | UPW water samples in PTFE containers | UPW water samples for testing | |
| **OR-MA-0450** | Support Materials | UPW water samples in glass bottles | UPW water in glass containers for testing | |

| | | | |
|---|---|---|---|
| **OR-MA-0451** | Support Materials | Teflon that shredded off the runners of the x-y-r stage, which accumulated at the end of the documentation chamber of the TAGSAM glovebox | Teflon with supporting materials used to collect the shreds from the x-y-r stage in the TAGSAM glovebox, supporting materials included aluminum cups, one pair of stainless-steel tweezers, and two Teflon brushes. |
| **OR-MA-0452** | Support Materials | OSIRIS-REx Sample Return Capsule area Utah soil-environmental sample collected on September 24, 2023, Log number: 1, Time (UTTR): 9:40 am, Map#2, Event type: Gas Sampling, Heat Shield, Tag#34989. | Soil |
| **OR-MA-0453** | Support Materials | OSIRIS-REx Sample Return Capsule area Utah soil-environmental sample collected on September 24, 2023, Log number: 2, Time (UTTR): 9:41 am, Map#2, Event type: Gas Sampling, Back shell, Tag#12848. | Soil |
| **OR-MA-0454** | Support Materials | OSIRIS-REx Sample Return Capsule area Utah soil-environmental sample collected on September 24, 2023, Log number: 3, Time (UTTR): 9:43 am, Map#2, Event type: Gas Sampling, East Vent#1, Tag#22378. Label incorrect, and corrected in field. | Soil |
| **OR-MA-0455** | Support Materials | OSIRIS-REx Sample Return Capsule area Utah soil-environmental sample collected on September 24, 2023, Log number: 4, Time (UTTR): 9:44 am, Map#2, Event type: Gas Sampling, West Vent#2, Tag#21190. | Soil |
| **OR-MA-0456** | Support Materials | OSIRIS-REx Sample Return Capsule area Utah soil-environmental sample collected on September 24, 2023, Log number: 5, Time (UTTR): 9:52 am, Map#1, Event type: Dust Sampling started, Tag#-. | Soil |
| **OR-MA-0457** | Support Materials | OSIRIS-REx Sample Return Capsule area Utah soil-environmental sample collected on September 24, 2023, Log number: 6, Time (UTTR): 9:54 am, Map#1, Event type: Dust Sampler#1, Tag#25218. didn't audibly verify hiss, was mislabeled, label corrected in field, resampled (see #7) (used spare bottle). | Soil |
| **OR-MA-0458** | Support Materials | OSIRIS-REx Sample Return Capsule area Utah soil-environmental sample collected on September 24, 2023, Log number: 7, Time (UTTR): 9:54 am, Map#1, Event type: Dust Sampler#2, Tag#21260. | Soil |
| **OR-MA-0459** | Support Materials | OSIRIS-REx Sample Return Capsule area Utah soil-environmental sample collected on September 24, 2023, Log number: 8, Time (UTTR): 10:07 am, Map#2, Event type: Soil Sample, under SRC#1, Tag#1007. | Soil |
| **OR-MA-0460** | Support Materials | OSIRIS-REx Sample Return Capsule area Utah soil-environmental sample collected on September 24, 2023, Log number: 9, Time (UTTR): 10:14 am, Map#2, Event type: Soil Sample, under SRC#2, Tag#1023. | Soil |
| **OR-MA-0461** | Support Materials | OSIRIS-REx Sample Return Capsule area Utah soil-environmental sample collected on September 24, 2023, Log number: 10, Time (UTTR): 10:15 am, Map#2, Event type: SRC lift image, Tag#-. | Soil |
| **OR-MA-0462** | Support Materials | OSIRIS-REx Sample Return Capsule area Utah soil-environmental sample collected on September 24, 2023, Log number: 11, Time (UTTR): 10:20 am, Map#2, Event type: Side imprint soil sample, Tag#1039. | Soil |
| **OR-MA-0463** | Support Materials | OSIRIS-REx Sample Return Capsule area Utah soil-environmental sample collected on September 24, 2023, Log number: 12, Time (UTTR): 10:25 am, Map#2, Event type: Context soil, sample control, Tag#1055. | Soil |
| **OR-MA-0464** | Support Materials | OSIRIS-REx Sample Return Capsule area Utah soil-environmental sample collected on September 24, 2023, Log number: 13, Time (UTTR): 10:28 am, Map#2, Event type: Context soil, sample control, Tag#1071. | Soil |
| **OR-MA-0465** | Support Materials | OSIRIS-REx Sample Return Capsule area Utah soil-environmental sample collected on September 24, 2023, Log number: 14, Time (UTTR): 10:32 am, Map#2, Event type: Context soil, sample control, Tag#1072. | Soil |
| **OR-MA-0466** | Support Materials | OSIRIS-REx Sample Return Capsule area Utah soil-environmental sample collected on September 24, 2023, Log number: 15, Time (UTTR): 10:35 am, Map#2, Event type: Context soil, sample control, Tag#1056. Small amount of biomaterial in sample. | Soil |
| **OR-MA-0467** | Support Materials | OSIRIS-REx Sample Return Capsule area Utah soil-environmental sample collected on September 24, 2023, Log number: 16, Time (UTTR): 10:37 am, Map#2, Event type: Trample sample, Tag#1040. | Soil |
| **OR-MA-0468** | Support Materials | OSIRIS-REx Sample Return Capsule area Utah soil-environmental sample collected on September 24, 2023, Log number: 17, Time (UTTR): 10:43 am, Map#1, Event type: Stop dust sampling, Tag#1008. | Soil |

**Table S2-3** Contamination knowledge witness plates archived (343 total).

| Catalog Number | Substrate | Process Monitored | Exposure Start | Exposure Stop | Days Exposed |
|---|---|---|---|---|---|
| **OR-GCKP-01-1-Al** | Al foil | SRC assembly and functional testing, TAGSAM assembly with clean qual head | 03/11/2015 | 04/14/2015 | 34 |
| **OR-GCKP-01-1-Si** | Si wafer | SRC assembly and functional testing, TAGSAM assembly with clean qual head | 03/11/2015 | 04/14/2015 | 34 |
| **OR-GCKP-01-2-Al** | Al foil | SRC assembly and functional testing, TAGSAM assembly with clean qual head | 03/11/2015 | 04/14/2015 | 34 |
| **OR-GCKP-01-2-Si** | Si wafer | SRC assembly and functional testing, TAGSAM assembly with clean qual head | 03/11/2015 | 04/14/2015 | 34 |
| **OR-GCKP-01-3-Al** | Al foil | SRC assembly and functional testing, TAGSAM assembly with clean qual head | 03/11/2015 | 04/14/2015 | 34 |
| **OR-GCKP-01-3-Si** | Si wafer | SRC assembly and functional testing, TAGSAM assembly with clean qual head | 03/11/2015 | 04/14/2015 | 34 |
| **OR-GCKP-01-4-Al** | Al foil | SRC assembly and functional testing, TAGSAM assembly with clean qual head | 03/11/2015 | 04/14/2015 | 34 |
| **OR-GCKP-01-4-Si** | Si wafer | SRC assembly and functional testing, TAGSAM assembly with clean qual head | 03/11/2015 | 04/14/2015 | 34 |
| **OR-GCKP-02-1-Al** | Al foil | TAGSAM functional deployment (not in same room) | 04/14/2015 | 05/11/2015 | 27 |
| **OR-GCKP-02-1-Si** | Si wafer | TAGSAM functional deployment (not in same room) | 04/14/2015 | 05/11/2015 | 27 |
| **OR-GCKP-02-2-Al** | Al foil | TAGSAM functional deployment (not in same room) | 04/14/2015 | 05/11/2015 | 27 |
| **OR-GCKP-02-2-Si** | Si wafer | TAGSAM functional deployment (not in same room) | 04/14/2015 | 05/11/2015 | 27 |
| **OR-GCKP-02-3-Al** | Al foil | TAGSAM functional deployment (not in same room) | 04/14/2015 | 05/11/2015 | 27 |
| **OR-GCKP-02-3-Si** | Si wafer | TAGSAM functional deployment (not in same room) | 04/14/2015 | 05/11/2015 | 27 |
| **OR-GCKP-02-4-Al** | Al foil | TAGSAM functional deployment (not in same room) | 04/14/2015 | 05/11/2015 | 27 |
| **OR-GCKP-02-4-Si** | Si wafer | TAGSAM functional deployment (not in same room) | 04/14/2015 | 05/11/2015 | 27 |
| **OR-GCKP-03-1-Al** | Al foil | Avionics box assembly, SRC functional post-vibe | 05/11/2015 | 06/10/2015 | 30 |
| **OR-GCKP-03-1-Si** | Si wafer | Avionics box assembly, SRC functional post-vibe | 05/11/2015 | 06/10/2015 | 30 |
| **OR-GCKP-03-2-Al** | Al foil | Avionics box assembly, SRC functional post-vibe | 05/11/2015 | 06/10/2015 | 30 |
| **OR-GCKP-03-2-Si** | Si wafer | Avionics box assembly, SRC functional post-vibe | 05/11/2015 | 06/10/2015 | 30 |
| **OR-GCKP-03-3-Al** | Al foil | Avionics box assembly, SRC functional post-vibe | 05/11/2015 | 06/10/2015 | 30 |
| **OR-GCKP-03-3-Si** | Si wafer | Avionics box assembly, SRC functional post-vibe | 05/11/2015 | 06/10/2015 | 30 |
| **OR-GCKP-03-4-Al** | Al foil | Avionics box assembly, SRC functional post-vibe | 05/11/2015 | 06/10/2015 | 30 |
| **OR-GCKP-03-4-Si** | Si wafer | Avionics box assembly, SRC functional post-vibe | 05/11/2015 | 06/10/2015 | 30 |
| **OR-GCKP-04-1-AL** | Al foil | SARA TVAC (RAL), TAGSAM deployment (high bay), TAGSAM functional post-vibe (high bay); OVIRS install (high bay) | 06/10/2015 | 07/17/2015 | 37 |
| **OR-GCKP-04-1-Si** | Si wafer | SARA TVAC (RAL), TAGSAM deployment (high bay), TAGSAM functional post-vibe (high bay); OVIRS install (high bay) | 06/10/2015 | 07/17/2015 | 37 |
| **OR-GCKP-04-2-Al** | Al foil | SARA TVAC (RAL), TAGSAM deployment (high bay), TAGSAM functional post-vibe (high bay); OVIRS install (high bay) | 06/10/2015 | 07/17/2015 | 37 |
| **OR-GCKP-04-2-Si** | Si wafer | SARA TVAC (RAL), TAGSAM deployment (high bay), TAGSAM functional post-vibe (high bay); OVIRS install (high bay) | 06/10/2015 | 07/17/2015 | 37 |
| **OR-GCKP-04-3-Al** | Al foil | SARA TVAC (RAL), TAGSAM deployment (high bay), TAGSAM functional post-vibe (high bay); OVIRS install (high bay) | 06/10/2015 | 07/17/2015 | 37 |
| **OR-GCKP-04-3-Si** | Si wafer | SARA TVAC (RAL), TAGSAM deployment (high bay), TAGSAM functional post-vibe (high bay); OVIRS install (high bay) | 06/10/2015 | 07/17/2015 | 37 |
| **OR-GCKP-04-4-Al** | Al foil | SARA TVAC (RAL), TAGSAM deployment (high bay), TAGSAM functional post-vibe (high bay); OVIRS install (high bay) | 06/10/2015 | 07/17/2015 | 37 |
| **OR-GCKP-04-4-Si** | Si wafer | SARA TVAC (RAL), TAGSAM deployment (high bay), TAGSAM functional post-vibe (high bay); OVIRS install (high bay) | 06/10/2015 | 07/17/2015 | 37 |
| **OR-GCKP-05-1-Al** | Al foil | TAGSAM install (high bay); TAGSAM deployment (high bay) | 07/17/2015 | 08/19/2015 | 33 |
| **OR-GCKP-05-1-Si** | Si wafer | TAGSAM install (high bay); TAGSAM deployment (high bay) | 07/17/2015 | 08/19/2015 | 33 |
| **OR-GCKP-05-2-Al** | Al foil | TAGSAM install (high bay); TAGSAM deployment (high bay) | 07/17/2015 | 08/19/2015 | 33 |
| **OR-GCKP-05-2-Si** | Si wafer | TAGSAM install (high bay); TAGSAM deployment (high bay) | 07/17/2015 | 08/19/2015 | 33 |
| **OR-GCKP-05-3-Al** | Al foil | TAGSAM install (high bay); TAGSAM deployment (high bay) | 07/17/2015 | 08/19/2015 | 33 |
| **OR-GCKP-05-3-Si** | Si wafer | TAGSAM install (high bay); TAGSAM deployment (high bay) | 07/17/2015 | 08/19/2015 | 33 |
| **OR-GCKP-05-4-Al** | Al foil | TAGSAM install (high bay); TAGSAM deployment (high bay) | 07/17/2015 | 08/19/2015 | 33 |
| **OR-GCKP-05-4-Si** | Si wafer | TAGSAM install (high bay); TAGSAM deployment (high bay) | 07/17/2015 | 08/19/2015 | 33 |
| **OR-GCKP-06-1-Al** | Al foil | OCAMS Installation, spacecraft lift to rotation fixture, SARA deployment | 08/19/2015 | 09/18/2015 | 30 |
| **OR-GCKP-06-1-Si** | Si wafer | OCAMS Installation, spacecraft lift to rotation fixture, SARA deployment | 08/19/2015 | 09/18/2015 | 30 |
| **OR-GCKP-06-2-Al** | Al foil | OCAMS Installation, spacecraft lift to rotation fixture, SARA deployment | 08/19/2015 | 09/18/2015 | 30 |
| **OR-GCKP-06-2-Si** | Si wafer | OCAMS Installation, spacecraft lift to rotation fixture, SARA deployment | 08/19/2015 | 09/18/2015 | 30 |
| **OR-GCKP-06-3-Al** | Al foil | OCAMS Installation, spacecraft lift to rotation fixture, SARA deployment | 08/19/2015 | 09/18/2015 | 30 |
| **OR-GCKP-06-3-Si** | Si wafer | OCAMS Installation, spacecraft lift to rotation fixture, SARA deployment | 08/19/2015 | 09/18/2015 | 30 |
| **OR-GCKP-06-4-Al** | Al foil | OCAMS Installation, spacecraft lift to rotation fixture, SARA deployment | 08/19/2015 | 09/18/2015 | 30 |
| **OR-GCKP-06-4-Si** | Si wafer | OCAMS Installation, spacecraft lift to rotation fixture, SARA deployment | 08/19/2015 | 09/18/2015 | 30 |
| **OR-GCKP-07-1-Al** | Al foil | SARA Deployment, SC move to RAL/Building 245, lift to chamber MOIT stand | 09/18/2015 | 11/04/2015 | 47 |
| **OR-GCKP-07-1-Si** | Si wafer | SARA Deployment, SC move to RAL/Building 245, lift to chamber MOIT stand | 09/18/2015 | 11/04/2015 | 47 |
| **OR-GCKP-07-2-Al** | Al foil | SARA Deployment, SC move to RAL/Building 245, lift to chamber MOIT stand | 09/18/2015 | 11/04/2015 | 47 |
| **OR-GCKP-07-2-Si** | Si wafer | SARA Deployment, SC move to RAL/Building 245, lift to chamber MOIT stand | 09/18/2015 | 11/04/2015 | 47 |
| **OR-GCKP-07-3-Al** | Al foil | SARA Deployment, SC move to RAL/Building 245, lift to chamber MOIT stand | 09/18/2015 | 11/04/2015 | 47 |
| **OR-GCKP-07-3-Si** | Si wafer | SARA Deployment, SC move to RAL/Building 245, lift to chamber MOIT stand | 09/18/2015 | 11/04/2015 | 47 |
| **OR-GCKP-07-4-Al** | Al foil | SARA Deployment, SC move to RAL/Building 245, lift to chamber MOIT stand | 09/18/2015 | 11/04/2015 | 47 |
| **OR-GCKP-07-4-Si** | Si wafer | SARA Deployment, SC move to RAL/Building 245, lift to chamber MOIT stand | 09/18/2015 | 11/04/2015 | 47 |
| **OR-GCKP-08-1-Al** | Al foil | Still in RAL; Sine vibe testing, SC move to SSB on 12/1; SC lift 12/2 | 11/04/2015 | 12/09/2015 | 35 |
| **OR-GCKP-08-1-Si** | Si wafer | Still in RAL; Sine vibe testing, SC move to SSB on 12/1; SC lift 12/2 | 11/04/2015 | 12/09/2015 | 35 |
| **OR-GCKP-08-2-Al** | Al foil | Still in RAL; Sine vibe testing, SC move to SSB on 12/1; SC lift 12/2 | 11/04/2015 | 12/09/2015 | 35 |
| **OR-GCKP-08-2-Si** | Si wafer | Still in RAL; Sine vibe testing, SC move to SSB on 12/1; SC lift 12/2 | 11/04/2015 | 12/09/2015 | 35 |
| **OR-GCKP-08-3-Al** | Al foil | Still in RAL; Sine vibe testing, SC move to SSB on 12/1; SC lift 12/2 | 11/04/2015 | 12/09/2015 | 35 |
| **OR-GCKP-08-3-Si** | Si wafer | Still in RAL; Sine vibe testing, SC move to SSB on 12/1; SC lift 12/2 | 11/04/2015 | 12/09/2015 | 35 |
| **OR-GCKP-08-4-Al** | Al foil | Still in RAL; Sine vibe testing, SC move to SSB on 12/1; SC lift 12/2 | 11/04/2015 | 12/09/2015 | 35 |
| **OR-GCKP-08-4-Si** | Si wafer | Still in RAL; Sine vibe testing, SC move to SSB on 12/1; SC lift 12/2 | 11/04/2015 | 12/09/2015 | 35 |
| **OR-GCKP-09-1-Al** | Al foil | SC move - science plates + EM plates went with SC inside shipping container | 12/09/2015 | 01/07/2016 | 29 |
| **OR-GCKP-09-1-Si** | Si wafer | SC move - science plates + EM plates went with SC inside shipping container | 12/09/2015 | 01/07/2016 | 29 |
| **OR-GCKP-09-2-Al** | Al foil | SC move - science plates + EM plates went with SC inside shipping container | 12/09/2015 | 01/07/2016 | 29 |
| **OR-GCKP-09-2-Si** | Si wafer | SC move - science plates + EM plates went with SC inside shipping container | 12/09/2015 | 01/07/2016 | 29 |
| **OR-GCKP-09-3-Al** | Al foil | SC move - science plates + EM plates went with SC inside shipping container | 12/09/2015 | 01/07/2016 | 29 |
| **OR-GCKP-09-3-Si** | Si wafer | SC move - science plates + EM plates went with SC inside shipping container | 12/09/2015 | 01/07/2016 | 29 |
| **OR-GCKP-09-4-Al** | Al foil | SC move - science plates + EM plates went with SC inside shipping container | 12/09/2015 | 01/07/2016 | 29 |
| **OR-GCKP-09-4-Si** | Si wafer | SC move - science plates + EM plates went with SC inside shipping container | 12/09/2015 | 01/07/2016 | 29 |
| **OR-GCKP-10-1-Al** | Al foil | Flight TAGSAM build; SC lift; moved SC to RAL; EMI/EMC testing; TVAC pre-cert. | 01/08/2016 | 02/05/2016 | 28 |
| **OR-GCKP-10-1-Si** | Si wafer | Flight TAGSAM build; SC lift; moved SC to RAL; EMI/EMC testing; TVAC pre-cert. | 01/08/2016 | 02/05/2016 | 28 |
| **OR-GCKP-10-2-Al** | Al foil | Flight TAGSAM build; SC lift; moved SC to RAL; EMI/EMC testing; TVAC pre-cert. | 01/08/2016 | 02/05/2016 | 28 |
| **OR-GCKP-10-2-Si** | Si wafer | SC lift; moved SC to RAL; EMI/EMC testing; TVAC pre-cert. | 01/08/2016 | 02/05/2016 | 28 |
| **OR-GCKP-10-3-Al** | Al foil | Flight TAGSAM build; SC lift; moved SC to RAL; EMI/EMC testing; TVAC pre-cert. | 01/08/2016 | 02/05/2016 | 28 |
| **OR-GCKP-10-3-Si** | Si wafer | Flight TAGSAM build; SC lift; moved SC to RAL; EMI/EMC testing; TVAC pre-cert. | 01/08/2016 | 02/05/2016 | 28 |
| **OR-GCKP-10-4-Al** | Al foil | Flight TAGSAM build; SC lift; moved SC to RAL; EMI/EMC testing; TVAC pre-cert. | 01/08/2016 | 02/05/2016 | 28 |
| **OR-GCKP-10-4-Si** | Si wafer | Flight TAGSAM build; SC lift; moved SC to RAL; EMI/EMC testing; TVAC pre-cert. | 01/08/2016 | 02/05/2016 | 28 |
| **OR-GCKP-11-1-Al** | Al foil | EM11 removed; EM13 deployed; EM12 removed; EM14 deployed | 02/05/2016 | 03/15/2016 | 39 |
| **OR-GCKP-11-1-Si** | Si wafer | EM11 removed; EM13 deployed; EM12 removed; EM14 deployed | 02/05/2016 | 03/15/2016 | 39 |
| **OR-GCKP-11-2-Al** | Al foil | EM11 removed; EM13 deployed; EM12 removed; EM14 deployed | 02/05/2016 | 03/15/2016 | 39 |
| **OR-GCKP-11-2-Si** | Si wafer | EM11 removed; EM13 deployed; EM12 removed; EM14 deployed | 02/05/2016 | 03/15/2016 | 39 |
| **OR-GCKP-11-3-Al** | Al foil | EM11 removed; EM13 deployed; EM12 removed; EM14 deployed | 02/05/2016 | 03/15/2016 | 39 |
| **OR-GCKP-11-3-Si** | Si wafer | EM11 removed; EM13 deployed; EM12 removed; EM14 deployed | 02/05/2016 | 03/15/2016 | 39 |
| **OR-GCKP-11-4-Al** | Al foil | EM11 removed; EM13 deployed; EM12 removed; EM14 deployed | 02/05/2016 | 03/15/2016 | 39 |
| **OR-GCKP-11-4-Si** | Si wafer | EM11 removed; EM13 deployed; EM12 removed; EM14 deployed | 02/05/2016 | 03/15/2016 | 39 |
| **OR-GCKP-12-1-Al** | Al foil | Launch container deployment; SARA deployment | 03/15/2016 | 04/25/2016 | 41 |
| **OR-GCKP-12-1-Si** | Si wafer | Launch container deployment; SARA deployment | 03/15/2016 | 04/25/2016 | 41 |
| **OR-GCKP-12-2-Al** | Al foil | Launch container deployment; SARA deployment | 03/15/2016 | 04/25/2016 | 41 |
| **OR-GCKP-12-2-Si** | Si wafer | Launch container deployment; SARA deployment | 03/15/2016 | 04/25/2016 | 41 |
| **OR-GCKP-12-3-Al** | Al foil | Launch container deployment; SARA deployment | 03/15/2016 | 04/25/2016 | 41 |
| **OR-GCKP-12-3-Si** | Si wafer | Launch container deployment; SARA deployment | 03/15/2016 | 04/25/2016 | 41 |
| **OR-GCKP-12-4-Al** | Al foil | Launch container deployment; SARA deployment | 03/15/2016 | 04/25/2016 | 41 |
| **OR-GCKP-12-4-Si** | Si wafer | Launch container deployment; SARA deployment | 03/15/2016 | 04/25/2016 | 41 |
| **OR-GCKP-13-1-Al** | Al foil | Flight TAGSAM head exposure | 04/25/2016 | 04/26/2016 | 1 |
| **OR-GCKP-13-1-Si** | Si wafer | Flight TAGSAM head exposure | 04/25/2016 | 04/26/2016 | 1 |
| **OR-GCKP-13-2-Al** | Al foil | Flight TAGSAM head exposure | 04/25/2016 | 04/26/2016 | 1 |
| **OR-GCKP-13-2-Si** | Si wafer | Flight TAGSAM head exposure | 04/25/2016 | 04/26/2016 | 1 |
| **OR-GCKP-13-3-Al** | Al foil | Flight TAGSAM head exposure | 04/25/2016 | 04/26/2016 | 1 |
| **OR-GCKP-13-3-Si** | Si wafer | Flight TAGSAM head exposure | 04/25/2016 | 04/26/2016 | 1 |
| **OR-GCKP-13-4-Al** | Al foil | Flight TAGSAM head exposure | 04/25/2016 | 04/26/2016 | 1 |
| **OR-GCKP-13-4-Si** | Si wafer | Flight TAGSAM head exposure | 04/25/2016 | 04/26/2016 | 1 |
| **OR-GCKP-14-1-Al** | Al foil | End of TAGSAM install; move to KSC; KSC | 04/25/2016 | 06/16/2016 | 52 |
| **OR-GCKP-14-1-Si** | Si wafer | TAGSAM install; move to KSC; KSC | 04/25/2016 | 06/16/2016 | 52 |
| **OR-GCKP-14-2-Al** | Al foil | End of TAGSAM install; move to KSC; KSC | 04/25/2016 | 06/16/2016 | 52 |
| **OR-GCKP-14-2-Si** | Si wafer | End of TAGSAM install; move to KSC; KSC | 04/25/2016 | 06/16/2016 | 52 |
| **OR-GCKP-14-3-Al** | Al foil | End of TAGSAM install; move to KSC; KSC | 04/25/2016 | 06/16/2016 | 52 |
| **OR-GCKP-14-3-Si** | Si wafer | End of TAGSAM install; move to KSC; KSC | 04/25/2016 | 06/16/2016 | 52 |
| **OR-GCKP-14-4-Al** | Al foil | End of TAGSAM install; move to KSC; KSC | 04/25/2016 | 06/16/2016 | 52 |
| **OR-GCKP-14-4-Si** | Si wafer | End of TAGSAM install; move to KSC; KSC | 04/25/2016 | 06/16/2016 | 52 |
| **OR-GCKP-15-1-Al** | Al foil | KSC | 06/16/2016 | 07/13/2016 | 27 |
| **OR-GCKP-15-1-Si** | Si wafer | KSC | 06/16/2016 | 07/13/2016 | 27 |
| **OR-GCKP-15-2-Al** | Al foil | KSC | 06/16/2016 | 07/13/2016 | 27 |
| **OR-GCKP-15-2-Si** | Si wafer | KSC | 06/16/2016 | 07/13/2016 | 27 |
| **OR-GCKP-15-3-Al** | Al foil | KSC | 06/16/2016 | 07/13/2016 | 27 |
| **OR-GCKP-15-3-Si** | Si wafer | KSC | 06/16/2016 | 07/13/2016 | 27 |
| **OR-GCKP-15-4-Al** | Al foil | KSC | 06/16/2016 | 07/13/2016 | 27 |
| **OR-GCKP-15-4-Si** | Si wafer | KSC | 06/16/2016 | 07/13/2016 | 27 |
| **OR-GCKP-16-1-Al** | Al foil | KSC | 07/13/2016 | 08/25/2016 | 43 |
| **OR-GCKP-16-1-Si** | Si wafer | KSC | 07/13/2016 | 08/25/2016 | 43 |
| **OR-GCKP-16-2-Al** | Al foil | KSC | 07/13/2016 | 08/25/2016 | 43 |
| **OR-GCKP-16-2-Si** | Si wafer | KSC | 07/13/2016 | 08/25/2016 | 43 |
| **OR-GCKP-16-3-Al** | Al foil | KSC | 07/13/2016 | 08/25/2016 | 43 |
| **OR-GCKP-16-3-Si** | Si wafer | KSC | 07/13/2016 | 08/25/2016 | 43 |
| **OR-GCKP-16-4-Al** | Al foil | KSC | 07/13/2016 | 08/25/2016 | 43 |
| **OR-GCKP-16-4-Si** | Si wafer | KSC | 07/13/2016 | 08/25/2016 | 43 |
| **OR-GCKP-17-1-Al** | Al foil | Process blank; Dec. 17 Messenger lab, Dec. 19 RSV; procedural blank for DNA analyses | 12/19/2017 | 12/19/2017 | 0 |
| **OR-GCKP-17-2-Al** | Al foil | Process blank; Dec. 17 Messenger lab, Dec. 19 RSV; | 12/19/2017 | 12/19/2017 | 0 |
| **OR-GCKP-17-3-Al** | Al foil | Procedural blank ; Dec. 17 in Messenger lab; Dec. 19 in RSV lab | 12/19/2017 | 12/19/2017 | 0 |
| **OR-GCKP-17-4-AL** | Al foil | Procedural blank ; Dec. 17 in Messenger lab; Dec. 19 in RSV lab | 12/19/2017 | 12/19/2017 | 0 |
| **OR-GCKP-20-1-AL** | Al foil | Curation cleanroom (Building 31 Rm 216) after completion of construction | 10/14/2021 | 11/15/2021 | 32 |
| **OR-GCKP-20-2-AL** | Al foil | Curation cleanroom (Building 31 Rm 216) after completion of construction | 10/14/2021 | 11/15/2021 | 32 |
| **OR-GCKP-20-3-AL** | Al foil | Curation cleanroom (Building 31 Rm 216) after completion of construction | 10/14/2021 | 11/15/2021 | 32 |
| **OR-GCKP-20-4-AL** | Al foil | Curation cleanroom (Building 31 Rm 216) after completion of construction | 10/14/2021 | 11/15/2021 | 32 |
| **OR-GCKP-21-1-AL** | Al foil | Curation cleanroom (Building 31 Rm 216) after completion of construction | 10/14/2021 | 11/15/2021 | 32 |
| **OR-GCKP-21-2-AL** | Al foil | Curation cleanroom (Building 31 Rm 216) after completion of construction | 10/14/2021 | 11/15/2021 | 32 |
| **OR-GCKP-21-3-AL** | Al foil | Curation cleanroom (Building 31 Rm 216) after completion of construction | 10/14/2021 | 11/15/2021 | 32 |
| **OR-GCKP-21-4-AL** | Al foil | Curation cleanroom (Building 31 Rm 216) after completion of construction | 10/14/2021 | 11/15/2021 | 32 |
| **OR-GCKP-22-1-AL** | Al foil | Curation cleanroom (Building 31 Rm 216) after completion of construction | 11/15/2021 | 12/15/2021 | 30 |
| **OR-GCKP-22-2-AL** | Al foil | Curation cleanroom (Building 31 Rm 216) after completion of construction | 11/15/2021 | 12/15/2021 | 30 |
| **OR-GCKP-22-3-AL** | Al foil | Curation cleanroom (Building 31 Rm 216) after completion of construction | 11/15/2021 | 12/15/2021 | 30 |
| **OR-GCKP-22-4-AL** | Al foil | Curation cleanroom (Building 31 Rm 216) after completion of construction | 11/15/2021 | 12/15/2021 | 30 |
| **OR-GCKP-23-1-AL** | Al foil | Curation cleanroom (Building 31 Rm 216) after completion of construction | 11/15/2021 | 12/15/2021 | 30 |
| **OR-GCKP-23-2-AL** | Al foil | Curation cleanroom (Building 31 Rm 216) after completion of construction | 11/15/2021 | 12/15/2021 | 30 |
| **OR-GCKP-23-3-AL** | Al foil | Curation cleanroom (Building 31 Rm 216) after completion of construction | 11/15/2021 | 12/15/2021 | 30 |
| **OR-GCKP-23-4-AL** | Al foil | Curation cleanroom (Building 31 Rm 216) after completion of construction | 11/15/2021 | 12/15/2021 | 30 |
| **OR-GCKP-24-1-AL** | Al foil | Curation cleanroom (Building 31 Rm 216) after completion of construction | 12/15/2021 | 01/13/2022 | 29 |
| **OR-GCKP-24-2-AL** | Al foil | Curation cleanroom (Building 31 Rm 216) after completion of construction | 12/15/2021 | 01/13/2022 | 29 |
| **OR-GCKP-24-3-AL** | Al foil | Curation cleanroom (Building 31 Rm 216) after completion of construction | 12/15/2021 | 01/13/2022 | 29 |
| **OR-GCKP-24-4-AL** | Al foil | Curation cleanroom (Building 31 Rm 216) after completion of construction | 12/15/2021 | 01/13/2022 | 29 |
| **OR-GCKP-25-1-AL** | Al foil | Curation cleanroom (Building 31 Rm 216) after completion of construction | 12/15/2021 | 01/13/2022 | 29 |
| **OR-GCKP-25-2-AL** | Al foil | Curation cleanroom (Building 31 Rm 216) after completion of construction | 12/15/2021 | 01/13/2022 | 29 |
| **OR-GCKP-25-3-AL** | Al foil | Curation cleanroom (Building 31 Rm 216) after completion of construction | 12/15/2021 | 01/13/2022 | 29 |
| **OR-GCKP-25-4-AL** | Al foil | Curation cleanroom (Building 31 Rm 216) after completion of construction | 12/15/2021 | 01/13/2022 | 29 |
| **OR-GCKP-26-1-AL** | Al foil | Curation cleanroom (Building 31 Rm 216) after completion of construction | 01/13/2022 | 02/14/2022 | 32 |
| **OR-GCKP-26-2-AL** | Al foil | Curation cleanroom (Building 31 Rm 216) after completion of construction | 01/13/2022 | 02/14/2022 | 32 |
| **OR-GCKP-26-3-AL** | Al foil | Curation cleanroom (Building 31 Rm 216) after completion of construction | 01/13/2022 | 02/14/2022 | 32 |
| **OR-GCKP-26-4-AL** | Al foil | Curation cleanroom (Building 31 Rm 216) after completion of construction | 01/13/2022 | 02/14/2022 | 32 |
| **OR-GCKP-27-1-AL** | Al foil | Curation cleanroom (Building 31 Rm 216) after completion of construction | 01/13/2022 | 02/14/2022 | 32 |
| **OR-GCKP-27-2-AL** | Al foil | Curation cleanroom (Building 31 Rm 216) after completion of construction | 01/13/2022 | 02/14/2022 | 32 |
| **OR-GCKP-27-3-AL** | Al foil | Curation cleanroom (Building 31 Rm 216) after completion of construction | 01/13/2022 | 02/14/2022 | 32 |
| **OR-GCKP-27-4-AL** | Al foil | Curation cleanroom (Building 31 Rm 216) after completion of construction | 01/13/2022 | 02/14/2022 | 32 |
| **OR-GCKP-28-1-AL** | Al foil | Curation cleanroom (Building 31 Rm 216) after completion of construction | 02/14/2022 | 03/16/2022 | 30 |
| **OR-GCKP-28-2-AL** | Al foil | Curation cleanroom (Building 31 Rm 216) after completion of construction | 02/14/2022 | 03/16/2022 | 30 |
| **OR-GCKP-28-3-AL** | Al foil | Curation cleanroom (Building 31 Rm 216) after completion of construction | 02/14/2022 | 03/16/2022 | 30 |
| **OR-GCKP-28-4-AL** | Al foil | Curation cleanroom (Building 31 Rm 216) after completion of construction | 02/14/2022 | 03/16/2022 | 30 |
| **OR-GCKP-29-1-AL** | Al foil | Curation cleanroom (Building 31 Rm 216) after completion of construction | 02/14/2022 | 03/16/2022 | 30 |
| **OR-GCKP-29-2-AL** | Al foil | Curation cleanroom (Building 31 Rm 216) after completion of construction | 02/14/2022 | 03/16/2022 | 30 |
| **OR-GCKP-29-3-AL** | Al foil | Curation cleanroom (Building 31 Rm 216) after completion of construction | 02/14/2022 | 03/16/2022 | 30 |
| **OR-GCKP-29-4-AL** | Al foil | Curation cleanroom (Building 31 Rm 216) after completion of construction | 02/14/2022 | 03/16/2022 | 30 |
| **OR-GCKP-30-1-AL** | Al foil | Curation cleanroom (Building 31 Rm 216) after completion of construction | 03/16/2022 | 04/15/2022 | 30 |
| **OR-GCKP-30-2-AL** | Al foil | Curation cleanroom (Building 31 Rm 216) after completion of construction | 03/16/2022 | 04/15/2022 | 30 |
| **OR-GCKP-30-3-AL** | Al foil | Curation cleanroom (Building 31 Rm 216) after completion of construction | 03/15/2022 | 04/16/2022 | 32 |
| **OR-GCKP-30-4-AL** | Al foil | Curation cleanroom (Building 31 Rm 216) after completion of construction | 03/16/2022 | 04/15/2022 | 30 |
| **OR-GCKP-31-1-AL** | Al foil | Curation cleanroom (Building 31 Rm 216) after completion of construction | 03/16/2022 | 04/15/2022 | 30 |
| **OR-GCKP-31-2-AL** | Al foil | Curation cleanroom (Building 31 Rm 216) after completion of construction | 03/16/2022 | 04/15/2022 | 30 |
| **OR-GCKP-31-3-AL** | Al foil | Curation cleanroom (Building 31 Rm 216) after completion of construction | 03/16/2022 | 04/15/2022 | 30 |
| **OR-GCKP-31-4-AL** | Al foil | Curation cleanroom (Building 31 Rm 216) after completion of construction | 03/16/2022 | 04/15/2022 | 30 |
| **OR-GCKP-32-1-AL** | Al foil | Curation cleanroom (Building 31 Rm 216) after completion of construction | 04/15/2022 | 05/16/2022 | 31 |
| **OR-GCKP-32-2-AL** | Al foil | Curation cleanroom (Building 31 Rm 216) after completion of construction | 04/15/2022 | 05/16/2022 | 31 |
| **OR-GCKP-32-3-AL** | Al foil | Curation cleanroom (Building 31 Rm 216) after completion of construction | 04/15/2022 | 05/16/2022 | 31 |
| **OR-GCKP-32-4-AL** | Al foil | Curation cleanroom (Building 31 Rm 216) after completion of construction | 04/15/2022 | 05/16/2022 | 31 |
| **OR-GCKP-33-1-AL** | Al foil | Curation cleanroom (Building 31 Rm 216) after completion of construction | 04/15/2022 | 05/16/2022 | 31 |
| **OR-GCKP-33-2-AL** | Al foil | Curation cleanroom (Building 31 Rm 216) after completion of construction | 04/15/2022 | 05/16/2022 | 31 |
| **OR-GCKP-33-3-AL** | Al foil | Curation cleanroom (Building 31 Rm 216) after completion of construction | 04/15/2022 | 05/16/2022 | 31 |
| **OR-GCKP-33-4-AL** | Al foil | Curation cleanroom (Building 31 Rm 216) after completion of construction | 04/15/2022 | 05/16/2022 | 31 |
| **OR-GCKP-34-1-AL** | Al foil | Curation cleanroom (Building 31 Rm 216) after completion of construction | 05/16/2022 | 06/14/2022 | 29 |
| **OR-GCKP-34-2-AL** | Al foil | Curation cleanroom (Building 31 Rm 216) after completion of construction | 05/16/2022 | 06/14/2022 | 29 |
| **OR-GCKP-34-3-AL** | Al foil | Curation cleanroom (Building 31 Rm 216) after completion of construction | 05/16/2022 | 06/14/2022 | 29 |
| **OR-GCKP-34-4-AL** | Al foil | Curation cleanroom (Building 31 Rm 216) after completion of construction | 05/16/2022 | 06/14/2022 | 29 |
| **OR-GCKP-35-1-AL** | Al foil | Curation cleanroom (Building 31 Rm 216) after completion of construction | 05/16/2022 | 06/14/2022 | 29 |
| **OR-GCKP-35-2-AL** | Al foil | Curation cleanroom (Building 31 Rm 216) after completion of construction | 05/16/2022 | 06/14/2022 | 29 |
| **OR-GCKP-35-4-AL** | Al foil | Curation cleanroom (Building 31 Rm 216) after completion of construction | 05/16/2022 | 06/14/2022 | 29 |
| **OR-GCKP-36-1-AL** | Al foil | Curation cleanroom (Building 31 Rm 216) after completion of construction | 06/14/2022 | 07/21/2022 | 37 |
| **OR-GCKP-36-2-AL** | Al foil | Curation cleanroom (Building 31 Rm 216) after completion of construction | 06/14/2022 | 07/21/2022 | 37 |
| **OR-GCKP-36-3-AL** | Al foil | Curation cleanroom (Building 31 Rm 216) after completion of construction | 06/14/2022 | 07/21/2022 | 37 |
| **OR-GCKP-36-4-AL** | Al foil | Curation cleanroom (Building 31 Rm 216) after completion of construction | 06/14/2022 | 07/21/2022 | 37 |
| **OR-GCKP-37-1-AL** | Al foil | Curation cleanroom (Building 31 Rm 216) after completion of construction | 06/14/2022 | 07/21/2022 | 37 |
| **OR-GCKP-37-2-AL** | Al foil | Curation cleanroom (Building 31 Rm 216) after completion of construction | 06/14/2022 | 07/21/2022 | 37 |
| **OR-GCKP-37-3-AL** | Al foil | Curation cleanroom (Building 31 Rm 216) after completion of construction | 06/14/2022 | 07/21/2022 | 37 |
| **OR-GCKP-37-4-AL** | Al foil | Curation cleanroom (Building 31 Rm 216) after completion of construction | 06/14/2022 | 07/21/2022 | 37 |
| **OR-GCKP-38-1-AL** | Al foil | Curation cleanroom (Building 31 Rm 216) after completion of construction | 07/21/2022 | 08/25/2022 | 35 |
| **OR-GCKP-38-2-AL** | Al foil | Curation cleanroom (Building 31 Rm 216) after completion of construction | 07/21/2022 | 08/25/2022 | 35 |
| **OR-GCKP-38-3-AL** | Al foil | Curation cleanroom (Building 31 Rm 216) after completion of construction | 07/21/2022 | 08/25/2022 | 35 |
| **OR-GCKP-38-4-AL** | Al foil | Curation cleanroom (Building 31 Rm 216) after completion of construction | 07/21/2022 | 08/25/2022 | 35 |
| **OR-GCKP-39-1-AL** | Al foil | Curation cleanroom (Building 31 Rm 216) after completion of construction | 07/21/2022 | 08/25/2022 | 35 |
| **OR-GCKP-39-2-AL** | Al foil | Curation cleanroom (Building 31 Rm 216) after completion of construction | 07/21/2022 | 08/25/2022 | 35 |
| **OR-GCKP-39-3-AL** | Al foil | Curation cleanroom (Building 31 Rm 216) after completion of construction | 07/21/2022 | 08/25/2022 | 35 |
| **OR-GCKP-39-4-AL** | Al foil | Curation cleanroom (Building 31 Rm 216) after completion of construction | 07/21/2022 | 08/25/2022 | 35 |
| **OR-GCKP-40-1-AL** | Al foil | Curation cleanroom (Building 31 Rm 216) after completion of construction | 08/25/2022 | 09/26/2022 | 32 |
| **OR-GCKP-40-2-AL** | Al foil | Curation cleanroom (Building 31 Rm 216) after completion of construction | 08/25/2022 | 09/26/2022 | 32 |
| **OR-GCKP-40-3-AL** | Al foil | Curation cleanroom (Building 31 Rm 216) after completion of construction | 08/25/2022 | 09/26/2022 | 32 |
| **OR-GCKP-40-4-AL** | Al foil | Curation cleanroom (Building 31 Rm 216) after completion of construction | 08/25/2022 | 09/26/2022 | 32 |
| **OR-GCKP-41-1-AL** | Al foil | Curation cleanroom (Building 31 Rm 216) after completion of construction | 08/25/2022 | 09/26/2022 | 32 |
| **OR-GCKP-41-2-AL** | Al foil | Curation cleanroom (Building 31 Rm 216) after completion of construction | 08/25/2022 | 09/26/2022 | 32 |
| **OR-GCKP-41-3-AL** | Al foil | Curation cleanroom (Building 31 Rm 216) after completion of construction | 08/25/2022 | 09/26/2022 | 32 |
| **OR-GCKP-41-4-AL** | Al foil | Curation cleanroom (Building 31 Rm 216) after completion of construction | 08/25/2022 | 09/26/2022 | 32 |
| **OR-GCKP-42-1-AL** | Al foil | Curation cleanroom (Building 31 Rm 216) after completion of construction | 09/26/2022 | 10/27/2022 | 31 |
| **OR-GCKP-42-2-AL** | Al foil | Curation cleanroom (Building 31 Rm 216) after completion of construction | 09/26/2022 | 10/27/2022 | 31 |
| **OR-GCKP-42-3-AL** | Al foil | Curation cleanroom (Building 31 Rm 216) after completion of construction | 09/26/2022 | 10/27/2022 | 31 |
| **OR-GCKP-42-4-AL** | Al foil | Curation cleanroom (Building 31 Rm 216) after completion of construction | 09/26/2022 | 10/27/2022 | 31 |
| **OR-GCKP-43-1-AL** | Al foil | Curation cleanroom (Building 31 Rm 216) after completion of construction | 09/26/2022 | 10/27/2022 | 31 |
| **OR-GCKP-43-2-AL** | Al foil | Curation cleanroom (Building 31 Rm 216) after completion of construction | 09/26/2022 | 10/27/2022 | 31 |
| **OR-GCKP-43-3-AL** | Al foil | Curation cleanroom (Building 31 Rm 216) after completion of construction | 09/26/2022 | 10/27/2022 | 31 |
| **OR-GCKP-43-4-AL** | Al foil | Curation cleanroom (Building 31 Rm 216) after completion of construction | 09/26/2022 | 10/27/2022 | 31 |
| **OR-GCKP-44-1-AL** | Al foil | Curation cleanroom (Building 31 Rm 216) after completion of construction | 10/27/2022 | 12/06/2022 | 40 |
| **OR-GCKP-44-2-AL** | Al foil | Curation cleanroom (Building 31 Rm 216) after completion of construction | 10/27/2022 | 12/06/2022 | 40 |
| **OR-GCKP-44-3-AL** | Al foil | Curation cleanroom (Building 31 Rm 216) after completion of construction | 10/27/2022 | 12/06/2022 | 40 |
| **OR-GCKP-44-4-AL** | Al foil | Curation cleanroom (Building 31 Rm 216) after completion of construction | 10/27/2022 | 12/06/2022 | 40 |
| **OR-GCKP-45-1-AL** | Al foil | Curation cleanroom (Building 31 Rm 216) after completion of construction | 10/27/2022 | 12/06/2022 | 40 |
| **OR-GCKP-45-2-AL** | Al foil | Curation cleanroom (Building 31 Rm 216) after completion of construction | 10/27/2022 | 12/06/2022 | 40 |
| **OR-GCKP-45-3-AL** | Al foil | Curation cleanroom (Building 31 Rm 216) after completion of construction | 10/27/2022 | 12/06/2022 | 40 |
| **OR-GCKP-45-4-AL** | Al foil | Curation cleanroom (Building 31 Rm 216) after completion of construction | 10/27/2022 | 12/06/2022 | 40 |
| **OR-GCKP-46-1-AL** | Al foil | Curation cleanroom (Building 31 Rm 216) after completion of construction | 12/06/2022 | 01/04/2023 | 29 |
| **OR-GCKP-46-2-AL** | Al foil | Curation cleanroom (Building 31 Rm 216) after completion of construction | 12/06/2022 | 01/04/2023 | 29 |
| **OR-GCKP-46-3-AL** | Al foil | Curation cleanroom (Building 31 Rm 216) after completion of construction | 12/06/2022 | 01/04/2023 | 29 |
| **OR-GCKP-46-4-AL** | Al foil | Curation cleanroom (Building 31 Rm 216) after completion of construction | 12/06/2022 | 01/04/2023 | 29 |
| **OR-GCKP-47-1-AL** | Al foil | Curation cleanroom (Building 31 Rm 216) after completion of construction | 12/06/2022 | 01/04/2023 | 29 |
| **OR-GCKP-47-2-AL** | Al foil | Curation cleanroom (Building 31 Rm 216) after completion of construction | 12/06/2022 | 01/04/2023 | 29 |
| **OR-GCKP-47-3-AL** | Al foil | Curation cleanroom (Building 31 Rm 216) after completion of construction | 12/06/2022 | 01/04/2023 | 29 |
| **OR-GCKP-47-4-AL** | Al foil | Curation cleanroom (Building 31 Rm 216) after completion of construction | 12/06/2022 | 01/04/2023 | 29 |
| **OR-GCKP-48-1-AL** | Al foil | Curation cleanroom (Building 31 Rm 216) after completion of construction | 01/04/2023 | 02/03/2023 | 30 |
| **OR-GCKP-48-2-AL** | Al foil | Curation cleanroom (Building 31 Rm 216) after completion of construction | 01/04/2023 | 02/03/2023 | 30 |
| **OR-GCKP-48-3-AL** | Al foil | Curation cleanroom (Building 31 Rm 216) after completion of construction | 01/04/2023 | 02/03/2023 | 30 |
| **OR-GCKP-48-4-AL** | Al foil | Curation cleanroom (Building 31 Rm 216) after completion of construction | 01/04/2023 | 02/03/2023 | 30 |
| **OR-GCKP-49-1-AL** | Al foil | Curation cleanroom (Building 31 Rm 216) after completion of construction | 01/04/2023 | 02/03/2023 | 30 |
| **OR-GCKP-49-2-AL** | Al foil | Curation cleanroom (Building 31 Rm 216) after completion of construction | 01/04/2023 | 02/03/2023 | 30 |
| **OR-GCKP-49-3-AL** | Al foil | Curation cleanroom (Building 31 Rm 216) after completion of construction | 01/04/2023 | 02/03/2023 | 30 |
| **OR-GCKP-49-4-AL** | Al foil | Curation cleanroom (Building 31 Rm 216) after completion of construction | 01/04/2023 | 02/03/2023 | 30 |
| **OR-GCKP-50-1-AL** | Al foil | Curation cleanroom (Building 31 Rm 216) after completion of construction | 02/03/2023 | 03/06/2023 | 31 |
| **OR-GCKP-50-2-AL** | Al foil | Curation cleanroom (Building 31 Rm 216) after completion of construction | 02/03/2023 | 03/06/2023 | 31 |
| **OR-GCKP-50-3-AL** | Al foil | Curation cleanroom (Building 31 Rm 216) after completion of construction | 02/03/2023 | 03/06/2023 | 31 |
| **OR-GCKP-50-4-AL** | Al foil | Curation cleanroom (Building 31 Rm 216) after completion of construction | 02/03/2023 | 03/06/2023 | 31 |
| **OR-GCKP-51-1-AL** | Al foil | Curation cleanroom (Building 31 Rm 216) after completion of construction | 02/03/2023 | 03/06/2023 | 31 |
| **OR-GCKP-51-2-AL** | Al foil | Curation cleanroom (Building 31 Rm 216) after completion of construction | 02/03/2023 | 03/06/2023 | 31 |
| **OR-GCKP-51-3-AL** | Al foil | Curation cleanroom (Building 31 Rm 216) after completion of construction | 02/03/2023 | 03/06/2023 | 31 |
| **OR-GCKP-51-4-AL** | Al foil | Curation cleanroom (Building 31 Rm 216) after completion of construction | 02/03/2023 | 03/06/2023 | 31 |
| **OR-GCKP-52-1-AL** | Al foil | Curation cleanroom (Building 31 Rm 216) after completion of construction | 03/06/2023 | 04/10/2023 | 35 |
| **OR-GCKP-52-2-AL** | Al foil | Curation cleanroom (Building 31 Rm 216) after completion of construction | 03/06/2023 | 04/10/2023 | 35 |
| **OR-GCKP-52-3-AL** | Al foil | Curation cleanroom (Building 31 Rm 216) after completion of construction | 03/06/2023 | 04/10/2023 | 35 |
| **OR-GCKP-52-4-AL** | Al foil | Curation cleanroom (Building 31 Rm 216) after completion of construction | 03/06/2023 | 04/10/2023 | 35 |
| **OR-GCKP-53-1-AL** | Al foil | Curation cleanroom (Building 31 Rm 216) after completion of construction | 03/06/2023 | 04/10/2023 | 35 |
| **OR-GCKP-53-2-AL** | Al foil | Curation cleanroom (Building 31 Rm 216) after completion of construction | 03/06/2023 | 04/10/2023 | 35 |
| **OR-GCKP-53-3-AL** | Al foil | Curation cleanroom (Building 31 Rm 216) after completion of construction | 03/06/2023 | 04/10/2023 | 35 |
| **OR-GCKP-53-4-AL** | Al foil | Curation cleanroom (Building 31 Rm 216) after completion of construction | 03/06/2023 | 04/10/2023 | 35 |
| **OR-GCKP-54-1-AL** | Al foil | Curation cleanroom (Building 31 Rm 216) after completion of construction | 04/10/2023 | 05/17/2023 | 37 |
| **OR-GCKP-54-2-AL** | Al foil | Curation cleanroom (Building 31 Rm 216) after completion of construction | 04/10/2023 | 05/17/2023 | 37 |
| **OR-GCKP-54-3-AL** | Al foil | Curation cleanroom (Building 31 Rm 216) after completion of construction | 04/10/2023 | 05/17/2023 | 37 |
| **OR-GCKP-54-4-AL** | Al foil | Curation cleanroom (Building 31 Rm 216) after completion of construction | 04/10/2023 | 05/17/2023 | 37 |
| **OR-GCKP-55-1-AL** | Al foil | Curation cleanroom (Building 31 Rm 216) after completion of construction | 04/10/2023 | 05/17/2023 | 37 |
| **OR-GCKP-55-2-AL** | Al foil | Curation cleanroom (Building 31 Rm 216) after completion of construction | 04/10/2023 | 05/17/2023 | 37 |
| **OR-GCKP-55-3-AL** | Al foil | Curation cleanroom (Building 31 Rm 216) after completion of construction | 04/10/2023 | 05/17/2023 | 37 |
| **OR-GCKP-55-4-AL** | Al foil | Curation cleanroom (Building 31 Rm 216) after completion of construction | 04/10/2023 | 05/17/2023 | 37 |
| **OR-GCKP-56-1-AL** | Al foil | Curation cleanroom (Building 31 Rm 216) after completion of construction | 05/17/2023 | 06/14/2023 | 28 |
| **OR-GCKP-56-2-AL** | Al foil | Curation cleanroom (Building 31 Rm 216) after completion of construction | 05/17/2023 | 06/14/2023 | 28 |
| **OR-GCKP-56-3-AL** | Al foil | Curation cleanroom (Building 31 Rm 216) after completion of construction | 05/17/2023 | 06/14/2023 | 28 |
| **OR-GCKP-56-4-AL** | Al foil | Curation cleanroom (Building 31 Rm 216) after completion of construction | 05/17/2023 | 06/14/2023 | 28 |
| **OR-GCKP-57-1-AL** | Al foil | Curation cleanroom (Building 31 Rm 216) after completion of construction | 05/17/2023 | 06/14/2023 | 28 |
| **OR-GCKP-57-2-AL** | Al foil | Curation cleanroom (Building 31 Rm 216) after completion of construction | 05/17/2023 | 06/14/2023 | 28 |
| **OR-GCKP-57-3-AL** | Al foil | Curation cleanroom (Building 31 Rm 216) after completion of construction | 05/17/2023 | 06/14/2023 | 28 |
| **OR-GCKP-57-4-AL** | Al foil | Curation cleanroom (Building 31 Rm 216) after completion of construction | 05/17/2023 | 06/14/2023 | 28 |
| **OR-GCKP-58-1-AL** | Al foil | Curation cleanroom (Building 31 Rm 216) after completion of construction | 06/14/2023 | 07/12/2023 | 28 |
| **OR-GCKP-58-2-AL** | Al foil | Curation cleanroom (Building 31 Rm 216) after completion of construction | 06/14/2023 | 07/12/2023 | 28 |
| **OR-GCKP-58-3-AL** | Al foil | Curation cleanroom (Building 31 Rm 216) after completion of construction | 06/14/2023 | 07/12/2023 | 28 |
| **OR-GCKP-58-4-AL** | Al foil | Curation cleanroom (Building 31 Rm 216) after completion of construction | 06/14/2023 | 07/12/2023 | 28 |
| **OR-GCKP-59-1-AL** | Al foil | Curation cleanroom (Building 31 Rm 216) after completion of construction | 06/14/2023 | 07/12/2023 | 28 |
| **OR-GCKP-59-2-AL** | Al foil | Curation cleanroom (Building 31 Rm 216) after completion of construction | 06/14/2023 | 07/12/2023 | 28 |
| **OR-GCKP-59-3-AL** | Al foil | Curation cleanroom (Building 31 Rm 216) after completion of construction | 06/14/2023 | 07/12/2023 | 28 |
| **OR-GCKP-59-4-AL** | Al foil | Curation cleanroom (Building 31 Rm 216) after completion of construction | 06/14/2023 | 07/12/2023 | 28 |
| **OR-GCKP-60-1-AL** | Al foil | Curation cleanroom (Building 31 Rm 216) after completion of construction | 06/14/2023 | 07/12/2023 | 28 |
| **OR-GCKP-60-2-AL** | Al foil | Curation cleanroom (Building 31 Rm 216) after completion of construction | 06/14/2023 | 07/12/2023 | 28 |
| **OR-GCKP-60-3-AL** | Al foil | Curation cleanroom (Building 31 Rm 216) after completion of construction | 06/14/2023 | 07/12/2023 | 28 |
| **OR-GCKP-60-4-AL** | Al foil | Curation cleanroom (Building 31 Rm 216) after completion of construction | 06/14/2023 | 07/12/2023 | 28 |
| **OR-GCKP-62-1-AL** | Al foil | Curation cleanroom (Building 31 Rm 216) after completion of construction | 06/14/2023 | 07/12/2023 | 28 |
| **OR-GCKP-62-2-AL** | Al foil | Curation cleanroom (Building 31 Rm 216) after completion of construction | 06/14/2023 | 07/12/2023 | 28 |
| **OR-GCKP-62-3-AL** | Al foil | Curation cleanroom (Building 31 Rm 216) after completion of construction | 06/14/2023 | 07/12/2023 | 28 |
| **OR-GCKP-62-4-AL** | Al foil | Curation cleanroom (Building 31 Rm 216) after completion of construction | 06/14/2023 | 07/12/2023 | 28 |
| **OR-GCKP-64-1-AL** | Al foil | Curation cleanroom (Building 31 Rm 216) after completion of construction | 08/10/2023 | 09/11/2023 | 32 |
| **OR-GCKP-64-2-AL** | Al foil | Curation cleanroom (Building 31 Rm 216) after completion of construction | 08/10/2023 | 09/11/2023 | 32 |
| **OR-GCKP-64-3-AL** | Al foil | Curation cleanroom (Building 31 Rm 216) after completion of construction | 08/10/2023 | 09/11/2023 | 32 |
| **OR-GCKP-64-4-AL** | Al foil | Curation cleanroom (Building 31 Rm 216) after completion of construction | 08/10/2023 | 09/11/2023 | 32 |
| **OR-GCKP-66-1-AL** | Al foil | Curation cleanroom (Building 31 Rm 216) | 09/11/2023 | 10/13/2023 | 32 |
| **OR-GCKP-66-2-AL** | Al foil | Curation cleanroom (Building 31 Rm 216) | 09/11/2023 | 10/13/2023 | 32 |
| **OR-GCKP-66-3-AL** | Al foil | Curation cleanroom (Building 31 Rm 216) | 09/11/2023 | 10/13/2023 | 32 |
| **OR-GCKP-66-4-AL** | Al foil | Curation cleanroom (Building 31 Rm 216) | 09/11/2023 | 10/13/2023 | 32 |
| **OR-GCKP-67-1-AL** | Al foil | Curation cleanroom (Building 31 Rm 216) | 10/13/2023 | 12/04/2023 | 52 |
| **OR-GCKP-67-2-AL** | Al foil | Curation cleanroom (Building 31 Rm 216) | 10/13/2023 | 12/04/2023 | 52 |
| **OR-GCKP-67-3-AL** | Al foil | Curation cleanroom (Building 31 Rm 216) | 10/13/2023 | 12/04/2023 | 52 |
| **OR-GCKP-67-4-AL** | Al foil | Curation cleanroom (Building 31 Rm 216) | 10/13/2023 | 12/04/2023 | 52 |
| **OR-GCKP-68-1-AL** | Al foil | Curation cleanroom (Building 31 Rm 216) | 12/04/2023 | 01/08/2024 | 35 |
| **OR-GCKP-68-2-AL** | Al foil | Curation cleanroom (Building 31 Rm 216) | 12/04/2023 | 01/08/2024 | 35 |
| **OR-GCKP-68-3-AL** | Al foil | Curation cleanroom (Building 31 Rm 216) | 12/04/2023 | 01/08/2024 | 35 |
| **OR-GCKP-68-4-AL** | Al foil | Curation cleanroom (Building 31 Rm 216) | 12/04/2023 | 01/08/2024 | 35 |
| **OR-GCKP-69-1-AL** | Al foil | Curation cleanroom (Building 31 Rm 216) | 01/08/2024 | 04/12/2024 | 95 |
| **OR-GCKP-69-2-AL** | Al foil | Curation cleanroom (Building 31 Rm 216) | 01/08/2024 | 04/12/2024 | 95 |
| **OR-GCKP-69-3-AL** | Al foil | Curation cleanroom (Building 31 Rm 216) | 01/08/2024 | 04/12/2024 | 95 |
| **OR-GCKP-69-4-AL** | Al foil | Curation cleanroom (Building 31 Rm 216) | 01/08/2024 | 04/12/2024 | 95 |
| **OR-GCKP-70-1-AL** | Al foil | Curation cleanroom (Building 31 Rm 216) | 04/12/2024 | 06/20/2024 | 69 |
| **OR-GCKP-70-2-AL** | Al foil | Curation cleanroom (Building 31 Rm 216) | 04/12/2024 | 06/20/2024 | 69 |
| **OR-GCKP-70-3-AL** | Al foil | Curation cleanroom (Building 31 Rm 216) | 04/12/2024 | 06/20/2024 | 69 |
| **OR-GCKP-70-4-AL** | Al foil | Curation cleanroom (Building 31 Rm 216) | 04/12/2024 | 06/20/2024 | 69 |
| **OR-GCKP-71-1-AL** | Al foil | Curation cleanroom (Building 31 Rm 216) | 06/20/2024 | 05/09/2025 | 323 |
| **OR-GCKP-71-2-AL** | Al foil | Curation cleanroom (Building 31 Rm 216) | 06/20/2024 | 05/09/2025 | 323 |
| **OR-GCKP-71-3-AL** | Al foil | Curation cleanroom (Building 31 Rm 216) | 06/20/2024 | 05/09/2025 | 323 |
| **OR-GCKP-71-4-AL** | Al foil | Curation cleanroom (Building 31 Rm 216) | 06/20/2024 | 05/09/2025 | 323 |
| **OR-GCKP-72-1-AL** | Al foil | Curation cleanroom (Building 31 Rm 216) | 05/09/2025 | 07/08/2025 | 60 |
| **OR-GCKP-72-2-AL** | Al foil | Curation cleanroom (Building 31 Rm 216) | 05/09/2025 | 07/08/2025 | 60 |
| **OR-GCKP-72-3-AL** | Al foil | Curation cleanroom (Building 31 Rm 216) | 05/09/2025 | 07/08/2025 | 60 |
| **OR-GCKP-72-4-AL** | Al foil | Curation cleanroom (Building 31 Rm 216) | 05/09/2025 | 07/08/2025 | 60 |
| **OR-GCKP-73-1-AL** | Al foil | Curation cleanroom (Building 31 Rm 216) | 07/08/2025 | 09/12/2025 | 66 |
| **OR-GCKP-73-2-AL** | Al foil | Curation cleanroom (Building 31 Rm 216) | 07/08/2025 | 09/12/2025 | 66 |
| **OR-GCKP-73-3-AL** | Al foil | Curation cleanroom (Building 31 Rm 216) | 07/08/2025 | 09/12/2025 | 66 |
| **OR-GCKP-73-4-AL** | Al foil | Curation cleanroom (Building 31 Rm 216) | 07/08/2025 | 09/12/2025 | 66 |
| **OR-GCKP-74-1-AL** | Al foil | Curation cleanroom (Building 31 Rm 216) | 09/12/2025 | 11/20/2025 | 69 |
| **OR-GCKP-74-2-AL** | Al foil | Curation cleanroom (Building 31 Rm 216) | 09/12/2025 | 11/20/2025 | 69 |
| **OR-GCKP-74-3-AL** | Al foil | Curation cleanroom (Building 31 Rm 216) | 09/12/2025 | 11/20/2025 | 69 |
| **OR-GCKP-74-4-AL** | Al foil | Curation cleanroom (Building 31 Rm 216) | 09/12/2025 | 11/20/2025 | 69 |
| **OR-GCKP-75-1-AL** | Al foil | Curation cleanroom (Building 31 Rm 216) | 11/20/2025 | 01/30/2026 | 71 |
| **OR-GCKP-75-2-AL** | Al foil | Curation cleanroom (Building 31 Rm 216) | 11/20/2025 | 01/30/2026 | 71 |
| **OR-GCKP-75-3-AL** | Al foil | Curation cleanroom (Building 31 Rm 216) | 11/20/2025 | 01/30/2026 | 71 |
| **OR-GCKP-75-4-AL** | Al foil | Curation cleanroom (Building 31 Rm 216) | 11/20/2025 | 01/30/2026 | 71 |

**Supplement 3:** SART report of archived flight hardware

# CCWG Report SART-5-040

Jason Dworkin

Updated From

30 April 2023

## Table of Contents

**Test Objective**
This SART activity consists of the investigation of three different components of potential concern in the sample return capsule (SRC) canister. Two of these were generated by the analysis of components from SART-4-003: the stainless steel convoluted tube (Fig. S3.1a) and Henkel Loctite EA9394, which was used inside of the canister air filter (Fig. S3.1b). The mineralogy and petrology working group (MAPWG) raised a concern about the mylar (biaxially-oriented polyethylene terephthalate polyester) on the touch and go sample acquisition mechanism (TAGSAM) head (Fig. S3.1c) so that was bundled with this activity. The objective is to become familiar with the signatures of contamination from these three materials to permit the sample analysis team to rapidly identify and discriminate against these contaminants.

A B C

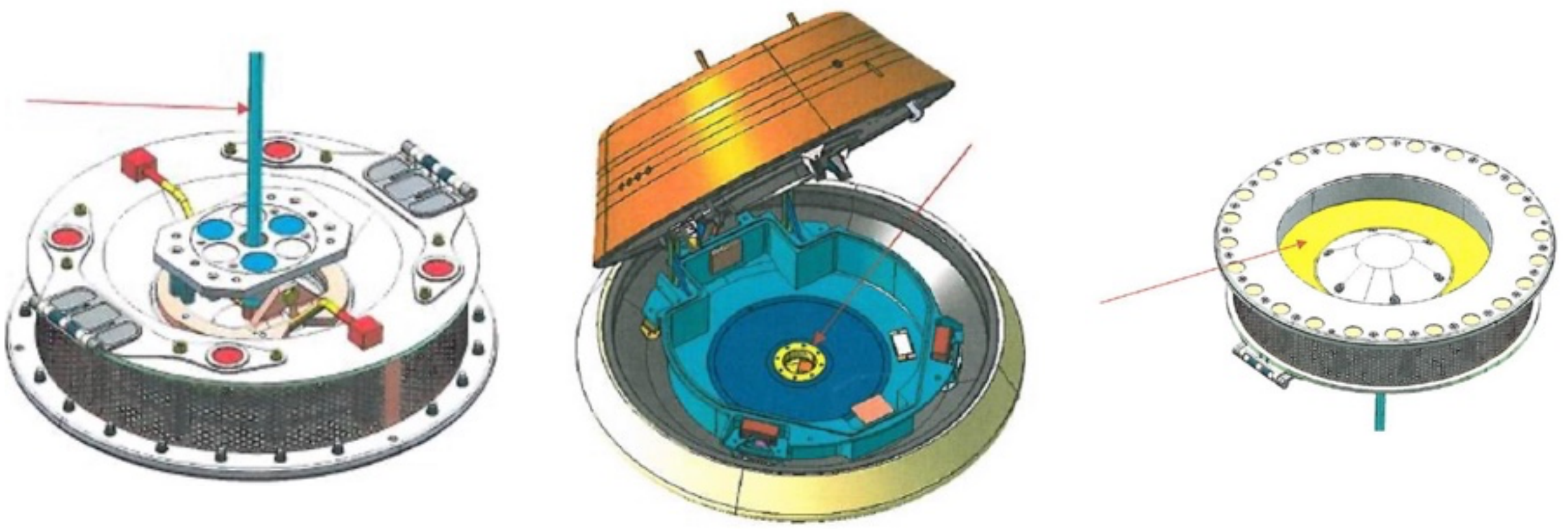

*Fig. S2-1 A. TAGSAM head with a red arrow pointing to the convoluted tube. B. SRC with a red arrow pointing to the location of Henkel Loctite EA9394 epoxy in the air filter. C. TAGSAM head with a red arrow pointing to the mylar flap.*

**Success Criteria**
Establish the contamination risk from these three materials.

**Summary of Results**
Thus, SART-5-040 determined that these materials are not a concern for sample analysis, since any contamination from these materials should be readily identifiable in the sample.

- Convoluted tube steel is unremarkable and recognizable. Care will need to be taken to look at elemental and isotopic ratios of elements in common with these steels.
- Henkel Loctite EA9394 epoxy is generally recognizable by microscopy, resistant to extraction, and obvious in pyrolysis.
- Henkel Loctite EA9394 epoxy, like in the case of steel contamination, by the presence of the two simplest amines–methylamine and ethylamine (but no other amines)–means that care needs to be taken in the interpretation of these two compounds.
- No evidence of Henkel Loctite EA9394 epoxy was seen in ATLO witness plates or in the air filter analyses conducted in parallel for SART-5-038.
- Mylar is distinct and readily recognizable.
- Some researchers work to (extended) deadlines and leave no margin if things go wrong.

## Procedure

Contamination and curation conferred to determine which samples from the materials archive were best suited to the analyses and in sufficient quantity. The contamination control/knowledge working group (CCWG) lead conferred with the relevant working groups and volunteer analysts to determine the analyses and sample requirements for each. CCWG assisted the analysts with the requisition of the materials from curation. The analysts studied the material with the relevant techniques and produced a report.

### *Selection of materials*

There was a wide selection of the Henkel Loctite EA9394 samples from the materials archive (Table S3.1). Sample OR-MA-0221 was selected for analysis since it was indicated in spacecraft drawing NFP30643142 of the SRC air filter so has a high probability of being one of the samples used for flight since it had the most commonly used batch and lot numbers, it is a large sample, and consuming part of it would still leave two large examples untouched.

The selection of convoluted tube and mylar were simpler since the traceability to the flight unit was clearer as only a single example was in the archive, Convoluted tube OR-MA-0303 and Mylar OR-MA-0266. The samples from the archive are shown in Fig. S3.2.

**Table S3-1** Selection of Henkel Loctite EA9394 samples used in the SRC air filter available for study. A chip of sample OR-MA-0221 (blue) was selected for analysis. The distinction between batch and lot was unclear.

| Curation # | Batch | Lot | Mass | Major Component | Drawing | Where |
|---|---|---|---|---|---|---|
| OR-MA-0218 | 950112 | JH4AAH5278 | 67 g | SRC | NFP30643142 | SRC Filter Assembly |
| OR-MA-0219 | 943012 | JH4AAH5278 | 56 g | SRC | – | SRC Filter Assembly |
| OR-MA-0221 | 943012 | JH4AAG6437 | 71 g | SRC | NFP30643142 | SRC Filter Assembly |
| OR-MA-0225 | 943012 | JH4AAG6437 | 70 g | SRC | NFP30643142 | SRC Filter Assembly |
| OR-MA-0227 | 950112 | JH4AAH5278 | 80 g | SRC | NFP30643142 | SRC Filter Assembly |
| OR-MA-0236 | 943012 | JH4AAG6437 | 70 g | SRC | NFP30643142 | Filter Assembly |
| OR-MA-0238 | 787384 | JH2DAG5892 | 67 g | SRC | NFP30643142 | Lap Shear Testing |

A B C

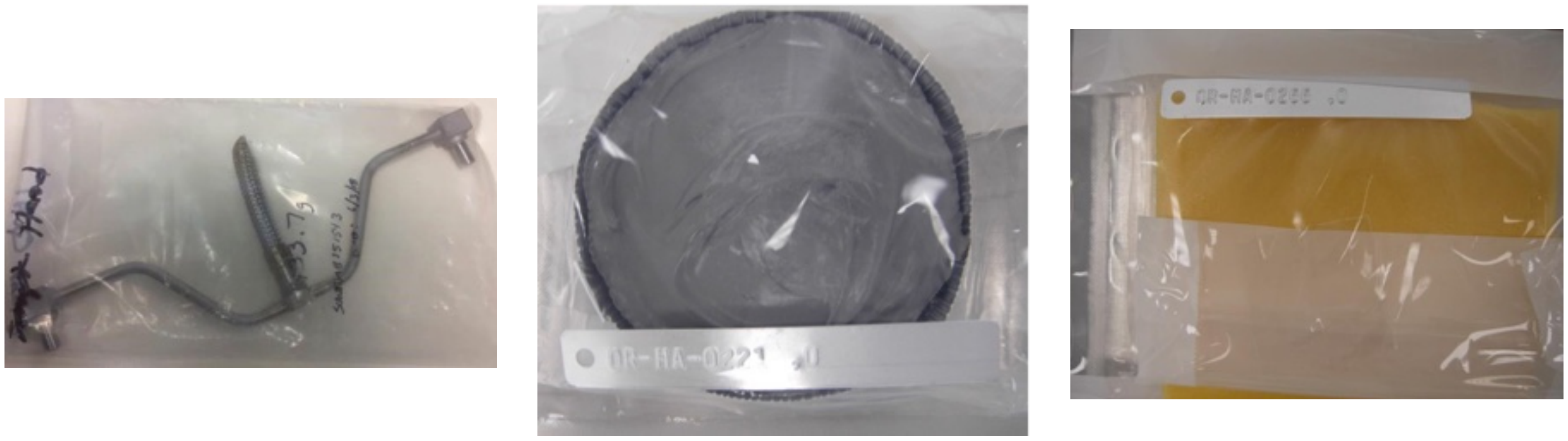

*Fig. S2-2 A. Convoluted tube OR-MA-0303. B. Henkel Loctite EA9394 epoxy OR-MA-0221. C. TAGSAM mylar flap OR-MA-0266.*

***Analyses***

The analyses of the convoluted tube were via two different inductively coupled plasma mass spectrometry (ICP-MS) instruments on behalf of Sample Elements and Isotopes Working Group (SEIWG). Greg Brennecka, Josh Wimpenny, and Jan Render (Lawrence Livermore National Laboratory; LLNL using a Thermo Scientific Element XR High-Resolution ICP-MS) and Dominique Weis, Vivian Lai, and Marg Amini (University of British Columbia, UBC using a Thermo Scientific Element 2 high-resolution sector-field-ICP-MS) and an Agilent 7700x Quadrupole CP-MS analyzed the wire braid of the convoluted tube as well as also the steel core. Both laboratories employed the methods described in Wang et al. (in press).

The analysis of Henkel Loctite EA9394 was conducted by multiple analysts and techniques within MAPWG and Sample Organic Analysis Working Group (SOAWG). Some techniques were of a fragment of Henkel Loctite EA9394 while others were of a solvent (water or dichloromethane) extract.

The methods were Scanning Electron Microscopy with Energy-Dispersive X-ray Spectroscopy (SEM/EDX) at Smithsonian Institution National Museum of Natural History (NMNH) on a Thermo Fisher FEI Quattro Analytical FE-SEM equipped with two opposing Thermo Fisher UltraDry 60 $mm^2$ EDS via the methods in McCoy & Russell et al. (2025) and a Hitachi S-4800 SEM located in the Kuiper-Arizona Laboratory for Astromaterials Analysis (K-ALFAA). BSE and WDS The Hitachi S-4800 cold-field emission gun (cold FEG) is a 0.5 to 30 keV SEM equipped with SE and BSE imaging detectors and an Oxford Instruments Aztec Live/x-stream/Ultimax 170 SDD EDX according to the methods in Lauretta & Connolly et al. (2024).

The elemental and isotopic analyses of C and N were carried out at the Carnegie Institution for Science (CIS) conducted by a Thermo Scientific Delta $V^{Plus}$ mass spectrometer that was interfaced with a Carlo Erba (NA 2500) elemental analyzer via the methods in Lauretta & Connolly et al. (2024).

Pyrolysis-gas chromatography mass spectrometry (GCMS) at CIS was performed with a CDS 1000 pyrolyzer interfaced with an Agilent 6890 GC and 5972 MS and at NASA’s Goddard Space Flight Center (GSFC) with a CDS Analytical 6200 pyroprobe feeding a Thermo Scientific TRACE 1610 gas chromatograph coupled to a Thermo Scientific 9610 triple quadrupole mass spectrometer system via the methods of Mojarro et al. (2025). Two-dimensional high-resolution GC-MS (GC×GC-HRMS) at GSFC using an Agilent 7890B gas chromatograph coupled to a LECO Pegasus HRT+ 4D time-of-flight mass spectrometer was used for an untargeted survey of dichloromethane extracts via the methods of Aponte et al. (2026). GCMS of hexane and dichloromethane extracts was performed at Japan Agency for Marine-Earth Science and Technology (JAMSTEC) for hydrocarbons using a Thermo Trace 13100 and Thermo TSQ via the methods of Naraoka et al. (2023). GCMS of dichloromethane extracts was performed at NASA’s Jet Propulsion Laboratory/California Institute of Technology for polycyclic aromatic hydrocarbons with a Thermo Fisher Trace 1310 GC and QExactive Orbitrap HRMS using the methods of Zeichner et al. (2023).

Liquid chromatography mass spectrometry (LCMS) of hot water extracts was performed at GSFC with a Waters ACQUITY with a Waters Xevo TQS-Micro triple quadrupole MS for quantification of amino acids via the methods of (Dworkin et al. 2018). LCMS at GSFC for amines with a Thermo

Fisher Scientific Vanquish Horizon LC, Vanquish fluorescence detector, and Q Exactive hybrid quadrupole-Orbitrap HRMS using the methods of Glavin & Dworkin et al. (2025). LCMS analysis of N-heterocycles was conducted at JAMSTEC using an Agilent 1100 series LC and 6460 Triple Quadrupole MS using the methods of Glavin & Dworkin et al. (2025).

The SEM/EDX analysis of mylar was conducted by Lindsay Keller and Kathie Thomas-Keprta at Johnson Space Center (JSC) on behalf of Tom Zega and Tim McCoy of MAPWG via a JEOL 7600 FESEM equipped with an Oxford Ultim Max 170 EDX spectrometer via the methods in Lauretta & Connolly et al. (2024). Hyperion 3000 IR microscope attached to a Bruker Vertex 70 IR was used to collect a transmission infrared spectrum at JSC using the method in Lauretta & Connolly et al. (2024).

## Results

### *Convoluted tube*

The convoluted tube was analyzed for major and trace elemental abundances by two slightly different ICP-MS instruments employing similar methods. The measured compositions are consistent with 304 stainless steel for the braid and 316L for the tube core. The composition of 304 stainless steel is expected to be Fe, Cr (18-20%), Ni (8-11%), with <2% Mn, Si, C, P, and S and the composition of 316L stainless steel is expected to be Fe, Cr (16-18%), Ni (10-12%), Mo (2-3%) with <1% Si, P, and S. A distribution of minor elements was also observed. These results are in Table S3.2. The results of the two methods agree broadly within a 10% uncertainty envelope. There are minor inconsistencies for a few elements, but these are mostly related to the extremely low abundance elements, high dilution, or potential laboratory artifacts. However, a cross comparison between the two laboratories

**Table S3-2** Elemental abundances (μg/g) provided by LLNL and UBC for the convoluted tube braid (one 11 mg to LLNL and two samples of 11.5 and 27.11 mg to UBC) and by UBC for the core tube (12.85 mg). LLNL data have a conservative relative uncertainty of 10% for trace elements, whereas uncertainties are 1σ for UBC data. For major elements are 25% due to the lack of isotope dilution

| | Braid LLNL | Braid UBC | Tube UBC | | Braid LLNL | Braid UBC | Tube UBC |
|---|---|---|---|---|---|---|---|
| Li | 0.39 | <LOD (0.008) | <LOD (0.008) | Ag | 0.74 | 0.67±0.19 | 1.67±0.06 |
| Be | 0.010 | | | Cd | 32.86 | 32.98±6.27 | 35.47±0.45 |
| B | | 4.2±0.5 | 36±0.6 | Sn | 47.09 | 122.4±10 | 121.74±3.20 |
| Na | 22.66 | | | Sb | | 28.3±5.8 | 17.43±0.41 |
| Mg | 3.97 | | | Cs | 0.36[b] | 0.055±0.008[b] | 0.164±0.005[b] |
| Al | 4.44 | | | Ba | 0.29 | <LOD (0.067) | <LOD (0.067) |
| Si | | 4242±845 | <LOD (0.001) | La | 0.016 | <LOD (0.002) | <LOD (0.002) |
| P | | 366±39 | | Ce | 0.056 | <LOD (0.009) | <LOD (0.009) |
| K | 24.81 | | | Pr | 0.22 | <LOD (0.007) | <LOD (0.007) |
| Ca | 26.54 | | | Nd | 1.07 | <LOD (0.008) | <LOD (0.008) |
| Sc | 0.008 | <LOD (0.01) | <LOD (0.01) | Sm | 1.61 | <LOD (0.002) | <LOD (0.002) |
| Ti | 6.23 | | | Eu | 0.059 | <LOD (0.009) | <LOD (0.009) |
| V | 755.2 | 636.7±11.5 | 903.81±7.1 | Tb | 0.008 | <LOD (0.005) | <LOD (0.005) |
| Cr | 176758 | 180063±8707 | 161168±2208 | Gd | 0.26 | <LOD (0.05) | <LOD (0.05) |
| Mn | 12201 | 13786±805 | 18826±350 | Dy | 0.014 | <LOD (0.02) | <LOD (0.02) |
| Fe | 667108 | 704187±36914 | 704405±10284 | Ho | 0.0017 | <LOD (0.007) | <LOD (0.007) |
| Co | 1790.0 | 1775±10 | 3914±25 | Er | 0.21 | <LOD (0.02) | <LOD (0.02) |
| Ni | 111883 | 111899±6880 | 97357±1373 | Tm | 0.27 | <LOD (0.007) | <LOD (0.007) |
| Cu | 8753 | 10751±1502 | 5148±51 | Yb | 0.005 | <LOD (0.03) | <LOD (0.03) |
| Zn | 4.81 | | 25.53±3.97 | Lu | 0.0004 | <LOD (0.006) | <LOD (0.006) |
| Ga | 31.11 | 30.59±0.29 | 30.66±0.43 | Hf | 0.0009 | <LOD (0.02) | <LOD (0.02) |
| As | 85.23 | | | Ta | 0.0045 | 0.03±0.01[c] | 0.10±0.01[c] |
| Rb | 10.70 | | 0.06±0.01 | W | 381.6 | 421.4±33.6 | 933.3±12.6 |
| Sr | 1.02 | | | Tl | 0.045 | 0.033±0.001 | 0.06±0.01 |
| Y | 0.044 | <LOD (0.00457) | <LOD (0.00457) | Pb | 2.65 | 2.4±0.2 | 0.11±0.005 |
| Zr | 0.29 | 0.7±0.2[a] | 1.31±0.02 [a] | Bi | | 0.016±0.001 | <LOD (0.001) |
| Nb | 52.94 | 60.0±2.8 | 165.70±2.10 | Th | 0.003 | <LOD (0.001) | <LOD (0.001) |
| Mo | 20007 | 21042±1524 | 21291±143 | U | 0.004 | <LOD (0.0003) | <LOD (0.0003) |

[a] Possible laboratory artifact due to routine zircon analyses.
[b] Disagreement may be due to matrix or memory effects.
[c] May be degraded due to W interference.

### *Henkel Loctite EA9394*

The analysis of the Henkel Loctite EA9394 by wavelength-dispersive spectroscopy (WDS), and electron backscatter diffraction (BSD) showed distinct phases where the larger particles will be easily distinguishable from Bennu samples by elemental and sub-mm morphology. Scanning electron microscopy (SEM) also showed that the particles should all be easily recognized as contaminants. It was noted that the silica component might not be so readily recognized as a contaminant except through association with either a resin or alumina component (Fig. S3.3).

Solvent extraction of the Henkel Loctite EA9394 for different LC-MS and GC-MS techniques showed very little. The low abundances of species observed were generally near the detection limits from extracts of large chips of pure Henkel Loctite EA9394. These trace detections are dianalines, plasticizers, siloxanes, and perhaps naphthalene.

Conversely, when a grain of Henkel Loctite EA9394 was subjected to pyrolysis-GC-MS vast quantities and diversities of compounds were seen, enough to saturate the instruments. Primarily phenols were seen, perhaps some cellulose (though it may have been an artifact from the initial sample preparation). It is important to emphasize that the abundance of compounds from Henkel Loctite EA9394 pyrolysis did not overlap with any species observed in any pyrolysis experiments on the flight-like air filter. It should be noted that the air filter tested in SART-5-038 was not subjected to thermovacuum (TVAC) cleaning, unlike the flight unit. Thus, concerns of Henkel Loctite EA9394 outgassing during curing in the air filter appear not to have manifested or these species are not present in the dried epoxy, even after pyrolysis.

The elemental ratios (EA-IRMS) of the C and N in the Henkel Loctite EA9394 (Epoxy) is in Table S3.3. These values appear distinct from relevant meteoritic values.

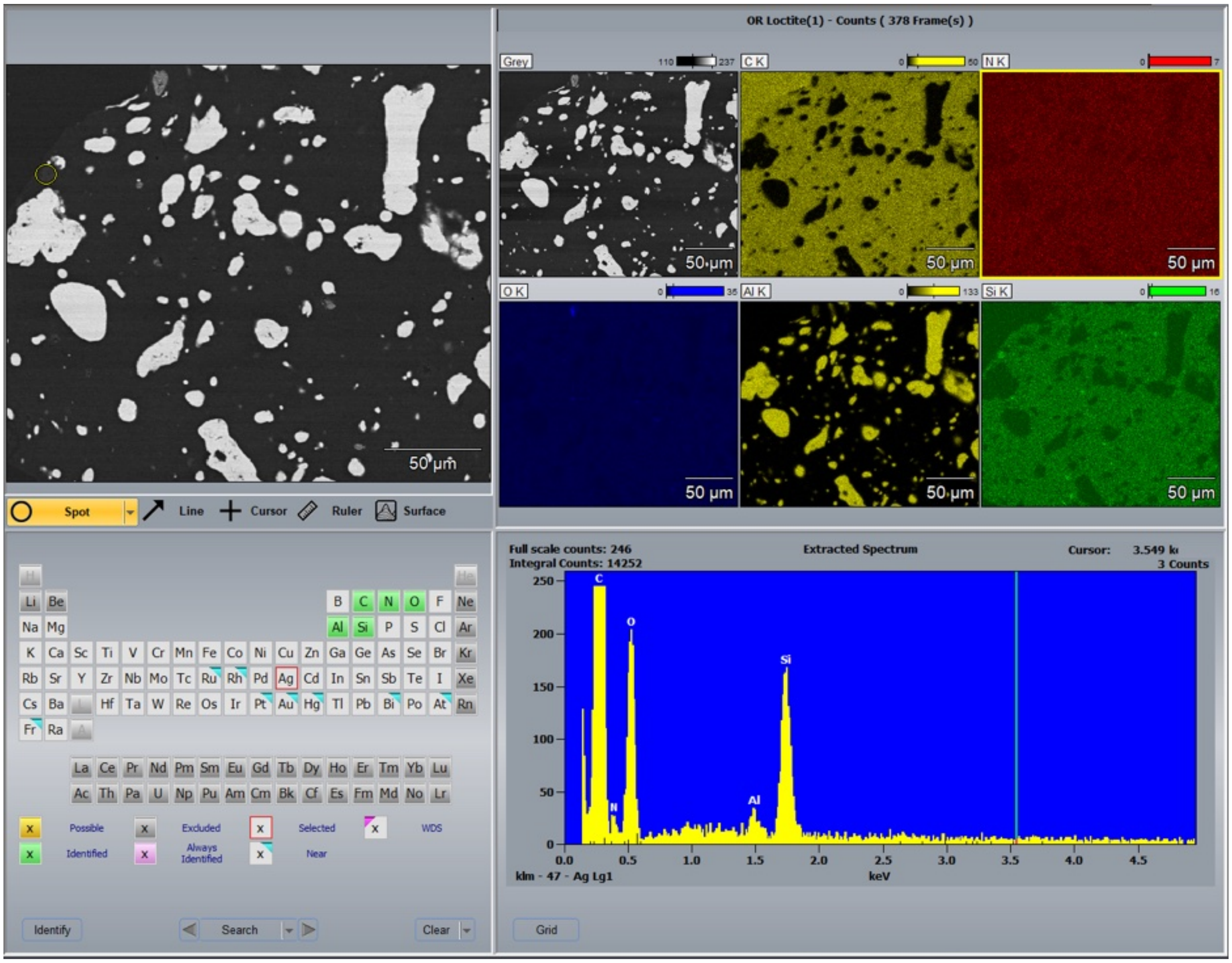

*Fig. S2-3. SEM (NMNH instrument) shows two phases of aluminum (light) and resin (dark) with Si-O hotspots embedded in the resin.*

**Table S3-3** Organic elemental and isotopic ratios in Henkel Loctite EA9394.

| | C (wt %) | N (wt %) | C/N | $\delta^{13}$C (‰, VPDB) | $\delta^{15}$N (‰, AIR) |
|---|---|---|---|---|---|
| Henkel Loctite EA9394 | 41.74±1.38 | 7.43±1.71 | 5.44±0.07 | -29.03 | -1.33±0.08 |
| Ryugu (A0106)[1] | 3.76±0.14 | 0.16±0.01 | 23.5±0.4 | -0.58±2.0 | +43.0±9.0 |
| Ryugu IOM[2] | 5.2 | 0.15 | 34.5 | -32±2 | +42±14 |
| Orgueil[3] | 3.52 | 0.149 | 23.6 | -14.7 | +44.1 |
| Orgueil (IOM)[4] | 67.0 | 0.192 | 28.7 | -17.1 | +30.7 |
| Murchison[3] | 2.08 | 0.105 | 19.7 | -2.9 | +45.6 |

[1]Naroaka et al. 2023; [2]Yabuta et al. 2023; [3]Alexander et al. 2012; [3]Alexander et al. 2007

Henkel Loctite EA9394, when hot water extracted and hydrolyzed, showed a suite of protein amino acids. They were absent in the unhydrolyzed sample and no non-protein amino acids were detected. This indicates that the amino acids derive from a protein. One likely source is latex or possibly fingerprints or sweat from the sample preparation. The lack of high sodium and nucleobases suggests that it may have been latex. The presence of potential cellulose may suggest that the witness mixture was stirred with a wooden splint or swab. Perhaps the preparation of this sample was done with less care than the application on the spacecraft, which would have prohibited latex and made exposure to fingerprints difficult.

The analysis of amines, the most likely contaminants from Henkel Loctite EA9394, showed methylamine and ethylamine. The presence of both methylamine and ethylamine are concerning, and the ratios are not inconsistent with hydrolyzed and unhydrolyzed extracts of relevant extraterrestrial samples (Table S3.4). However, Henkel Loctite EA9394 did not have any larger amines (propylamines, butylamines, pentylamines, and hexylamine). With the exception of Stardust aerogel exposed to comet Wild 2 (Glavin et al. 2008) all extraterrestrial samples containing methylamine and ethylamine also contained at least both isomers of propylamine (Naraoka et al. 2023; Parker et al. 2023) as has been observed in all other CI, CM, CR, CV CO, CK, and Tagish Lake meteorites (Pizzarello et al. 2001; Aponte et al. 2016; Aponte et al. 2017).

It should also be emphasized that unlike the flight hardware the materials archive samples, including OR-MA-0221 sample of Henkel Loctite EA9394, was not subjected to TVAC. Thus, the abundances of volatiles, like methylamine and ethylamine would be expected to be lower in Henkel Loctite EA9394 on the spacecraft.

**Table S3-4** Methylamine/ethylamine ratio for the aqueous different extraction and work-up conditions to be employed on Bennu samples.

| Methylamine/Ethylamine | Sonication Extraction | | Hot Water Extraction | |
|---|---|---|---|---|
| | Unhydrolyzed | Hydrolyzed | Unhydrolyzed | Hydrolyzed |
| Henkel Loctite EA9394 | >30 | 6.8±1.5 | 2.9±0.1 | 3.6±0.5 |
| Ryugu (A0106)[1] | | | 1.45±0.06 | |
| Ryugu (C0107)[1] | | | 2.2±0.4 | |
| Wild 2 (C2054,4)[2] | | | 0.5±0.5 | 1.0±0.3 |
| Wild 2 (C2086,1)[2] | | | 0.9±1 | 1.7±0.2 |
| Orgueil[3] | | | | 3.9±0.1 |
| Aguas Zarcas[4] | | | 29.5±5.0 | |
| Murchison[5] | | | 5.7±0.4 | 3.9±0.1 |

[1]Parker et al. 2023 and Parker et al. 2026; [2]Glavin et al. 2008; [3]Aponte et al. 2015; [4]Aponte et al. 2020; [5]Aponte et al. 2014

The results from the pyrolysis GCMS and LCMS for amines and amino acids were used as a reference to search for unidentified trace peaks in data collected from assembly, test, and launch operations (ATLO) contamination knowledge witness plates in 2015-2016 (Dworkin et al. 2018). The witness plates selected are listed in Table S3.5. These were selected in pairs early before TVAC to look for trends and after TVAC. An additional witness was selected as a control which was exposed while the SRC was closed.

It is notable that these species were absent in the later exposures, including after TVAC. This could mean that the source of the trace levels of these amines was due the same contamination source as the amino acids in the beginning of ATLO. It could mean that the levels of amines from the open SRC were decreasing with time, or that the factor of 3 lower exposure time of the open SRC to the foils pushed the abundances below the detection limit. These compounds were very small, previously unidentified peaks in the ATLO chromatograms.

**Table S3-5** ATLO witness plate data compared against Henkel Loctite EA9394 data.

| Witness Plate | Witness Exposure dates | SRC Exposure time | Rationale |
|---|---|---|---|
| OR-GC-KP-02 | 4/14/15-5/11/15 | 20 hr | Recently cured, should be highest amount |
| OR-GC-KP-03 | 5/11/15-6/10/15 | 19.5 hr | Similar exposure to OR-GC-KP-02 |
| OR-GC-KP-05 | 7/14/15-8/19/15 | 6.3 hr | Similar exposure to OR-GC-KP-14 but pre-TVAC |
| OR-GC-KP-13 | 4/26/16-4/27/16 | 0 hr | Control |
| OR-GC-KP-14 | 4/27/16-6/17/16 | 5.5 hr | Post-TVAC, should lowest amounts and most flight-like |

There were no peaks in common with the pyrolysis-GC-MS of Henkel Loctite EA9394 and pyrolysis-GC-MS of any of the ATLO witness plates. The LC-MS showed only protein amino acids, some of which were found in OR-GC-KP-02, though in different ratios. The earliest ATLO contamination knowledge and contamination control witness plates (Dworkin et al. 2018) showed higher protein amino acids than later witness plates, likely due to the cleanroom technicians becoming more familiar with amino acid-clean techniques.

Similarly, there were trace levels of methylamine and ethylamine observed in the first two exposures in ATLO (Table S3.6). The first exposure shows a methylamine/ethylamine ratio similar to Henkel Loctite EA9394 (and extraterrestrial samples, see Table S3.6). This ratio significantly changed in the following month's exposure, OR-GC-KP-02. It is possible that it is due to the 60% lower vapor pressure of ethylamine, coincidence, or the error from the very small peaks observed in the ATLO witness plates. If the former, then the methylamine/ethylamine ratio should continue to decrease as the spacecraft outgases, taking it further from meteoritic and Ryugu values.

**Table S3-6** Blank-corrected amine abundances analyzed in hot water extracted hydrolyzed samples of Henkel Loctite EA9394 and the ATLO samples from Table S3.5. Since the small amine peaks from ATLO had not been previously identified, quantitation was not performed. Instead, peak areas in arbitrary units are shown.

| Amines | OR-CKP-02 | OR-CKP-03 | OR-CKP-05 | OR-CKP-13 | OR-CKP-14 | Henkel Loctite EA9394 (nmol/g) |
|---|---|---|---|---|---|---|
| Methylamine | 3.78 | 0.17 | – | – | – | 31.14±2.42 |
| Ethylamine | 1.04 | 0.25 | – | – | – | 8.61±0.93 |
| Methylamine/ Ethylamine | 3.6 | 0.68 | – | – | – | 3.6±0.5 |

***Mylar***

To discriminate possible Mylar contamination from indigenous C-bearing phases in the returned sample, optical microscopy, FTIR, and field-emission SEM were performed. The sample was unremarkable mylar from the measurements. The sample exhibited electron beam damage exhibited by a change in texture with longer exposure gradually transforms into dark spots and channels. The onset of damage is proportional to the electron beam current density so that acquisition of spot EDS spectra produces greater damage compared to area measurements. Organic matter typically associated with extraterrestrial samples does not display the same beam damage behavior and so under similar conditions should be discernable from possible Mylar contaminants.

**Supplement 4:** Gas analysis reports from SRC recovery operations, 22 September 2022

Gas samples were collected in pre-cleaned evacuated stainless steel 480 mL valved Grab Sample Containers (Sandford et al. 2010) provided by the JSC Toxicology and Environmental Chemistry (TEC) group (Fig. S4-1). These were then promptly analyzed by TEC via gas chromatography mass spectrometry for target species (Wallace et al. 2024; Sandford et al. 2026).

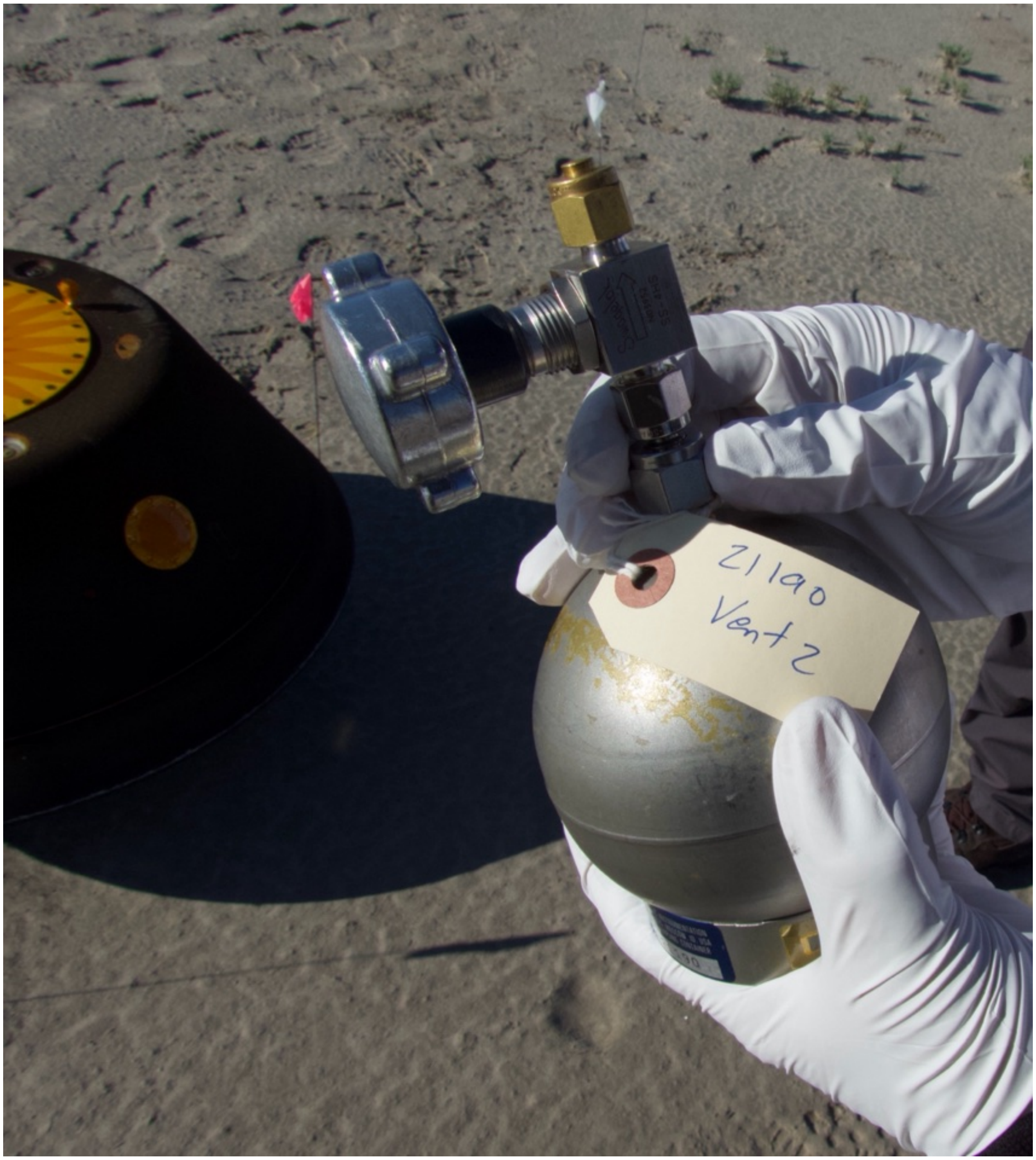

**Fig. S4-1**. A Grab Sample Container to be used in the collection of gasses from the OSIRIS-REx SRC shortly after landing in Utah. Image credit: NASA/Francis McCubbin.

# Air Analysis Report

**Project: OSIRIS-REx Cleanroom Monitoring - JSC-XI2-001**
**Order ID: 231003004**

| | | | |
|---|---|---|---|
| **Sample Description:** | 21189 | **JSC Sample No.:** | AQ230782 |
| **Sample Location:** | Purge Back Up / UTTR Cleanroom 1012 | **Collection Date/Time:** | 09/22/2023 12:35 |
| **Site:** | JSC B31 | **Submitted by:** | N. Lunning |

| Test | Conc. | Units | Analyst |
|---|---|---|---|
| **Volatiles Targets GCMS (TO-15 mod)** | | | |
| 1,1,1,2-Tetrafluoroethane | < 5.0 | ppbv | IAH |
| 1,1,1-Trichloroethane | < 3.0 | ppbv | IAH |
| 1,1,2,2-Tetrachloroethane | < 3.0 | ppbv | IAH |
| 1,1,2-Trichloroethane | < 3.0 | ppbv | IAH |
| 1,1-Dichloroethane | < 3.0 | ppbv | IAH |
| 1,1-Dichloroethene | < 3.0 | ppbv | IAH |
| 1,2,4-Trichlorobenzene | < 5.0 | ppbv | IAH |
| 1,2,4-Trimethylbenzene | < 3.0 | ppbv | IAH |
| 1,2-Dibromoethane (EDB) | < 3.0 | ppbv | IAH |
| 1,2-Dichlorobenzene | < 3.0 | ppbv | IAH |
| 1,2-Dichloroethane | < 3.0 | ppbv | IAH |
| 1,2-Dichloropropane | < 3.0 | ppbv | IAH |
| 1,3,5-Trimethylbenzene | < 3.0 | ppbv | IAH |
| 1,3-Butadiene | < 3.0 | ppbv | IAH |
| 1,3-Dichlorobenzene | < 3.0 | ppbv | IAH |
| 1,4-Dichlorobenzene | < 3.0 | ppbv | IAH |
| 1,4-Dioxane | < 3.0 | ppbv | IAH |
| 1-Butanol | < 3.0 | ppbv | IAH |
| 1-Propanol | < 3.0 | ppbv | IAH |
| 2,3-Dimethylpentane | < 3.0 | ppbv | IAH |
| 2,5-Dimethylfuran | < 3.0 | ppbv | IAH |
| 2-Butanone (Methyl ethyl ketone) | < 3.0 | ppbv | IAH |
| 2-Butenal | < 3.0 | ppbv | IAH |
| 2-Heptanone | < 3.0 | ppbv | IAH |
| 2-Methyl-1-propene | < 3.0 | ppbv | IAH |
| 2-Methyl-2-propanol | < 3.0 | ppbv | IAH |
| 2-Methylfuran | < 3.0 | ppbv | IAH |
| 2-Methylhexane | < 3.0 | ppbv | IAH |
| 2-Pentanone | < 3.0 | ppbv | IAH |
| 2-Pentenal | < 3.0 | ppbv | IAH |
| 2-Propanol (Isopropanol) | 75 | ppbv | IAH |
| 3-Chloropropene (Allyl chloride) | < 3.0 | ppbv | IAH |
| 3-Methylhexane | < 3.0 | ppbv | IAH |

# Air Analysis Report

**Project: OSIRIS-REx Cleanroom Monitoring - JSC-XI2-001**
**Order ID: 231003004**

| | | | |
|---|---|---|---|
| **Sample Description:** | 21189 | **JSC Sample No.:** | AQ230782 |
| **Sample Location:** | Purge Back Up / UTTR Cleanroom 1012 | **Collection Date/Time:** | 09/22/2023 12:35 |
| **Site:** | JSC B31 | **Submitted by:** | N. Lunning |

| Test | Conc. | Units | Analyst |
|---|---|---|---|
| **Volatiles Targets GCMS (TO-15 mod)** | | | |
| 4-Methyl-2-pentanone (MIBK) | < 3.0 | ppbv | IAH |
| Acetaldehyde | < 5.0 | ppbv | IAH |
| Acetone | 4.6 | ppbv | IAH |
| Acetonitrile | < 3.0 | ppbv | IAH |
| Acrylonitrile | < 3.0 | ppbv | IAH |
| Benzene | < 3.0 | ppbv | IAH |
| Bromomethane | < 3.0 | ppbv | IAH |
| Butanal (Butyraldehyde) | < 3.0 | ppbv | IAH |
| Butane | < 3.0 | ppbv | IAH |
| Butyl acetate | < 3.0 | ppbv | IAH |
| Carbon disulfide | < 3.0 | ppbv | IAH |
| Carbon tetrachloride | < 3.0 | ppbv | IAH |
| Carbonyl sulfide (Carbon oxide sulfide) | < 3.0 | ppbv | IAH |
| Chlorobenzene | < 3.0 | ppbv | IAH |
| Chloroethane | < 3.0 | ppbv | IAH |
| Chloroform | < 3.0 | ppbv | IAH |
| Chloromethane | < 3.0 | ppbv | IAH |
| cis-1,2-Dichloroethene | < 3.0 | ppbv | IAH |
| cis-1,3-Dichloropropene | < 3.0 | ppbv | IAH |
| Cyclohexanone | < 3.0 | ppbv | IAH |
| Decamethylcyclopentasiloxane (DMCPS) | < 5.0 | ppbv | IAH |
| Dimethyl sulfide | < 3.0 | ppbv | IAH |
| Ethanol | < 3.0 | ppbv | IAH |
| Ethyl acetate | < 3.0 | ppbv | IAH |
| Ethylbenzene | < 3.0 | ppbv | IAH |
| Freon 11 (Trichlorofluoromethane) | < 3.0 | ppbv | IAH |
| Freon 113 (1,1,2-Trichloro-1,2,2-trifluoroethane) | < 3.0 | ppbv | IAH |
| Freon 114 (1,2-Dichloro-1,1,2,2-tetrafluoroethane) | < 3.0 | ppbv | IAH |
| Freon 12 (Dichlorodifluoromethane) | < 3.0 | ppbv | IAH |
| Furan | < 3.0 | ppbv | IAH |
| Heptanal | < 3.0 | ppbv | IAH |
| Hexachlorobutadiene | < 3.0 | ppbv | IAH |
| Hexanal | < 3.0 | ppbv | IAH |

# Air Analysis Report

**Project:** **OSIRIS-REx Cleanroom Monitoring - JSC-XI2-001**
**Order ID:** **231003004**

| | | | |
|---|---|---|---|
| **Sample Description:** | 21189 | **JSC Sample No.:** | AQ230782 |
| **Sample Location:** | Purge Back Up / UTTR Cleanroom 1012 | **Collection Date/Time:** | 09/22/2023 12:35 |
| **Site:** | JSC B31 | **Submitted by:** | N. Lunning |

| Test | Conc. | Units | Analyst |
|---|---|---|---|
| **Volatiles Targets GCMS (TO-15 mod)** | | | |
| Hexane | < 3.0 | ppbv | IAH |
| Isobutane | < 3.0 | ppbv | IAH |
| Isoprene (2-Methyl-1,3-butadiene) | < 3.0 | ppbv | IAH |
| m and p-Xylene | < 6.0 | ppbv | IAH |
| Mesityl oxide (4-Methyl-3-penten-2-one) | < 3.0 | ppbv | IAH |
| Methanol | < 5.0 | ppbv | IAH |
| Methyl acetate | < 3.0 | ppbv | IAH |
| Methylene chloride (Dichloromethane) | < 3.0 | ppbv | IAH |
| n-Heptane | < 3.0 | ppbv | IAH |
| Nonane | < 3.0 | ppbv | IAH |
| Octamethylcyclotetrasiloxane (OMCTS) | < 5.0 | ppbv | IAH |
| Octane | < 3.0 | ppbv | IAH |
| o-Xylene | < 3.0 | ppbv | IAH |
| Pentanal | < 3.0 | ppbv | IAH |
| Pentane | < 3.0 | ppbv | IAH |
| Perfluoro(2-methylpentane) | < 3.0 | ppbv | IAH |
| Propanal (Propionaldehyde) | < 3.0 | ppbv | IAH |
| Propane | < 3.0 | ppbv | IAH |
| Propenal (Acrolein) | < 3.0 | ppbv | IAH |
| Propene | < 3.0 | ppbv | IAH |
| Styrene (Ethenylbenzene) | < 3.0 | ppbv | IAH |
| Tetrachloroethene (Perchloroethene) | < 3.0 | ppbv | IAH |
| Toluene | < 3.0 | ppbv | IAH |
| trans-1,3-Dichloropropene | < 3.0 | ppbv | IAH |
| Trichloroethene | < 3.0 | ppbv | IAH |
| Trimethylsilanol | < 3.0 | ppbv | IAH |
| Vinyl chloride | < 3.0 | ppbv | IAH |
| **Volatiles SICs GCMS (estimated conc.)** | | | |
| Hexamethylcyclotrisiloxane (HMCTS) | < 10.0 | ppbv | IAH |

**Comments:**

*Samples:* AQ230782: Collected inline with quick disconnect fitting and line pressure near 30 psi

*Results:* None

# Air Analysis Report

**Project: OSIRIS-REx Cleanroom Monitoring - JSC-XI2-001**
**Order ID: 231003004**

| | | | |
|---|---|---|---|
| **Sample Description:** | 23681 | **JSC Sample No.:** | AQ230783 |
| **Sample Location:** | Purge / UTTR Cleanroom 1012 | **Collection Date/Time:** | 09/22/2023 12:38 |
| **Site:** | JSC B31 | **Submitted by:** | N. Lunning |

| Test | Conc. | Units | Analyst |
|---|---|---|---|
| **Volatiles Targets GCMS (TO-15 mod)** | | | |
| 1,1,1,2-Tetrafluoroethane | < 5.0 | ppbv | IAH |
| 1,1,1-Trichloroethane | < 3.0 | ppbv | IAH |
| 1,1,2,2-Tetrachloroethane | < 3.0 | ppbv | IAH |
| 1,1,2-Trichloroethane | < 3.0 | ppbv | IAH |
| 1,1-Dichloroethane | < 3.0 | ppbv | IAH |
| 1,1-Dichloroethene | < 3.0 | ppbv | IAH |
| 1,2,4-Trichlorobenzene | < 5.0 | ppbv | IAH |
| 1,2,4-Trimethylbenzene | < 3.0 | ppbv | IAH |
| 1,2-Dibromoethane (EDB) | < 3.0 | ppbv | IAH |
| 1,2-Dichlorobenzene | < 3.0 | ppbv | IAH |
| 1,2-Dichloroethane | < 3.0 | ppbv | IAH |
| 1,2-Dichloropropane | < 3.0 | ppbv | IAH |
| 1,3,5-Trimethylbenzene | < 3.0 | ppbv | IAH |
| 1,3-Butadiene | < 3.0 | ppbv | IAH |
| 1,3-Dichlorobenzene | < 3.0 | ppbv | IAH |
| 1,4-Dichlorobenzene | < 3.0 | ppbv | IAH |
| 1,4-Dioxane | < 3.0 | ppbv | IAH |
| 1-Butanol | < 3.0 | ppbv | IAH |
| 1-Propanol | < 3.0 | ppbv | IAH |
| 2,3-Dimethylpentane | < 3.0 | ppbv | IAH |
| 2,5-Dimethylfuran | < 3.0 | ppbv | IAH |
| 2-Butanone (Methyl ethyl ketone) | < 3.0 | ppbv | IAH |
| 2-Butenal | < 3.0 | ppbv | IAH |
| 2-Heptanone | < 3.0 | ppbv | IAH |
| 2-Methyl-1-propene | < 3.0 | ppbv | IAH |
| 2-Methyl-2-propanol | < 3.0 | ppbv | IAH |
| 2-Methylfuran | < 3.0 | ppbv | IAH |
| 2-Methylhexane | < 3.0 | ppbv | IAH |
| 2-Pentanone | < 3.0 | ppbv | IAH |
| 2-Pentenal | < 3.0 | ppbv | IAH |
| 2-Propanol (Isopropanol) | 130 | ppbv | IAH |
| 3-Chloropropene (Allyl chloride) | < 3.0 | ppbv | IAH |
| 3-Methylhexane | < 3.0 | ppbv | IAH |

# Air Analysis Report

**Project: OSIRIS-REx Cleanroom Monitoring - JSC-XI2-001**
**Order ID: 231003004**

| | | | |
|---|---|---|---|
| **Sample Description:** | 23681 | **JSC Sample No.:** | AQ230783 |
| **Sample Location:** | Purge / UTTR Cleanroom 1012 | **Collection Date/Time:** | 09/22/2023 12:38 |
| **Site:** | JSC B31 | **Submitted by:** | N. Lunning |

| Test | Conc. | Units | Analyst |
|---|---|---|---|
| **Volatiles Targets GCMS (TO-15 mod)** | | | |
| 4-Methyl-2-pentanone (MIBK) | < 3.0 | ppbv | IAH |
| Acetaldehyde | < 5.0 | ppbv | IAH |
| Acetone | 6.1 | ppbv | IAH |
| Acetonitrile | < 3.0 | ppbv | IAH |
| Acrylonitrile | < 3.0 | ppbv | IAH |
| Benzene | < 3.0 | ppbv | IAH |
| Bromomethane | < 3.0 | ppbv | IAH |
| Butanal (Butyraldehyde) | < 3.0 | ppbv | IAH |
| Butane | < 3.0 | ppbv | IAH |
| Butyl acetate | < 3.0 | ppbv | IAH |
| Carbon disulfide | < 3.0 | ppbv | IAH |
| Carbon tetrachloride | < 3.0 | ppbv | IAH |
| Carbonyl sulfide (Carbon oxide sulfide) | < 3.0 | ppbv | IAH |
| Chlorobenzene | < 3.0 | ppbv | IAH |
| Chloroethane | < 3.0 | ppbv | IAH |
| Chloroform | < 3.0 | ppbv | IAH |
| Chloromethane | < 3.0 | ppbv | IAH |
| cis-1,2-Dichloroethene | < 3.0 | ppbv | IAH |
| cis-1,3-Dichloropropene | < 3.0 | ppbv | IAH |
| Cyclohexanone | < 3.0 | ppbv | IAH |
| Decamethylcyclopentasiloxane (DMCPS) | < 5.0 | ppbv | IAH |
| Dimethyl sulfide | < 3.0 | ppbv | IAH |
| Ethanol | 4.9 | ppbv | IAH |
| Ethyl acetate | < 3.0 | ppbv | IAH |
| Ethylbenzene | < 3.0 | ppbv | IAH |
| Freon 11 (Trichlorofluoromethane) | < 3.0 | ppbv | IAH |
| Freon 113 (1,1,2-Trichloro-1,2,2-trifluoroethane) | < 3.0 | ppbv | IAH |
| Freon 114 (1,2-Dichloro-1,1,2,2-tetrafluoroethane) | < 3.0 | ppbv | IAH |
| Freon 12 (Dichlorodifluoromethane) | < 3.0 | ppbv | IAH |
| Furan | < 3.0 | ppbv | IAH |
| Heptanal | < 3.0 | ppbv | IAH |
| Hexachlorobutadiene | < 3.0 | ppbv | IAH |
| Hexanal | < 3.0 | ppbv | IAH |

# Air Analysis Report

**Project:** **OSIRIS-REx Cleanroom Monitoring - JSC-XI2-001**
**Order ID:** **231003004**

| | | | |
|---|---|---|---|
| **Sample Description:** | 23681 | **JSC Sample No.:** | AQ230783 |
| **Sample Location:** | Purge / UTTR Cleanroom 1012 | **Collection Date/Time:** | 09/22/2023 12:38 |
| **Site:** | JSC B31 | **Submitted by:** | N. Lunning |

| Test | Conc. | Units | Analyst |
|---|---|---|---|
| **Volatiles Targets GCMS (TO-15 mod)** | | | |
| Hexane | < 3.0 | ppbv | IAH |
| Isobutane | 23 | ppbv | IAH |
| Isoprene (2-Methyl-1,3-butadiene) | < 3.0 | ppbv | IAH |
| m and p-Xylene | < 6.0 | ppbv | IAH |
| Mesityl oxide (4-Methyl-3-penten-2-one) | < 3.0 | ppbv | IAH |
| Methanol | 5.8 | ppbv | IAH |
| Methyl acetate | < 3.0 | ppbv | IAH |
| Methylene chloride (Dichloromethane) | < 3.0 | ppbv | IAH |
| n-Heptane | < 3.0 | ppbv | IAH |
| Nonane | < 3.0 | ppbv | IAH |
| Octamethylcyclotetrasiloxane (OMCTS) | < 5.0 | ppbv | IAH |
| Octane | < 3.0 | ppbv | IAH |
| o-Xylene | < 3.0 | ppbv | IAH |
| Pentanal | < 3.0 | ppbv | IAH |
| Pentane | < 3.0 | ppbv | IAH |
| Perfluoro(2-methylpentane) | < 3.0 | ppbv | IAH |
| Propanal (Propionaldehyde) | < 3.0 | ppbv | IAH |
| Propane | < 3.0 | ppbv | IAH |
| Propenal (Acrolein) | < 3.0 | ppbv | IAH |
| Propene | < 3.0 | ppbv | IAH |
| Styrene (Ethenylbenzene) | < 3.0 | ppbv | IAH |
| Tetrachloroethene (Perchloroethene) | < 3.0 | ppbv | IAH |
| Toluene | < 3.0 | ppbv | IAH |
| trans-1,3-Dichloropropene | < 3.0 | ppbv | IAH |
| Trichloroethene | < 3.0 | ppbv | IAH |
| Trimethylsilanol | < 3.0 | ppbv | IAH |
| Vinyl chloride | < 3.0 | ppbv | IAH |
| **Volatiles SICs GCMS (estimated conc.)** | | | |
| Hexamethylcyclotrisiloxane (HMCTS) | < 10.0 | ppbv | IAH |

**Comments:**

*Samples:* AQ230783: Collected inline with quick disconnect fitting and line pressure near 30 psi

*Results:* None

# Air Analysis Report

**Project: OSIRIS-REx Cleanroom Monitoring - JSC-XI2-001**
**Order ID: 231003004**

| | | | |
|---|---|---|---|
| **Sample Description:** | 34990 | **JSC Sample No.:** | AQ230784 |
| **Sample Location:** | Cleanroom ambient air / UTTR Cleanroom 1012 | **Collection Date/Time:** | 09/22/2023 12:41 |
| **Site:** | JSC B31 | **Submitted by:** | N. Lunning |

| Test | Conc. | Units | Analyst |
|---|---|---|---|
| **Volatiles Targets GCMS (TO-15 mod)** | | | |
| 1,1,1,2-Tetrafluoroethane | < 10 | ppbv | IAH |
| 1,1,1-Trichloroethane | < 6.0 | ppbv | IAH |
| 1,1,2,2-Tetrachloroethane | < 6.0 | ppbv | IAH |
| 1,1,2-Trichloroethane | < 6.0 | ppbv | IAH |
| 1,1-Dichloroethane | < 6.0 | ppbv | IAH |
| 1,1-Dichloroethene | < 6.0 | ppbv | IAH |
| 1,2,4-Trichlorobenzene | < 10 | ppbv | IAH |
| 1,2,4-Trimethylbenzene | < 6.0 | ppbv | IAH |
| 1,2-Dibromoethane (EDB) | < 6.0 | ppbv | IAH |
| 1,2-Dichlorobenzene | < 6.0 | ppbv | IAH |
| 1,2-Dichloroethane | < 6.0 | ppbv | IAH |
| 1,2-Dichloropropane | < 6.0 | ppbv | IAH |
| 1,3,5-Trimethylbenzene | < 6.0 | ppbv | IAH |
| 1,3-Butadiene | < 6.0 | ppbv | IAH |
| 1,3-Dichlorobenzene | < 6.0 | ppbv | IAH |
| 1,4-Dichlorobenzene | < 6.0 | ppbv | IAH |
| 1,4-Dioxane | < 6.0 | ppbv | IAH |
| 1-Butanol | < 6.0 | ppbv | IAH |
| 1-Propanol | < 6.0 | ppbv | IAH |
| 2,3-Dimethylpentane | < 6.0 | ppbv | IAH |
| 2,5-Dimethylfuran | < 6.0 | ppbv | IAH |
| 2-Butanone (Methyl ethyl ketone) | < 6.0 | ppbv | IAH |
| 2-Butenal | < 6.0 | ppbv | IAH |
| 2-Heptanone | < 6.0 | ppbv | IAH |
| 2-Methyl-1-propene | < 6.0 | ppbv | IAH |
| 2-Methyl-2-propanol | < 6.0 | ppbv | IAH |
| 2-Methylfuran | < 6.0 | ppbv | IAH |
| 2-Methylhexane | < 6.0 | ppbv | IAH |
| 2-Pentanone | < 6.0 | ppbv | IAH |
| 2-Pentenal | < 6.0 | ppbv | IAH |
| 2-Propanol (Isopropanol) | See GC-FID | ppbv | IAH |
| 3-Chloropropene (Allyl chloride) | < 6.0 | ppbv | IAH |

# Air Analysis Report

**Project:** **OSIRIS-REx Cleanroom Monitoring - JSC-XI2-001**
**Order ID:** **231003004**

| | | | |
|---|---|---|---|
| **Sample Description:** | 34990 | **JSC Sample No.:** | AQ230784 |
| **Sample Location:** | Cleanroom ambient air / UTTR Cleanroom 1012 | **Collection Date/Time:** | 09/22/2023 12:41 |
| **Site:** | JSC B31 | **Submitted by:** | N. Lunning |

| Test | Conc. | Units | Analyst |
|---|---|---|---|
| **Volatiles Targets GCMS (TO-15 mod)** | | | |
| 3-Methylhexane | < 6.0 | ppbv | IAH |
| 4-Methyl-2-pentanone (MIBK) | < 6.0 | ppbv | IAH |
| Acetaldehyde | < 10 | ppbv | IAH |
| Acetone | 46 | ppbv | IAH |
| Acetonitrile | < 6.0 | ppbv | IAH |
| Acrylonitrile | < 6.0 | ppbv | IAH |
| Benzene | < 6.0 | ppbv | IAH |
| Bromomethane | < 6.0 | ppbv | IAH |
| Butanal (Butyraldehyde) | < 6.0 | ppbv | IAH |
| Butane | < 6.0 | ppbv | IAH |
| Butyl acetate | < 6.0 | ppbv | IAH |
| Carbon disulfide | < 6.0 | ppbv | IAH |
| Carbon tetrachloride | < 6.0 | ppbv | IAH |
| Carbonyl sulfide (Carbon oxide sulfide) | < 6.0 | ppbv | IAH |
| Chlorobenzene | < 6.0 | ppbv | IAH |
| Chloroethane | < 6.0 | ppbv | IAH |
| Chloroform | < 6.0 | ppbv | IAH |
| Chloromethane | < 6.0 | ppbv | IAH |
| cis-1,2-Dichloroethene | < 6.0 | ppbv | IAH |
| cis-1,3-Dichloropropene | < 6.0 | ppbv | IAH |
| Cyclohexanone | < 6.0 | ppbv | IAH |
| Decamethylcyclopentasiloxane (DMCPS) | < 10 | ppbv | IAH |
| Dimethyl sulfide | < 6.0 | ppbv | IAH |
| Ethanol | 24 | ppbv | IAH |
| Ethyl acetate | < 6.0 | ppbv | IAH |
| Ethylbenzene | < 6.0 | ppbv | IAH |
| Freon 11 (Trichlorofluoromethane) | < 6.0 | ppbv | IAH |
| Freon 113 (1,1,2-Trichloro-1,2,2-trifluoroethane) | < 6.0 | ppbv | IAH |
| Freon 114 (1,2-Dichloro-1,1,2,2-tetrafluoroethane) | < 6.0 | ppbv | IAH |
| Freon 12 (Dichlorodifluoromethane) | < 6.0 | ppbv | IAH |
| Furan | < 6.0 | ppbv | IAH |
| Heptanal | < 6.0 | ppbv | IAH |

# Air Analysis Report

**Project:** **OSIRIS-REx Cleanroom Monitoring - JSC-XI2-001**
**Order ID:** **231003004**

| | | | |
|---|---|---|---|
| **Sample Description:** | 34990 | **JSC Sample No.:** | AQ230784 |
| **Sample Location:** | Cleanroom ambient air / UTTR Cleanroom 1012 | **Collection Date/Time:** | 09/22/2023 12:41 |
| **Site:** | JSC B31 | **Submitted by:** | N. Lunning |

| Test | Conc. | Units | Analyst |
|---|---|---|---|
| **Volatiles Targets GCMS (TO-15 mod)** | | | |
| Hexachlorobutadiene | < 6.0 | ppbv | IAH |
| Hexanal | < 6.0 | ppbv | IAH |
| Hexane | < 6.0 | ppbv | IAH |
| Isobutane | < 6.0 | ppbv | IAH |
| Isoprene (2-Methyl-1,3-butadiene) | < 6.0 | ppbv | IAH |
| m and p-Xylene | < 12 | ppbv | IAH |
| Mesityl oxide (4-Methyl-3-penten-2-one) | < 6.0 | ppbv | IAH |
| Methanol | 20 | ppbv | IAH |
| Methyl acetate | < 6.0 | ppbv | IAH |
| Methylene chloride (Dichloromethane) | < 6.0 | ppbv | IAH |
| n-Heptane | < 6.0 | ppbv | IAH |
| Nonane | < 6.0 | ppbv | IAH |
| Octamethylcyclotetrasiloxane (OMCTS) | < 10 | ppbv | IAH |
| Octane | < 6.0 | ppbv | IAH |
| o-Xylene | < 6.0 | ppbv | IAH |
| Pentanal | < 6.0 | ppbv | IAH |
| Pentane | < 6.0 | ppbv | IAH |
| Perfluoro(2-methylpentane) | < 6.0 | ppbv | IAH |
| Propanal (Propionaldehyde) | 9.6 | ppbv | IAH |
| Propane | 250 | ppbv | IAH |
| Propenal (Acrolein) | < 6.0 | ppbv | IAH |
| Propene | 32 | ppbv | IAH |
| Styrene (Ethenylbenzene) | < 6.0 | ppbv | IAH |
| Tetrachloroethene (Perchloroethene) | 6.2 | ppbv | IAH |
| Toluene | < 6.0 | ppbv | IAH |
| trans-1,3-Dichloropropene | < 6.0 | ppbv | IAH |
| Trichloroethene | < 6.0 | ppbv | IAH |
| Trimethylsilanol | < 6.0 | ppbv | IAH |
| Vinyl chloride | < 6.0 | ppbv | IAH |
| **Volatiles SICs GCMS (estimated conc.)** | | | |
| Hexamethylcyclotrisiloxane (HMCTS) | < 20 | ppbv | IAH |
| **Volatiles Targets GCFID** | | | |
| 2-Propanol (Isopropanol) | 13000 | ppbv | CMM |

# Air Analysis Report

**Project:** **OSIRIS-REx Cleanroom Monitoring - JSC-XI2-001**
**Order ID:** **231003004**

| | | | |
|---|---|---|---|
| **Sample Description:** | 34990 | **JSC Sample No.:** | AQ230784 |
| **Sample Location:** | Cleanroom ambient air / UTTR Cleanroom 1012 | **Collection Date/Time:** | 09/22/2023 12:41 |
| **Site:** | JSC B31 | **Submitted by:** | N. Lunning |

| Test | Conc. | Units | Analyst |
|---|---|---|---|
| **Volatiles Targets GCFID** | | | |
| Octafluoropropane (Perfluoropropane) | < 200 | ppbv | CMM |

**Comments:**

*Samples:* AQ230784: Ambient air pressure

*Results:* None

# Air Analysis Report

**Project: OSIRIS-REx Cleanroom Monitoring - JSC-XI2-001**
**Order ID: 231003004**

| | | | |
|---|---|---|---|
| **Sample Description:** | 22378 | **JSC Sample No.:** | AQ230785 |
| **Sample Location:** | Vent 1 / Recovery site | **Collection Date/Time:** | 09/24/2023 09:43 |
| **Site:** | JSC B31 | **Submitted by:** | N. Lunning |

| Test | Conc. | Units | Analyst |
|---|---|---|---|
| **Volatiles Targets GCMS (TO-15 mod)** | | | |
| 1,1,1,2-Tetrafluoroethane | < 10 | ppbv | IAH |
| 1,1,1-Trichloroethane | < 6.0 | ppbv | IAH |
| 1,1,2,2-Tetrachloroethane | < 6.0 | ppbv | IAH |
| 1,1,2-Trichloroethane | < 6.0 | ppbv | IAH |
| 1,1-Dichloroethane | < 6.0 | ppbv | IAH |
| 1,1-Dichloroethene | < 6.0 | ppbv | IAH |
| 1,2,4-Trichlorobenzene | < 10 | ppbv | IAH |
| 1,2,4-Trimethylbenzene | < 6.0 | ppbv | IAH |
| 1,2-Dibromoethane (EDB) | < 6.0 | ppbv | IAH |
| 1,2-Dichlorobenzene | < 6.0 | ppbv | IAH |
| 1,2-Dichloroethane | < 6.0 | ppbv | IAH |
| 1,2-Dichloropropane | < 6.0 | ppbv | IAH |
| 1,3,5-Trimethylbenzene | < 6.0 | ppbv | IAH |
| 1,3-Butadiene | < 6.0 | ppbv | IAH |
| 1,3-Dichlorobenzene | < 6.0 | ppbv | IAH |
| 1,4-Dichlorobenzene | < 6.0 | ppbv | IAH |
| 1,4-Dioxane | < 6.0 | ppbv | IAH |
| 1-Butanol | < 6.0 | ppbv | IAH |
| 1-Propanol | < 6.0 | ppbv | IAH |
| 2,3-Dimethylpentane | < 6.0 | ppbv | IAH |
| 2,5-Dimethylfuran | < 6.0 | ppbv | IAH |
| 2-Butanone (Methyl ethyl ketone) | < 6.0 | ppbv | IAH |
| 2-Butenal | < 6.0 | ppbv | IAH |
| 2-Heptanone | < 6.0 | ppbv | IAH |
| 2-Methyl-1-propene | < 6.0 | ppbv | IAH |
| 2-Methyl-2-propanol | < 6.0 | ppbv | IAH |
| 2-Methylfuran | < 6.0 | ppbv | IAH |
| 2-Methylhexane | < 6.0 | ppbv | IAH |
| 2-Pentanone | < 6.0 | ppbv | IAH |
| 2-Pentenal | < 6.0 | ppbv | IAH |
| 2-Propanol (Isopropanol) | < 10 | ppbv | IAH |
| 3-Chloropropene (Allyl chloride) | < 6.0 | ppbv | IAH |
| 3-Methylhexane | < 6.0 | ppbv | IAH |

# Air Analysis Report

**Project:** **OSIRIS-REx Cleanroom Monitoring - JSC-XI2-001**
**Order ID:** **231003004**

| | | | |
|---|---|---|---|
| **Sample Description:** | 22378 | **JSC Sample No.:** | AQ230785 |
| **Sample Location:** | Vent 1 / Recovery site | **Collection Date/Time:** | 09/24/2023 09:43 |
| **Site:** | JSC B31 | **Submitted by:** | N. Lunning |

| Test | Conc. | Units | Analyst |
|---|---|---|---|
| **Volatiles Targets GCMS (TO-15 mod)** | | | |
| 4-Methyl-2-pentanone (MIBK) | < 6.0 | ppbv | IAH |
| Acetaldehyde | 11 | ppbv | IAH |
| Acetone | 18 | ppbv | IAH |
| Acetonitrile | < 6.0 | ppbv | IAH |
| Acrylonitrile | < 6.0 | ppbv | IAH |
| Benzene | < 6.0 | ppbv | IAH |
| Bromomethane | < 6.0 | ppbv | IAH |
| Butanal (Butyraldehyde) | < 6.0 | ppbv | IAH |
| Butane | < 6.0 | ppbv | IAH |
| Butyl acetate | < 6.0 | ppbv | IAH |
| Carbon disulfide | < 6.0 | ppbv | IAH |
| Carbon tetrachloride | < 6.0 | ppbv | IAH |
| Carbonyl sulfide (Carbon oxide sulfide) | < 6.0 | ppbv | IAH |
| Chlorobenzene | < 6.0 | ppbv | IAH |
| Chloroethane | < 6.0 | ppbv | IAH |
| Chloroform | < 6.0 | ppbv | IAH |
| Chloromethane | < 6.0 | ppbv | IAH |
| cis-1,2-Dichloroethene | < 6.0 | ppbv | IAH |
| cis-1,3-Dichloropropene | < 6.0 | ppbv | IAH |
| Cyclohexanone | < 6.0 | ppbv | IAH |
| Decamethylcyclopentasiloxane (DMCPS) | < 10 | ppbv | IAH |
| Dimethyl sulfide | < 6.0 | ppbv | IAH |
| Ethanol | < 6.0 | ppbv | IAH |
| Ethyl acetate | < 6.0 | ppbv | IAH |
| Ethylbenzene | < 6.0 | ppbv | IAH |
| Freon 11 (Trichlorofluoromethane) | < 6.0 | ppbv | IAH |
| Freon 113 (1,1,2-Trichloro-1,2,2-trifluoroethane) | < 6.0 | ppbv | IAH |
| Freon 114 (1,2-Dichloro-1,1,2,2-tetrafluoroethane) | < 6.0 | ppbv | IAH |
| Freon 12 (Dichlorodifluoromethane) | < 6.0 | ppbv | IAH |
| Furan | < 6.0 | ppbv | IAH |
| Heptanal | < 6.0 | ppbv | IAH |
| Hexachlorobutadiene | < 6.0 | ppbv | IAH |
| Hexanal | < 6.0 | ppbv | IAH |

# Air Analysis Report

**Project: OSIRIS-REx Cleanroom Monitoring - JSC-XI2-001**
**Order ID: 231003004**

| | | | |
|---|---|---|---|
| **Sample Description:** | 22378 | **JSC Sample No.:** | AQ230785 |
| **Sample Location:** | Vent 1 / Recovery site | **Collection Date/Time:** | 09/24/2023 09:43 |
| **Site:** | JSC B31 | **Submitted by:** | N. Lunning |

| Test | Conc. | Units | Analyst |
|---|---|---|---|
| **Volatiles Targets GCMS (TO-15 mod)** | | | |
| Hexane | < 6.0 | ppbv | IAH |
| Isobutane | < 6.0 | ppbv | IAH |
| Isoprene (2-Methyl-1,3-butadiene) | < 6.0 | ppbv | IAH |
| m and p-Xylene | < 12 | ppbv | IAH |
| Mesityl oxide (4-Methyl-3-penten-2-one) | < 6.0 | ppbv | IAH |
| Methanol | 16 | ppbv | IAH |
| Methyl acetate | < 6.0 | ppbv | IAH |
| Methylene chloride (Dichloromethane) | < 6.0 | ppbv | IAH |
| n-Heptane | < 6.0 | ppbv | IAH |
| Nonane | < 6.0 | ppbv | IAH |
| Octamethylcyclotetrasiloxane (OMCTS) | < 10 | ppbv | IAH |
| Octane | < 6.0 | ppbv | IAH |
| o-Xylene | < 6.0 | ppbv | IAH |
| Pentanal | < 6.0 | ppbv | IAH |
| Pentane | < 6.0 | ppbv | IAH |
| Perfluoro(2-methylpentane) | < 6.0 | ppbv | IAH |
| Propanal (Propionaldehyde) | < 6.0 | ppbv | IAH |
| Propane | < 6.0 | ppbv | IAH |
| Propenal (Acrolein) | < 6.0 | ppbv | IAH |
| Propene | < 6.0 | ppbv | IAH |
| Styrene (Ethenylbenzene) | < 6.0 | ppbv | IAH |
| Tetrachloroethene (Perchloroethene) | < 6.0 | ppbv | IAH |
| Toluene | < 6.0 | ppbv | IAH |
| trans-1,3-Dichloropropene | < 6.0 | ppbv | IAH |
| Trichloroethene | < 6.0 | ppbv | IAH |
| Trimethylsilanol | < 6.0 | ppbv | IAH |
| Vinyl chloride | < 6.0 | ppbv | IAH |
| **Volatiles SICs GCMS (estimated conc.)** | | | |
| Hexamethylcyclotrisiloxane (HMCTS) | < 20 | ppbv | IAH |

**Comments:**

*Samples:* AQ230785: Ambient air pressure

*Results:* None

# Air Analysis Report

**Project: OSIRIS-REx Cleanroom Monitoring - JSC-XI2-001**
**Order ID: 231003004**

| | | | |
|---|---|---|---|
| **Sample Description:** | 34989 | **JSC Sample No.:** | AQ230786 |
| **Sample Location:** | Heat Shield / Recovery site | **Collection Date/Time:** | 09/24/2023 09:40 |
| **Site:** | JSC B31 | **Submitted by:** | N. Lunning |

| Test | Conc. | Units | Analyst |
|---|---|---|---|
| **Volatiles Targets GCMS (TO-15 mod)** | | | |
| 1,1,1,2-Tetrafluoroethane | < 10 | ppbv | IAH |
| 1,1,1-Trichloroethane | < 6.0 | ppbv | IAH |
| 1,1,2,2-Tetrachloroethane | < 6.0 | ppbv | IAH |
| 1,1,2-Trichloroethane | < 6.0 | ppbv | IAH |
| 1,1-Dichloroethane | < 6.0 | ppbv | IAH |
| 1,1-Dichloroethene | < 6.0 | ppbv | IAH |
| 1,2,4-Trichlorobenzene | < 10 | ppbv | IAH |
| 1,2,4-Trimethylbenzene | < 6.0 | ppbv | IAH |
| 1,2-Dibromoethane (EDB) | < 6.0 | ppbv | IAH |
| 1,2-Dichlorobenzene | < 6.0 | ppbv | IAH |
| 1,2-Dichloroethane | < 6.0 | ppbv | IAH |
| 1,2-Dichloropropane | < 6.0 | ppbv | IAH |
| 1,3,5-Trimethylbenzene | < 6.0 | ppbv | IAH |
| 1,3-Butadiene | < 6.0 | ppbv | IAH |
| 1,3-Dichlorobenzene | < 6.0 | ppbv | IAH |
| 1,4-Dichlorobenzene | < 6.0 | ppbv | IAH |
| 1,4-Dioxane | < 6.0 | ppbv | IAH |
| 1-Butanol | < 6.0 | ppbv | IAH |
| 1-Propanol | < 6.0 | ppbv | IAH |
| 2,3-Dimethylpentane | < 6.0 | ppbv | IAH |
| 2,5-Dimethylfuran | < 6.0 | ppbv | IAH |
| 2-Butanone (Methyl ethyl ketone) | < 6.0 | ppbv | IAH |
| 2-Butenal | < 6.0 | ppbv | IAH |
| 2-Heptanone | < 6.0 | ppbv | IAH |
| 2-Methyl-1-propene | < 6.0 | ppbv | IAH |
| 2-Methyl-2-propanol | < 6.0 | ppbv | IAH |
| 2-Methylfuran | < 6.0 | ppbv | IAH |
| 2-Methylhexane | < 6.0 | ppbv | IAH |
| 2-Pentanone | < 6.0 | ppbv | IAH |
| 2-Pentenal | < 6.0 | ppbv | IAH |
| 2-Propanol (Isopropanol) | < 10 | ppbv | IAH |
| 3-Chloropropene (Allyl chloride) | < 6.0 | ppbv | IAH |
| 3-Methylhexane | < 6.0 | ppbv | IAH |

# Air Analysis Report

**Project: OSIRIS-REx Cleanroom Monitoring - JSC-XI2-001**
**Order ID: 231003004**

| | | | |
|---|---|---|---|
| **Sample Description:** | 34989 | **JSC Sample No.:** | AQ230786 |
| **Sample Location:** | Heat Shield / Recovery site | **Collection Date/Time:** | 09/24/2023 09:40 |
| **Site:** | JSC B31 | **Submitted by:** | N. Lunning |

| Test | Conc. | Units | Analyst |
|---|---|---|---|
| **Volatiles Targets GCMS (TO-15 mod)** | | | |
| 4-Methyl-2-pentanone (MIBK) | < 6.0 | ppbv | IAH |
| Acetaldehyde | < 10 | ppbv | IAH |
| Acetone | 13 | ppbv | IAH |
| Acetonitrile | < 6.0 | ppbv | IAH |
| Acrylonitrile | < 6.0 | ppbv | IAH |
| Benzene | < 6.0 | ppbv | IAH |
| Bromomethane | < 6.0 | ppbv | IAH |
| Butanal (Butyraldehyde) | < 6.0 | ppbv | IAH |
| Butane | < 6.0 | ppbv | IAH |
| Butyl acetate | < 6.0 | ppbv | IAH |
| Carbon disulfide | < 6.0 | ppbv | IAH |
| Carbon tetrachloride | < 6.0 | ppbv | IAH |
| Carbonyl sulfide (Carbon oxide sulfide) | < 6.0 | ppbv | IAH |
| Chlorobenzene | < 6.0 | ppbv | IAH |
| Chloroethane | < 6.0 | ppbv | IAH |
| Chloroform | < 6.0 | ppbv | IAH |
| Chloromethane | < 6.0 | ppbv | IAH |
| cis-1,2-Dichloroethene | < 6.0 | ppbv | IAH |
| cis-1,3-Dichloropropene | < 6.0 | ppbv | IAH |
| Cyclohexanone | < 6.0 | ppbv | IAH |
| Decamethylcyclopentasiloxane (DMCPS) | < 10 | ppbv | IAH |
| Dimethyl sulfide | < 6.0 | ppbv | IAH |
| Ethanol | < 6.0 | ppbv | IAH |
| Ethyl acetate | < 6.0 | ppbv | IAH |
| Ethylbenzene | < 6.0 | ppbv | IAH |
| Freon 11 (Trichlorofluoromethane) | < 6.0 | ppbv | IAH |
| Freon 113 (1,1,2-Trichloro-1,2,2-trifluoroethane) | < 6.0 | ppbv | IAH |
| Freon 114 (1,2-Dichloro-1,1,2,2-tetrafluoroethane) | < 6.0 | ppbv | IAH |
| Freon 12 (Dichlorodifluoromethane) | < 6.0 | ppbv | IAH |
| Furan | < 6.0 | ppbv | IAH |
| Heptanal | < 6.0 | ppbv | IAH |
| Hexachlorobutadiene | < 6.0 | ppbv | IAH |
| Hexanal | < 6.0 | ppbv | IAH |

# Air Analysis Report

**Project:** **OSIRIS-REx Cleanroom Monitoring - JSC-XI2-001**
**Order ID:** **231003004**

| | | | |
|---|---|---|---|
| **Sample Description:** | 34989 | **JSC Sample No.:** | AQ230786 |
| **Sample Location:** | Heat Shield / Recovery site | **Collection Date/Time:** | 09/24/2023 09:40 |
| **Site:** | JSC B31 | **Submitted by:** | N. Lunning |

| Test | Conc. | Units | Analyst |
|---|---|---|---|
| **Volatiles Targets GCMS (TO-15 mod)** | | | |
| Hexane | < 6.0 | ppbv | IAH |
| Isobutane | < 6.0 | ppbv | IAH |
| Isoprene (2-Methyl-1,3-butadiene) | < 6.0 | ppbv | IAH |
| m and p-Xylene | < 12 | ppbv | IAH |
| Mesityl oxide (4-Methyl-3-penten-2-one) | < 6.0 | ppbv | IAH |
| Methanol | < 10 | ppbv | IAH |
| Methyl acetate | < 6.0 | ppbv | IAH |
| Methylene chloride (Dichloromethane) | < 6.0 | ppbv | IAH |
| n-Heptane | < 6.0 | ppbv | IAH |
| Nonane | < 6.0 | ppbv | IAH |
| Octamethylcyclotetrasiloxane (OMCTS) | < 10 | ppbv | IAH |
| Octane | < 6.0 | ppbv | IAH |
| o-Xylene | < 6.0 | ppbv | IAH |
| Pentanal | < 6.0 | ppbv | IAH |
| Pentane | < 6.0 | ppbv | IAH |
| Perfluoro(2-methylpentane) | < 6.0 | ppbv | IAH |
| Propanal (Propionaldehyde) | < 6.0 | ppbv | IAH |
| Propane | < 6.0 | ppbv | IAH |
| Propenal (Acrolein) | < 6.0 | ppbv | IAH |
| Propene | < 6.0 | ppbv | IAH |
| Styrene (Ethenylbenzene) | < 6.0 | ppbv | IAH |
| Tetrachloroethene (Perchloroethene) | < 6.0 | ppbv | IAH |
| Toluene | < 6.0 | ppbv | IAH |
| trans-1,3-Dichloropropene | < 6.0 | ppbv | IAH |
| Trichloroethene | < 6.0 | ppbv | IAH |
| Trimethylsilanol | < 6.0 | ppbv | IAH |
| Vinyl chloride | < 6.0 | ppbv | IAH |
| **Volatiles SICs GCMS (estimated conc.)** | | | |
| Hexamethylcyclotrisiloxane (HMCTS) | < 20 | ppbv | IAH |

**Comments:**

*Samples:* AQ230786: Ambient air pressure

*Results:* None

# Air Analysis Report

**Project: OSIRIS-REx Cleanroom Monitoring - JSC-XI2-001**
**Order ID: 231003004**

| | | | |
|---|---|---|---|
| **Sample Description:** | 25218 | **JSC Sample No.:** | AQ230787 |
| **Sample Location:** | Dust Sampler 1 / Recovery site | **Collection Date/Time:** | 09/24/2023 09:54 |
| **Site:** | JSC B31 | **Submitted by:** | N. Lunning |

| Test | Conc. | Units | Analyst |
|---|---|---|---|
| **Volatiles Targets GCMS (TO-15 mod)** | | | |
| 1,1,1,2-Tetrafluoroethane | < 10 | ppbv | IAH |
| 1,1,1-Trichloroethane | < 6.0 | ppbv | IAH |
| 1,1,2,2-Tetrachloroethane | < 6.0 | ppbv | IAH |
| 1,1,2-Trichloroethane | < 6.0 | ppbv | IAH |
| 1,1-Dichloroethane | < 6.0 | ppbv | IAH |
| 1,1-Dichloroethene | < 6.0 | ppbv | IAH |
| 1,2,4-Trichlorobenzene | < 10 | ppbv | IAH |
| 1,2,4-Trimethylbenzene | < 6.0 | ppbv | IAH |
| 1,2-Dibromoethane (EDB) | < 6.0 | ppbv | IAH |
| 1,2-Dichlorobenzene | < 6.0 | ppbv | IAH |
| 1,2-Dichloroethane | < 6.0 | ppbv | IAH |
| 1,2-Dichloropropane | < 6.0 | ppbv | IAH |
| 1,3,5-Trimethylbenzene | < 6.0 | ppbv | IAH |
| 1,3-Butadiene | < 6.0 | ppbv | IAH |
| 1,3-Dichlorobenzene | < 6.0 | ppbv | IAH |
| 1,4-Dichlorobenzene | < 6.0 | ppbv | IAH |
| 1,4-Dioxane | < 6.0 | ppbv | IAH |
| 1-Butanol | < 6.0 | ppbv | IAH |
| 1-Propanol | < 6.0 | ppbv | IAH |
| 2,3-Dimethylpentane | < 6.0 | ppbv | IAH |
| 2,5-Dimethylfuran | < 6.0 | ppbv | IAH |
| 2-Butanone (Methyl ethyl ketone) | < 6.0 | ppbv | IAH |
| 2-Butenal | < 6.0 | ppbv | IAH |
| 2-Heptanone | < 6.0 | ppbv | IAH |
| 2-Methyl-1-propene | < 6.0 | ppbv | IAH |
| 2-Methyl-2-propanol | < 6.0 | ppbv | IAH |
| 2-Methylfuran | < 6.0 | ppbv | IAH |
| 2-Methylhexane | < 6.0 | ppbv | IAH |
| 2-Pentanone | < 6.0 | ppbv | IAH |
| 2-Pentenal | < 6.0 | ppbv | IAH |
| 2-Propanol (Isopropanol) | See GC-FID | ppbv | IAH |
| 3-Chloropropene (Allyl chloride) | < 6.0 | ppbv | IAH |
| 3-Methylhexane | < 6.0 | ppbv | IAH |

# Air Analysis Report

**Project: OSIRIS-REx Cleanroom Monitoring - JSC-XI2-001**
**Order ID: 231003004**

| | | | |
|---|---|---|---|
| **Sample Description:** | 25218 | **JSC Sample No.:** | AQ230787 |
| **Sample Location:** | Dust Sampler 1 / Recovery site | **Collection Date/Time:** | 09/24/2023 09:54 |
| **Site:** | JSC B31 | **Submitted by:** | N. Lunning |

| Test | Conc. | Units | Analyst |
|---|---|---|---|
| **Volatiles Targets GCMS (TO-15 mod)** | | | |
| 4-Methyl-2-pentanone (MIBK) | < 6.0 | ppbv | IAH |
| Acetaldehyde | < 10 | ppbv | IAH |
| Acetone | 22 | ppbv | IAH |
| Acetonitrile | < 6.0 | ppbv | IAH |
| Acrylonitrile | < 6.0 | ppbv | IAH |
| Benzene | < 6.0 | ppbv | IAH |
| Bromomethane | < 6.0 | ppbv | IAH |
| Butanal (Butyraldehyde) | < 6.0 | ppbv | IAH |
| Butane | 12 | ppbv | IAH |
| Butyl acetate | < 6.0 | ppbv | IAH |
| Carbon disulfide | < 6.0 | ppbv | IAH |
| Carbon tetrachloride | < 6.0 | ppbv | IAH |
| Carbonyl sulfide (Carbon oxide sulfide) | < 6.0 | ppbv | IAH |
| Chlorobenzene | < 6.0 | ppbv | IAH |
| Chloroethane | < 6.0 | ppbv | IAH |
| Chloroform | < 6.0 | ppbv | IAH |
| Chloromethane | < 6.0 | ppbv | IAH |
| cis-1,2-Dichloroethene | < 6.0 | ppbv | IAH |
| cis-1,3-Dichloropropene | < 6.0 | ppbv | IAH |
| Cyclohexanone | < 6.0 | ppbv | IAH |
| Decamethylcyclopentasiloxane (DMCPS) | < 10 | ppbv | IAH |
| Dimethyl sulfide | < 6.0 | ppbv | IAH |
| Ethanol | 57 | ppbv | IAH |
| Ethyl acetate | < 6.0 | ppbv | IAH |
| Ethylbenzene | < 6.0 | ppbv | IAH |
| Freon 11 (Trichlorofluoromethane) | < 6.0 | ppbv | IAH |
| Freon 113 (1,1,2-Trichloro-1,2,2-trifluoroethane) | < 6.0 | ppbv | IAH |
| Freon 114 (1,2-Dichloro-1,1,2,2-tetrafluoroethane) | < 6.0 | ppbv | IAH |
| Freon 12 (Dichlorodifluoromethane) | < 6.0 | ppbv | IAH |
| Furan | < 6.0 | ppbv | IAH |
| Heptanal | < 6.0 | ppbv | IAH |
| Hexachlorobutadiene | < 6.0 | ppbv | IAH |
| Hexanal | < 6.0 | ppbv | IAH |

# Air Analysis Report

**Project:** **OSIRIS-REx Cleanroom Monitoring - JSC-XI2-001**
**Order ID:** **231003004**

| | | | |
|---|---|---|---|
| **Sample Description:** | 25218 | **JSC Sample No.:** | AQ230787 |
| **Sample Location:** | Dust Sampler 1 / Recovery site | **Collection Date/Time:** | 09/24/2023 09:54 |
| **Site:** | JSC B31 | **Submitted by:** | N. Lunning |

| Test | Conc. | Units | Analyst |
|---|---|---|---|
| **Volatiles Targets GCMS (TO-15 mod)** | | | |
| Hexane | < 6.0 | ppbv | IAH |
| Isobutane | 330 | ppbv | IAH |
| Isoprene (2-Methyl-1,3-butadiene) | < 6.0 | ppbv | IAH |
| m and p-Xylene | < 12 | ppbv | IAH |
| Mesityl oxide (4-Methyl-3-penten-2-one) | < 6.0 | ppbv | IAH |
| Methanol | 33 | ppbv | IAH |
| Methyl acetate | < 6.0 | ppbv | IAH |
| Methylene chloride (Dichloromethane) | < 6.0 | ppbv | IAH |
| n-Heptane | < 6.0 | ppbv | IAH |
| Nonane | < 6.0 | ppbv | IAH |
| Octamethylcyclotetrasiloxane (OMCTS) | < 10 | ppbv | IAH |
| Octane | < 6.0 | ppbv | IAH |
| o-Xylene | < 6.0 | ppbv | IAH |
| Pentanal | < 6.0 | ppbv | IAH |
| Pentane | < 6.0 | ppbv | IAH |
| Perfluoro(2-methylpentane) | < 6.0 | ppbv | IAH |
| Propanal (Propionaldehyde) | < 6.0 | ppbv | IAH |
| Propane | 11 | ppbv | IAH |
| Propenal (Acrolein) | < 6.0 | ppbv | IAH |
| Propene | < 6.0 | ppbv | IAH |
| Styrene (Ethenylbenzene) | < 6.0 | ppbv | IAH |
| Tetrachloroethene (Perchloroethene) | < 6.0 | ppbv | IAH |
| Toluene | < 6.0 | ppbv | IAH |
| trans-1,3-Dichloropropene | < 6.0 | ppbv | IAH |
| Trichloroethene | < 6.0 | ppbv | IAH |
| Trimethylsilanol | < 6.0 | ppbv | IAH |
| Vinyl chloride | < 6.0 | ppbv | IAH |
| **Volatiles SICs GCMS (estimated conc.)** | | | |
| Hexamethylcyclotrisiloxane (HMCTS) | < 20 | ppbv | IAH |
| **Volatiles Targets GCFID** | | | |
| 2-Propanol (Isopropanol) | 2100 | ppbv | CMM |
| Octafluoropropane (Perfluoropropane) | < 200 | ppbv | CMM |

# Air Analysis Report

**Project:** **OSIRIS-REx Cleanroom Monitoring - JSC-XI2-001**
**Order ID:** **231003004**

| | | | |
|---|---|---|---|
| **Sample Description:** | 25218 | **JSC Sample No.:** | AQ230787 |
| **Sample Location:** | Dust Sampler 1 / Recovery site | **Collection Date/Time:** | 09/24/2023 09:54 |
| **Site:** | JSC B31 | **Submitted by:** | N. Lunning |

| Test | Conc. | Units | Analyst |
|---|---|---|---|

**Comments:**

*Samples:* AQ230787: Ambient air pressure

*Results:* None

# Air Analysis Report

**Project: OSIRIS-REx Cleanroom Monitoring - JSC-XI2-001**
**Order ID: 231003004**

| | | | |
|---|---|---|---|
| **Sample Description:** | 21190 | **JSC Sample No.:** | AQ230788 |
| **Sample Location:** | Vent 2 / Recovery site | **Collection Date/Time:** | 09/24/2023 09:44 |
| **Site:** | JSC B31 | **Submitted by:** | N. Lunning |

| Test | Conc. | Units | Analyst |
|---|---|---|---|
| **Volatiles Targets GCMS (TO-15 mod)** | | | |
| 1,1,1,2-Tetrafluoroethane | < 10 | ppbv | IAH |
| 1,1,1-Trichloroethane | < 6.0 | ppbv | IAH |
| 1,1,2,2-Tetrachloroethane | < 6.0 | ppbv | IAH |
| 1,1,2-Trichloroethane | < 6.0 | ppbv | IAH |
| 1,1-Dichloroethane | < 6.0 | ppbv | IAH |
| 1,1-Dichloroethene | < 6.0 | ppbv | IAH |
| 1,2,4-Trichlorobenzene | < 10 | ppbv | IAH |
| 1,2,4-Trimethylbenzene | < 6.0 | ppbv | IAH |
| 1,2-Dibromoethane (EDB) | < 6.0 | ppbv | IAH |
| 1,2-Dichlorobenzene | < 6.0 | ppbv | IAH |
| 1,2-Dichloroethane | < 6.0 | ppbv | IAH |
| 1,2-Dichloropropane | < 6.0 | ppbv | IAH |
| 1,3,5-Trimethylbenzene | < 6.0 | ppbv | IAH |
| 1,3-Butadiene | < 6.0 | ppbv | IAH |
| 1,3-Dichlorobenzene | < 6.0 | ppbv | IAH |
| 1,4-Dichlorobenzene | < 6.0 | ppbv | IAH |
| 1,4-Dioxane | < 6.0 | ppbv | IAH |
| 1-Butanol | < 6.0 | ppbv | IAH |
| 1-Propanol | < 6.0 | ppbv | IAH |
| 2,3-Dimethylpentane | < 6.0 | ppbv | IAH |
| 2,5-Dimethylfuran | < 6.0 | ppbv | IAH |
| 2-Butanone (Methyl ethyl ketone) | < 6.0 | ppbv | IAH |
| 2-Butenal | < 6.0 | ppbv | IAH |
| 2-Heptanone | < 6.0 | ppbv | IAH |
| 2-Methyl-1-propene | < 6.0 | ppbv | IAH |
| 2-Methyl-2-propanol | < 6.0 | ppbv | IAH |
| 2-Methylfuran | < 6.0 | ppbv | IAH |
| 2-Methylhexane | < 6.0 | ppbv | IAH |
| 2-Pentanone | < 6.0 | ppbv | IAH |
| 2-Pentenal | < 6.0 | ppbv | IAH |
| 2-Propanol (Isopropanol) | < 10 | ppbv | IAH |
| 3-Chloropropene (Allyl chloride) | < 6.0 | ppbv | IAH |
| 3-Methylhexane | < 6.0 | ppbv | IAH |

# Air Analysis Report

**Project:** **OSIRIS-REx Cleanroom Monitoring - JSC-XI2-001**
**Order ID:** **231003004**

| | | | |
|---|---|---|---|
| **Sample Description:** | 21190 | **JSC Sample No.:** | AQ230788 |
| **Sample Location:** | Vent 2 / Recovery site | **Collection Date/Time:** | 09/24/2023 09:44 |
| **Site:** | JSC B31 | **Submitted by:** | N. Lunning |

| Test | Conc. | Units | Analyst |
|---|---|---|---|
| **Volatiles Targets GCMS (TO-15 mod)** | | | |
| 4-Methyl-2-pentanone (MIBK) | < 6.0 | ppbv | IAH |
| Acetaldehyde | < 10 | ppbv | IAH |
| Acetone | 15 | ppbv | IAH |
| Acetonitrile | < 6.0 | ppbv | IAH |
| Acrylonitrile | < 6.0 | ppbv | IAH |
| Benzene | < 6.0 | ppbv | IAH |
| Bromomethane | < 6.0 | ppbv | IAH |
| Butanal (Butyraldehyde) | < 6.0 | ppbv | IAH |
| Butane | < 6.0 | ppbv | IAH |
| Butyl acetate | < 6.0 | ppbv | IAH |
| Carbon disulfide | < 6.0 | ppbv | IAH |
| Carbon tetrachloride | < 6.0 | ppbv | IAH |
| Carbonyl sulfide (Carbon oxide sulfide) | < 6.0 | ppbv | IAH |
| Chlorobenzene | < 6.0 | ppbv | IAH |
| Chloroethane | < 6.0 | ppbv | IAH |
| Chloroform | < 6.0 | ppbv | IAH |
| Chloromethane | < 6.0 | ppbv | IAH |
| cis-1,2-Dichloroethene | < 6.0 | ppbv | IAH |
| cis-1,3-Dichloropropene | < 6.0 | ppbv | IAH |
| Cyclohexanone | < 6.0 | ppbv | IAH |
| Decamethylcyclopentasiloxane (DMCPS) | < 10 | ppbv | IAH |
| Dimethyl sulfide | < 6.0 | ppbv | IAH |
| Ethanol | < 6.0 | ppbv | IAH |
| Ethyl acetate | < 6.0 | ppbv | IAH |
| Ethylbenzene | < 6.0 | ppbv | IAH |
| Freon 11 (Trichlorofluoromethane) | < 6.0 | ppbv | IAH |
| Freon 113 (1,1,2-Trichloro-1,2,2-trifluoroethane) | < 6.0 | ppbv | IAH |
| Freon 114 (1,2-Dichloro-1,1,2,2-tetrafluoroethane) | < 6.0 | ppbv | IAH |
| Freon 12 (Dichlorodifluoromethane) | < 6.0 | ppbv | IAH |
| Furan | < 6.0 | ppbv | IAH |
| Heptanal | < 6.0 | ppbv | IAH |
| Hexachlorobutadiene | < 6.0 | ppbv | IAH |
| Hexanal | < 6.0 | ppbv | IAH |

# Air Analysis Report

**Project:** **OSIRIS-REx Cleanroom Monitoring - JSC-XI2-001**
**Order ID:** **231003004**

| | | | |
|---|---|---|---|
| **Sample Description:** | 21190 | **JSC Sample No.:** | AQ230788 |
| **Sample Location:** | Vent 2 / Recovery site | **Collection Date/Time:** | 09/24/2023 09:44 |
| **Site:** | JSC B31 | **Submitted by:** | N. Lunning |

| Test | Conc. | Units | Analyst |
|---|---|---|---|
| **Volatiles Targets GCMS (TO-15 mod)** | | | |
| Hexane | < 6.0 | ppbv | IAH |
| Isobutane | < 6.0 | ppbv | IAH |
| Isoprene (2-Methyl-1,3-butadiene) | < 6.0 | ppbv | IAH |
| m and p-Xylene | < 12 | ppbv | IAH |
| Mesityl oxide (4-Methyl-3-penten-2-one) | < 6.0 | ppbv | IAH |
| Methanol | 35 | ppbv | IAH |
| Methyl acetate | < 6.0 | ppbv | IAH |
| Methylene chloride (Dichloromethane) | < 6.0 | ppbv | IAH |
| n-Heptane | < 6.0 | ppbv | IAH |
| Nonane | < 6.0 | ppbv | IAH |
| Octamethylcyclotetrasiloxane (OMCTS) | < 10 | ppbv | IAH |
| Octane | < 6.0 | ppbv | IAH |
| o-Xylene | < 6.0 | ppbv | IAH |
| Pentanal | < 6.0 | ppbv | IAH |
| Pentane | < 6.0 | ppbv | IAH |
| Perfluoro(2-methylpentane) | < 6.0 | ppbv | IAH |
| Propanal (Propionaldehyde) | < 6.0 | ppbv | IAH |
| Propane | < 6.0 | ppbv | IAH |
| Propenal (Acrolein) | < 6.0 | ppbv | IAH |
| Propene | < 6.0 | ppbv | IAH |
| Styrene (Ethenylbenzene) | < 6.0 | ppbv | IAH |
| Tetrachloroethene (Perchloroethene) | < 6.0 | ppbv | IAH |
| Toluene | < 6.0 | ppbv | IAH |
| trans-1,3-Dichloropropene | < 6.0 | ppbv | IAH |
| Trichloroethene | < 6.0 | ppbv | IAH |
| Trimethylsilanol | 9.9 | ppbv | IAH |
| Vinyl chloride | < 6.0 | ppbv | IAH |
| **Volatiles SICs GCMS (estimated conc.)** | | | |
| Hexamethylcyclotrisiloxane (HMCTS) | < 20 | ppbv | IAH |

**Comments:**

*Samples:* AQ230788: Ambient air pressure

*Results:* None

# Air Analysis Report

**Project: OSIRIS-REx Cleanroom Monitoring - JSC-XI2-001**
**Order ID: 231003004**

| | | | |
|---|---|---|---|
| **Sample Description:** | 12848 | **JSC Sample No.:** | AQ230789 |
| **Sample Location:** | Backshell / Recovery site | **Collection Date/Time:** | 09/24/2023 09:41 |
| **Site:** | JSC B31 | **Submitted by:** | N. Lunning |

| Test | Conc. | Units | Analyst |
|---|---|---|---|
| **Volatiles Targets GCMS (TO-15 mod)** | | | |
| 1,1,1,2-Tetrafluoroethane | < 10 | ppbv | IAH |
| 1,1,1-Trichloroethane | < 6.0 | ppbv | IAH |
| 1,1,2,2-Tetrachloroethane | < 6.0 | ppbv | IAH |
| 1,1,2-Trichloroethane | < 6.0 | ppbv | IAH |
| 1,1-Dichloroethane | < 6.0 | ppbv | IAH |
| 1,1-Dichloroethene | < 6.0 | ppbv | IAH |
| 1,2,4-Trichlorobenzene | < 10 | ppbv | IAH |
| 1,2,4-Trimethylbenzene | < 6.0 | ppbv | IAH |
| 1,2-Dibromoethane (EDB) | < 6.0 | ppbv | IAH |
| 1,2-Dichlorobenzene | < 6.0 | ppbv | IAH |
| 1,2-Dichloroethane | < 6.0 | ppbv | IAH |
| 1,2-Dichloropropane | < 6.0 | ppbv | IAH |
| 1,3,5-Trimethylbenzene | < 6.0 | ppbv | IAH |
| 1,3-Butadiene | < 6.0 | ppbv | IAH |
| 1,3-Dichlorobenzene | < 6.0 | ppbv | IAH |
| 1,4-Dichlorobenzene | < 6.0 | ppbv | IAH |
| 1,4-Dioxane | < 6.0 | ppbv | IAH |
| 1-Butanol | < 6.0 | ppbv | IAH |
| 1-Propanol | < 6.0 | ppbv | IAH |
| 2,3-Dimethylpentane | < 6.0 | ppbv | IAH |
| 2,5-Dimethylfuran | < 6.0 | ppbv | IAH |
| 2-Butanone (Methyl ethyl ketone) | < 6.0 | ppbv | IAH |
| 2-Butenal | < 6.0 | ppbv | IAH |
| 2-Heptanone | < 6.0 | ppbv | IAH |
| 2-Methyl-1-propene | < 6.0 | ppbv | IAH |
| 2-Methyl-2-propanol | < 6.0 | ppbv | IAH |
| 2-Methylfuran | < 6.0 | ppbv | IAH |
| 2-Methylhexane | < 6.0 | ppbv | IAH |
| 2-Pentanone | < 6.0 | ppbv | IAH |
| 2-Pentenal | < 6.0 | ppbv | IAH |
| 2-Propanol (Isopropanol) | < 10 | ppbv | IAH |
| 3-Chloropropene (Allyl chloride) | < 6.0 | ppbv | IAH |
| 3-Methylhexane | < 6.0 | ppbv | IAH |

# Air Analysis Report

**Project: OSIRIS-REx Cleanroom Monitoring - JSC-XI2-001**
**Order ID: 231003004**

| | | | |
|---|---|---|---|
| **Sample Description:** | 12848 | **JSC Sample No.:** | AQ230789 |
| **Sample Location:** | Backshell / Recovery site | **Collection Date/Time:** | 09/24/2023 09:41 |
| **Site:** | JSC B31 | **Submitted by:** | N. Lunning |

| Test | Conc. | Units | Analyst |
|---|---|---|---|
| **Volatiles Targets GCMS (TO-15 mod)** | | | |
| 4-Methyl-2-pentanone (MIBK) | < 6.0 | ppbv | IAH |
| Acetaldehyde | < 10 | ppbv | IAH |
| Acetone | 14 | ppbv | IAH |
| Acetonitrile | < 6.0 | ppbv | IAH |
| Acrylonitrile | < 6.0 | ppbv | IAH |
| Benzene | < 6.0 | ppbv | IAH |
| Bromomethane | < 6.0 | ppbv | IAH |
| Butanal (Butyraldehyde) | < 6.0 | ppbv | IAH |
| Butane | < 6.0 | ppbv | IAH |
| Butyl acetate | < 6.0 | ppbv | IAH |
| Carbon disulfide | < 6.0 | ppbv | IAH |
| Carbon tetrachloride | < 6.0 | ppbv | IAH |
| Carbonyl sulfide (Carbon oxide sulfide) | < 6.0 | ppbv | IAH |
| Chlorobenzene | < 6.0 | ppbv | IAH |
| Chloroethane | < 6.0 | ppbv | IAH |
| Chloroform | < 6.0 | ppbv | IAH |
| Chloromethane | < 6.0 | ppbv | IAH |
| cis-1,2-Dichloroethene | < 6.0 | ppbv | IAH |
| cis-1,3-Dichloropropene | < 6.0 | ppbv | IAH |
| Cyclohexanone | < 6.0 | ppbv | IAH |
| Decamethylcyclopentasiloxane (DMCPS) | < 10 | ppbv | IAH |
| Dimethyl sulfide | < 6.0 | ppbv | IAH |
| Ethanol | < 6.0 | ppbv | IAH |
| Ethyl acetate | < 6.0 | ppbv | IAH |
| Ethylbenzene | < 6.0 | ppbv | IAH |
| Freon 11 (Trichlorofluoromethane) | < 6.0 | ppbv | IAH |
| Freon 113 (1,1,2-Trichloro-1,2,2-trifluoroethane) | < 6.0 | ppbv | IAH |
| Freon 114 (1,2-Dichloro-1,1,2,2-tetrafluoroethane) | < 6.0 | ppbv | IAH |
| Freon 12 (Dichlorodifluoromethane) | < 6.0 | ppbv | IAH |
| Furan | < 6.0 | ppbv | IAH |
| Heptanal | < 6.0 | ppbv | IAH |
| Hexachlorobutadiene | < 6.0 | ppbv | IAH |
| Hexanal | < 6.0 | ppbv | IAH |

# Air Analysis Report

**Project:** **OSIRIS-REx Cleanroom Monitoring - JSC-XI2-001**
**Order ID:** **231003004**

| | | | |
|---|---|---|---|
| **Sample Description:** | 12848 | **JSC Sample No.:** | AQ230789 |
| **Sample Location:** | Backshell / Recovery site | **Collection Date/Time:** | 09/24/2023 09:41 |
| **Site:** | JSC B31 | **Submitted by:** | N. Lunning |

| Test | Conc. | Units | Analyst |
|---|---|---|---|
| **Volatiles Targets GCMS (TO-15 mod)** | | | |
| Hexane | < 6.0 | ppbv | IAH |
| Isobutane | < 6.0 | ppbv | IAH |
| Isoprene (2-Methyl-1,3-butadiene) | < 6.0 | ppbv | IAH |
| m and p-Xylene | < 12 | ppbv | IAH |
| Mesityl oxide (4-Methyl-3-penten-2-one) | < 6.0 | ppbv | IAH |
| Methanol | < 10 | ppbv | IAH |
| Methyl acetate | < 6.0 | ppbv | IAH |
| Methylene chloride (Dichloromethane) | < 6.0 | ppbv | IAH |
| n-Heptane | < 6.0 | ppbv | IAH |
| Nonane | < 6.0 | ppbv | IAH |
| Octamethylcyclotetrasiloxane (OMCTS) | < 10 | ppbv | IAH |
| Octane | < 6.0 | ppbv | IAH |
| o-Xylene | < 6.0 | ppbv | IAH |
| Pentanal | < 6.0 | ppbv | IAH |
| Pentane | < 6.0 | ppbv | IAH |
| Perfluoro(2-methylpentane) | < 6.0 | ppbv | IAH |
| Propanal (Propionaldehyde) | < 6.0 | ppbv | IAH |
| Propane | < 6.0 | ppbv | IAH |
| Propenal (Acrolein) | < 6.0 | ppbv | IAH |
| Propene | < 6.0 | ppbv | IAH |
| Styrene (Ethenylbenzene) | < 6.0 | ppbv | IAH |
| Tetrachloroethene (Perchloroethene) | < 6.0 | ppbv | IAH |
| Toluene | < 6.0 | ppbv | IAH |
| trans-1,3-Dichloropropene | < 6.0 | ppbv | IAH |
| Trichloroethene | < 6.0 | ppbv | IAH |
| Trimethylsilanol | < 6.0 | ppbv | IAH |
| Vinyl chloride | < 6.0 | ppbv | IAH |
| **Volatiles SICs GCMS (estimated conc.)** | | | |
| Hexamethylcyclotrisiloxane (HMCTS) | < 20 | ppbv | IAH |

**Comments:**

*Samples:* AQ230789: Ambient air pressure

*Results:* None

# Air Analysis Report

**Project:** **OSIRIS-REx Cleanroom Monitoring - JSC-XI2-001**
**Order ID:** **231003004**

| | | | |
|---|---|---|---|
| **Sample Description:** | 35336 | **JSC Sample No.:** | AQ230790 |
| **Sample Location:** | UTTR Cleanroom 1012; Vent 1 | **Collection Date/Time:** | 09/24/2023 11:20 |
| **Site:** | JSC B31 | **Submitted by:** | N. Lunning |

| Test | Conc. | Units | Analyst |
|---|---|---|---|
| **Volatiles Targets GCMS (TO-15 mod)** | | | |
| 1,1,1,2-Tetrafluoroethane | < 10 | ppbv | IAH |
| 1,1,1-Trichloroethane | < 6.0 | ppbv | IAH |
| 1,1,2,2-Tetrachloroethane | < 6.0 | ppbv | IAH |
| 1,1,2-Trichloroethane | < 6.0 | ppbv | IAH |
| 1,1-Dichloroethane | < 6.0 | ppbv | IAH |
| 1,1-Dichloroethene | < 6.0 | ppbv | IAH |
| 1,2,4-Trichlorobenzene | < 10 | ppbv | IAH |
| 1,2,4-Trimethylbenzene | < 6.0 | ppbv | IAH |
| 1,2-Dibromoethane (EDB) | < 6.0 | ppbv | IAH |
| 1,2-Dichlorobenzene | < 6.0 | ppbv | IAH |
| 1,2-Dichloroethane | < 6.0 | ppbv | IAH |
| 1,2-Dichloropropane | < 6.0 | ppbv | IAH |
| 1,3,5-Trimethylbenzene | < 6.0 | ppbv | IAH |
| 1,3-Butadiene | < 6.0 | ppbv | IAH |
| 1,3-Dichlorobenzene | < 6.0 | ppbv | IAH |
| 1,4-Dichlorobenzene | < 6.0 | ppbv | IAH |
| 1,4-Dioxane | < 6.0 | ppbv | IAH |
| 1-Butanol | < 6.0 | ppbv | IAH |
| 1-Propanol | < 6.0 | ppbv | IAH |
| 2,3-Dimethylpentane | < 6.0 | ppbv | IAH |
| 2,5-Dimethylfuran | < 6.0 | ppbv | IAH |
| 2-Butanone (Methyl ethyl ketone) | < 6.0 | ppbv | IAH |
| 2-Butenal | < 6.0 | ppbv | IAH |
| 2-Heptanone | < 6.0 | ppbv | IAH |
| 2-Methyl-1-propene | < 6.0 | ppbv | IAH |
| 2-Methyl-2-propanol | < 6.0 | ppbv | IAH |
| 2-Methylfuran | < 6.0 | ppbv | IAH |
| 2-Methylhexane | < 6.0 | ppbv | IAH |
| 2-Pentanone | < 6.0 | ppbv | IAH |
| 2-Pentenal | < 6.0 | ppbv | IAH |
| 2-Propanol (Isopropanol) | See GC-FID | ppbv | IAH |
| 3-Chloropropene (Allyl chloride) | < 6.0 | ppbv | IAH |
| 3-Methylhexane | < 6.0 | ppbv | IAH |

# Air Analysis Report

**Project: OSIRIS-REx Cleanroom Monitoring - JSC-XI2-001**
**Order ID: 231003004**

| | | | |
|---|---|---|---|
| **Sample Description:** | 35336 | **JSC Sample No.:** | AQ230790 |
| **Sample Location:** | UTTR Cleanroom 1012; Vent 1 | **Collection Date/Time:** | 09/24/2023 11:20 |
| **Site:** | JSC B31 | **Submitted by:** | N. Lunning |

| Test | Conc. | Units | Analyst |
|---|---|---|---|
| **Volatiles Targets GCMS (TO-15 mod)** | | | |
| 4-Methyl-2-pentanone (MIBK) | < 6.0 | ppbv | IAH |
| Acetaldehyde | 14 | ppbv | IAH |
| Acetone | 38 | ppbv | IAH |
| Acetonitrile | < 6.0 | ppbv | IAH |
| Acrylonitrile | < 6.0 | ppbv | IAH |
| Benzene | 19 | ppbv | IAH |
| Bromomethane | < 6.0 | ppbv | IAH |
| Butanal (Butyraldehyde) | < 6.0 | ppbv | IAH |
| Butane | 7.1 | ppbv | IAH |
| Butyl acetate | < 6.0 | ppbv | IAH |
| Carbon disulfide | < 6.0 | ppbv | IAH |
| Carbon tetrachloride | < 6.0 | ppbv | IAH |
| Carbonyl sulfide (Carbon oxide sulfide) | < 6.0 | ppbv | IAH |
| Chlorobenzene | < 6.0 | ppbv | IAH |
| Chloroethane | < 6.0 | ppbv | IAH |
| Chloroform | < 6.0 | ppbv | IAH |
| Chloromethane | < 6.0 | ppbv | IAH |
| cis-1,2-Dichloroethene | < 6.0 | ppbv | IAH |
| cis-1,3-Dichloropropene | < 6.0 | ppbv | IAH |
| Cyclohexanone | < 6.0 | ppbv | IAH |
| Decamethylcyclopentasiloxane (DMCPS) | < 10 | ppbv | IAH |
| Dimethyl sulfide | < 6.0 | ppbv | IAH |
| Ethanol | 62 | ppbv | IAH |
| Ethyl acetate | < 6.0 | ppbv | IAH |
| Ethylbenzene | < 6.0 | ppbv | IAH |
| Freon 11 (Trichlorofluoromethane) | < 6.0 | ppbv | IAH |
| Freon 113 (1,1,2-Trichloro-1,2,2-trifluoroethane) | < 6.0 | ppbv | IAH |
| Freon 114 (1,2-Dichloro-1,1,2,2-tetrafluoroethane) | < 6.0 | ppbv | IAH |
| Freon 12 (Dichlorodifluoromethane) | < 6.0 | ppbv | IAH |
| Furan | < 6.0 | ppbv | IAH |
| Heptanal | < 6.0 | ppbv | IAH |
| Hexachlorobutadiene | < 6.0 | ppbv | IAH |
| Hexanal | < 6.0 | ppbv | IAH |

# Air Analysis Report

**Project: OSIRIS-REx Cleanroom Monitoring - JSC-XI2-001**
**Order ID: 231003004**

| | | | |
|---|---|---|---|
| **Sample Description:** | 35336 | **JSC Sample No.:** | AQ230790 |
| **Sample Location:** | UTTR Cleanroom 1012; Vent 1 | **Collection Date/Time:** | 09/24/2023 11:20 |
| **Site:** | JSC B31 | **Submitted by:** | N. Lunning |

| Test | Conc. | Units | Analyst |
|---|---|---|---|
| **Volatiles Targets GCMS (TO-15 mod)** | | | |
| Hexane | < 6.0 | ppbv | IAH |
| Isobutane | 18 | ppbv | IAH |
| Isoprene (2-Methyl-1,3-butadiene) | < 6.0 | ppbv | IAH |
| m and p-Xylene | < 12 | ppbv | IAH |
| Mesityl oxide (4-Methyl-3-penten-2-one) | < 6.0 | ppbv | IAH |
| Methanol | 310 | ppbv | IAH |
| Methyl acetate | < 6.0 | ppbv | IAH |
| Methylene chloride (Dichloromethane) | < 6.0 | ppbv | IAH |
| n-Heptane | < 6.0 | ppbv | IAH |
| Nonane | < 6.0 | ppbv | IAH |
| Octamethylcyclotetrasiloxane (OMCTS) | 18 | ppbv | IAH |
| Octane | < 6.0 | ppbv | IAH |
| o-Xylene | < 6.0 | ppbv | IAH |
| Pentanal | < 6.0 | ppbv | IAH |
| Pentane | < 6.0 | ppbv | IAH |
| Perfluoro(2-methylpentane) | < 6.0 | ppbv | IAH |
| Propanal (Propionaldehyde) | 9.6 | ppbv | IAH |
| Propane | 380 | ppbv | IAH |
| Propenal (Acrolein) | 13 | ppbv | IAH |
| Propene | 60 | ppbv | IAH |
| Styrene (Ethenylbenzene) | < 6.0 | ppbv | IAH |
| Tetrachloroethene (Perchloroethene) | < 6.0 | ppbv | IAH |
| Toluene | 22 | ppbv | IAH |
| trans-1,3-Dichloropropene | < 6.0 | ppbv | IAH |
| Trichloroethene | < 6.0 | ppbv | IAH |
| Trimethylsilanol | 44 | ppbv | IAH |
| Vinyl chloride | < 6.0 | ppbv | IAH |
| **Volatiles SICs GCMS (estimated conc.)** | | | |
| Hexamethylcyclotrisiloxane (HMCTS) | 370 | ppbv | IAH |
| **Volatiles Targets GCFID** | | | |
| 2-Propanol (Isopropanol) | 15000 | ppbv | CMM |
| Octafluoropropane (Perfluoropropane) | < 200 | ppbv | CMM |

# Air Analysis Report

**Project:** OSIRIS-REx Cleanroom Monitoring - JSC-XI2-001
**Order ID:** 231003004

| | | | |
|---|---|---|---|
| **Sample Description:** | 35336 | **JSC Sample No.:** | AQ230790 |
| **Sample Location:** | UTTR Cleanroom 1012; Vent 1 | **Collection Date/Time:** | 09/24/2023 11:20 |
| **Site:** | JSC B31 | **Submitted by:** | N. Lunning |

| Test | Conc. | Units | Analyst |
|---|---|---|---|

**Comments:**

*Samples:* AQ230790: Ambient air pressure

*Results:* None

# Air Analysis Report

**Project: OSIRIS-REx Cleanroom Monitoring - JSC-XI2-001**
**Order ID: 231003004**

| | | | |
|---|---|---|---|
| **Sample Description:** | 22383 | **JSC Sample No.:** | AQ230791 |
| **Sample Location:** | UTTR Cleanroom 1012; Filter | **Collection Date/Time:** | 09/24/2023 12:55 |
| **Site:** | JSC B31 | **Submitted by:** | N. Lunning |

| Test | Conc. | Units | Analyst |
|---|---|---|---|
| **Volatiles Targets GCMS (TO-15 mod)** | | | |
| 1,1,1,2-Tetrafluoroethane | < 10 | ppbv | IAH |
| 1,1,1-Trichloroethane | < 6.0 | ppbv | IAH |
| 1,1,2,2-Tetrachloroethane | < 6.0 | ppbv | IAH |
| 1,1,2-Trichloroethane | < 6.0 | ppbv | IAH |
| 1,1-Dichloroethane | < 6.0 | ppbv | IAH |
| 1,1-Dichloroethene | < 6.0 | ppbv | IAH |
| 1,2,4-Trichlorobenzene | < 10 | ppbv | IAH |
| 1,2,4-Trimethylbenzene | < 6.0 | ppbv | IAH |
| 1,2-Dibromoethane (EDB) | < 6.0 | ppbv | IAH |
| 1,2-Dichlorobenzene | < 6.0 | ppbv | IAH |
| 1,2-Dichloroethane | < 6.0 | ppbv | IAH |
| 1,2-Dichloropropane | < 6.0 | ppbv | IAH |
| 1,3,5-Trimethylbenzene | < 6.0 | ppbv | IAH |
| 1,3-Butadiene | < 6.0 | ppbv | IAH |
| 1,3-Dichlorobenzene | < 6.0 | ppbv | IAH |
| 1,4-Dichlorobenzene | < 6.0 | ppbv | IAH |
| 1,4-Dioxane | < 6.0 | ppbv | IAH |
| 1-Butanol | < 6.0 | ppbv | IAH |
| 1-Propanol | < 6.0 | ppbv | IAH |
| 2,3-Dimethylpentane | < 6.0 | ppbv | IAH |
| 2,5-Dimethylfuran | < 6.0 | ppbv | IAH |
| 2-Butanone (Methyl ethyl ketone) | < 6.0 | ppbv | IAH |
| 2-Butenal | < 6.0 | ppbv | IAH |
| 2-Heptanone | < 6.0 | ppbv | IAH |
| 2-Methyl-1-propene | < 6.0 | ppbv | IAH |
| 2-Methyl-2-propanol | < 6.0 | ppbv | IAH |
| 2-Methylfuran | < 6.0 | ppbv | IAH |
| 2-Methylhexane | < 6.0 | ppbv | IAH |
| 2-Pentanone | < 6.0 | ppbv | IAH |
| 2-Pentenal | < 6.0 | ppbv | IAH |
| 2-Propanol (Isopropanol) | See GC-FID | ppbv | IAH |
| 3-Chloropropene (Allyl chloride) | < 6.0 | ppbv | IAH |
| 3-Methylhexane | < 6.0 | ppbv | IAH |

# Air Analysis Report

**Project:** **OSIRIS-REx Cleanroom Monitoring - JSC-XI2-001**
**Order ID:** **231003004**

| | | | |
|---|---|---|---|
| **Sample Description:** | 22383 | **JSC Sample No.:** | AQ230791 |
| **Sample Location:** | UTTR Cleanroom 1012; Filter | **Collection Date/Time:** | 09/24/2023 12:55 |
| **Site:** | JSC B31 | **Submitted by:** | N. Lunning |

| Test | Conc. | Units | Analyst |
|---|---|---|---|
| **Volatiles Targets GCMS (TO-15 mod)** | | | |
| 4-Methyl-2-pentanone (MIBK) | < 6.0 | ppbv | IAH |
| Acetaldehyde | < 10 | ppbv | IAH |
| Acetone | 67 | ppbv | IAH |
| Acetonitrile | < 6.0 | ppbv | IAH |
| Acrylonitrile | < 6.0 | ppbv | IAH |
| Benzene | < 6.0 | ppbv | IAH |
| Bromomethane | < 6.0 | ppbv | IAH |
| Butanal (Butyraldehyde) | < 6.0 | ppbv | IAH |
| Butane | 6.0 | ppbv | IAH |
| Butyl acetate | < 6.0 | ppbv | IAH |
| Carbon disulfide | < 6.0 | ppbv | IAH |
| Carbon tetrachloride | < 6.0 | ppbv | IAH |
| Carbonyl sulfide (Carbon oxide sulfide) | < 6.0 | ppbv | IAH |
| Chlorobenzene | < 6.0 | ppbv | IAH |
| Chloroethane | < 6.0 | ppbv | IAH |
| Chloroform | < 6.0 | ppbv | IAH |
| Chloromethane | < 6.0 | ppbv | IAH |
| cis-1,2-Dichloroethene | < 6.0 | ppbv | IAH |
| cis-1,3-Dichloropropene | < 6.0 | ppbv | IAH |
| Cyclohexanone | < 6.0 | ppbv | IAH |
| Decamethylcyclopentasiloxane (DMCPS) | < 10 | ppbv | IAH |
| Dimethyl sulfide | < 6.0 | ppbv | IAH |
| Ethanol | 51 | ppbv | IAH |
| Ethyl acetate | < 6.0 | ppbv | IAH |
| Ethylbenzene | < 6.0 | ppbv | IAH |
| Freon 11 (Trichlorofluoromethane) | < 6.0 | ppbv | IAH |
| Freon 113 (1,1,2-Trichloro-1,2,2-trifluoroethane) | < 6.0 | ppbv | IAH |
| Freon 114 (1,2-Dichloro-1,1,2,2-tetrafluoroethane) | < 6.0 | ppbv | IAH |
| Freon 12 (Dichlorodifluoromethane) | < 6.0 | ppbv | IAH |
| Furan | < 6.0 | ppbv | IAH |
| Heptanal | < 6.0 | ppbv | IAH |
| Hexachlorobutadiene | < 6.0 | ppbv | IAH |
| Hexanal | < 6.0 | ppbv | IAH |

# Air Analysis Report

**Project: OSIRIS-REx Cleanroom Monitoring - JSC-XI2-001**
**Order ID: 231003004**

| | | | |
|---|---|---|---|
| **Sample Description:** | 22383 | **JSC Sample No.:** | AQ230791 |
| **Sample Location:** | UTTR Cleanroom 1012; Filter | **Collection Date/Time:** | 09/24/2023 12:55 |
| **Site:** | JSC B31 | **Submitted by:** | N. Lunning |

| Test | Conc. | Units | Analyst |
|---|---|---|---|
| **Volatiles Targets GCMS (TO-15 mod)** | | | |
| Hexane | < 6.0 | ppbv | IAH |
| Isobutane | 6.4 | ppbv | IAH |
| Isoprene (2-Methyl-1,3-butadiene) | < 6.0 | ppbv | IAH |
| m and p-Xylene | < 12 | ppbv | IAH |
| Mesityl oxide (4-Methyl-3-penten-2-one) | < 6.0 | ppbv | IAH |
| Methanol | 19 | ppbv | IAH |
| Methyl acetate | < 6.0 | ppbv | IAH |
| Methylene chloride (Dichloromethane) | < 6.0 | ppbv | IAH |
| n-Heptane | < 6.0 | ppbv | IAH |
| Nonane | < 6.0 | ppbv | IAH |
| Octamethylcyclotetrasiloxane (OMCTS) | < 10 | ppbv | IAH |
| Octane | < 6.0 | ppbv | IAH |
| o-Xylene | < 6.0 | ppbv | IAH |
| Pentanal | < 6.0 | ppbv | IAH |
| Pentane | < 6.0 | ppbv | IAH |
| Perfluoro(2-methylpentane) | < 6.0 | ppbv | IAH |
| Propanal (Propionaldehyde) | 12 | ppbv | IAH |
| Propane | 310 | ppbv | IAH |
| Propenal (Acrolein) | < 6.0 | ppbv | IAH |
| Propene | 41 | ppbv | IAH |
| Styrene (Ethenylbenzene) | < 6.0 | ppbv | IAH |
| Tetrachloroethene (Perchloroethene) | 7.6 | ppbv | IAH |
| Toluene | < 6.0 | ppbv | IAH |
| trans-1,3-Dichloropropene | < 6.0 | ppbv | IAH |
| Trichloroethene | < 6.0 | ppbv | IAH |
| Trimethylsilanol | < 6.0 | ppbv | IAH |
| Vinyl chloride | < 6.0 | ppbv | IAH |
| **Volatiles SICs GCMS (estimated conc.)** | | | |
| Hexamethylcyclotrisiloxane (HMCTS) | < 20 | ppbv | IAH |
| **Volatiles Targets GCFID** | | | |
| 2-Propanol (Isopropanol) | 10000 | ppbv | CMM |
| Octafluoropropane (Perfluoropropane) | < 200 | ppbv | CMM |

# Air Analysis Report

**Project:** **OSIRIS-REx Cleanroom Monitoring - JSC-XI2-001**
**Order ID:** **231003004**

| | | | |
|---|---|---|---|
| **Sample Description:** | 22383 | **JSC Sample No.:** | AQ230791 |
| **Sample Location:** | UTTR Cleanroom 1012; Filter | **Collection Date/Time:** | 09/24/2023 12:55 |
| **Site:** | JSC B31 | **Submitted by:** | N. Lunning |

| Test | Conc. | Units | Analyst |
|---|---|---|---|

**Comments:**

*Samples:* AQ230791: Ambient air pressure

*Results:* None

# Air Analysis Report

**Project: OSIRIS-REx Cleanroom Monitoring - JSC-XI2-001**
**Order ID: 231003004**

| | | | |
|---|---|---|---|
| **Sample Description:** | 24021 | **JSC Sample No.:** | AQ230792 |
| **Sample Location:** | UTTR Cleanroom 1012; Vent 2 | **Collection Date/Time:** | 09/24/2023 11:20 |
| **Site:** | JSC B31 | **Submitted by:** | N. Lunning |

| Test | Conc. | Units | Analyst |
|---|---|---|---|
| **Volatiles Targets GCMS (TO-15 mod)** | | | |
| 1,1,1,2-Tetrafluoroethane | < 10 | ppbv | IAH |
| 1,1,1-Trichloroethane | < 6.0 | ppbv | IAH |
| 1,1,2,2-Tetrachloroethane | < 6.0 | ppbv | IAH |
| 1,1,2-Trichloroethane | < 6.0 | ppbv | IAH |
| 1,1-Dichloroethane | < 6.0 | ppbv | IAH |
| 1,1-Dichloroethene | < 6.0 | ppbv | IAH |
| 1,2,4-Trichlorobenzene | < 10 | ppbv | IAH |
| 1,2,4-Trimethylbenzene | < 6.0 | ppbv | IAH |
| 1,2-Dibromoethane (EDB) | < 6.0 | ppbv | IAH |
| 1,2-Dichlorobenzene | < 6.0 | ppbv | IAH |
| 1,2-Dichloroethane | < 6.0 | ppbv | IAH |
| 1,2-Dichloropropane | < 6.0 | ppbv | IAH |
| 1,3,5-Trimethylbenzene | < 6.0 | ppbv | IAH |
| 1,3-Butadiene | < 6.0 | ppbv | IAH |
| 1,3-Dichlorobenzene | < 6.0 | ppbv | IAH |
| 1,4-Dichlorobenzene | < 6.0 | ppbv | IAH |
| 1,4-Dioxane | < 6.0 | ppbv | IAH |
| 1-Butanol | < 6.0 | ppbv | IAH |
| 1-Propanol | < 6.0 | ppbv | IAH |
| 2,3-Dimethylpentane | < 6.0 | ppbv | IAH |
| 2,5-Dimethylfuran | < 6.0 | ppbv | IAH |
| 2-Butanone (Methyl ethyl ketone) | < 6.0 | ppbv | IAH |
| 2-Butenal | < 6.0 | ppbv | IAH |
| 2-Heptanone | < 6.0 | ppbv | IAH |
| 2-Methyl-1-propene | < 6.0 | ppbv | IAH |
| 2-Methyl-2-propanol | < 6.0 | ppbv | IAH |
| 2-Methylfuran | < 6.0 | ppbv | IAH |
| 2-Methylhexane | < 6.0 | ppbv | IAH |
| 2-Pentanone | < 6.0 | ppbv | IAH |
| 2-Pentenal | < 6.0 | ppbv | IAH |
| 2-Propanol (Isopropanol) | <125 | ppbv | IAH |
| 3-Chloropropene (Allyl chloride) | < 6.0 | ppbv | IAH |
| 3-Methylhexane | < 6.0 | ppbv | IAH |

# Air Analysis Report

**Project: OSIRIS-REx Cleanroom Monitoring - JSC-XI2-001**
**Order ID: 231003004**

| | | | |
|---|---|---|---|
| **Sample Description:** | 24021 | **JSC Sample No.:** | AQ230792 |
| **Sample Location:** | UTTR Cleanroom 1012; Vent 2 | **Collection Date/Time:** | 09/24/2023 11:20 |
| **Site:** | JSC B31 | **Submitted by:** | N. Lunning |

| Test | Conc. | Units | Analyst |
|---|---|---|---|
| **Volatiles Targets GCMS (TO-15 mod)** | | | |
| 4-Methyl-2-pentanone (MIBK) | < 6.0 | ppbv | IAH |
| Acetaldehyde | < 10 | ppbv | IAH |
| Acetone | 35 | ppbv | IAH |
| Acetonitrile | < 6.0 | ppbv | IAH |
| Acrylonitrile | < 6.0 | ppbv | IAH |
| Benzene | < 6.0 | ppbv | IAH |
| Bromomethane | < 6.0 | ppbv | IAH |
| Butanal (Butyraldehyde) | < 6.0 | ppbv | IAH |
| Butane | < 6.0 | ppbv | IAH |
| Butyl acetate | < 6.0 | ppbv | IAH |
| Carbon disulfide | < 6.0 | ppbv | IAH |
| Carbon tetrachloride | < 6.0 | ppbv | IAH |
| Carbonyl sulfide (Carbon oxide sulfide) | < 6.0 | ppbv | IAH |
| Chlorobenzene | < 6.0 | ppbv | IAH |
| Chloroethane | < 6.0 | ppbv | IAH |
| Chloroform | < 6.0 | ppbv | IAH |
| Chloromethane | < 6.0 | ppbv | IAH |
| cis-1,2-Dichloroethene | < 6.0 | ppbv | IAH |
| cis-1,3-Dichloropropene | < 6.0 | ppbv | IAH |
| Cyclohexanone | < 6.0 | ppbv | IAH |
| Decamethylcyclopentasiloxane (DMCPS) | < 10 | ppbv | IAH |
| Dimethyl sulfide | < 6.0 | ppbv | IAH |
| Ethanol | 61 | ppbv | IAH |
| Ethyl acetate | < 6.0 | ppbv | IAH |
| Ethylbenzene | < 6.0 | ppbv | IAH |
| Freon 11 (Trichlorofluoromethane) | < 6.0 | ppbv | IAH |
| Freon 113 (1,1,2-Trichloro-1,2,2-trifluoroethane) | < 6.0 | ppbv | IAH |
| Freon 114 (1,2-Dichloro-1,1,2,2-tetrafluoroethane) | < 6.0 | ppbv | IAH |
| Freon 12 (Dichlorodifluoromethane) | < 6.0 | ppbv | IAH |
| Furan | < 6.0 | ppbv | IAH |
| Heptanal | < 6.0 | ppbv | IAH |
| Hexachlorobutadiene | < 6.0 | ppbv | IAH |
| Hexanal | < 6.0 | ppbv | IAH |

# Air Analysis Report

**Project:** **OSIRIS-REx Cleanroom Monitoring - JSC-XI2-001**
**Order ID:** **231003004**

| | | | |
|---|---|---|---|
| **Sample Description:** | 24021 | **JSC Sample No.:** | AQ230792 |
| **Sample Location:** | UTTR Cleanroom 1012; Vent 2 | **Collection Date/Time:** | 09/24/2023 11:20 |
| **Site:** | JSC B31 | **Submitted by:** | N. Lunning |

| Test | Conc. | Units | Analyst |
|---|---|---|---|
| **Volatiles Targets GCMS (TO-15 mod)** | | | |
| Hexane | < 6.0 | ppbv | IAH |
| Isobutane | 11 | ppbv | IAH |
| Isoprene (2-Methyl-1,3-butadiene) | < 6.0 | ppbv | IAH |
| m and p-Xylene | < 12 | ppbv | IAH |
| Mesityl oxide (4-Methyl-3-penten-2-one) | < 6.0 | ppbv | IAH |
| Methanol | 58 | ppbv | IAH |
| Methyl acetate | < 6.0 | ppbv | IAH |
| Methylene chloride (Dichloromethane) | < 6.0 | ppbv | IAH |
| n-Heptane | < 6.0 | ppbv | IAH |
| Nonane | < 6.0 | ppbv | IAH |
| Octamethylcyclotetrasiloxane (OMCTS) | < 10 | ppbv | IAH |
| Octane | < 6.0 | ppbv | IAH |
| o-Xylene | < 6.0 | ppbv | IAH |
| Pentanal | < 6.0 | ppbv | IAH |
| Pentane | < 6.0 | ppbv | IAH |
| Perfluoro(2-methylpentane) | < 6.0 | ppbv | IAH |
| Propanal (Propionaldehyde) | 8.8 | ppbv | IAH |
| Propane | < 6.0 | ppbv | IAH |
| Propenal (Acrolein) | < 6.0 | ppbv | IAH |
| Propene | 60 | ppbv | IAH |
| Styrene (Ethenylbenzene) | < 6.0 | ppbv | IAH |
| Tetrachloroethene (Perchloroethene) | 6.4 | ppbv | IAH |
| Toluene | 6.9 | ppbv | IAH |
| trans-1,3-Dichloropropene | < 6.0 | ppbv | IAH |
| Trichloroethene | < 6.0 | ppbv | IAH |
| Trimethylsilanol | 11 | ppbv | IAH |
| Vinyl chloride | < 6.0 | ppbv | IAH |
| **Volatiles SICs GCMS (estimated conc.)** | | | |
| Hexamethylcyclotrisiloxane (HMCTS) | 40 | ppbv | IAH |
| **Volatiles Targets GCFID** | | | |
| 2-Propanol (Isopropanol) | 16000 | ppbv | CMM |
| Octafluoropropane (Perfluoropropane) | < 200 | ppbv | CMM |

# Air Analysis Report

**Project:** **OSIRIS-REx Cleanroom Monitoring - JSC-XI2-001**
**Order ID:** **231003004**

| | | | |
|---|---|---|---|
| **Sample Description:** | 24021 | **JSC Sample No.:** | AQ230792 |
| **Sample Location:** | UTTR Cleanroom 1012; Vent 2 | **Collection Date/Time:** | 09/24/2023 11:20 |
| **Site:** | JSC B31 | **Submitted by:** | N. Lunning |

| Test | Conc. | Units | Analyst |
|---|---|---|---|

**Comments:**
*Samples:* AQ230792: Ambient air pressure
*Results:* None

# Air Analysis Report

**Project: OSIRIS-REx Cleanroom Monitoring - JSC-XI2-001**
**Order ID: 231003004**

| | | | |
|---|---|---|---|
| **Sample Description:** | 21260 | **JSC Sample No.:** | AQ230793 |
| **Sample Location:** | Dust Sampler 2 / Recovery site | **Collection Date/Time:** | 09/24/2023 09:57 |
| **Site:** | JSC B31 | **Submitted by:** | N. Lunning |

| Test | Conc. | Units | Analyst |
|---|---|---|---|
| **Volatiles Targets GCMS (TO-15 mod)** | | | |
| 1,1,1,2-Tetrafluoroethane | < 10 | ppbv | IAH |
| 1,1,1-Trichloroethane | < 6.0 | ppbv | IAH |
| 1,1,2,2-Tetrachloroethane | < 6.0 | ppbv | IAH |
| 1,1,2-Trichloroethane | < 6.0 | ppbv | IAH |
| 1,1-Dichloroethane | < 6.0 | ppbv | IAH |
| 1,1-Dichloroethene | < 6.0 | ppbv | IAH |
| 1,2,4-Trichlorobenzene | < 10 | ppbv | IAH |
| 1,2,4-Trimethylbenzene | < 6.0 | ppbv | IAH |
| 1,2-Dibromoethane (EDB) | < 6.0 | ppbv | IAH |
| 1,2-Dichlorobenzene | < 6.0 | ppbv | IAH |
| 1,2-Dichloroethane | < 6.0 | ppbv | IAH |
| 1,2-Dichloropropane | < 6.0 | ppbv | IAH |
| 1,3,5-Trimethylbenzene | < 6.0 | ppbv | IAH |
| 1,3-Butadiene | < 6.0 | ppbv | IAH |
| 1,3-Dichlorobenzene | < 6.0 | ppbv | IAH |
| 1,4-Dichlorobenzene | < 6.0 | ppbv | IAH |
| 1,4-Dioxane | < 6.0 | ppbv | IAH |
| 1-Butanol | < 6.0 | ppbv | IAH |
| 1-Propanol | < 6.0 | ppbv | IAH |
| 2,3-Dimethylpentane | < 3.0 | ppbv | IAH |
| 2,5-Dimethylfuran | < 6.0 | ppbv | IAH |
| 2-Butanone (Methyl ethyl ketone) | < 6.0 | ppbv | IAH |
| 2-Butenal | < 6.0 | ppbv | IAH |
| 2-Heptanone | < 6.0 | ppbv | IAH |
| 2-Methyl-1-propene | < 6.0 | ppbv | IAH |
| 2-Methyl-2-propanol | < 6.0 | ppbv | IAH |
| 2-Methylfuran | < 6.0 | ppbv | IAH |
| 2-Methylhexane | < 6.0 | ppbv | IAH |
| 2-Pentanone | < 6.0 | ppbv | IAH |
| 2-Pentenal | < 6.0 | ppbv | IAH |
| 2-Propanol (Isopropanol) | < 10 | ppbv | IAH |
| 3-Chloropropene (Allyl chloride) | < 6.0 | ppbv | IAH |
| 3-Methylhexane | < 6.0 | ppbv | IAH |

# Air Analysis Report

**Project:** **OSIRIS-REx Cleanroom Monitoring - JSC-XI2-001**
**Order ID:** **231003004**

| | | | |
|---|---|---|---|
| **Sample Description:** | 21260 | **JSC Sample No.:** | AQ230793 |
| **Sample Location:** | Dust Sampler 2 / Recovery site | **Collection Date/Time:** | 09/24/2023 09:57 |
| **Site:** | JSC B31 | **Submitted by:** | N. Lunning |

| Test | Conc. | Units | Analyst |
|---|---|---|---|
| **Volatiles Targets GCMS (TO-15 mod)** | | | |
| 4-Methyl-2-pentanone (MIBK) | < 6.0 | ppbv | IAH |
| Acetaldehyde | < 10 | ppbv | IAH |
| Acetone | 13 | ppbv | IAH |
| Acetonitrile | < 6.0 | ppbv | IAH |
| Acrylonitrile | < 6.0 | ppbv | IAH |
| Benzene | < 6.0 | ppbv | IAH |
| Bromomethane | < 6.0 | ppbv | IAH |
| Butanal (Butyraldehyde) | < 6.0 | ppbv | IAH |
| Butane | < 6.0 | ppbv | IAH |
| Butyl acetate | < 6.0 | ppbv | IAH |
| Carbon disulfide | < 6.0 | ppbv | IAH |
| Carbon tetrachloride | < 6.0 | ppbv | IAH |
| Carbonyl sulfide (Carbon oxide sulfide) | < 6.0 | ppbv | IAH |
| Chlorobenzene | < 6.0 | ppbv | IAH |
| Chloroethane | < 6.0 | ppbv | IAH |
| Chloroform | < 6.0 | ppbv | IAH |
| Chloromethane | < 6.0 | ppbv | IAH |
| cis-1,2-Dichloroethene | < 6.0 | ppbv | IAH |
| cis-1,3-Dichloropropene | < 6.0 | ppbv | IAH |
| Cyclohexanone | < 6.0 | ppbv | IAH |
| Decamethylcyclopentasiloxane (DMCPS) | < 10 | ppbv | IAH |
| Dimethyl sulfide | < 6.0 | ppbv | IAH |
| Ethanol | < 6.0 | ppbv | IAH |
| Ethyl acetate | < 6.0 | ppbv | IAH |
| Ethylbenzene | < 6.0 | ppbv | IAH |
| Freon 11 (Trichlorofluoromethane) | < 6.0 | ppbv | IAH |
| Freon 113 (1,1,2-Trichloro-1,2,2-trifluoroethane) | < 6.0 | ppbv | IAH |
| Freon 114 (1,2-Dichloro-1,1,2,2-tetrafluoroethane) | < 6.0 | ppbv | IAH |
| Freon 12 (Dichlorodifluoromethane) | < 6.0 | ppbv | IAH |
| Furan | < 6.0 | ppbv | IAH |
| Heptanal | < 6.0 | ppbv | IAH |
| Hexachlorobutadiene | < 6.0 | ppbv | IAH |
| Hexanal | < 6.0 | ppbv | IAH |

# Air Analysis Report

**Project:** OSIRIS-REx Cleanroom Monitoring - JSC-XI2-001
**Order ID:** 231003004

| | | | |
|---|---|---|---|
| **Sample Description:** | 21260 | **JSC Sample No.:** | AQ230793 |
| **Sample Location:** | Dust Sampler 2 / Recovery site | **Collection Date/Time:** | 09/24/2023 09:57 |
| **Site:** | JSC B31 | **Submitted by:** | N. Lunning |

| Test | Conc. | Units | Analyst |
|---|---|---|---|
| **Volatiles Targets GCMS (TO-15 mod)** | | | |
| Hexane | < 6.0 | ppbv | IAH |
| Isobutane | < 6.0 | ppbv | IAH |
| Isoprene (2-Methyl-1,3-butadiene) | < 6.0 | ppbv | IAH |
| m and p-Xylene | < 12 | ppbv | IAH |
| Mesityl oxide (4-Methyl-3-penten-2-one) | < 6.0 | ppbv | IAH |
| Methanol | < 10 | ppbv | IAH |
| Methyl acetate | < 6.0 | ppbv | IAH |
| Methylene chloride (Dichloromethane) | < 6.0 | ppbv | IAH |
| n-Heptane | < 6.0 | ppbv | IAH |
| Nonane | < 6.0 | ppbv | IAH |
| Octamethylcyclotetrasiloxane (OMCTS) | < 10 | ppbv | IAH |
| Octane | < 6.0 | ppbv | IAH |
| o-Xylene | < 6.0 | ppbv | IAH |
| Pentanal | < 6.0 | ppbv | IAH |
| Pentane | < 6.0 | ppbv | IAH |
| Perfluoro(2-methylpentane) | < 6.0 | ppbv | IAH |
| Propanal (Propionaldehyde) | < 6.0 | ppbv | IAH |
| Propane | < 6.0 | ppbv | IAH |
| Propenal (Acrolein) | < 6.0 | ppbv | IAH |
| Propene | < 6.0 | ppbv | IAH |
| Styrene (Ethenylbenzene) | < 6.0 | ppbv | IAH |
| Tetrachloroethene (Perchloroethene) | < 6.0 | ppbv | IAH |
| Toluene | < 6.0 | ppbv | IAH |
| trans-1,3-Dichloropropene | < 6.0 | ppbv | IAH |
| Trichloroethene | < 6.0 | ppbv | IAH |
| Trimethylsilanol | < 6.0 | ppbv | IAH |
| Vinyl chloride | < 6.0 | ppbv | IAH |
| **Volatiles SICs GCMS (estimated conc.)** | | | |
| Hexamethylcyclotrisiloxane (HMCTS) | < 20 | ppbv | IAH |

**Comments:**

*Samples:* AQ230793: Ambient air pressure

*Results:* None

# Air Analysis Report

**Project:** **OSIRIS-REx Cleanroom Monitoring - JSC-XI2-001**
**Order ID:** **231003004**

NOTES MI = Matrix Interference | NA = Not Analyzed | NR = Not Reported

**Remarks:** *Sample Login:* None.

**Reviewed by:** LTW

**Approved by:** Ed Hudson, TEC Laboratory Lead on 10/19/23 08:33

# TOXICOLOGY & ENVIRONMENTAL CHEMISTRY (TEC) CHAIN OF CUSTODY FORM

Project Name: OSIRIS REx Clean Room Monitoring
Mission (if applicable): OSIRIS REx
Project Site/Location: B31, R216
Project No.: JSC-XI2-001

Order ID: 231003004

Customer Contact: Nicole Lunning
Contact Phone No.: 281-244-9923
Report Results to: Nicole Lunning

R. 10/17/23

| | | Sample Identification | | Sample Collection | | | | | | Comments | Analyses |
|---|---|---|---|---|---|---|---|---|---|---|---|
| Item No. | JSC Sample No. | Sample Identification | Location | Date | Time (24 hr) | Volume | Qty | Matrix Code* | Sample Type** | Sample Remarks | Tests Requested |
| 1 | AQ230782 | 21189 | purge back up / UTTR cleanroom 1012 | 9/22/2023 | 12:35 | 528cc | 1 | A | G | collected inline with quick disconnect fitting and line pressure near 30 psi | Volatiles: GC-MS Targets, SICs, Non-Targets; TGA |
| 2 | 83 | 23681 | purge / UTTR cleanroom 1012 | 9/22/2023 | 12:38 | 528cc | 1 | A | G | collected inline with quick disconnect fitting and line pressure near 30 psi | Volatiles: GC-MS Targets, SICs, Non-Targets; TGA |
| 3 | 84 | 34990 | cleanroom ambient air / UTTR cleanroom 1012 | 9/22/2023 | 12:41 | 528cc | 1 | A | G | ambient air pressure | Volatiles: GC-MS Targets, SICs, Non-Targets; TGA |
| 4 | 85 | 22378 | vent 1 / recovery site | 9/24/2023 | 9:43 | 528cc | 1 | A | G | ambient air pressure | Volatiles: GC-MS Targets, SICs, Non-Targets; TGA |
| 5 | 86 | 34989 | heat shield / recovery site | 9/24/2023 | 9:40 | 528cc | 1 | A | G | ambient air pressure | Volatiles: GC-MS Targets, SICs, Non-Targets; TGA |
| 6 | 87 | 25218 | dust sampler 1 / recovery site | 9/24/2023 | 9:54 | 528cc | 1 | A | G | ambient air pressure | Volatiles: GC-MS Targets, SICs, Non-Targets; TGA |
| 7 | 88 | 21190 | vent 2 / recovery site | 9/24/2023 | 9:44 | 528cc | 1 | A | G | ambient air pressure | Volatiles: GC-MS Targets, SICs, Non-Targets; TGA |
| 8 | 89 | 12848 | backshell / recovery site | 9/24/2023 | 9:41 | 528cc | 1 | A | G | ambient air pressure | Volatiles: GC-MS Targets, SICs, Non-Targets; TGA |

*Matrix Codes: A=Air, BA=Badge, CG=Compressed Gas, U=Urine (raw, preserved & brine), W=Water (potable, product & groundwater) & WW=Wastewater (condensate,extract, ww & urine distillate)
**Sample Type : G=Grab , C=Composite, O=OGT Test Article

| Sampler (Print): | Ian Harper |
|---|---|

| | Signature | Date/Time |
|---|---|---|
| Relinquished by: | Ian A. Harper | 10/3/2023 |
| Relinquished by: | | |

| | Signature | Date/Time |
|---|---|---|
| Received by: | | |
| Received by: | Anne Brown | 10/3/23 1305 |

FRM-TEC-002
Rev. 1

Verify this is the correct version before use

04/03/2018
Page 1 of 2

# TOXICOLOGY & ENVIRONMENTAL CHEMISTRY (TEC) CHAIN OF CUSTODY FORM

Project Name: OSIRIS REx Clean Room Monitoring
Mission (if applicable): OSIRIS REx
Project Site/Location: B31, R216
Project No.: JSC-XI2-001

Order ID: 23100 3004

Customer Contact: Nicole Lunning
Contact Phone No.: 281-244-9923
Report Results to: Nicole Lunning

R - 10/17/23

| | | Sample Identification | | Sample Collection | | | | | | Comments | Analyses |
|---|---|---|---|---|---|---|---|---|---|---|---|
| Item No. | JSC Sample No. | Description/ID | Location | Date | Time (24 hr) | Volume | Qty | Matrix Code* | Sample Type** | Sample Remarks | Tests Requested |
| 9 | AQ 230790 | 35336 | UTTR cleanroom 1012; vent 1 | 9/24/2023 | 11:20 | 528cc | 1 | A | G | ambient air pressure | Volatiles: GC-MS Targets, SICs, Non-Targets; TGA |
| 10 | 91 | 22383 | UTTR cleanroom 1012; filter | 9/24/2023 | 12:55 | 528cc | 1 | A | G | ambient air pressure | Volatiles: GC-MS Targets, SICs, Non-Targets; TGA |
| 11 | 92 | 24021 | UTTR cleanroom 1012; vent 2 | 9/24/2023 | 11:20 | 528cc | 1 | A | G | ambient air pressure | Volatiles: GC-MS Targets, SICs, Non-Targets; TGA |
| 12 | 93 | 21260 | dust sampler 2 / recovery site | 9/24/2023 | 9:57 | 528cc | 1 | A | G | ambient air pressure | Volatiles: GC-MS Targets, SICs, Non-Targets; TGA |
| | | | | | | | | | | | |
| | | | | | | | | | | | |
| | | | | | | | | | | | |
| 8 | | | | | | | | | | | |

*Matrix Codes: A=Air, BA=Badge, CG=Compressed Gas, U=Urine (raw, preserved & brine), W=Water (potable, product & groundwater) & WW=Wastewater (condensate,extract, ww & urine distillate)
**Sample Type : G=Grab , C=Composite, O=OGT Test Article

| Sampler (Print): | Ian Harper |
|---|---|

| | Signature | Date/Time |
|---|---|---|
| Relinquished by: | Ian A. Harper | 10/3/2023 |
| Relinquished by: | | |

| | Signature | Date/Time |
|---|---|---|
| Received by: | | |
| Received by: | Bruce Brown | 10/3/23 1305 |

FRM-TEC-002
Rev. 1

Verify this is the correct version before use

04/03/2018

## **Supplement 5:** Material restrictions within the OSIRIS-REx curation lab

***Forbidden Materials (material waivers will not be considered for these materials):***

- Nylon
- Silicones
- 3D printed materials

**Materials allowed in gloveboxes (all must undergo OSIRIS-REx cleaning procedure and triple bagging in PFTE).**

Glass:

- Schott Amiran Low Iron Laminated (glovebox windows)
- Pilkington OpticWhite (uncoated glovebox windows only)
- Fused Quartz/Fused Silica
- Borosilicate Crown Glass (BK7)
- Soda-Lime silicate glass (above without Boron)
- Sapphire/Sapphire Glass

Plastics (In order of preference):

- Polytetrafluoroethylene (PTFE; Any metal with PTFE coating is acceptable w/approval)
- Fluorinated Ethylene Propylene (FEP)
- Polyvinylidene Fluoride (PVDF; For Tubing)
- Perfluoroalkyl Alkanes (PFA; For Tubing)

Metals (In order of preference):

- 6061 and 6063 Aluminum (Unfinished or Clear anodized)
- 316/316l Stainless Steel
- 304 Stainless Steel
- 2024 Aluminum (Clear anodized)
- 301, 302, 303 Stainless Steel

Finishing for metals:

- Clear and Hard Anodized Aluminum
- Pickled and Passivated
- Electro-polished
- Gold Plated (For electrical contacts only)
- Satin/Pebble/media blast

Elastomers:

- Viton (FKM)
- Chlorosulfonated Polyethylene (CSM, Hypalon)

**Materials allowed in OSIRIS-REx Curation Lab (not in gloveboxes). All other materials will require a materials waiver review to be allowed in the curation lab:**

- All of the materials allowed in gloveboxes (above)
- Cleanroom Polyester (gowning and similar uses)
- Teflon and Kapton tape
- Nitrile (gloves and similar uses)
- Polycarbonate (Lexan)

- Additional stainless-steel alloys beyond those approved for glovebox use (e.g., hand tools, carts, tables)
- Computers, cameras, and microscopes that live permanently in the curation lab (minimize contamination that would “hitchhike” on this type of equipment coming and going from the lab)
- LED light sources for above

## **Supplement 6:** Contamination memos

Contamination Assessment Memo

22 March 2023

# Glovebox Pycnometer

Jason Dworkin
C. Chris Lorentson

This memo is a **conditional approval** and evaluation of the materials and implementation of the custom pycnometer proposed to be used on "curation-pristine" Bennu samples in the sample glovebox in the ISO 5 OSIRIS-REx cleanroom at NASA JSC. Curation-pristine is a condition where the sample is unimpacted by the process to be considered unchanged by the process and is retained as the unallocated NASA collection.

The pycnometer team should be commended for their responsiveness to contamination control requests over an extended period of discussion and iteration. They have done an excellent job of ensuring that all components and gases directly exposed to the sample and glovebox interior (wetted components) are compliant and will not contaminate the sample or environment. Even so, it is strongly recommended (and in the baseline plan) **to place representative wetted components into the materials archive**, as insurance against unexpected issues.

**Contamination knowledge information for particulates** should be collected inside the sample chamber. The recommended procedure is to place a substrate in the sample chamber and perform one measurement and repeat it with a different substrate with 10 measurements performed sequentially on a single substrate to allow for improved confidence for the SEM/EDS detection. The substrates would then be measured by SEM/EDS at JSC. This examination could be conducted in parallel with Curation's collection of Balazs witness plates for assessment of the glovebox interior.

There are components in the electronics and plumbing boxes which are known to be suspect/non-compliant to the ISO 5 environment. Some components have been proposed to be and **should be replaced** with compliant materials, such as the unknown plastic zip-ties. The use of Kapton tape to cover problematic components is effective for particulates, but only delays outgassing. Thus, the plumbing box, electronics box, and cables between **should be bagged** in PTFE according to JSC Cleanroom procedures. The metering valve on the plumbing box is of potential concern. It has been stated that it only rarely needs to be adjusted and can be adjusted through the bag. As with other computers in the ISO 5 space, it must not contain a fan and cannot return to the cleanroom if it is removed.

**Slow abrasion of the rotating handle is a known potential source of steel particles**. Other options for mitigation have been explored and been unsatisfactory. The rotating handle should be periodically inspected for wear after each day the pycnometer is used. The steel in the handle used should be characterized by SEM/EDS at JSC to enable quick identification on cleanroom contamination knowledge witness plates prior to cleanroom commissioning. In addition to the archive of the wetted components above, >1g of the same steel used on the rotating handle should also be archived.

**Additional procedural recommendations:**

1. The molding of the foil liner to the mold should not be done by gloved hand or must be cleaned by heating to 500 °C in air for 3-16 hours prior to use.
2. There is a tortuous path from the sample chamber to the interior wetted components of the pycnometer. As such, it is unlikely that dust from one Bennu sample can mix with a subsequent Bennu or calibration sample. Nevertheless, since the wetted areas of the pycnometer outside of the sample chamber are sealed, a visual inspection for dust outside the foil liner, but inside the sample chamber, should be conducted before adding a sample for analysis.

2 November 2023

## Glovebox Pycnometer

Jason Dworkin

This memo is an assessment of the contamination testing of the pycnometer as installed in the canister glovebox is sufficiently non-contaminating to analyze Bennu samples while maintaining them as "Curation Pristine." The tests performed suggest the pycnometer can be operated as designed without a detectable impact on the pristinity of the sample.

***Procedure***

The test, described in cleanroom Step B3, was conducted by placing a 0.9x0.9-inch foil in the pycnometer chamber and running a cycle (sample foil). An identical foil was left in the glovebox during the procedure as a control. This process was repeated a total of four times. These eight foils had been cut in flow bench to prevent the introduction of ambient particles. These foils were wrapped in a foil envelope and baked at 500 °C overnight in air to combust any organic material. Three pairs of samples were analyzed independently by SEM/EDS for particles, pyrolysis-GCMS for broad organic survey, and LCMS for amino acids. The procedures for the instruments are described in appendices 1, 2, and 3, respectively. The fourth foil was held in reserve to be used either for an archive or a repeat analysis. Analysts were told which sample was which and asked to look for differences between the pair of foils. Since the canister glovebox has been used for the processing of hardware and Bennu material, it is reasonable that traces of either would be in the glovebox.

***Results***

Particles: SEM/EDS analysis arbitrarily examined one side of each foil (the interior of the fold) and avoided the edges where the foils were cut since tool shavings are likely. An exhaustive search (see Appendix 1) showed no visible particles on the control and four particles 2-3 μm on the sample in pits. One particle is Fe-Cr-Ni metal (interpreted as stainless steel), two of Ti metal, and one is composed of a mixture of Fe, Si, and Ti.

Bulk organics: Pyrolysis GCMS showed that the control and sample are both indistinguishable and exceptionally clean.

Amino acids: LC-MS analysis of extracted acid-hydrolyzed foil generally detected lower amino acids than in the control. This indicates that the pycnometer has 3% of the amino acids than the ambient environment in the glovebox. A slight excess of serine was seen but may not be statistically significant. Also, a slight excess of asparagine was seen in the sample over the control, but it should not have survived the hydrolysis step in the workup (converted to aspartic acid). The ISO 5 lab space measured in July and August 2023 during the same LCMS session do not show detectable asparagine (or glutamine). As a note, these show a higher abundance of amino acids, except for serine and asparagine than the pycnometer, but generally lower than the glovebox interior. A witness sample from the canister glovebox collected after final cleaning but before the canister was introduced on 25 September 2023 has not been analyzed yet. It is likely that the low levels of amino acids observed on the control are from Bennu or the SRC.

***Conclusions***

Particles: First of it is notable how clean the control foil is, reinforcing the effectiveness of both the processor techniques and glovebox. Second, the presence of only four particles found only in pits suggests that they may be indigenous to the foil (and random chance saw the other side of the control foil analyzed). *This hypothesis could be verified by the examination of both sides of the reserve control foil, if requested.*

The existence of only four particles on the surface is remarkably clean. Yet, if the four particles are contamination from the pycnometer, it is unclear where all but the presumed stainless steel one could have originated, as there are no known titanium components in the pycnometer assembly. Thus, there is no recommendation on what corrective action could be taken to reduce their abundance. The composition of these particles is unlikely to be confused with astromaterials and their low abundance is unlikely to be significant in bulk analyses.

Bulk organics: The absence of any detectable material in the control or sample speaks to the exceptional organic cleanliness of both the pycnometer and the glovebox.

Amino acids: The low levels of amino acids, nearly all below the control, suggest that the pycnometer will not violate the pristinity of samples introduced.

## Appendix S6-A1

### *Analysis of a Blank and Deployed Pycnometer Aluminum Foil Samples by FESEM/EDX*

Kathie Thomas-Keprta and Loan Le, JSC

### *Introduction*

Provided by JSC Curation as part of the cleanroom activity Step B3, deployed pycnometer aluminum (Al) foil witness plates were analyzed using Field Emission Scanning Electron Microscopy (FESEM) and Energy Dispersive X-ray spectrometry (EDX). One foil (OR-MA-0443,0; X2183) served as a blank while the other (Or-MA-0442,0; X2182) was exposed within the pycnometer. The goal was to identify and characterize particles collected on and/or near the center regions of each of the Al foil surfaces.

### *Samples Analyzed and Technical Approach*

Both Al foils we analyzed using a JEOL 7600 FESEM equipped with an Oxford Ultim Max 170 EDX spectrometer. Each sample was received folded within separate sealed glass vials. After removal the foils were first unfolded using cleaned stainless steel tweezers before being mounted onto a cleaned 40 mm ∅ sample holder. In each case the foils were attached using a small piece of C adhesive tape under the foil center and four strips of conductive Cu tape over the corners as illustrated in Fig. S6-A1-1. Since the foils are conductive no surface sputter coating was necessary to dissipated surface charging. The FESEM was operated at 15 kV with a 1 nA beam current and spot/area EDX spectra acquired with collection times ranging from 20 – 30 s. at a process time of 4 (dead time <15%).

**Fig. S7-A1-1** Optical view of the two rectangular Al foils received for FESEM analysis from JSC Curation (Blank at left; Exposed Sample at right). Each was mounted on a large SEM holder and held in place using one small rectangular piece of double-sided C tape underneath the foil center and a piece of Cu tape at each corner of the foil rectangle.

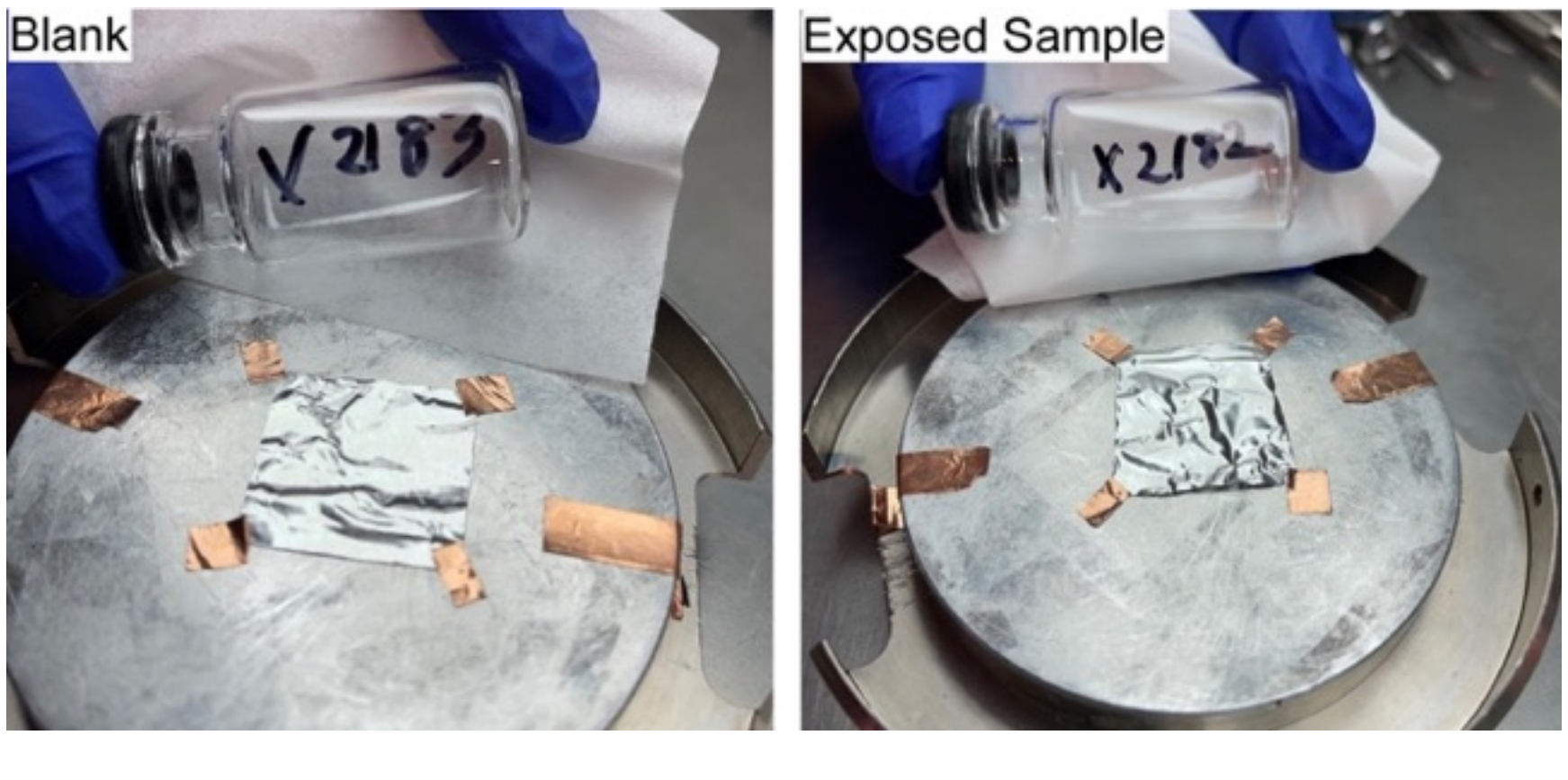

### *SEM Results*

*Blank Foil*: Low magnification secondary electron (SEI) and low angle backscatter electron (LABE) images of the blank surface are shown in Fig. S6-2. The surface has a striated texture containing numerous defects (*e.g.,* irregular shaped depression/pits) but no surface particulates were identified. At higher magnification the surface defects can divided into two groupings: (1) irregular shaped pits and (2) embedded bright features, ranging from ~ 0.1 – 2 µm in size, that are heterogeneously distributed across the foil surface (Fig. S6-A1-3 and S6-A1-4). Some regions within the pits contain enrichments of Fe ± Si ± Ti ± Cu (Fig. S6-A1-3) while the bright features are composed primarily of Fe and Si with minor Cu (Fig. S6-A1-4). In both cases these are interpreted as products of the manufacturing/milling process.

*Exposed Foil*: Background textures/matrix composition of the exposed foil are identical to the blank sample, with the exception of four surface particles that are shown in Fig. S6-A1-5. One particle is composed of Fe-Cr-Ni metal (interpreted as stainless steel), two of Ti metal, and one is composed of a mixture of Fe, Si (overlap with a bright particle; see Fig. S6-A1-4) and Ti.

**Fig. S6-A1-2** Low magnification SEI and LABE images of the blank foil illustrating the oriented parallel striations in the surface produced during the milling/manufacturing process.

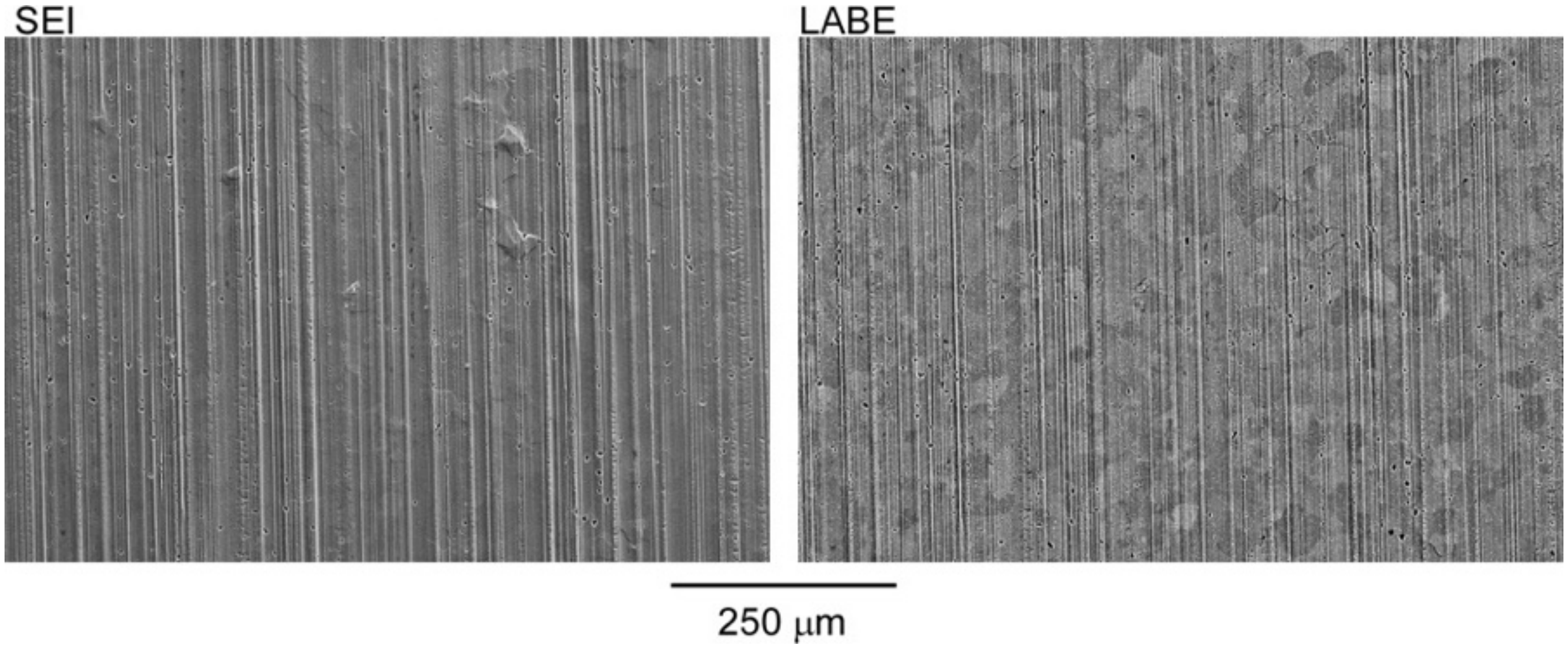

***Conclusions***

Based on the above analyses of both blank and exposed Al foils the conclusions of this study can be summarized as:

(1) Both blank and exposed foils contained features embedded within the surface Al matrix that are interpreted as artifacts acquired during the manufacturing process.
(2) The foil blank was devoid of any exogenous / foreign surface particles in the areas imaged, *i.e.*, it was considered clean.
(3) The exposed foil was identical to the blank with the exception of four foreign particles identified residing within surface depressions/pits. Based on element compositions, these are interpreted as being sourced from curation sample processing handling tools (*e.g.*, stainless steel, Ti metal).

**Fig. S6-A1-3 Upper views:** Higher magnification SEI view (left) and LABE view (right) of the same region on the blank Al foil. Vertical milling ridges are visible along with irregularly shaped pits. Note most of the dark regions in the SEI view correspond to pits, as shown in the LABE image.
**Lower view**: EDX spectra of six locations (circles in SEI view; spectra coordinated with circle colors) show compositional variations of highlighted regions. The bright features in the LABE view are heterogeneously distributed within the Al matrix and are shown at higher magnification in Figure 4.

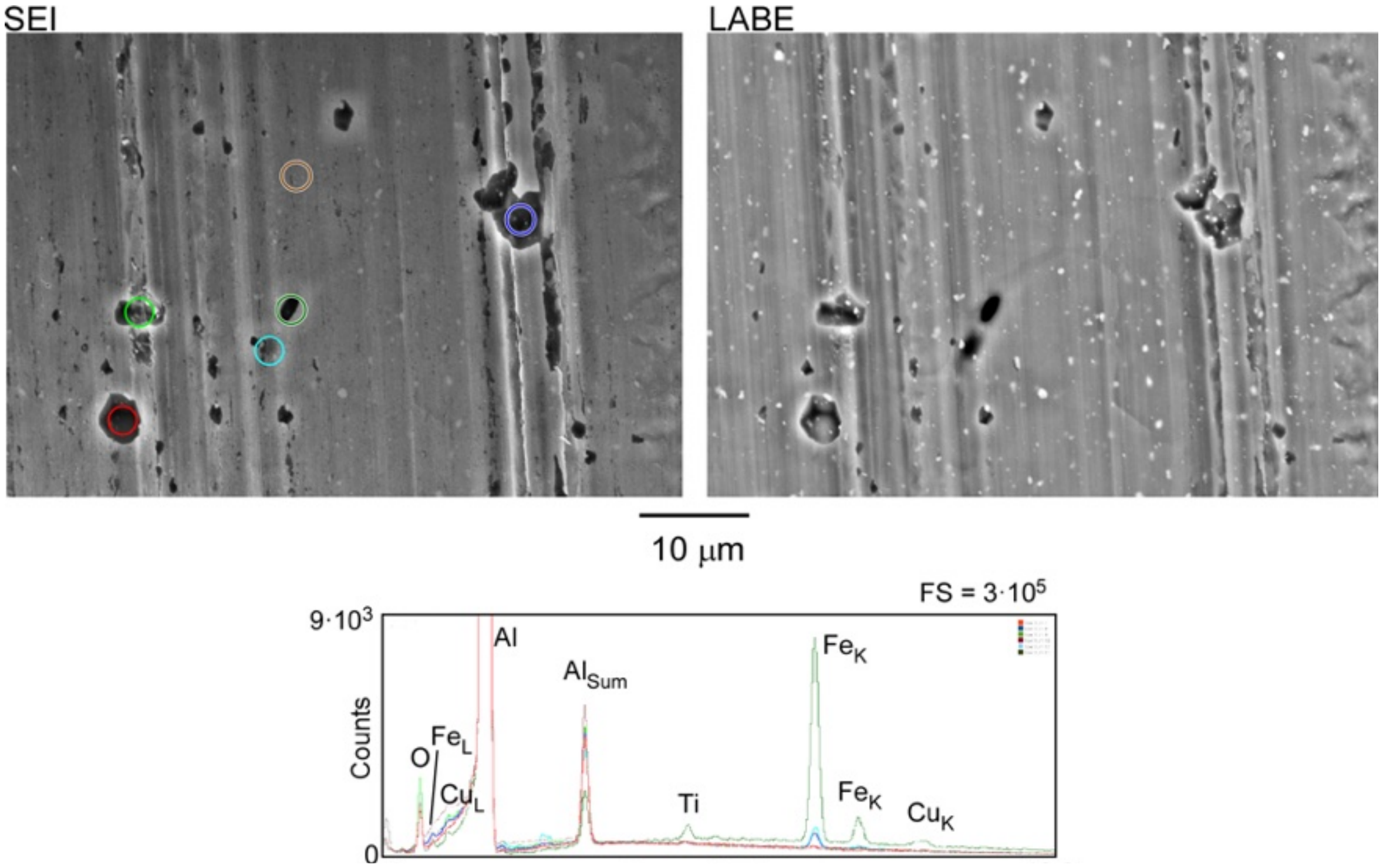

**Fig. S6-A1-4 Left views**: LABE images of the blank Al matrix showing defects in the form of irregular shaped pits. The bright features are embedded in the foil matrix and range in size from ~ 0.1 to 2 μm. **Right views**: EDX spectra of highlighted regions in the views at left. The white features are composed of major Fe and Si with minor Cu. The foil surface has been slightly oxidized as evidenced by the small O peak.

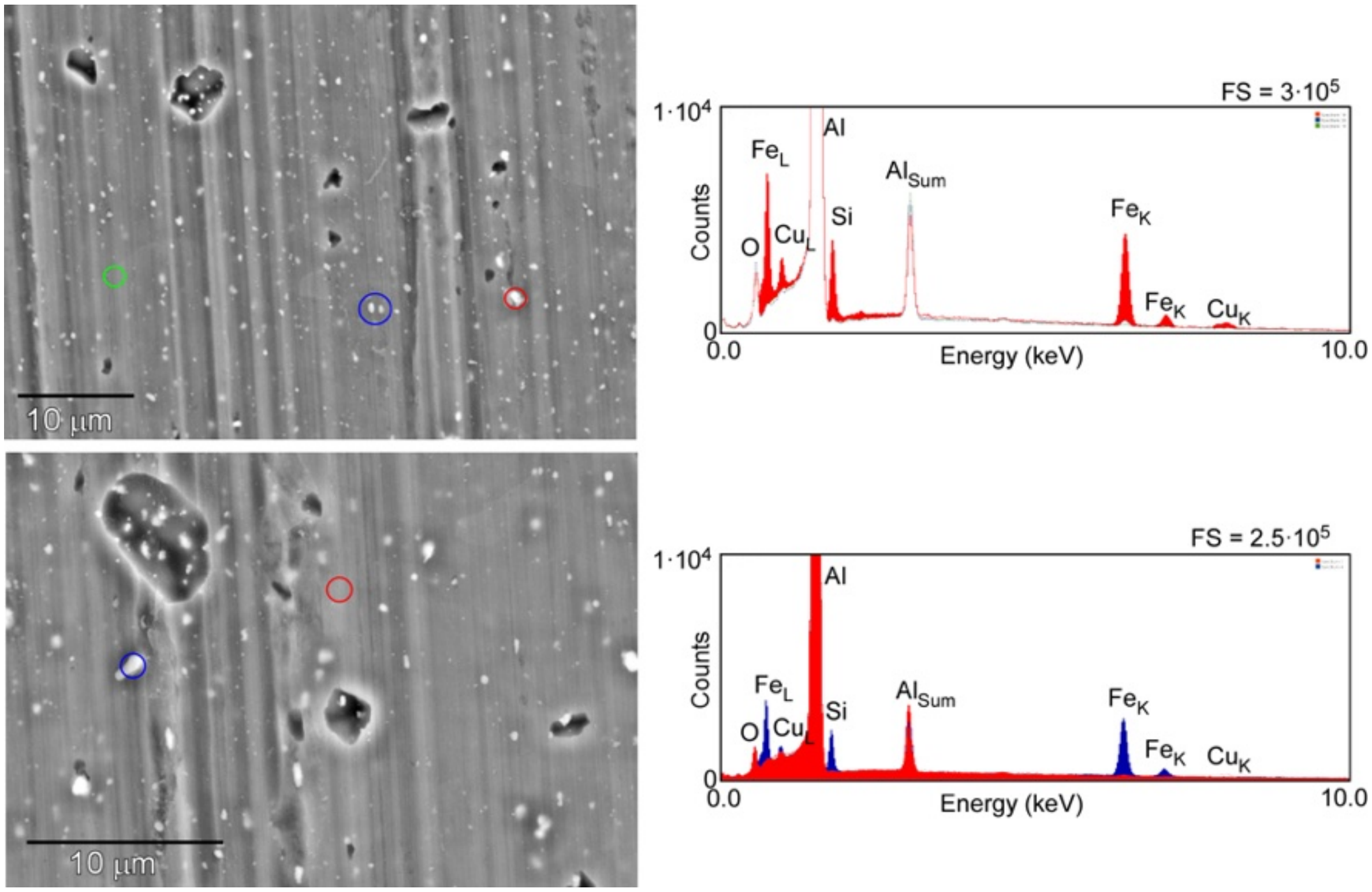

**Fig.S6-A1-5. Left views**: LABE images of particles identified on the exposed foil surface. All appear to be located in depressions or pits in the foil surface. Particles are highlighted by red arrows with corresponding EDX spectra at right. **Right views**: EDX spectra of particles at left ranging in size from ~ 2 - 3 µm. The uppermost particle is composed of stainless steel (Fe-Cr-Ni). The middle two particles are composed of Ti metal and the bottom particle is a mixture of Ti metal and Fe and Si (contributed by overlap with a bright feature embedded in the matrix; also see Fig. S6-A1-4).

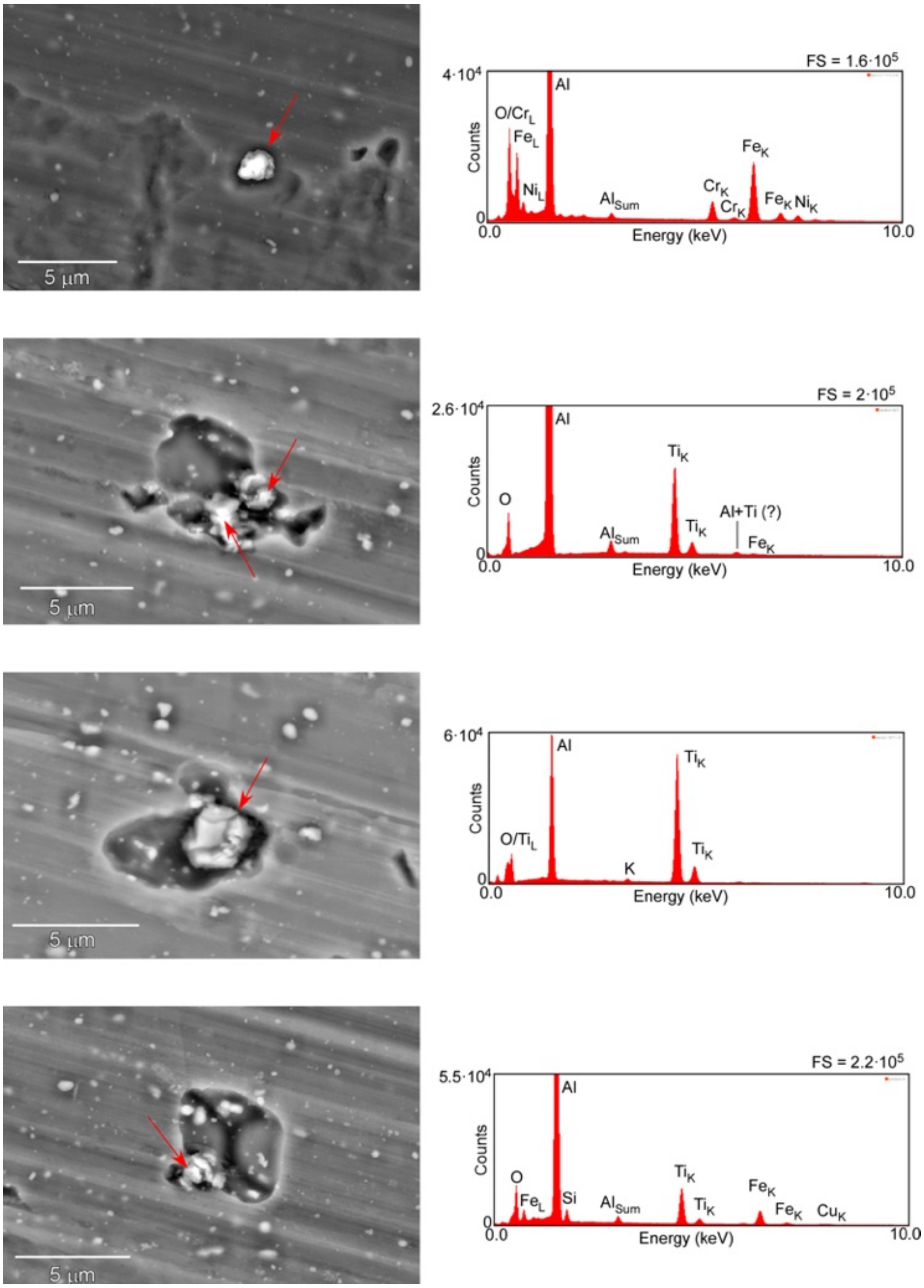

## Appendix S6-A2

### *Analysis of a Blank and Deployed Pycnometer Aluminum Foil Samples by py-GC-MS*

Angel Mojarro, GSFC

**(1) Pyrolysis of pycnometer witness foils**

Samples:
OR-MA-0440,0: Pycnometer witness foil # 1 container X2180 Goddard organics
OR-MA-0441,0: Pycnometer witness foil # 1 blank container X2181 Goddard organics

**(2) Methods**

Foils were rolled into a thin cigar shape using solvent-cleaned tweezers (Fig. S6-A2-1). The prepared foils were then placed directly into the pyrolysis chamber for analysis at 600 °C. The GC was programed to start at 40 °C and ramp to 300 °C at 3.5 °C/min while the MS simultaneous conducted full scans between *m/z* 50 - 500 and multiple reaction monitoring for targeted analysis of insoluble organic matter (IOM) derived compounds.

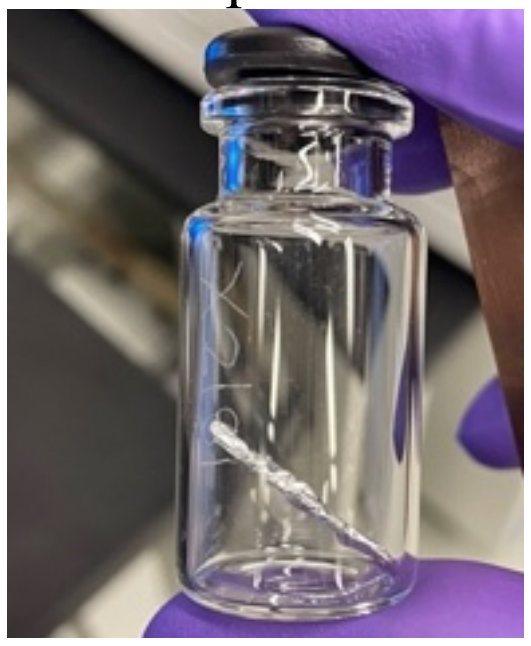

**Fig. S6-A2-1** A foil ready for pyrolysis.

**(3) Results**

No organic compounds in excess of the procedural blank were detected in either foil (Fig. S6-A-2).

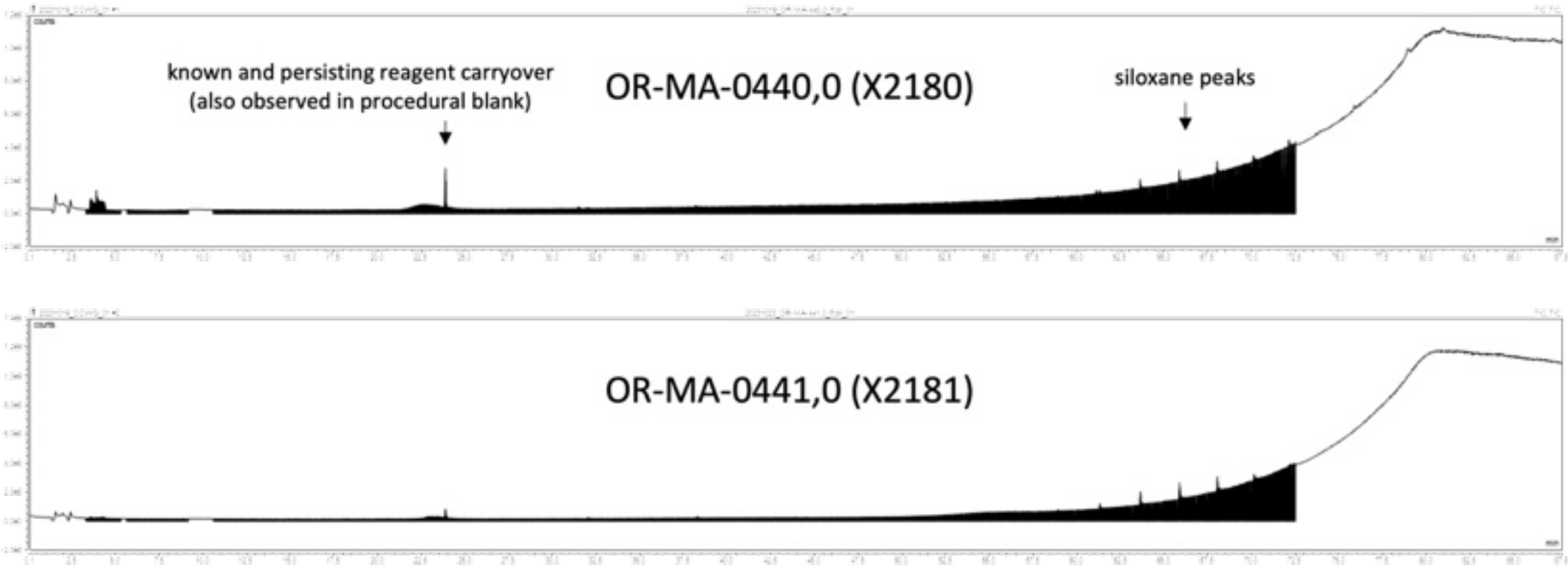

**Fig. S6-A2-2** Combined fullscan and multiple reaction monitoring (MRM) trace of witness foils do not show positive detection of organic compounds. *The reagent carryover is from prior analyses and is unrelated to this study.*

## Appendix S6-A3

***Analysis of a Blank and Deployed Pycnometer Aluminum Foil Samples by LC-MS***

Eve Berger, JSC

**Table S6-A3-1** Amino acid abundances (ng/cm$^2$) from pycnometer witness foils via AccQ·TAG derivatization[a].

| | | ***Pycnometer X2186 Sample*** [b] | ***Pycnometer X2187 Control*** [b] | ***Ratio*** |
|---|---|---|---|---|
| Glycine | avg. | 0.00 | 0.20 | 0 |
| β-Alanine | avg. | 0.05 | 0.12 | 0.42 |
| Alanine | avg. | 0.09 | 0.27 | 0.33 |
| ɣ-Amino-*n*-butyric acid | avg. | 0.00 | 0.20 | 0 |
| α- Amino-isobutyric acid | avg. | 0.03 | 0.08 | 0.38 |
| Serine | avg. | 0.06 | 0.04 | **1.50** |
| Proline | avg. | 0.04 | 0.12 | 0.33 |
| 5-Amino-*n*-pentanoic acid | avg. | 0.00 | 0.00 | 0 |
| Isovaline | avg. | 0.00 | 0.00 | 0 |
| Valine | avg. | 0.00 | 0.00 | 0 |
| Threonine | avg. | 0.01 | 0.04 | 0.25 |
| Cysteine | avg. | 0.00 | 0.00 | 0 |
| Isoleucine + Leucine | avg. | 0.00 | 9.63 | 0 |
| Asparagine | avg. | 0.10 | 0.07 | **1.43** |
| Aspartic acid | avg. | 0.00 | 0.10 | 0 |
| Glutamine | avg. | 0.00 | 0.00 | 0 |
| Lysine | avg. | 0.00 | 0.12 | 0 |
| Glutamine | avg. | 0.02 | 0.86 | 0.02 |
| Methionine | avg. | 0.00 | 0.00 | 0 |
| Histidine | avg. | 0.00 | 0.00 | 0 |
| Phenylalanine | avg. | 0.01 | 0.04 | 0.25 |
| Arginine | avg. | 0.00 | 0.00 | 0 |
| Tyrosine | avg. | 0.00 | 0.00 | 0 |
| Tryptophan | avg. | 0.00 | 0.00 | 0 |
| Total ng/cm$^2$ | | 0.40 | 11.90 | 0.03 |

[a] Dworkin JP et al. (2018) OSIRIS-REx contamination control strategy and implementation. Space Sci Rev 214:1-53. https://doi.org/10.1007/s11214-017-0439-4

[b] Standard deviation not reported since values are average of two runs.

Nicole Lunning Kevin Righter Christopher Snead Jason Dworkin

This is memo is a **conditional approval** and evaluation of the materials and implementation of the custom QRIS reflectance standard proposed to be used adjacent to "curation-pristine" Bennu samples in the sample glovebox in the ISO 5 OSIRIS-REx cleanroom at NASA JSC. Curation-pristine is a condition where the sample is considered unchanged by the process and is retained in the unallocated NASA collection.

The QRIS team has done an excellent job of ensuring that all components in the reflectance standard are redundantly isolated via compliant materials for direct exposure to the sample and glovebox interior. Based on the design presented on 10 May 2023 (Fig. S6-1), there are several recommendations and observations necessary for the design.

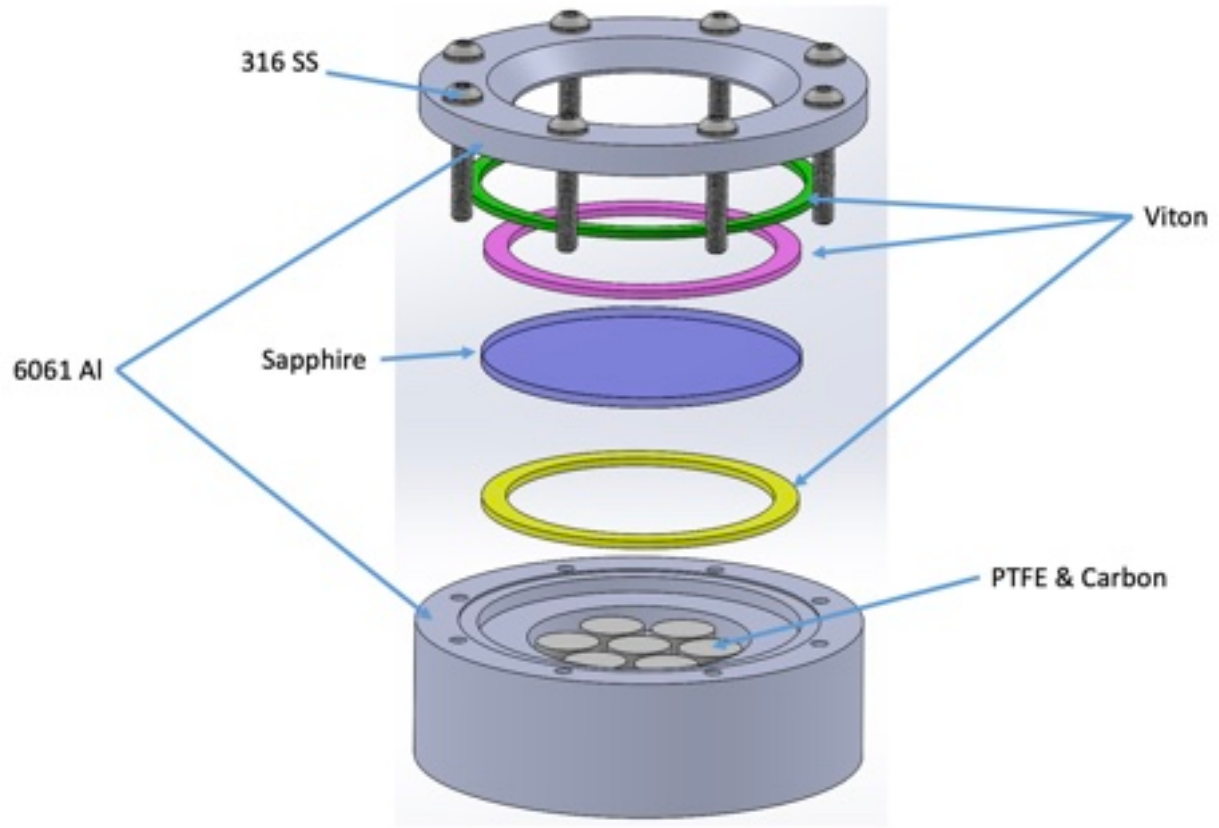

**Fig. S6-1** QRIS reflectance standard encapsulation design.

1. Ensure that surfaces in contact with the gasket are mirror smooth.
2. Ensure that the green and magenta gaskets are compressed equally. This will impact the tolerances of the machining.
3. Change the design to use captured screws in the base.
4. Change the design to use a chamfered edge.
5. Ensure that there is no lubricant used on the gaskets or fasteners.
6. Construct at least two, preferably three units: one for use, one for testing, and one spare.
7. Avoid using a waterjet cutter with garnet abrasive in the manufacturing.
8. It is permissible to roughen (e.g., bead-blast) the surfaces not in contact with the gaskets to reduce reflections.
9. It is permissible to change the fasteners to be countersunk flat to reduce reflections.
10. An aluminum protective cover is permitted, but unlikely to be necessary.

We agree that the unit will be assembled and cleaned in a $GN_2$ glovebox by NASA JSC personnel. For characterizing the units, one will be fitted with an $O_2$ sensor to measure the rate of $O_2$ leak in. This value will be used to determine the maximum duration the unit can remain inside the sample glovebox.

16 June 2023

# Glovebox Exhaust Characterization

Nicole Lunning Kevin Righter Christopher Snead Wayland Connolly Jason Dworkin

This memo is **approval** and evaluation of the materials and implementation of the exhaust sample collection from the TAGSAM glovebox. A design was presented for the collection of compounds which may be entrained in exhaust nitrogen with no impact to the sample and minimal impact to the curation workload. This design would collect gases on a best-effort basis downstream from the glovebox. The collection would be on adsorbent materials inside an airlock made of 2-inch 316 stainless steel and Viton sanitary fittings with no opportunity for exposure to the sample nor the cleanroom environment (Fig. S6-2, Fig. S6-3) even in the event of a nitrogen gas ($GN_2$) failure, glove leak, or power outage.

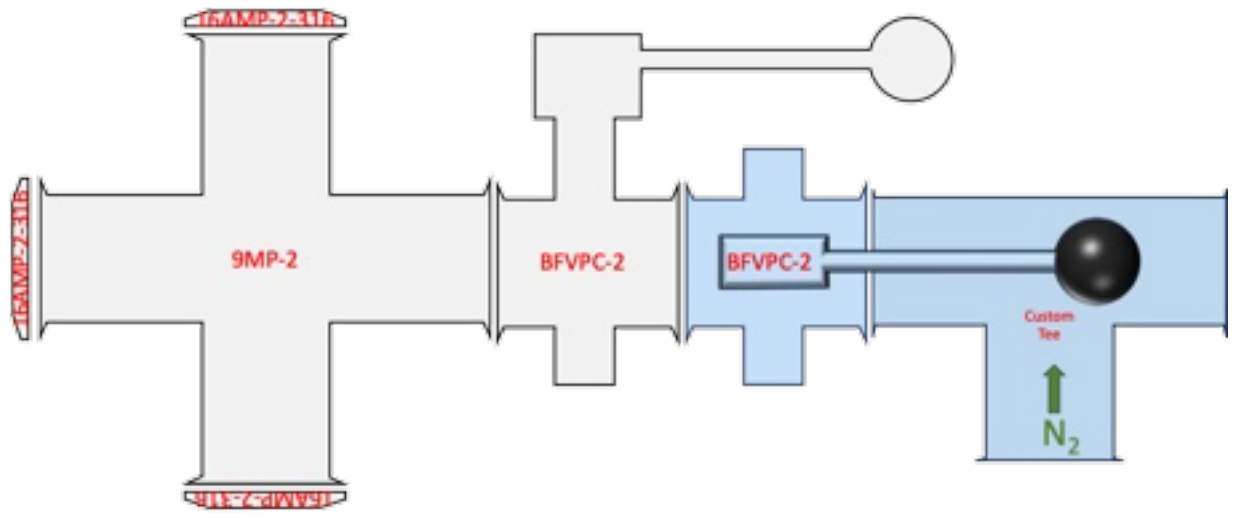

**Fig. S6-2** The blue components, a custom tee and valve would be fitted to the TAGSAM glovebox exhaust line after the check valve. The white components would contain the adsorbents and can be attached and removed via the two valves without disrupting the exhaust flow or exposing the interior to the cleanroom. Red text indicates sanitaryfittings.us part numbers, gaskets and clamps are omitted for clarity.

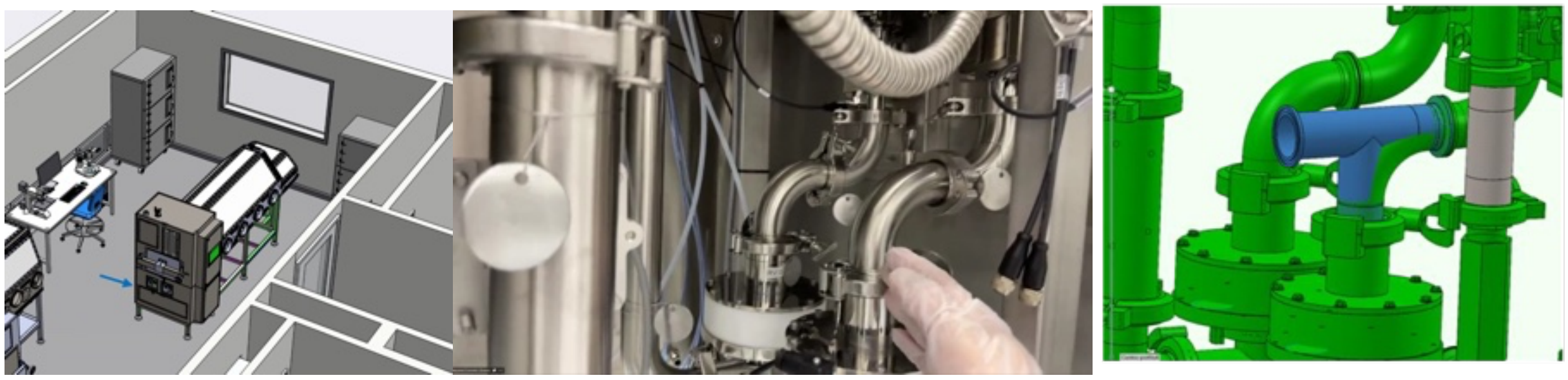

**Fig. S6-3** Left, the arrow points to the approximate location of the tee. Center photo of location of the tee to be added after the exhaust check valve at the far end of the glovebox from the exhaust connection to the sample chamber. Right, the design of the custom tee. Some clamps and gaskets are omitted for clarity.

All parts will be cleaned and installed by curation personnel. One custom tee from DF Siena (dfsiena.it), the same vendor that produced the exhaust plumbing on the glovebox, would be procured and installed with an off-the-shelf approved valve and end cap for when not in use (sanitaryfittings.us). Three collectors would be assembled and supplied by GSFC from the same approved vendor (sanitaryfittings.us). The first would be attached in August prior to the start of the cleanroom quiescent period, to collect a blank. The second would be attached prior to the arrival of the sample and valved off until the TAGSAM head is placed in the glovebox. The third is a backup but can be installed partway through the TAGSAM disassembly. The gas collection operation would conclude after sample containerization. However, the valved off and capped custom tee can remain on the glovebox.

Curation accepts on a best effort basis that the collector valve be closed when the glovebox airlock is used to minimize the introduction of room air into the collector, since there can be backflow into the collector from the airlock. Curation accepts that the valve should be closed when the airlock is cleaned to avoid the introduction of cleaning solvents (hydrogen peroxide solution and/or isopropanol) from the airlock into the collector.

Sample collectors would be shipped or hand carried to GSFC for disassembly and analysis, since the design has significant dead space, and it is understood that the material collected will be relative at best and qualitative at worst.

**Supplement 7:** Environmental sample plan UA-PRO-4.3.3-0001

# SAMPLE RETURN CAPSULE RECOVERY ENVIRONMENTAL SAMPLING PROCEDURE

## OSIRIS-REx Document UA-PRO-4.3.3-0001

Version 2a

## REVIEWERS

| 1. Reviewers |
|---|
| Scott A. Sandford<br>Science Team to SRC Interface |
| E. B. Bierhaus<br>Science Team to Spacecraft Interface |
| Sandra Freund<br>Lockheed Martin |
| Nicole Lunning<br>OSIRIS-REx Sample Curator |
| Jason P. Dworkin<br>Project Scientist |

| 3. Reviewers |
|---|
| Dante S. Lauretta<br>OSIRIS-REx Principal Investigator |
| Harold C. Connolly Jr.<br>OSIRIS-REx Mission Sample Scientist |
| Richard Burns<br>OSIRIS-REx Project Manager |
| Jeffrey N. Grossman<br>OSIRIS-REx Program Scientist |
| Francis McCubbin<br>Astromaterials Curator |

### REVIEWER APPROVALS

| 4. Approvals |
|---|
| Digitally Signed<br>9/8/23 |
| Scott A. Sandford<br>Science Team to SRC Interface |
| Digitally Signed<br>9/13/23 |
| E. B. Bierhaus<br>Science Team to Spacecraft Interface |
| Digitally Signed<br>9/15/23 |
| Sandra Freund<br>Lockheed Martin |
| Digitally Signed<br>9/19/23 |
| Nicole Lunning<br>OSIRIS-REx Sample Curator |
| Digitally Signed<br>9/8/23 |
| Jason P. Dworkin<br>Project Scientist |

| 6. Approvals |
|---|
| Digitally Signed<br>9/8/23 |
| Dante S. Lauretta<br>OSIRIS-REx Principal Investigator |
| Digitally Signed<br>9/15/23 |
| Harold C. Connolly Jr.<br>OSIRIS-REx Mission Sample Scientist |
| Digitally Signed<br>9/15/23 |
| Richard Burns<br>OSIRIS-REx Project Manager |
| Digitally Signed<br>9/8/23 |
| Jeffrey N. Grossman<br>OSIRIS-REx Program Scientist |
| |
| |

## CM Forward

This document is the Origins, Spectral Interpretation, Resource Identification, and Security-Regolith Explorer (OSIRIS-REx) project-controlled document. Changes to this document require prior approval of the OSIRIS-REx PI (Principal Investigator).

Questions or comments concerning this document should be addressed to:

OSIRIS-REx Principal Investigator's Office

## REVISION LOG

| Version | Date mm/dd/yyyy | Affected Section(s) | Engineering Change # | Reason/Initiation/Remarks |
|---|---|---|---|---|
| 1.0 | 04/21/2023 | all | | Initial release |
| 2.0 | 8/31/2023 | Appendix S2.1, 2, 3, 4 | | Added field procedure, camera procedure, and notebook |
| 2.0 | 8/31/2023 | 3, 4 | | Removed soil profile and added anemometer, increased volume of soil to be collected, changed date of JSC gas testing, added additional tools. |
| 2.0 | 8/31/2023 | 6 | | Samples will not be chilled on the Range and cold storage is the responsibility of Aaron Regberg |
| 2.0 | 8/31/2023 | 4 | | Reduced the number of gas bottles |
| 2.0 | 9/13/2023 | all | | Updated the verb tense for actions that have already occurred |
| 2.0 | 9/19/2023 | 4.5 | | Updated the CK deployment schedule |
| 2a | 9/8/2026 | all | | Updated formatting for publication |
| | | | | |

## LIST OF TBDS/TBRS

| Item No. | Location | Summary | Ind./Org. | Due Date |
|---|---|---|---|---|
| | | | | |
| | | | | |

# Table of Contents

# OSIRIS-REx SAMPLE RETURN CAPSULE RECOVERY ENVIRONMENTAL SAMPLING PROCEDURE

## S7-1.0 Governing Documents

Curation of Institutional Scientific Collections, Office of the Chief Scientist, NPD 7100.10F
Appendix F to the New Frontiers Program Plan: Program Level Requirements for the OSIRIS-REx Project
OSIRIS-REx Mission Requirements Document (OSIRIS-REx-RQMT-0001)
OSIRIS-REx Mission Curation Plan (OSIRIS-REx-PLAN-0041, Revision A)
OSIRIS-REx Mission Contamination Control Plan (NFP3-PN-11-MA13-1)
OSIRIS-REx Sample Analysis Plan (UA-PLN 4.5.4-001)
OSIRIS-REx Sample Return Capsule Components Science Analysis Plan – Part 2 (Science and Engineering Analyses) (OREX-DOCS-04.00-00027)
OSIRIS-REx Sample Return Capsule (SRC) Recovery Plan (NFP3-PN-12-SE-13)
OSIRIS-REx SRC Recovery Procedure (NFP3-PR-16-0491)
OSIRIS-REx UTTR Curation Procedure (due June 2023)
Dworkin JP et al. (2018) OSIRIS-REx contamination control strategy and implementation. Space Sci Rev 214:1-53. https://doi.org/10.1007/s11214-017-0439-4

## S7-2.0 Purpose

The purpose of OSIRIS-REx Sample Return Capsule Recovery Environmental Sampling Procedure is to establish how we will meet the mission Level-1 requirements:

> *PLRA 4.1.1.2 Document the contamination of the sample acquired from collection, transport, curation, and distribution.*
>
> *PLRA 4.7.4: The OSIRIS-REx Project shall analyze the SRC for assessment of contamination and capsule performance.*

This requirement is met by analyses of key Sample Return Capsule (SRC) components or sample in comparison with the environmental conditions present during the recovery and transport of the SRC into the controlled curation environment. A record of these environments serves to complement the chemical record of the Assembly, Test, and Launch Operations (ATLO) environment already obtained to ensure an end-to-end chain of evidence for potential contamination. Environmental samples also provide a physical record of the environments to which the SRC, and thus sample, may have been exposed in the event of an anomaly discovered anytime in the recovery sequence. Although the focus of the OSIRIS-REx Sample Analysis Team will be on the detailed analysis of the returned samples in the sample canister and Touch-And-Go Sample Acquisition Mechanism (TAGSAM) head, the returned hardware and environmental samples collected at the Utah Test and Training Range (UTTR) will be analyzed to establish the nature of potential terrestrial contaminants. In summary, the study of environmental samples collected at UTTR will address issues of scientific, engineering, and curatorial importance.

## S7-3.0 Team Roles, Responsibilities, and Governance

Lockheed Martin Space (LMS) personnel will perform gas sampling near the SRC on the UTTR Range. NASA's Johnson Space Center (JSC) Curation personnel assisted by Sample Analysis Team members will collect other gas samples. JSC Curation personnel assisted by Sample Analysis Team members will collect relevant samples of UTTR soil, water (if present), precipitation (if present), or debris (if present and declared safe to collect by UTTR personnel) on the UTTR Range.

JSC Curation personnel will assess the sufficiency of materials collected with the verbal concurrence of the PI (or designee) on the Range. JSC Curation will provide all sampling equipment and storage materials (Table S7.3-1), assign nomenclature, store and distribute samples for analysis, and arrange with the JSC Toxicology Lab for the gas collection bottles and their analysis. The roles of individual members of the environmental sampling team on the Range are listed in Appendix S7.1.

**Table S7.3-1** Total expected and reserve equipment needed for all sampling of all sample types to be supplied by JSC Curation. This list is in addition to equipment for video or photo documentation, SRC recovery, and off-nominal sample recovery.

| | Item | Minimum (Nominal) | Reserve (Off-nominal) | Purpose |
|---|---|---|---|---|
| Shared | Gloves (not black) in baggies by size (S,M,L,XL) | 21 pair | 100 pair | Protection of personnel and samples |
| | Sample tags | 75 | 125 | Documentation |
| | Writing implement | 4 | 10 | Documentation in the field |
| | Small towel | 1 | 1 | Dry hands for glove changes |
| | Flags (color 1, red) | 50 | 100 | Indicate collection locations |
| | Flags (color 2, white) | 10 | 10 | Indicate walking locations |
| | Flags (color 3, pink) | 3 | 3 | Indicate SRC position |
| | Flags (color 4, yellow) | 0 | 50 | Grid markers |
| | Scissors | 2 | 4 | Cut tape |
| | Camera (Appendix S2.2) | 2 | 2 | Documentation |
| | Spare camera battery | 1 | 1 | Documentation |
| | Compass and GPS (digital) | 2 | 2 | Documentation |
| | Spare battery bank & cables | 1 | 1 | Supplemental power |
| | Bucket with lid | 2 | 4 | Sample carrying/table/off-nominal supplies |
| | Personal bag(s) | 1/person | 1/person | Tool carrying and hydration |
| | Dry-bag | 2 | 2 | Off-nominal (25 extra, each: foil, bags, tags) |
| | ANSMET cube | 2 | 4 | Scale and orientation for photos |
| | Notebooks (Appendix S2.4) | 3 | 6 | Documentation in the field |
| Gas | Gas bottles (Appendix S2.3) | 15 | 21 | Holds the gas (Range, 1012, JSC) |
| | NVR/particle collector[1] | 1 | 1 | Verification of gas manifold cleanliness |
| | Dry-bag | 2 | 2 | Holds all gas bottle containers |
| Soil | PTFE bags[2] | 40 | 80 | Holds the samples |
| | Bag tape | 2 rolls | 4 rolls | Contain the samples |
| | Trowel[2] | 2 | 6 | Collect solid samples |
| Water | Cold shipper | 1[1] | 2 | Preserves soil and water samples for transport |
| | PTFE flasks with lids[2] | 4 | 8 | Holds the samples |
| | Transfer pipette[2] | 4 | 20 | Collect liquid samples |
| Other | Dry filter air sampler | 1 | 1 | Collects dust |
| | Polypropylene disks[2] | 2 | 4 | Holds the dust |
| | Aluminum jars[2] | 2 | 20 | Holds the samples |
| | Aluminum foil | 25[2] | 2 rolls | Holds the samples |
| | Jar lids[2] | 2 | 20 | Contain the samples |
| | Tweezers[2] | 2 | 8 | Collect foreign objects |
| | Whistle | 1/person | 1/person | Call for help |
| | Body camera | 1/person | 1/person | Documentation |

[1] Kept in building 1012 or 1010.
[2] Pre-cleaned and sterilized.

## S7-4.0 Samples

The numbers, times, and locations of samples are listed in Table S7.4-1 and described in the subsequent sections. Fig. S7.4-1 shows the nominal timeline for all collections and a comparison with key events in the preparation for and following recovery.

**Table S7.4-1** Total required samples to be collected.

| Sample | Type | Location | Date | Number | Responsible |
|---|---|---|---|---|---|
| SRC vents (2) | Gas | UTTR Range | 24 Sept | 2 | Curation |
| Heatshield and backshell | Gas | UTTR Range | 24 Sept | 2 | Curation |
| Ambient air upwind from SRC | Gas | UTTR Range | 24 Sept | 1 | Curation |
| Rotorwash | Gas | UTTR Range | ~29 Aug | 1[1] | Curation |
| Ambient air upwind from building 1012 | Gas | UTTR Building 1012 exterior | 24 Sept | 1 | Curation |
| Ambient air prior to use | Gas | UTTR Cleanroom | ~22 Sept | 1 | Curation |
| $GN_2$ Purge manifold verification[2] | NVR & particles | UTTR Building 1012 | ~11 July | 1 | LMS |
| $GN_2$ Purge manifold verification[2] | Gas | UTTR Building 1012 | ~11 July | 1 | Curation |
| $GN_2$ Purge gas verification[2] | Gas | JSC Building 31 Cleaning Lab | Aug | 1 | Curation |
| $GN_2$ Purge gas verification[2] | Gas | JSC building 31 Curation Lab | Aug | 1 | Curation |
| $GN_2$ Purge[2] | Gas | UTTR Building 1012 | 22 Sept | 1 | Curation |
| SRC Vents (2) | Gas | UTTR Cleanroom | 24 Sept | 2 | Curation |
| Near canister air filter | Gas | UTTR Cleanroom | 24 Sept | 1 | Curation |
| Contingency | Gas | As needed | As needed | 6[1] | Curation |
| CK witness plates | Witness | UTTR Cleanroom | 18 July-25 Sept | Various | Curation |
| Contingency | Witness | As needed | As needed | 2 | Curation |
| Soil under SRC divots | Solid | UTTR Range | 24 Sept | 7-24 | Curation |
| Undisturbed soil | Solid | UTTR Range | 24 Sept | 4 | Curation |
| Water | Liquid | UTTR Range | 24 Sept | 0-4 | Curation |
| Debris | Solid | UTTR Range | 24 Sept | 0-4 | Curation |
| Precipitation | Liquid | UTTR Range | 24 Sept | 0-1 | Curation |
| Dust | Solid | UTTR Range | 24 Sept | 1 | Curation |

[1] One of the contingency bottles was used in the rotorwash collection at ORT 10.

[2] Requires quick connect fitting.

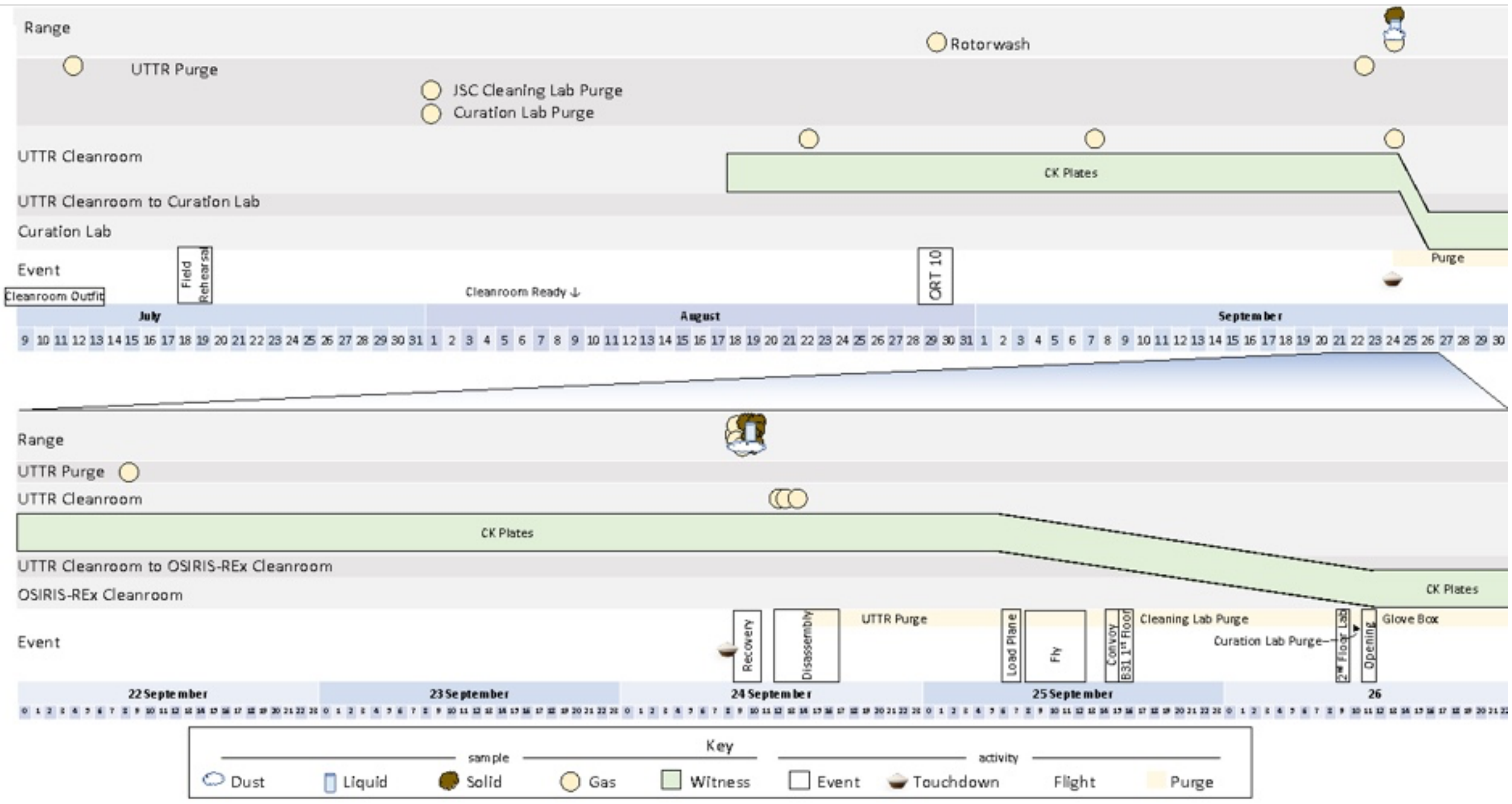

**Fig. S7.4-1** Notional timeline of environmental sample collection and witness exposures assuming earliest recovery and transportation schedule.

Personnel will be clothed as appropriate for the environment (Range or cleanroom). On the Range, required clothing and personal protective equipment (PPE) is based on operational and safety considerations as well as weather and field conditions, as described in the UTTR Curation Procedure document. Attire for operating in the UTTR cleanroom is described in UTTR Curation Procedure document. Nitrile gloves will be worn for collection activities, black gloves should not be used due to their tendency to get hot in direct sun. Curation and Sample Analysis Team personnel on the field team will not have assigned duties in the cleanroom to limit potential contamination from the Range into the cleanroom (i.e., dust and mud that a cleanroom suit may not effectively contain). If a field team member needs to enter the cleanroom due to unforeseen circumstances, they must shower and change into a separate set of clothing and shoes prior to entering the cleanroom. The number of field team members that can have dual roles on the field team is strictly capped at the number of immediately available showers (three).

Prior to sampling, the sample analysis team and curation field team members will photo document the landing site and describe (written and/or recording) local conditions. This description should include separate areas (e.g., bounce 1, bounce 2, beginning of roll, object struck, etc.) of the landing site that will be used in sample descriptions and will be used to determine when fresh clean tools should be deployed. Before initiating sampling, sample analysis team and curation field team members will establish concurrence, so the sample locations are consistently described and with nomenclature as defined by the OSIRIS-REx UTTR Curation Procedure. Key elements of this procedure are summarized in the recovery checklist in Appendix S2.1 and the recovery notebook in Appendix S2.4.

## S7-4.1 Gas

### S7-4.1.1 Collection procedure

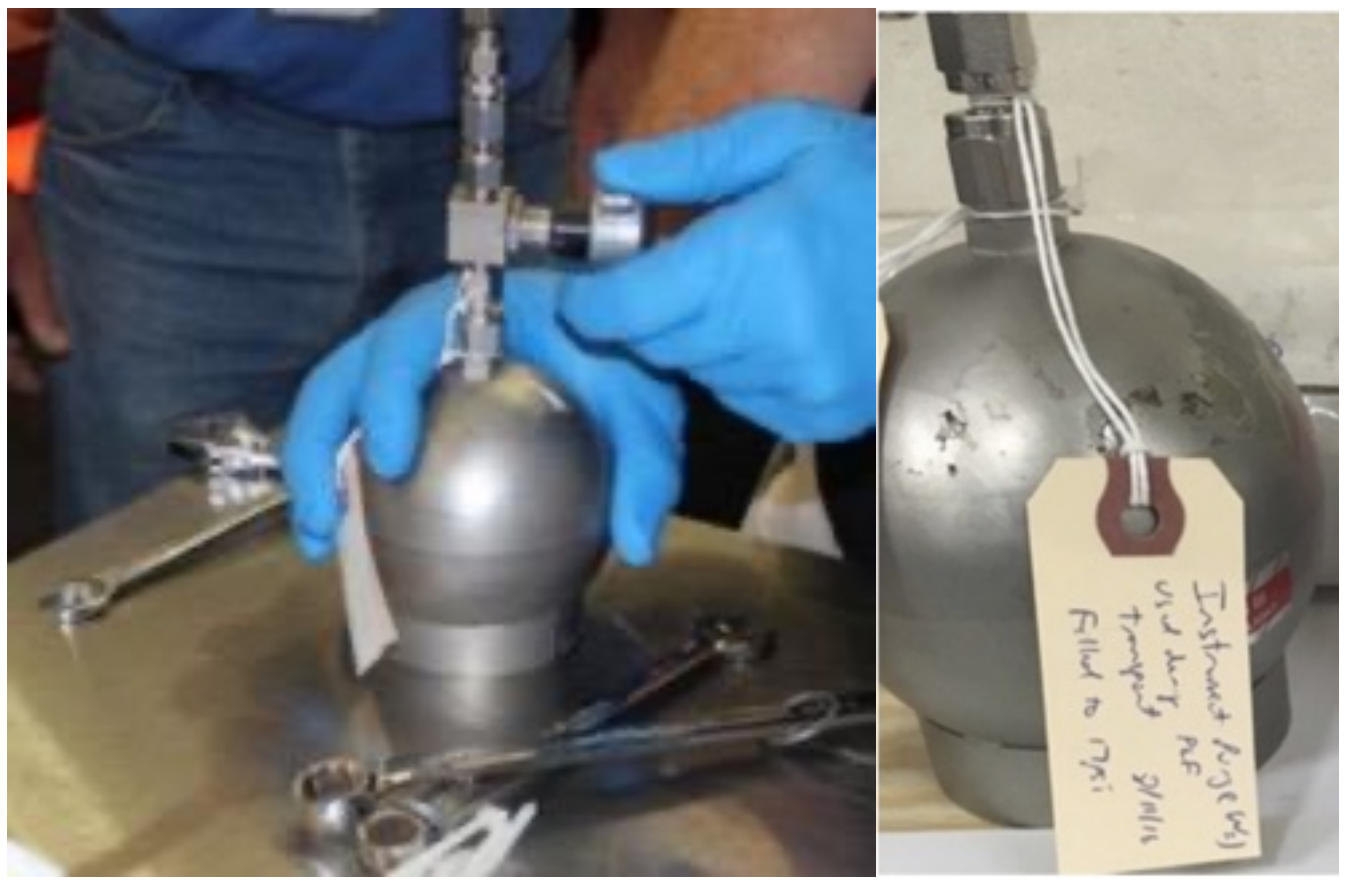

**Fig. S7.4-2** Two 500-mL evacuated gas bottles used in ATLO gas sampling.

Gas bottles will be supplied by the JSC Toxicology Lab as was done for gas sampling during ATLO (Fig. S7.4-2; Dworkin et al. 2018). JSC will arrange for the availability of sufficient bottles plus margin for off-nominal collections. JSC Toxicology provides instruction for use of the bottles. The bottles are pre-evacuated and calibrated. Sampling involves opening the provided valve to allow air to enter for approximately 30 seconds, then closing the valve. There will be a swish sound as the bottle fills; if there is not a sound then the bottle may be defective and a new collection should be made. This is a rare event. For bottles collected under pressure, note the maximum bottle pressure and use an appropriate regulator to prevent over-pressurization. Each should be labeled with the date, time, location, purpose of collection, and the name of the collector. Bottles collected on the Range have an assigned collector, so noting the person is not required. Bottles are terminated with a ¼ inch AN female fitting. However, they are capped with hand-tight knurled caps for use on the Range.

### S7-4.1.2 Gas collection on the UTTR Range

There are three gas sources of interest on the UTTR Range to be collected at the time of SRC recovery.

1. SRC ablation outgassing and potential products from the SRC battery
2. Bennu sample outgassing
3. Gases evolving from the ground, from recovery activities, or transported in the air

#### *S7-4.1.2.1 SRC gas samples*

LMS safety is concerned with gases from source 1. Operations will remain in a "No Go" state if gases are measured at levels above the instantaneous Threshold Limit Value (TLV) or above the 15-min average Short Term Exposure Limit (STEL). Plausible gases from the SRC battery and ablation products include sulfur dioxide ($SO_2$), hydrogen cyanide (HCN), and carbon monoxide (CO). LMS will monitor and document $SO_2$, HCN, and CO for safety with a hand-held device prior to permitting personnel to approach the SRC according to the limits in Table S7.4-2 then permit the collection of samples as described in OSIRIS-REx SRC Recovery Plan (NFP3-PN-12-SE-13):

G + 20 min - Gas samples collected from vicinity of SRC backshell vents
G + 40 min - Gas samples collected from vicinity of SRC heatshield

**Table S7.4-2.** SRC Gases monitored by LMS safety.

| Gas | TLV (ppm) | STEL (ppm) |
|---|---|---|
| Sulfur Dioxide ($SO_2$) | 2 | 5 |
| Hydrogen Cyanide (HCN) | - | 4.7 |
| Carbon Monoxide (CO) | 25 | - |

ppm – parts per million by volume (µL/L)

After pronouncing the SRC safe, the state of the SRC will be assessed to ensure a nominal SRC landing has been made. After verifying a nominal SRC, gas samples will be taken from the backshell vents following the SRC Recovery Plan (NFP3-PN-12-SE-13). Gases from the SRC will be collected once the SRC has been declared safe. Two samples will be collected from the two backshell vents and two samples will be collected from the heatshield and backshell area, requiring a **total of 4 bottles**.

Samples acquired at the SRC vents and heatshield may also be mixed with Bennu material that are off gassing as a result of the heat soak from the SRC or exposure to the terrestrial atmosphere. It is expected that any Bennu gases will be trapped by the sample canister air filter but escaping products are possible.

#### *S7-4.1.2.2 Environmental gas samples*

Gas samples of the ambient air will also be collected for comparison to the SRC gas samples. One sample will be collected from a location upwind of the SRC. One sample and one contingency sample were collected from a representative helicopter rotorwash during Operational Readiness Test (ORT) 10 in August 2023. One sample will be collected upwind of the SRC at Building 1012. A total of **2 bottles were used in August** (results posted on Confluence at [URL redacted] with an additional **2 bottles required in September**.

### S7-4.1.3 Purge gas collection

A $GN_2$ purge system for the SRC is required to maintain the pristine state of the sample. The system consists of a purge source and a purge cart (Fig. S7.4-3). One purge source will start at UTTR attached to the purge cart. This will be fixed to the sample canister from the time it is attached to the canister in the UTTR cleanroom through transport to JSC. In the JSC Building 31 first floor Cleaning Lab, room 110, the purge source attached to the purge cart will be switched from the UTTR source to the Cleaning Lab source. The source will be switched from the Cleaning Lab source to the Curation Lab source after the canister and purge cart are moved to the second floor Curation Lab, room 216, until the sample canister is sealed in its glovebox in room 216.

### S7-4.1.3.1 Gas source and manifold verification

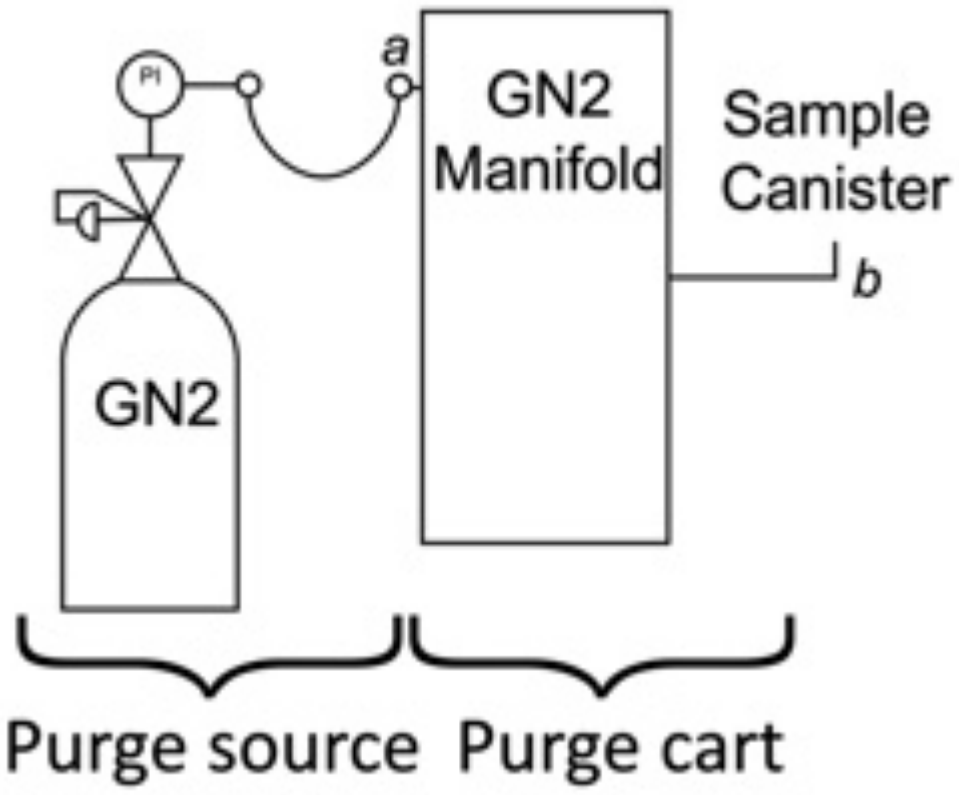

**Fig. S7.4-3** Schematic of the $GN_2$ source and cart for the sample canister purge system. Samples are collected at point a or b.

In July 2023, one sample of $GN_2$ from each of the purge sources from point *a* on Fig. S7.4-3 was taken to verify the gas and regulator cleanliness. One sample of $GN_2$ was collected from the combined purge source and purge cart in UTTR at point *b* on Fig. S7.4-3. These samples are to verify the proper assembly of the purge system and quality of the $GN_2$, and they must be analyzed promptly as part of the certification process of the complete manifold and sources to provide sufficient time for any corrective actions, such as re-cleaning, leak checking, or replacing a contaminated $GN_2$ bottle. The results of these tests were be compared with the purge gas reports from ATLO (Dworkin et al. 2018) to determine if corrective action would have been necessary (results posted on Confluence at [URL redacted]). An additional sample of gas for NVR and particulate verification will be collected in an LMS-supplied container and analyzed by LMS. A total of **3 bottles** (2 at JSC) are required as well as **1 container for NVR and particulate collection** supplied and analyzed by LMS. These bottles must have a quick-connect fitting compatible with the purge system. Note the bottle maximum pressure, a regulator may be required.

### S7-4.1.3.2 Purge contamination knowledge

The sample of the complete purge system must be collected to establish the purge baseline of the UTTR purge system at point *b* no earlier two days before SRC arrival. A total **of 1 bottle** is required. This bottle must have a quick-connect fitting compatible with the purge system. Note the bottle maximum pressure printed on the bottle; a regulator may be required.

### S7-4.1.4 Gas collection in the UTTR cleanroom

After the SRC is recovered, it will be partially disassembled under controlled conditions in a temporary cleanroom in Building 1012, as described in the Curation Plan (OSIRIS-REx-PLAN-0041, Revision A). To verify the condition of the cleanroom a sample of gas will be taken in August and in early September. Another gas sample will be collected to establish the baseline from inside this cleanroom no more than two days before SRC arrival. After SRC arrival in the cleanroom, two more gas samples will be taken:

G + 135 min - Gas samples taken from near heat shield and from backshell vents
- Gas sample taken near sample canister filter after purge is established

A total of **3 bottles** are required, one in around 22 September, and two on 24 September 2023.

## S7-4.2 Soil and water

An appropriate number of 8-24 soil samples (and water samples if water is present) at both the touchdown and recovery site (if not the same) will be collected at varying depths for contamination control purposes (NFP3-PN-12-SE-13).

### S7-4.2.1 Soil collection

Soil samples should be approximately 10x10 cm with a depth of about 2 cm deeper than the deepest depression made by the SRC. Soil will be collected by sterile scoops and transferred into containers provided by JSC Curation. Soil samples will be stored temporarily on ice at 4 °C or frozen at -20 °C as soon as practicable after arrival at Building 1012. The containers will be labeled with the date, time, location, purpose of collection. A fresh sterile scoop tool will be used when changing regions or by excavating and discarding three scoops of soil prior to use. The location of samples before and after collection will be photographed. Such photographs are in addition to the photo documentation of the SRC tracks.

**At least seven samples** of soil will be collected from areas exposed to the SRC: at least four samples of soil from points along the depressions or tracks made by the SRC if it rolled to its final position and at least two samples of soil from the location of the SRC after it has been moved from its post-touchdown resting location. For comparison at least four samples of soil from an undisturbed location at the recovery site will also be collected. Samples should be collected in areas without footprints and if multiple soil horizons are exposed an additional set of four samples of each will be collected. One additional collection of a footprint (trample-sample) will be collected.

### S7-4.2.2 Water collection

If the collection of standing water is necessary, water will be added to PTFE flasks provided by JSC Curation by dipping the container in the standing water depending on the depth and practical considerations or by using a transfer pipette. Approximately 4 mL of water (or a best effort volume of snow/ice) will be collected. Containers will be labeled with the date, time, location, purpose of collection, and the name of the collector. If a transfer pipette is used it should be filled three times with standing water before collection. The location of samples collected will be photographed.

If there is standing water at the recovery site, **at least one sample** will be collected. If there is accumulated snow or ice at the recovery site, **one sample** will also be collected. If there is active precipitation, a sample tube should be left open during the operation on the Range to make **one sample best effort attempt** to also collect precipitation. Water samples will be stored on ice at 4 °C or frozen at –20 °C as soon as practicable after arrival at Building 1012.

## S7-4.3 Windblown dust

Windblown dust will be collected by a dry filter air sampler on polypropylene disks for analysis with an InnovaPrep® portable Cub sampler (Fig. S7.4-4). The manufacturer's directions will be followed. The air sampler will be erected on the Range and operated at 200 L/min for the duration of the Range operation to collect 0.01 µm to >10 µm particles on a 52 mm polypropylene filter. **One filter** will be collected on the Range. The filter and a blank filter will be stored in a Curation-provided container.

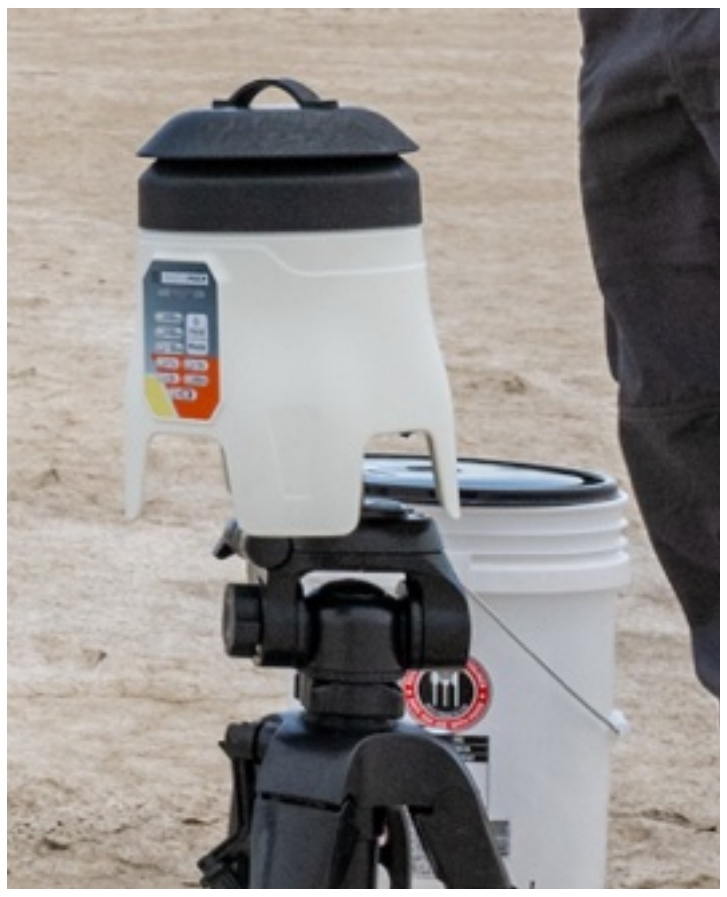

**Fig. S7.4-4** A commercial portable (3.5 kg) airborne dust collector deployed.

### S7-4.4 Other environmental materials

If other materials need to be sampled, the procedures for sampling soil (for solids) or water (for liquids) will be followed. Such materials could include plants (solid) or debris as permitted if declared safe by UTTR (solid). Collection will be made by scoop or forceps as appropriate. The location of samples before and after collection will be photographed. New sterile implements should be used for each collection if necessary.

### S7-4.5 Witness plates

Cleanroom contamination knowledge (CK) witness plates in addition to contamination control monitoring plates will be deployed in the UTTR Cleanroom according to the OSIRIS-REx Curation Plan for Utah Test & Training Range Operations plan. It states, "the monitoring of the UTTR cleanroom will utilize a variety of methods already used to monitor the OSIRIS-REx Curation cleanroom in building 31 at Johnson Space Center. These include daily particle counting when curation technical personnel are present at UTTR, Balazs air sampling, Balazs wafer sampling for organics and inorganics, microbial sampling, aluminum witness foils for amino acid measurements, and air sampling for analysis by JSC toxicology group. The cadence of this monitoring will leverage times that curation personnel need to be at UTTR for rehearsals and preparations." The CK plates include Balazs wafers deployed twice (7/18-19 and 9/18-20) microbial sampling four times (7/20, 8/31, ~9/20-23, and after completion of work on 9/24 or 25), and continuous exposure of aluminum witness foils (7/19-9/25, with foil changes on 8/30 and ~9/18-20).

To monitor contamination accumulation between the UTTR cleanroom and the OSIRIS-REx Curation Lab cleanroom (JSC building 31 room 216) the aluminum foils will be exposed with the sample canister during transport as practical. An additional set of CK aluminum plates will be brought for off-nominal deployments, for example if there is an extended delay between recovery and transport.

## S7-5.0 Off-Nominal Collection Locations

A range of off-nominal conditions is unlikely, but possible. Flexibility in the collection of environmental samples is necessary as conditions may be dynamic and uncertain. The prime responsibility is to ensure the recovery and protection of Bennu samples. However, environmental samples are critical to provide scientific information about potential contaminants. *All environmental sample collections are secondary to procedures to collect and secure Bennu samples.* Some examples follow:

- If there is a canister breach UTTR soil samples that contacted Bennu or SRC material will be collected. For comparison UTTR soil that is certain not to contain Bennu or SRC material should also be collected.
- If there is a battery or other SRC fire or smoke, gas from the plume should be collected, as safety permits.
- If the SRC lands in water, water draining from the SRC (at least 40 mL) will be collected. Standing water from the SRC's pool as well as a non-SRC pool should also be collected.

- If the SRC experiences an uncontrolled fall during recovery or transport to the helicopter, samples from the impacted surface will also be collected.
- If the SRC emits an odor when placed on purge, an effort to collect the gas should be made.
- If there is a substantial delay in placing the canister in its glovebox in JSC Curation Lab (building 31 room 216) a gas sample will be taken for each change in purge gas source and the CK witness plates accompanying the canister may need to be refreshed.
- If the sample canister is off purge in any location other than the scheduled durations in the UTTR cleanroom and JSC Cleaning and Curation Labs (building 31 room 110 complex and room 216), a gas sample will be taken.

Occasionally a gas bottle fails and needs to be resampled. A total of **6 extra bottles** is needed for these off-nominal events. One of these was consumed in the rotorwash collection during ORT 10.

## S7-6.0 Sample Storage and Analysis

All environmental samples will be returned to JSC Curation and stored away from uncontaminated Bennu samples or hardware. Soil, moist, or wet samples collected under nominal conditions will be frozen at -20˚C by A. Regberg at JSC, to impede microbial growth until samples are requested for analysis. Samples will be cooled as quickly as practicable. Samples will be refrigerated or frozen until they are shipped frozen to JSC. If water entered the SRC, then at least 40 mL of the recovered SRC water will be frozen in separate small vials for distribution and analysis. Gas samples will be delivered for prompt analysis by JSC Toxicology who will deliver their report to JSC Curation. Solid samples will be archived and up to ¼ available to the OSIRIS-REx Sample Analysis Team by request. CK witness plates will be archived (¾) and analyzed (¼) following the same procedure as during ATLO as described in Dworkin et al. (2017). These analyses include SEM/EDS at JSC and amino acid and GC-MS survey at GSFC or JSC.

## S7-7.0 Abbreviations and Acronyms

| | |
|---|---|
| ATLO | Assembly, Test, and Launch Operations |
| Aug | August |
| CK | Contamination knowledge |
| CM | Configuration Management |
| cm | Centimeter |
| CO | Carbon monoxide |
| e.g. | Exempli gratia |
| et al. | Et alia |
| etc. | Et cetera |
| ft. | Feet |
| G+ | Time after SRC ground contact |
| $GN_2$ | Nitrogen gas |
| GC-MS | Gas Chromatography Mass Spectrometry |
| GPS | Global Positioning System |
| GSFC | NASA Goddard Space Flight Center |
| HCN | Hydrogen cyanide |
| in. | Inch |
| Jr. | Junior |
| JSC | NASA Johnson Space Center |
| L | Liter |
| LMS | Lockheed Martin Space |
| m | Meter |
| MAAF | Michael Army Airfield |
| mL | Milliliter |
| µL | Microliter |
| NASA | National Aeronautics and Space Administration |
| NPD | NASA Policy Directive |
| NVR | Nonvolatile residue |
| ORT | Operational Readiness Test |
| OSCR | On Scene Commander |
| OSIRIS-REx | Origins, Spectral Interpretation, Resource Identification, and Security–Regolith |
| PI | Principal Investigator |
| PPE | Personal protective equipment |
| ppm | Parts per million (µL/L) |
| PTFE | Polytetrafluoroethylene (e.g. Teflon) |
| SEM/EDS | Scanning Electron Microscopy/Energy Dispersive X-Ray Spectroscopy |
| Sept | September |
| $SO_2$ | Sulfur dioxide |
| SRC | Sample Return Capsule |
| STEL | Short Term Exposure Limit |
| TAGSAM | Touch-And-Go Sample Acquisition Mechanism |
| TBD | To Be Determined |
| TBR | To Be Reviewed |
| TLV | Threshold Limit Value |
| UTTR | Utah Test and Training Range |
| WIG | Weapons Impact Grid |

# Appendix S7.1. Roles and checklist on the Range, Environmental Team

## Roles

| Title | PI | Curator | Co-I |
|---|---|---|---|
| **Prime** | Dante Lauretta | Francis McCubbin | Scott Sandford |
| **Backup** | Anjani Polit | Eve Berger | Jason Dworkin |
| **Role** | Tag coordinator<br>Dust sampler<br>Gas sampler<br>Flag Sample sites | Primary photographer<br>Solid/liquid sampler<br>Flag sample sites | Note taker<br>Mapper<br>Secondary photographer |
| **Bagging** | Bag holder and roller | Bag taper | Scissor steward |
| **Supplies** | Personal bag<br>Dry-bag (1)<br>Bucket with lid (1)<br>Gas bottles (5)<br>Dry filter air sampler (1)<br>Polypropylene disks (2)<br>Scissors (1)<br>PTFE bags (40)<br>Sample tags (75)<br>PTFE flasks with lids (4)<br>Aluminum jars (2)<br>Jar lids (2)<br>Body camera (1)<br>Whistle (1)<br>Gloves (7 pair) | Personal bag<br>Dry-bag (1)<br>Camera (1)<br>Spare camera battery (1)<br>Compass and GPS (1)<br>Trowel (2)<br>ANSMET cube (2)<br>Tape (2)<br>Tweezers (1)<br>Transfer pipette (4)<br>Foil (25 squares)<br>Body camera (1)<br>Whistle (1)<br>Gloves (7 pair) | Personal bag<br>Bucket with lid (1)<br>Notebook (3)<br>Writing implement (4)<br>Scissors (1)<br>Tweezers (1)<br>Spare compass and GPS (1)<br>Spare camera (1)<br>Flags (50 red/10 white/3 pink)<br>Battery bank & cables (1)<br>Towel (1)<br>Body camera (1)<br>Whistle (1)<br>Gloves (7 pair)<br>Heat gloves (1 pair) |

It is recommended that tools be on lanyards, sling packs, or hip packs as makes sense.

## Checklist

**All activities occur only with the clearance of OCSR**

### Staging sequence

1. Everyone verifies they have their equipment and that all batteries are fully charged.
2. Co-I record the departure time from MAAF.
3. Co-I record the arrival time at WIG.
4. Co-I record the departure time from WIG.

### Sequence at the landing zone

Upon leaving the helicopters, meet with the Recovery Field Lead, OSCR, and the entire team for a safety debrief. This will reveal the nominal vs. off nominal condition, the direction of SRC bounce and roll, gas safety levels, and capsule temperature. It will review the stop commands and other directions of note.

Co-I will sketch out the site in the field notebook and take notes on the entire process. The procedure below follows a nominal case. The lab notebook should be pre-filled with check boxes for each of the required photographs during sampling (Appendix S2.4). All members of environmental field team must wear nitrile gloves (§4.0).

## Arrival

After exiting the 4th helicopter, containing the environmental sampling team follow the below steps.

A1. Curator states the GPS location of where the helicopter landed and take a photograph of the GPS reading and Co-I records it in the notebook with the time of arrival.

A2. Turn on body camera.

A3. Huddle with Recovery Field Lead, OSCR, and the entire team for a safety debrief.

A4. The environmental field team will huddle to identify location for the dust sampler and preliminary identification of locations for other sampling.

A5. Co-I will begin to map the area and take notes. This will continue throughout the procedure, including the recording of all subsequent steps in the logbook and recording locations on the logbook maps.

A6. Curator will take preliminary context photographs of area.

A7. PI will use white flags (color 2) to mark the operational area for LMS technicians to approach and secure the SRC on the handling fixture. Team will inspect the SRC to assess if the breach team is required.

A8. PI will use two (2) pink flags (color 3) at SRC to note its position to know where to sample after it has been moved. Curator will mark the GPS coordinates of the SRC and take a photograph of the GPS reading and Co-I will record it in the notebook.

A9. PI will collect one gas sample in vicinity of heat shield. Curator photograph.

A10. PI will collect one gas sample in vicinity of backshell. Curator photograph.

A11. Co-I will remove the covers on the two SRC vents.

A12. PI will collect one gas sample in vicinity of each vent. Curator photograph.

A13. Co-I will replace the covers on the two SRC vents.

A14. PI will use red flags (color 1) to mark 4 sample locations along each bounce/roll path & 4 positions of control sample collection locations past the SRC in alignment with bounce and roll direction.

A15. Curator will take overview photographs that include the SRC and all flags, from 1-2 orientations.

A16. After sampling flags set, the environmental sampling team heads back to the Recovery Field Lead and the main huddle to recommend if the breach team can be called off and describe meaning behind flags. The group huddle decides on the sequence of sampling to ensure that environmental sampling does not interfere with securing the SRC.

A17. PI will set dust sampler 30-100 m (100-300 ft.) up wind from SRC.

A18. Curator will state the GPS location of the dust sampler and take a photograph of the GPS reading and Co-I will record the coordinates in the notebook.

A19. If there is active precipitation

- A19.1 Curator will place an open collection tube or jar near the dust sampler.
- A19.2 PI will place the tube or jar lid in a bag.
- A19.3 Curator take photos.
- A19.4 Co-I notes the time and location.

A20. Curator will use color 1 (red) flags to mark location of dust sampler.

A21. Curator will take context picture of dust sampler with SRC for context.

A22. Curator will take context picture of dust sampler in opposite direction of SRC for context.

A23. PI will collect gas sample in vicinity of dust sampler.

## Solid Sampling

The collection of soil, vegetation, or debris follows these steps.

S1. PI pulls tag and calls out tag number.
S2. Curator photographs the tag.
S3. Co-I records the tag number and each photo taken.
S4. Curator photographs the sampling location before sampling (one close up and one with SRC for context if possible).
S5. Curator hands camera to Co-I
S6. Curator scoops soil (if reusing shovel, scoop dirt three times and empty three times before collecting sample) or collects other solid material with forceps.
S7. Co-I photographs act of sampling.
S8. PI readies a bag (or jar if necessary for non-soil solids).
S9. Curator places the soil in the bag (or other solid in the jar).
S10. Co-I photograph the sample in the bag (or jar) being held by PI.
S11. Co-I photographs sampling location after sampling.
S12. Co-I return the camera to Curator.
S13. Seal the bag.
  - S13.1 PI folds the bag with the tag visible in the top fold.
  - S13.2 Curator pulls tape to seal bag against PI's knee.
  - S13.3 Co-I cuts tape
  - S13.4 PI folds ends of tape over back side of bag,

S14. Curator photographs sealed sample with tag facing up.
S15. Curator photographs relevant page of the field notebook being held by Co-I.
S16. PI or Curator will remove flag from location, everybody drinks water.
S17. Repeat S1-S16 until all solid samples are collected.
S18. After the SRC has been moved collect an additional sample from underneath a trample mark and 2 from underneath where the SRC was.

All relevant steps during this procedure are recorded in the logbook by Co-I.

## Water Sampling

For a wet landing site with standing water and SRC containing water that is below vents (not at risk of getting Bennu sample wet) follow the below steps.

W1. PI pulls tag and calls out tag number.

W2. Curator photographs the tag.

W3. Co-I records the tag number and each photo taken.

W4. Curator photographs the sampling location before sampling (one close up and one with SRC for context if possible).

W5. Curator hands camera to Co-I

W6. Collect water, less than ~4 cm (1.5 in.) deep.

- W6.1 PI will use a syringe to retrieve the water from the site (if reusing syringe, load and empty it three times before collecting sample).
- W6.2 Curator readies a PTFE tube.
- W6.3 PI adds water to the tube.

W7. Collect water, greater than ~4 cm (1.5 in.) deep.

- W7.1. Curator will directly sample water with PTFE tube by submerging open end of tube to collect a water sample.

W8. Collect water draining from the SRC.

- W8.1 LMS lifts the SRC.
- W8.2 Curator places the Teflon vial directly under the stream as the SRC is lifted and collects until the vial is full or the water stops exiting the capsule, being sure not to come into contact with live wires protruding from heatshield.
- W8.3 LMS places SRC on its side and drives a wedge into top of heat shield to allow water to drain from capsule.
- W8.4 Curator sample water with PTFE tube by bending down and collecting water dripping from SRC wedge point while Lockheed rolls SRC back and forth until the vial is full or the water stops exiting the capsule, being sure not to come into contact with live wires protruding from heatshield.
- W8.5 PI place an empty bucket under the Teflon vial to capture any extra water that does not fit in the vial or misses the vial.
- W8.6 If mud is stuck on bottom of SRC, collect a mud sample following the same procedure as the soil sample described above, being sure not to come into contact with live wires protruding from heatshield.

W9. Co-I photographs act of sampling.

W11. Co-I photograph the sample in the vial being held by PI.

W12. Co-I photographs sampling location after sampling.

W13. Seal the vial in a bag.

- W13.1 PI folds the bag with the tag visible in the top fold.
- W13.2 Curator pulls tape to seal bag against PI's knee.
- W13.3 Co-I cuts tape.
- W13.4 PI folds ends of tape over back side of bag.

W14. Co-I photograph sealed sample with tag facing up.

W15. Co-I return the camera to Curator.

W16. Curator photograph relevant page of the field notebook being held by Co-I.

W17. PI or Curator will remove flag from location, everybody drinks (clean) water.

W18. Repeat W1-W16 until all liquid samples are collected. Flag any soil sampling locations revealed or impacted by the collection of water from the SRC.

All relevant steps during this procedure are recorded in the logbook by Co-I.

## Departure

Once all samples are collected follow the below steps.

D1. The environmental field team will huddle to confirm that all sampling is done.
D2. Relate the completion to the Recovery Field Lead, or designee.
D3. Collect any remaining flags or dropped materials.
D4. If there was active precipitation collect the vial
  - D4.1 Curator photograph the collection tube/jar on the ground.
  - D4.2 PI remove the lid from the bag.
  - D4.3 Co-I note the time.
  - D4.4 PI folds the bag with the tag visible in the top fold.
  - D4.5 Curator pulls tape to seal bag against PI's knee.
  - D4.6 Co-I cuts tape.
  - D4.7 PI folds ends of tape over back side of bag
  - D4.8 Curator photograph sealed sample with tag facing up.
  - D4.9 Curator photograph relevant page of the field notebook being held by Co-I.
  - D4.10 PI or Curator will remove flag from location.

D5. PI remove the dust sampler.
D6. PI or Curator will remove flag from location.
D7. Report readiness to depart to the Recovery Field Lead, or designee.
D8. Turn off body camera.
D9. Record departure time.
D10. Record arrival time at MAAF.
D11. Record transfer time of samples to curation in building 1010.

All relevant steps during this procedure are recorded in the logbook by Co-I.

## Appendix S7.2. Camera settings

Two identical Nikon COOLPIX P950 cameras equipped with a lens hood, UV filter, and 512 GB memory card will be used as well as an additional backup battery.

[Instructions redacted for publication]

## Appendix S7.3. Gas collections

Since the gas collection bottles are in limited supply and bulky it may be necessary to make decisions to descope one collection over another if sufficient bottles are not available when needed. This is a prioritized list of collections by location for sampling required after ORT 10. Highest priority is A and lowest is C. X indicates that the sample has already been collected. Note that one spare was consumed in the collection of the ORT 10 rotorwash.

| Section | Event | Range Priority | 1012 Priority | Completed by 30 Aug 2023 |
|---|---|---|---|---|
| 4.1.2.1 | SRC vent 1 | A | | |
| 4.1.2.1 | SRC vent 2 | B | | |
| 4.1.2.1 | Backshell | C | | |
| 4.1.2.1 | Heatshield | A | | |
| 4.1.2.2 | Upwind of SRC | A | | |
| 4.1.2.2 | ORT 10 rotor wash | | | X |
| 4.1.2.2 | Upwind of 1012 | | C | |
| 4.1.3.1 | $GN_2$ manifold verification | | | X |
| 4.1.3.1 | JSC manifold 1st floor | | | X |
| 4.1.3.1 | JSC manifold 2nd floor | | | X |
| 4.1.3.2 | $GN_2$ manifold | | B | |
| 4.1.4 | Cleanroom, 9/23 | | A | |
| 4.1.4 | Cleanroom, SRC vents pre-purge | | B | |
| 4.1.4 | Cleanroom, SRC vents pre-purge | | C | |
| 4.1.4 | Cleanroom, SRC on purge | | A | |
| 5.0 | Spare 1 | | | X |
| 5.0 | Spare 2 | | A | |
| 5.0 | Spare 3 | | A | |
| 5.0 | Spare 4 | | B | |
| 5.0 | Spare 5 | | B | |
| 5.0 | Spare 6 | | C | |
| | Total bottles | 5 | 11 | 5 |

## Appendix S7.4. Notebook

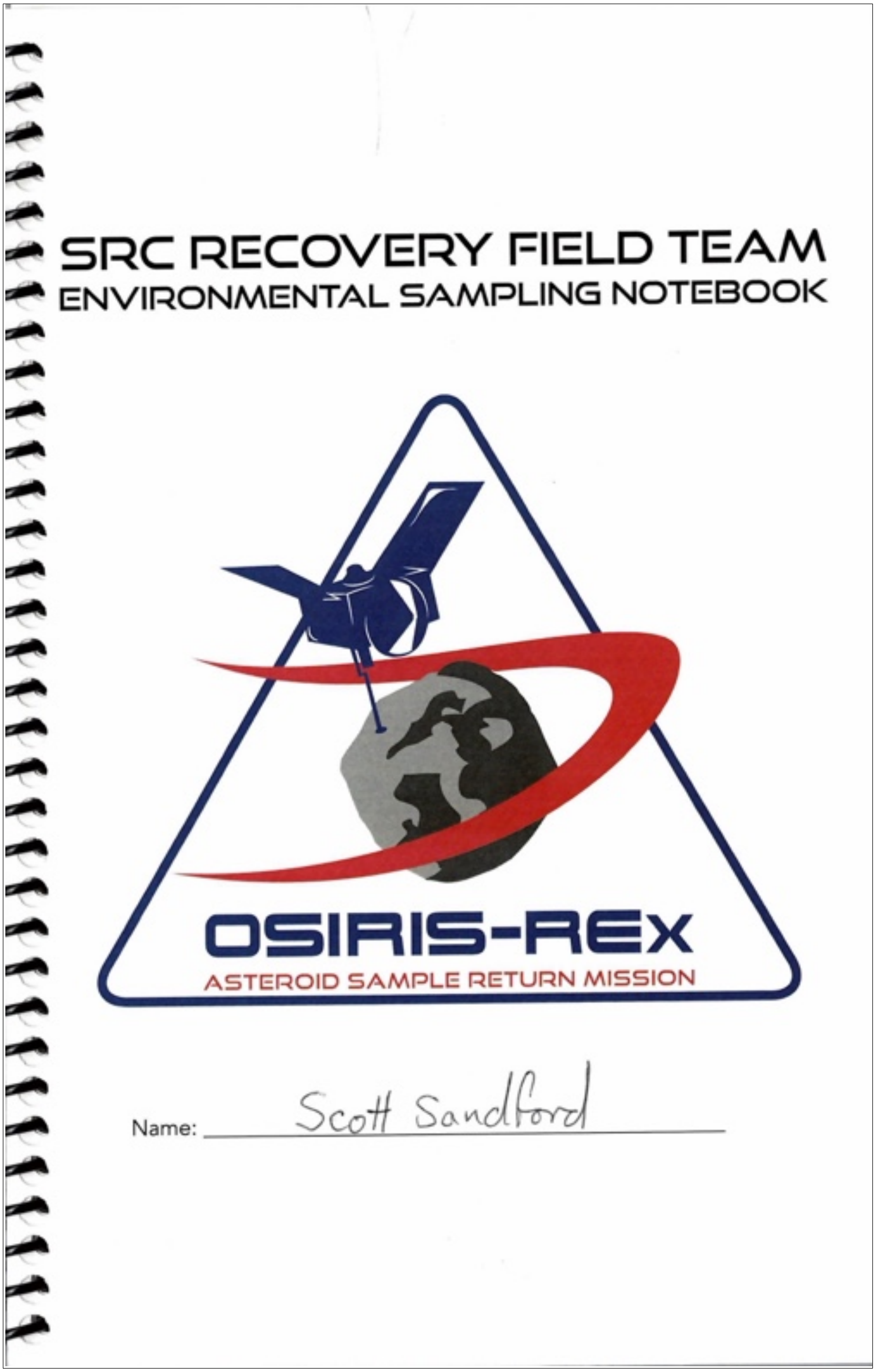

2 x 10 cm

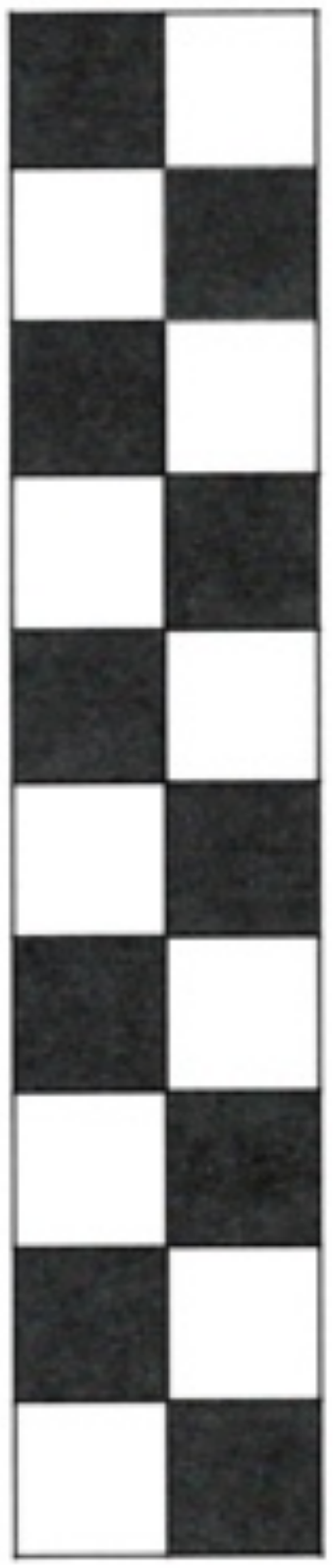

| Time (T) | Activity |
|---|---|
| **SUNDAY 9/24** | |
| 1:55 AM | Go/No-Go poll for SRC release (including UTTR) |
| 2:55 AM | Spacecraft SRC release command uplinked to s/c |
| 3:00 AM | Weather balloon #1 release |
| 4:55 AM | SRC Release |
| 5:30 AM | Confirmation of release, confirmation of divert communicated to Recovery Team (ORX-COORD) |
| 6:15 AM | Recovery Team Work Day Starts + Hill range control + MCC staff |
| 6:20 AM | Weather Briefing (In-Person/Teams) |
| 6:30 AM | RCS Authorizes Recovery Team to Execute Recovery Operations |
| 6:55 AM | MCC staff on-site (time tbd) |
| 7:00 AM | Team Briefing in building 1010 |
| 7:21 AM | Sunrise at Dugway |
| 7:35 AM | Breach personnel depart 1010 for WIG |
| 7:35 AM | Recovery personnel depart 1010 for MAAF |
| 7:45 AM | Personnel Loading @ MAAF & Helicopters Run-Up |
| 7:50 AM | Helicopters depart MAAF for WIG |
| 7:55 AM | Range protected by road guards |
| 8:00 AM | Arrival & Final Preparations @ WIG - Helicopters / Ground Vehicles |
| 8:24 AM | Personnel Loading - Helicopters / Ground Vehicles |
| 8:29 AM | Helicopter 1-2 spin up & depart for holding location (Earliest Time)<br>Helicopter 3-4 spin up & hold at WIG |
| 8:39 AM | Helicopter 1-2 arrive at hold point outside ellipse |
| 8:42 AM | Entry Interface |
| 8:44 AM | Drogue Deploy - ~102,300 ft |
| 8:44 AM | Closest Approach of OSIRIS-REx to the Earth |
| 8:49 AM | Main Deploy - ~5,050 ft |
| 8:50 AM | Helicopters #3 & 4 depart for landing area (per OSCR)<br>Helicopters 1-2 given permission to enter ellipse and go to landing co-ordinates<br>Helicopters 3-4 given permission to spin up and leave WIG for landing site<br>Breach team & support vehicles given permission to leave WIG and go to landing co-ordinates |
| 8:53 AM | SRC Landing - Earliest Time |
| 8:55 AM | SRC Landing - Nominal Time |
| 8:57 AM | SRC Landing - Latest Time |
| 8:56 AM | Helicopter #1 & 2 Arrives At Scene and Locates SRC (earliest time) |
| 8:59 AM | Helicopter #3 & 4 Arrive At Scene(earliest time) |
| 9:15 AM | Weather balloon #2 release from near landing site |
| 8:56 AM | Recovery Site Assessment Begin |
| | Helicopter #1 and #2 Landing (#2 departs) |
| | Approach Preparations (OSCAR & LM Safety) |
| | Perform Unexploded Ordnance (UXO) survey of helo landing zone (OSCAR) |
| | Clear helo #3 & #4 to land (OSCAR) |
| | Perform Unexploded Ordnance (UXO) survey (OSCAR) |
| | Parachute Attached safing and storage - ASSUMES PARACHUTE RELEASED |
| | Initial Assessment (LM Safety) |
| | Inspect Vents (LM Safety) |
| | Gas Safety Survey (LM Safety) |
| | Secure Cut Harness Cable Ends (LM Safety) |
| | Install Vent & Parachute Covers (LM Safety) |
| 9:21 AM | Recovery Operations Begin |
| | Site status huddle (whole team) |
| | Science/Curation team site assessment |
| | Science/Curation status huddle |
| | Science/Curation team operations in parallel |
| | Prepare SRC handling fixture and bagging material |
| | Secure the SRC in Handling Fixture |
| | SRC ready for long-line operations |
| | Move SRC to helo #1 for longline operations |
| | Hookup cargo net for longline operations |
| | Load recovery personnel into helo #3 (science team remain in field) |
| 10:20 AM | Recovery Operations Complete |
| 10:20 AM | Helicopter #1 & #3 - Departure to Building 1012<br>Helicopter #2 - Record departure and then land to pick up personnel |
| 10:30 AM | Helicopter #3 drop off personnel (parking lot of 1012) |
| 10:50 AM | Helicopter #1 drop off SRC (parking lot of 1012) |
| 10:55 AM | Helicopter #1 & 2 drop off personnel (MAAF) |
| 11:00 AM | Activity huddle |
| 11:05 AM | Gas Safety Survey (LM Safety) |
| 11:07 AM | Transfer into cleanroom (multipe steps) |
| 11:22 AM | Margin |
| 11:00 AM | Helicopter #4 Return to MAAF (predicted earliest) |
| 11:22 AM | Dis-Assembly Operations Begin |
| | Preparations & Photographs |
| | Gas Sample Vents |
| | Parachute Canister |
| | Remove Parachute Canister Fasteners |
| | Remove Parachute Canister |
| | SRC Backshell |
| | Remove Backshell Fasteners |
| | Remove MLI Blanketing Around Battery |
| | De-Mate Battery |
| | Remove Backshell |
| | Purge Final Preparations |
| | Purge Final Preparations |
| | Purge Margin |
| | Margin |
| 1:00 PM | Helicopter #4 Return to MAAF (predicted latest) |
| 1:22 PM | Initiate Purge - (w/in 2 hours of cleanroom entry) |

## Table of Contents

## GPS Locations at Recovery Site

Helicopter 4 N 40° 22.266'
W 113° 14.465'

SRC N 40° 22.336'
W 113° 14.401'

Dust Sampler N 40° 22.346'
W 113° 14.383'

1

## Timeline

| Event | Time (UTTR) |
|---|---|
| Loading helicopter | 07:40 AM |
| Departure from MAAF | 07:48 AM |
| Arrival at WIG | 07:56 AM |
| Helicopter departure from WIG | 08:55 AM |
| Arrival at recovery site | 09:21 AM |
| Placement of air sampler | see pages 10-29 |
| Placement of rain collection beakers | see pages 10-29 |
| Placement of sampling site flags | see pages 10-29 |
| Sampling | see pages 10-29 |
| Departure from field site | 10:57 AM |
| Arrival at helipad, post-recovery | 11:09 AM |
| Transfer of sample Bldg 1010, Avery | 11:30 AM |

## Sample/Event Types

- air sampling
- dust sampling
- soil sampling (surface)
- water sampling (surface)
- water sampling (rain)
- water sampling (surface)
- water sampling (SRC interior)
- ice/snow sampling
- other sampling (vegetation, etc.)

## Sequence at the landing zone

Upon leaving the helicopters meet with the Recovery Field Lead and the entire team for a safety debrief. This will reveal the nominal vs. off nominal condition, the direction of SRC bounce and roll, gas safety levels, and capsule temperature. It will review the stop commands and other directions of note.

Co-I will sketch out the site in this notebook and take notes on the entire process. All members of environmental field team must wear nitrile gloves.

## Checklist: Arrival

*After exiting the 4th helicopter, follow the below steps.*

A1. Curator states the GPS location of where the helicopter landed and take a photograph of the GPS reading and Co-I record in notebook with the arrival time.
A2. Turn on body cameras.
A3. Huddle with Recovery Field Lead, OSCR, and the entire team for a safety debrief.
A4. The environmental field team will huddle to identify location for the dust sampler and preliminary identification of locations for other sampling.
A5. Co-I will begin to map the area and take notes. This will continue throughout the procedure, including the recording of all subsequent steps in the logbook and recording locations on the logbook maps.
A6. Curator will take preliminary context photographs of area.
A7. PI will use white flags to mark the operational area for LM technicians to approach and secure the SRC on the handling fixture. Team will assess if the breach team is needed.
A8. PI will use two (2) pink flags at SRC to note its position to know where to sample after it has been moved. Curator will mark the GPS coordinates of the SRC and take a photograph of the GPS reading and Co-I will record it in the notebook.
A9. PI will collect one gas sample in vicinity of heat shield. Curator photograph.
A10. PI will collect one gas sample in vicinity of backshell. Curator photograph.
A11. Co-I will remove the covers on the two SRC vents.
A12. PI will collect one gas sample in vicinity of each vent. Curator photograph.
A13. Co-I will replace the covers on the two SRC vents.
A14. PI will use red flags to mark 4 sample locations along each bounce/roll path & 4 positions of control sample collection locations past the SRC in alignment with bounce and roll direction.
A15. Curator will take overview photographs that include the SRC and all flags, from 1-2 orientations.
A16. After sampling flags set, the environmental sampling team heads back to the Recovery Field Lead and the main huddle to recommend if the breach team can be called off and describe meaning behind flags. The group huddle decides on the sequence of sampling to ensure that environmental sampling does not interfere with securing the SRC.
A17. PI will set up dust sampler 30-100 m (100-300 ft.) up wind from SRC.
A18. Curator will state the GPS location of the dust sampler and take a photograph of the GPS reading and Co-I will record the coordinates in the notebook.
A19. If there is active precipitation
  A19.1 Curator will place an open collection tube or jar near the dust sampler.
  A19.2 PI will place the tube or jar lid in a bag.
  A19.3 Curator take photos.
  A19.4 Co-I notes the time and location.
A20. Curator will use red flags to mark location of dust sampler.
A21. Curator will take context picture of dust sampler with SRC.

A22. Curator will take context picture of dust sampler in opposite direction of SRC.
A23. PI will collect gas sample in vicinity of dust sampler.

## Checklist: Solid sampling

*The collection of soil, vegetation, or debris follows the below steps.*

S1. PI pulls tag and calls out tag number.
S2. Curator photographs the tag.
S3. Co-I records the tag number and each photo taken.
S4. Curator photographs the sampling location before sampling (one close up and one with SRC for context if possible).
S5. Curator hands camera to Co-I
S6. Curator scoops soil (if reusing shovel, scoop dirt three times and empty three times before collecting sample) or collects other solid material with forceps.
S7. Co-I photographs act of sampling.
S8. PI readies a bag (or jar if necessary for non-soil solids).
S9. Curator places the soil in the bag (or other solid in the jar).
S10. Co-I photograph the sample in the bag (or jar) being held by PI.
S11. Co-I photographs sampling location after sampling.
S12. Co-I return the camera to Curator.
S13. Seal the bag.
  S13.1 PI folds the bag with the tag visible in the top fold.
  S13.2 Curator pulls tape to seal bag against PI's knee.
  S13.3 Co-I cuts tape.
  S13.4 PI folds ends of tape over back side of bag.
S14. Curator photographs sealed sample with tag facing up.
S15. Curator photographs relevant page of the field notebook being held by Co-I.
S16. PI or Curator will remove flag from location, drink water.
S17. Repeat S1-S16 until all solid samples are collected.
S18. After the SRC has been moved collect an additional sample from underneath a trample mark and 2 from underneath where the SRC was.

All relevant steps during this procedure are recorded in the logbook by Co-I.

## Checklist: Liquid sampling

*For a wet landing site with standing water and SRC containing water that is below vents (not at risk of getting Bennu sample wet) follow the below steps.*

W1. PI pulls tag and calls out tag number.
W2. Curator photographs the tag.
W3. Co-I records the tag number and each photo taken.
W4. Curator photographs the sampling location before sampling (one close up and one with SRC for context if possible).
W5. Curator hands camera to Co-I
W6. Collect water, less than ~4 cm (1.5 in.) deep.

CoI photographs these steps

- W6.1 PI will use a syringe to retrieve the water from the site (if reusing syringe, load and empty it three times before collecting sample).
- W6.2 Curator readies a PTFE tube.
- W6.3 PI adds water to the tube.

W7. Collect water, greater than ~4 cm (1.5 in.) deep.

- W7.1 Curator will directly sample water with PTFE tube by submerging open end of tube to collect a water sample.

W8. Collect water draining from the SRC.

- W8.1 LM lifts the SRC.
- W8.2 Curator places the Teflon vial directly under the stream as the SRC is lifted and collects until the vial is full or the water stops exiting the capsule, being sure not to come into contact with live wires protruding from heatshield.
- W8.3 LM places SRC on its side and drives a wedge into top of heat shield to allow water to drain from capsule.
- W8.4 Curator sample water with PTFE tube by bending down and collecting water dripping from SRC wedge point while Lockheed rolls SRC back and forth until the vial is full or the water stops exiting the capsule, being sure not to come into contact with live wires protruding from heatshield.
- W8.5 PI place an empty bucket under the Teflon vial to capture any extra water that does not fit in the vial or misses the vial.
- W8.6 If mud is stuck on bottom of SRC, collect a mud sample following the same procedure as the soil sample described above, being sure not to come into contact with live wires protruding from heatshield.

W9. Co-I photographs act of sampling.
W10. Co-I return the camera to Curator.
W11. Co-I photograph the sample in the vial being held by PI.
W12. Co-I photographs sampling location after sampling.
W13. Seal the vial in a bag.

- W13.1 PI folds the bag with the tag visible in the top fold.
- W13.2 Curator pulls tape to seal bag against PI's knee.
- W13.3 Co-I cuts tape.
- W13.4 PI folds ends of tape over back side of bag.

W14. Co-I photograph sealed sample with tag facing up.

6

W15. Co-I returns the camera to Curator
W16. Curator photograph relevant page of the field notebook being held by Co-I.
W17. PI or Curator will remove flag from location, drink (clean) water
W18. Repeat W1-W16 until all liquid samples are collected. Flag any soil sampling locations revealed or impacted by the collection of water from the SRC.

All relevant steps during this procedure are recorded in the logbook by Co-I.

## Checklist: Departure

*Once all samples are collected follow the below steps.*

D1. The environmental field team will huddle to confirm that all sampling is done.
D2. Relate the completion to the Recovery Field Lead, or designee.
D3. Collect any remaining flags or dropped materials.
D4. If there was active precipitation collect the vial
    D4.1 Curator photograph the collection tube/jar on the ground.
    D4.2 PI remove the lid from the bag.
    D4.3 Co-I note the time.
    D4.4 PI folds the bag with the tag visible in the top fold.
    D4.5 Curator pulls tape to seal bag against PI's knee.
    D4.6 Co-I cuts tape.
    D4.7 PI folds ends of tape over back side of bag
    D4.8 Curator photograph sealed sample with tag facing up.
    D4.9 Curator photograph relevant page of the field notebook being held by Co-I.
    D4.10 PI or Curator will remove flag from location.
D5. PI turns off and packs the dust sampler
D6. PI or Curator will remove flag from location.
D7. Report readiness to depart to the Recovery Field Lead, or designee.
D8. Turn off body camera.
D9. Record departure time.
D10. Record arrival time at MAAF.
D11. Record transfer time of samples to curation in building 1010.

All relevant steps during this procedure are recorded in the logbook by Co-I.

## Field Map 1: Entire recovery site

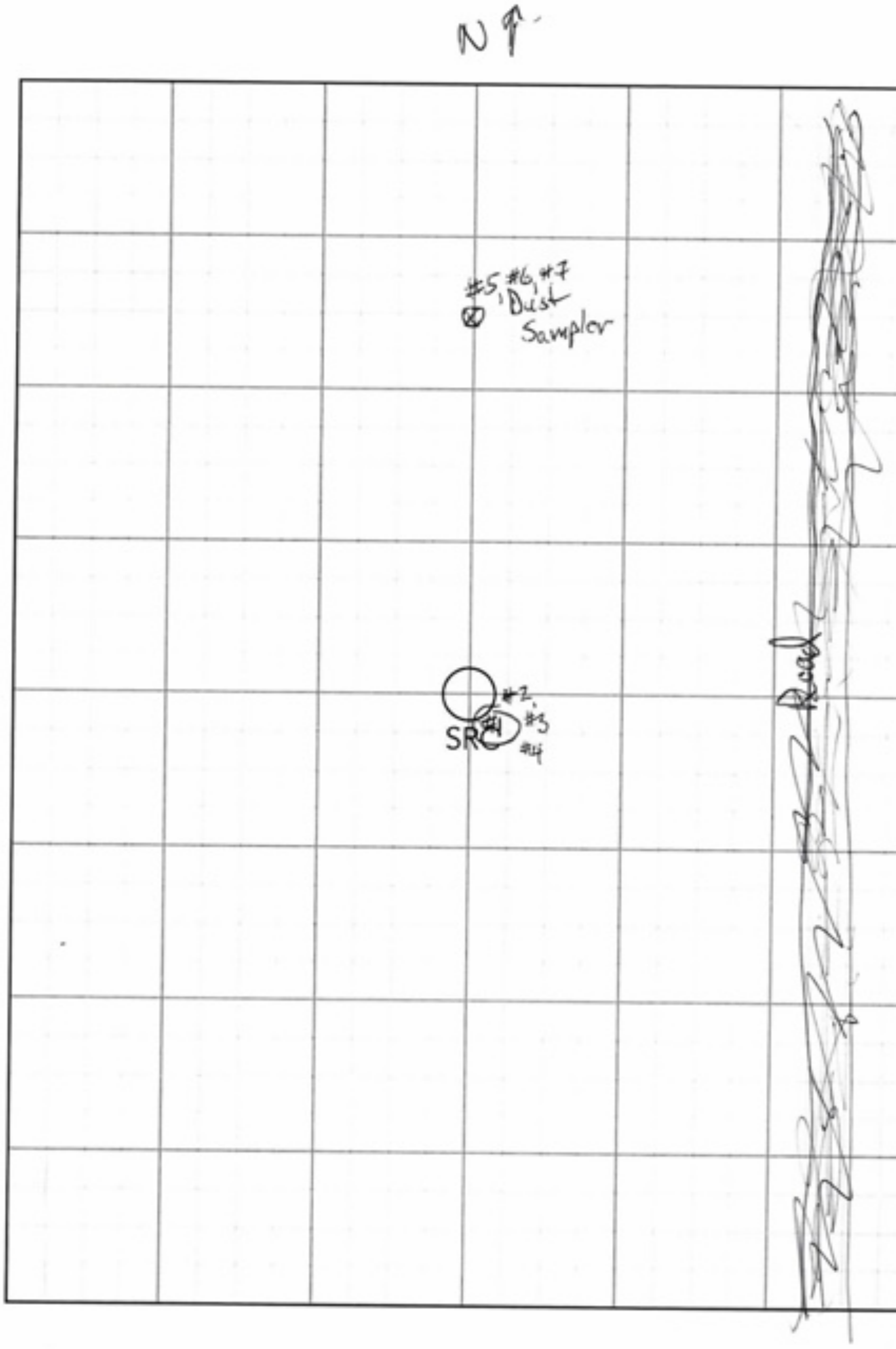

8

## Field Map 2: Immediate area around SRC

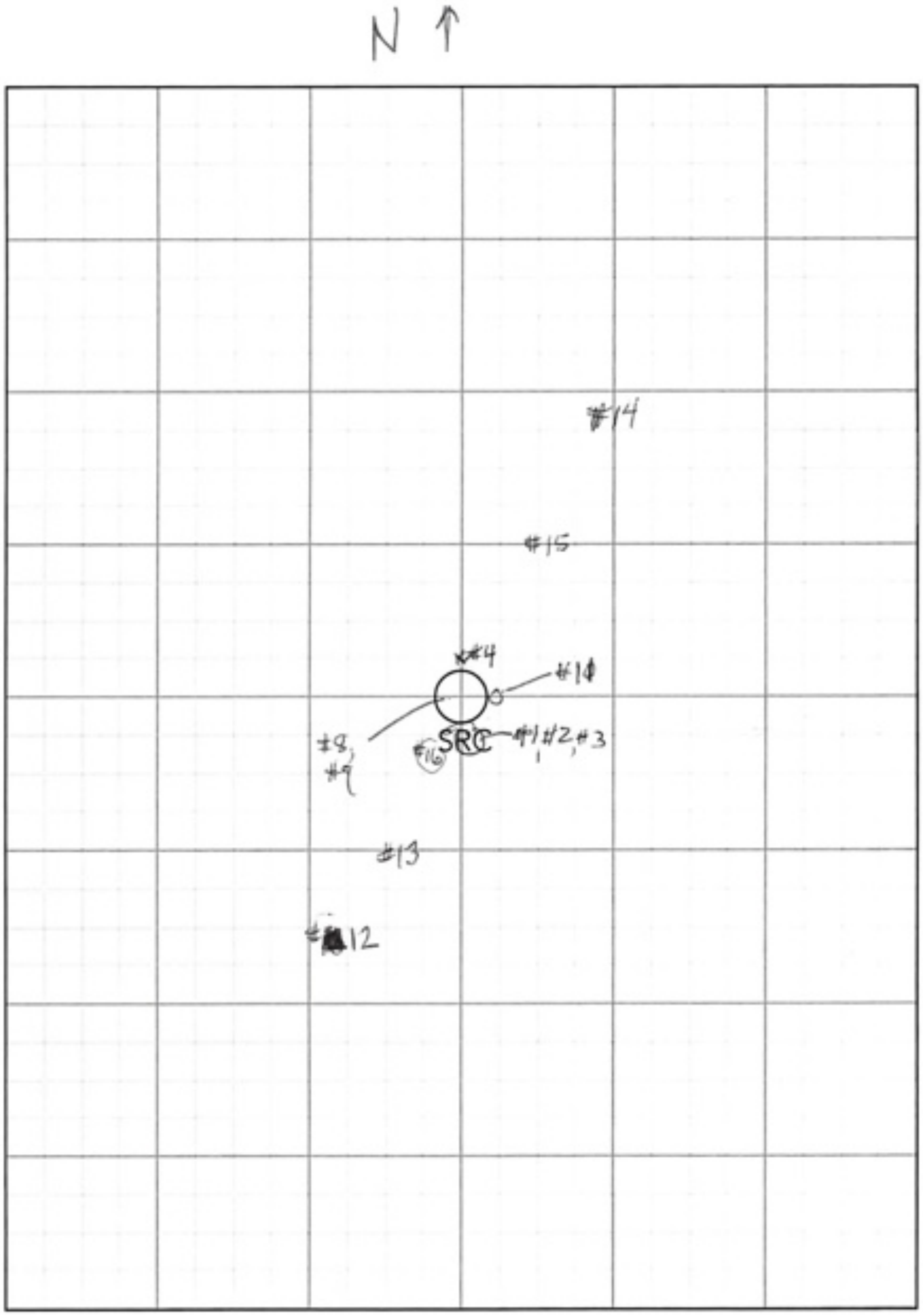

9

| Log Number | | 1 | 2 | 3 |
|---|---|---|---|---|
| Time (UTTR) | | Time 09:40 AM | Time 09:41 AM | Time 09:43A |
| Map number (pages 4/5) | | Map # 2 | Map # 2 | Map # 2 |
| Event type | | Event Gas sampling heat shield | Event Gas sampling bookshell | Event Gas sampling East vent #1 |
| Tag Number | | Tag # 34989 ✓ | Tag # 12848 ✓ | Tag # 22378 ✓ |
| Photos | Tag | Tag ☑ | Tag ☑ | Tag ☑ |
| | Sample site | Close ☑ | Close ☑ | Close ☑ |
| | Site w/SRC | Far ☑ | Far ☑ | Far ☑ |
| | Sampling activity | During ☑ | During ☑ | During ☑ |
| | Site post-collection | After – | After – | After – |
| | Open container | Open – | Open – | Open – |
| | Sealed container | Sealed – | Sealed ☐ | Sealed – |
| | Notebook entry | Book ☑ | Book ☑ | Book ☑ |
| | Extra | Other ☐ | Other ☐ | Other ☐ |
| | Extra | Other ☐ | Other ☐ | Other ☐ |
| | | | | Label incorrect and corrected in field |

10

| | 4 | 5 | 6 | 7 |
|---|---|---|---|---|
| Time | 09:44AM | 09:52 AM | 09:54AM | 09:57 AM |
| Map # | 2 | 1 | 1 | 1 |
| Event | Gas Sampling Vent #2 (west) | Dust sampling started | Dust Sampler #1 | Dust Sampler #2 |
| Tag # | 21190 | — | 25218 | 21260 |
| Tag | ☑ | ☑ | ☑ | ☑ |
| Close | ☑ | ☑ | ☑ | ☑ |
| Far | ☑ | ☑ | ☑ | ☑ |
| During | ☑ | ☑ | ☑ | ☑ |
| After | – | – | – | – |
| Open | – | ☐ | – | – |
| Sealed | – | ☐ | – | – |
| Book | ☑ | ☑ | ☑ | ☑ |
| Other | ☐ | ☐ | ☐ | ☐ |
| Other | ☐ | ☐ | ☐ | ☐ |
| | | | Did not audibly verify hiss. Was mislabeled - label corrected in field. resampled (see #7) (used spare bottle) | |

| Log Number | | 8 | 9 | 10 |
|---|---|---|---|---|
| Time (UTTR) | | Time 10:07AM | Time 10:14 AM | Time 10:15AM |
| Map number (pages 4/5) | | Map # 2 | Map # 2 | Map # 2 |
| Event type | | Event Soil Sample under SRC #1 | Event Soil Sample under SRC #2 | Event SRC lift image |
| Tag Number | | Tag # 1007 ✓ | Tag # 1023 ✓ | Tag # ——— |
| Photos | Tag | Tag ☑ | Tag ☑ | Tag ☐ |
| | Sample site | Close ☑ | Close ☑ | Close ☐ |
| | Site w/SRC | Far ☑ | Far ☑ | Far ☐ |
| | Sampling activity | During ☑ | During ☑ | During ☐ |
| | Site post-collection | After ☑ | After ☑ | After ☐ |
| | Open container | Open ☑ | Open ☑ | Open ☐ |
| | Sealed container | Sealed ☑ | Sealed ☑ | Sealed ☐ |
| | Notebook entry | Book ☑ | Book ☑ | Book ☐ |
| | Extra | Other ☐ | Other ☐ | Other ☐ |
| | Extra | Other ☐ | Other ☐ | Other ☐ |
| | | | | SRC lift image |

12

| 11 | 12 | 13 | 14 |
|---|---|---|---|
| Time 10:20 AM | Time 10:25 AM | Time 10:28 AM | Time 10:32 |
| Map # 2 | Map # 2 | Map # 2 | Map # 2 |
| Event Side imprint soil sample | Event Context soil Sample ~~[illegible]~~ control | Event Context soil sample control | Event Context soil Sample control |
| Tag # 1039 ✓ | Tag # 1055 ✓ | Tag # 1071 ✓ | Tag # 1072 ✓ |
| Tag ☑ | Tag ☑ | Tag ☑ | Tag ☑ |
| Close ☑ | Close ☑ | Close ☑ | Close ☑ |
| Far ☑ | Far ☑ | Far ☑ | Far ☑ |
| During ☑ | During ☑ | During ☑ | During ☑ |
| After ☑ | After ☑ | After ☑ | After ☑ |
| Open ☑ | Open ☑ | Open ☑ | Open ☑ |
| Sealed ☑ | Sealed ☑ | Sealed ☑ | Sealed ☑ |
| Book ☑ | Book ☑ | Book ☑ | Book ☑ |
| Other ☐ | Other ☐ | Other ☐ | Other ☐ |
| Other ☐ | Other ☐ | Other ☐ | Other ☐ |

| Log Number | 15 | 16 | 17 |
| --- | --- | --- | --- |
| Time (UTTR) | Time 10:35 AM | Time 10:37 AM | Time 10:43 |
| Map number (pages 4/5) | Map # 2 | Map # 2 | Map # 1 |
| Event type | Event Control soil sample control | Event Trample sample | Event Stop dust sampling |
| Tag Number | Tag # 1056 ✓ | Tag # 1040 ✓ | Tag # 1008 ✓ |
| Photos: Tag | Tag ☑ | Tag ☑ | Tag ☐ |
| Photos: Sample site | Close ☑ | Close ☑ | Close ☑ |
| Photos: Site w/SRC | Far ☑ | Far ☑ | Far ☑ |
| Photos: Sampling activity | During ☑ | During ☑ | During ☑ |
| Photos: Site post-collection | After ☑ | After ☑ | After ☐ |
| Photos: Open container | Open ☑ | Open ☑ | Open ☑ |
| Photos: Sealed container | Sealed ☑ | Sealed ☑ | Sealed ☑ |
| Photos: Notebook entry | Book ☑ | Book ☑ | Book ☑ |
| Photos: Extra | Other ☐ | Other ☐ | Other ☐ |
| Photos: Extra | Other ☐ | Other ☐ | Other ☐ |
| | Small amount of biomaterial in sample | | |

14

[Blank pages redacted for publication]

## Personnel and Equipment Check List

| | PI | | Curator | | Co-I |
|---|---|---|---|---|---|
| | Dante Lauretta or Anjani Polit | | Francis McCubbin or Eve Berger | | Scott Sandford or Jason Dworkin |
| | Personal bag | | Personal bag | | Personal bag |
| | Dry-bag (1) | | Dry-bag (1) | | Bucket with lid (1) |
| | Bucket with lid (1) | | Camera (1) | | Notebook (3) |
| | Gas bottles (5) | | Spare camera battery (1) | | Writing implement (4) |
| | Dry filter air sampler (1) | | Compass and GPS (1) | | Scissors (1) |
| | Polypropylene disks (2) | | Trowel (2) | | Tweezers (1) |
| | Scissors (1) | | ANSMET cube (2) | | Spare compass and GPS (1) |
| | PTFE bags (40) | | Tape (2) | | Spare camera (1) |
| | Sample tags (75) | | Tweezers (1) | | Flags (50 red/10 white/3 pink) |
| | PTFE flasks with lids (4) | | Transfer pipette (4) | | Battery bank & cables (1) |
| | Aluminum jars (2) | | Foil (25 squares) | | Towel (1) |
| | Jar lids (2) | | Body camera (1) | | Body camera (1) |
| | Body camera (1) | | Whistle (1) | | Whistle (1) |
| | Whistle (1) | | Gloves (7 pair) | | Gloves (7 pair) |
| | Gloves (7 pair) | | Smile | | Heat gloves (1 pair) |

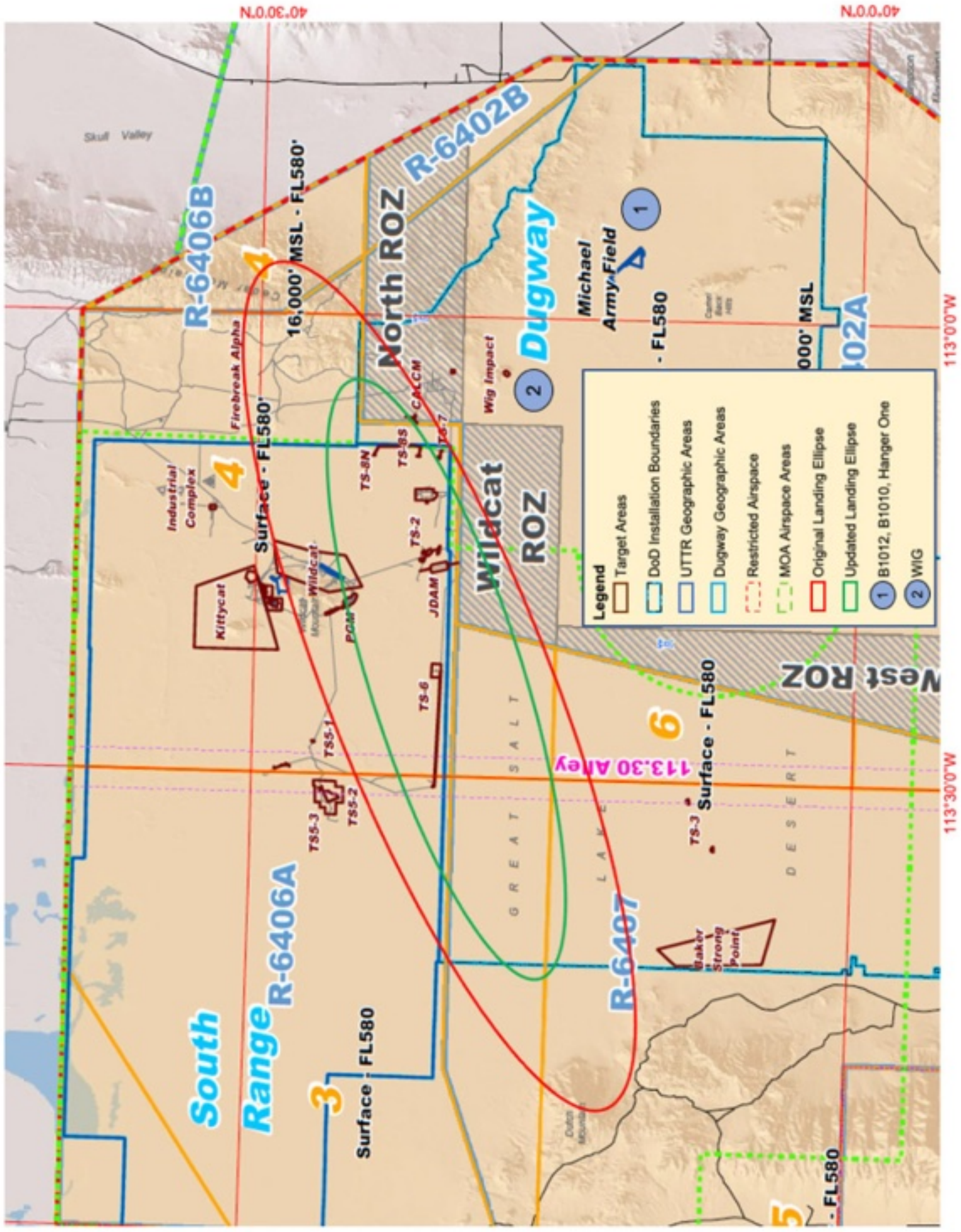

| | |
|---|---|
| A | Alfa |
| B | Bravo |
| C | Charlie |
| D | Delta |
| E | Echo |
| F | Foxtrot |
| G | Golf |
| H | Hotel |
| I | India |
| J | Juliett |
| K | Kilo |
| L | Lima |
| M | Mike |
| N | November |
| O | Oscar |
| P | Papa |
| Q | Quebec |
| R | Romeo |
| S | Sierra |
| T | Tango |
| U | Uniform |
| V | Victor |
| W | Whiskey |
| X | X-ray |
| Y | Yankee |
| Z | Zulu |

OSIRIS-REx 2023

**Supplement 8:** SRC environmental tracking

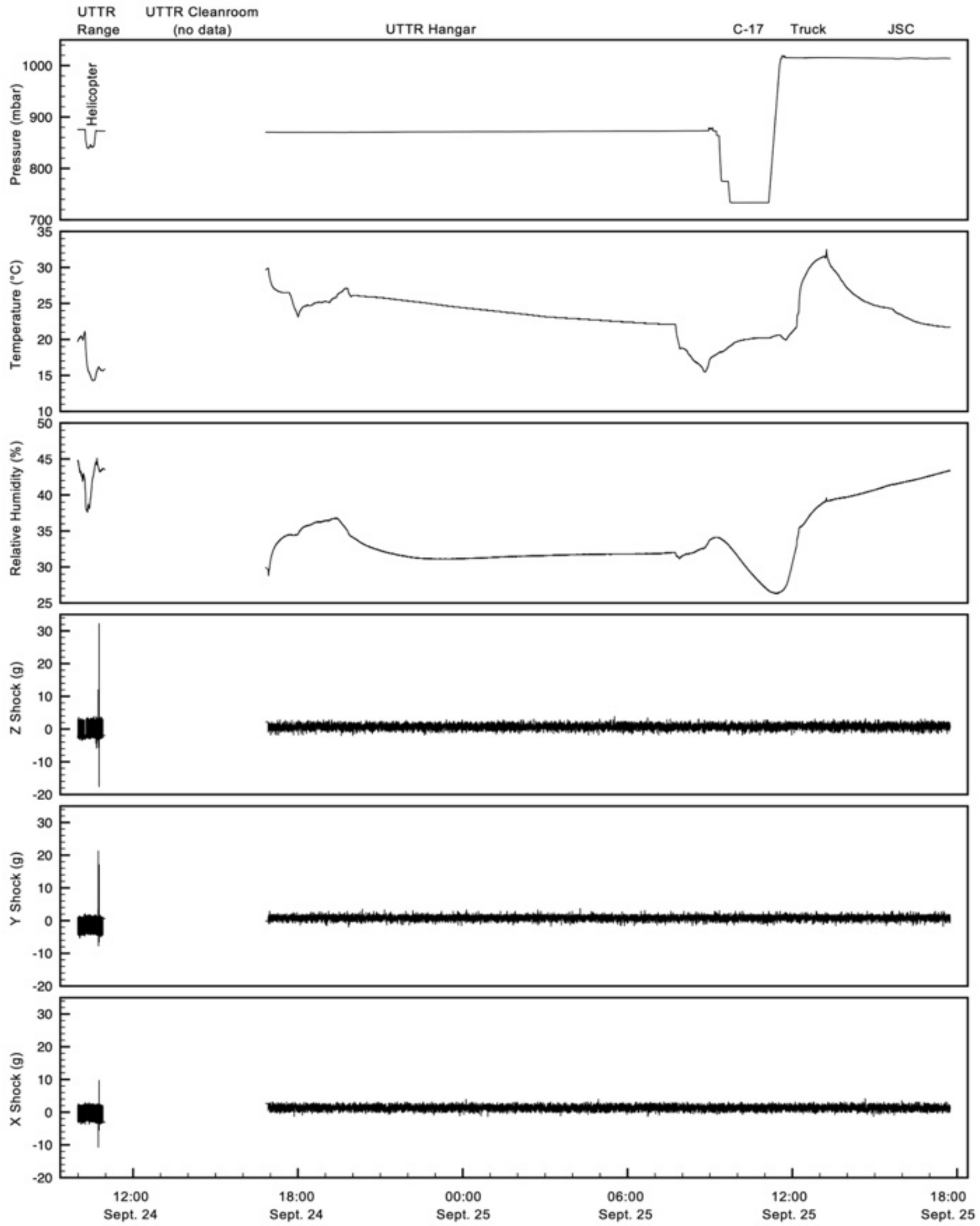

**Fig. S8-1** The transportation environment of the SRC from recovery on the range at UTTR to arrival at JSC was monitored via a MadgeTech Ultrashock 3-axis shock, temperature, pressure, and humidity data logger affixed to the SRC handling fixture. The logger recorded shock, temperature, relative humidity, and pressure at a resolution of 0.2 g, 0.1 °C, 0.1%, and 0.05 mbar, respectively, every 10 s. The pressure, temperature, and humidity traces show the environments the SRC traveled through though the sample was under $GN_2$ purge 2 hours after re-entry; thus the humidity and pressure values reflect the ambient environment, not the sample environment. However, they usefully delineate the different locations the SRC was in, including at altitude at UTTR and on the helicopter long line, on board the C-17 aircraft during transit from Utah to Texas, on the truck from Arlington Field to JSC, and inside at JSC. All times are Mountain Daylight Time (MDT), which was local time for UTTR. These environments are shown in **Figs. S8-2 to S8-9.**

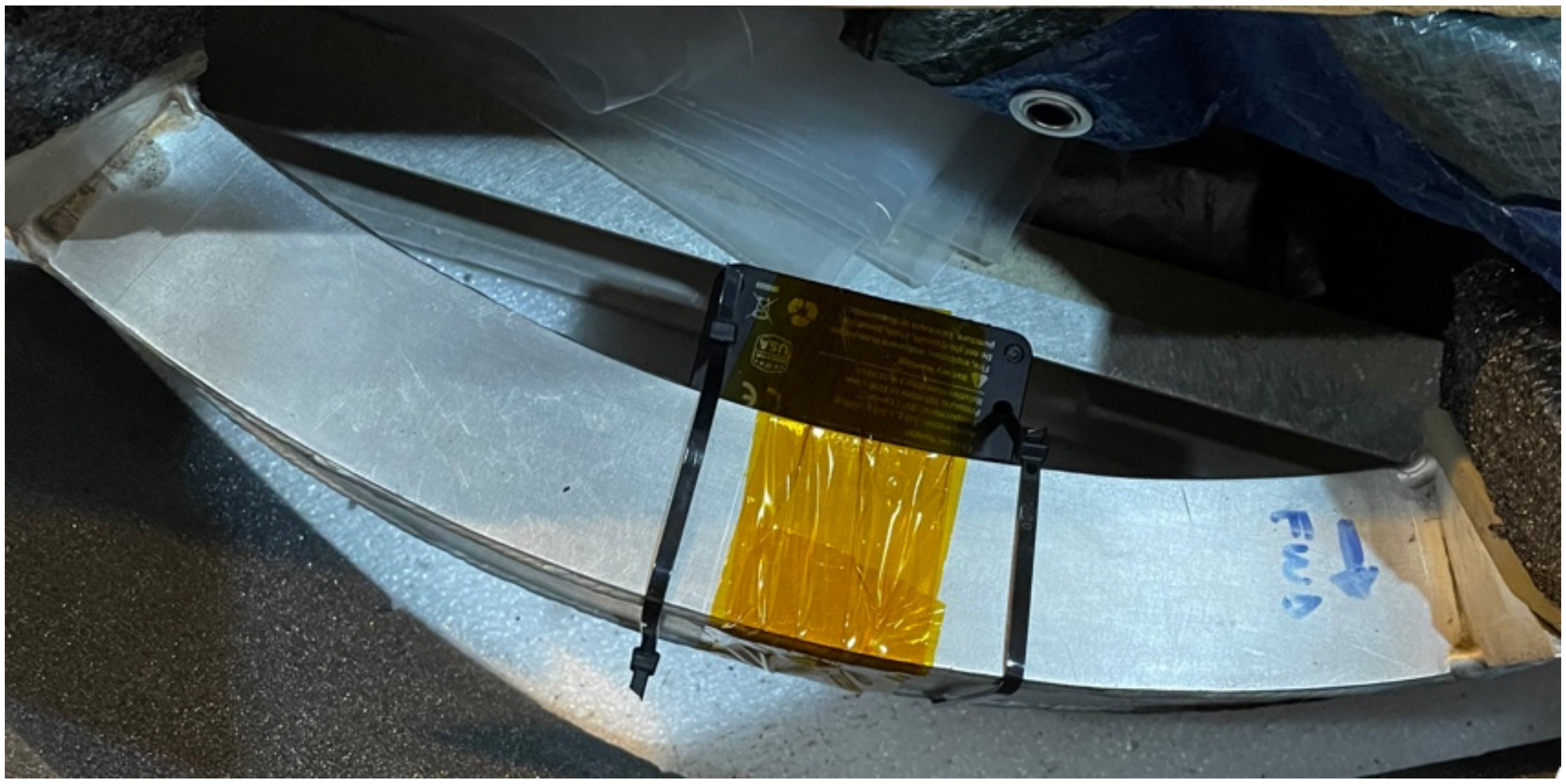

**Fig. S8-2** The data logger on the SRC handling fixture.

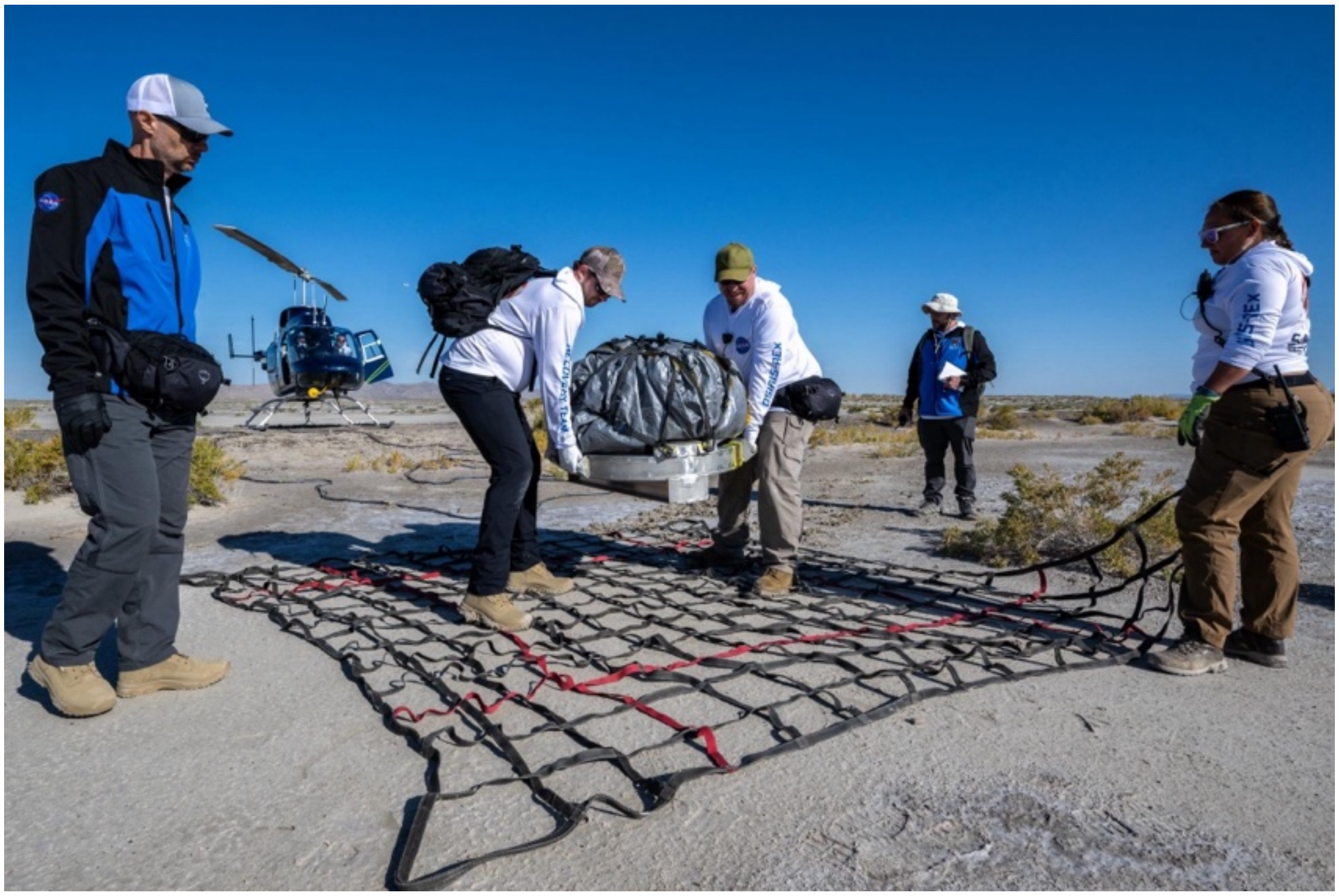

**Fig. S8-3** At the landing site, the SRC double wrapped in PTFE, a plastic tarpaulin, then attached to the handling fixture being is being placed on the netting to be transported by helicopter to the temporary cleanroom at the UTTR. Image credit: NASA/Keegan Barber, https://www.flickr.com/photos/gsfc/albums/72177720310727975/

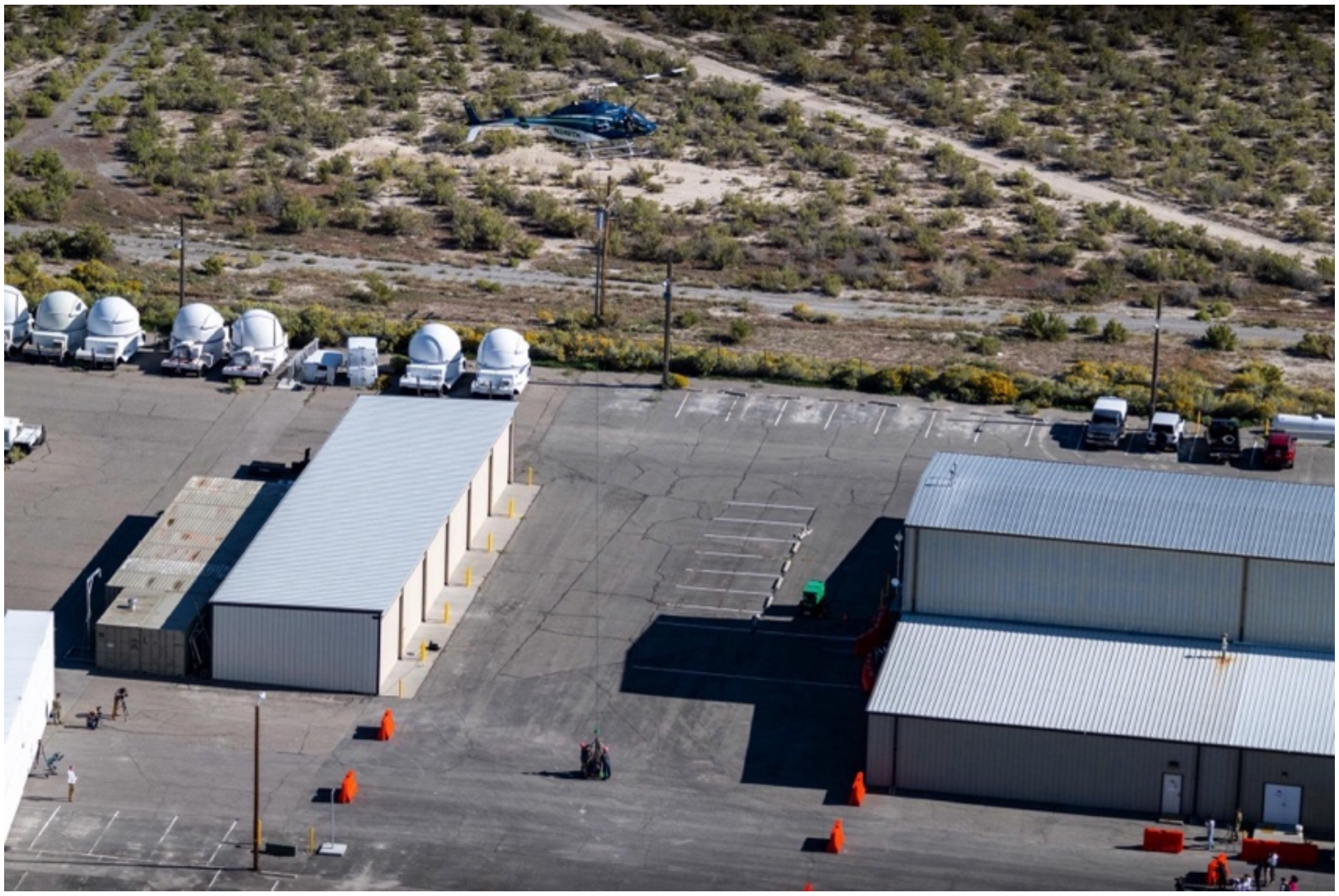

**Fig. S8-4** The helicopter delivering the SRC on longline to the temporary cleanroom inside Dugway building 1012, Avery Complex, Michael Army Airfield at UTTR, on the right. Image credit: NASA/Keegan Barber, https://www.flickr.com/photos/gsfc/albums/72177720310727975/

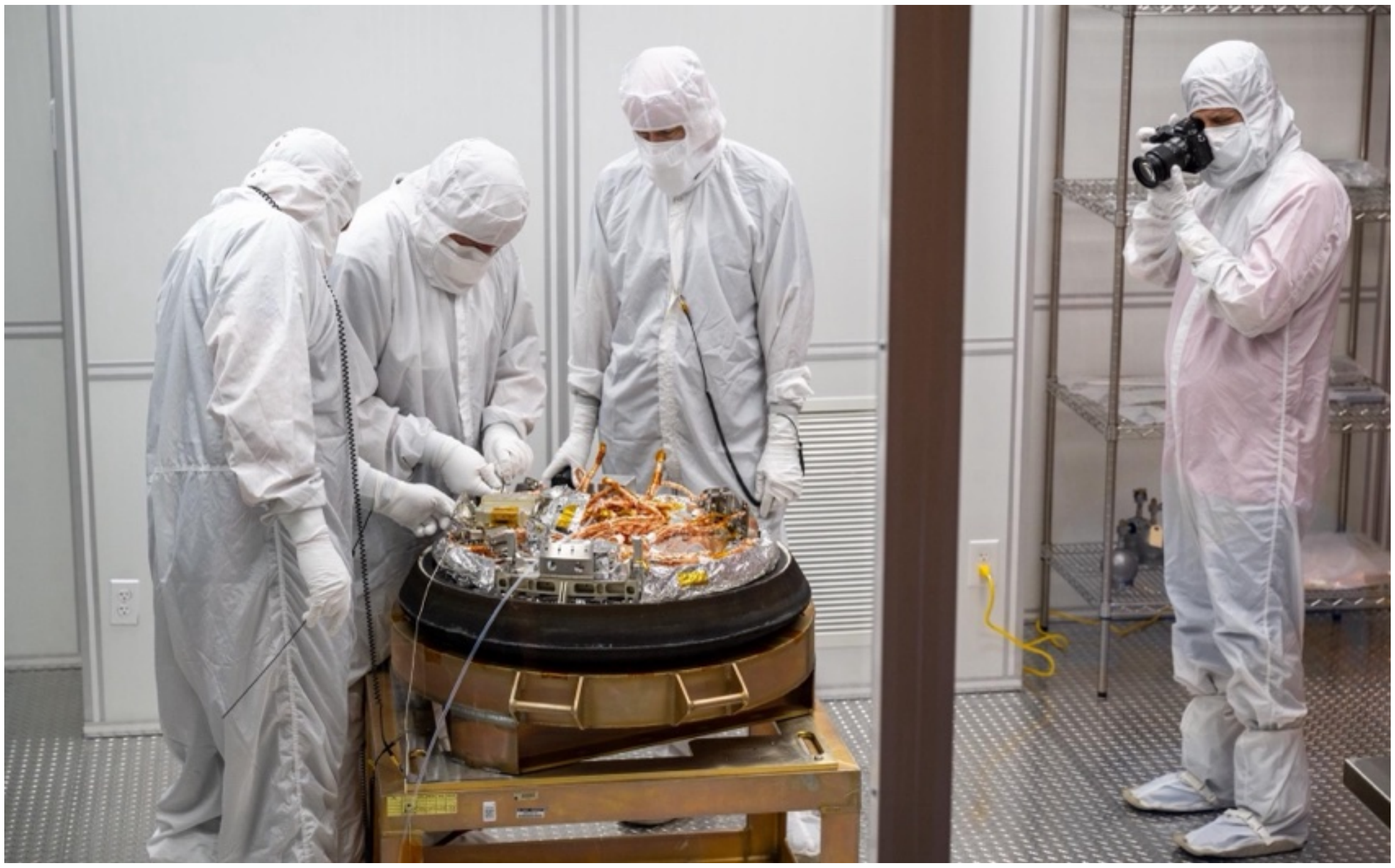

**Fig. S8-5** The SRC being partially disassembled in the UTTR temporary cleanroom. Image credit: NASA/Keegan Barber, https://www.flickr.com/photos/gsfc/albums/72177720310727975/

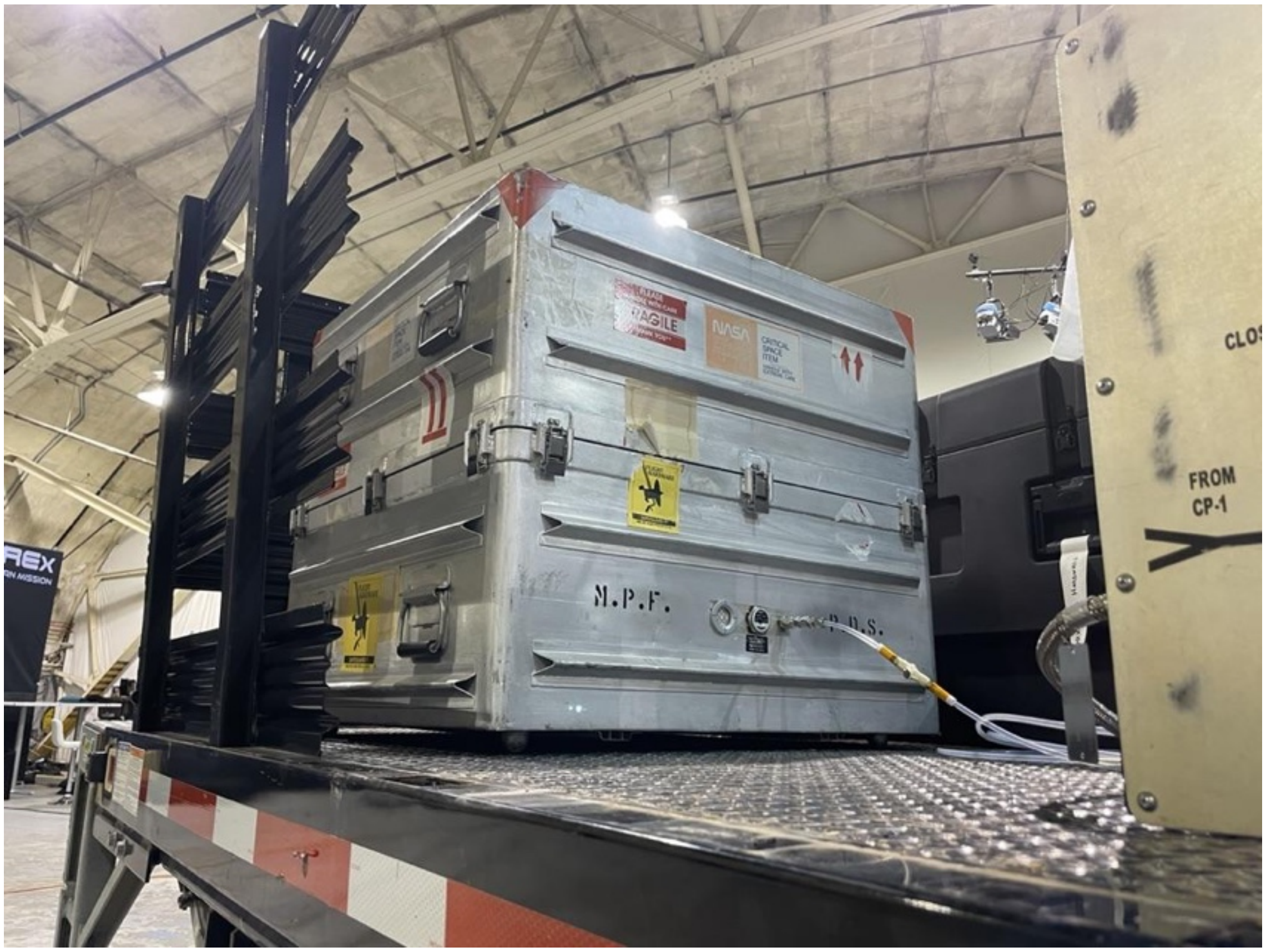

**Fig. S8-6** After partial disassembly, the sample canister was wrapped in PTFE and kept under $GN_2$ purge inside a shipping container. The heatshield and backshell were stored in separate containers.

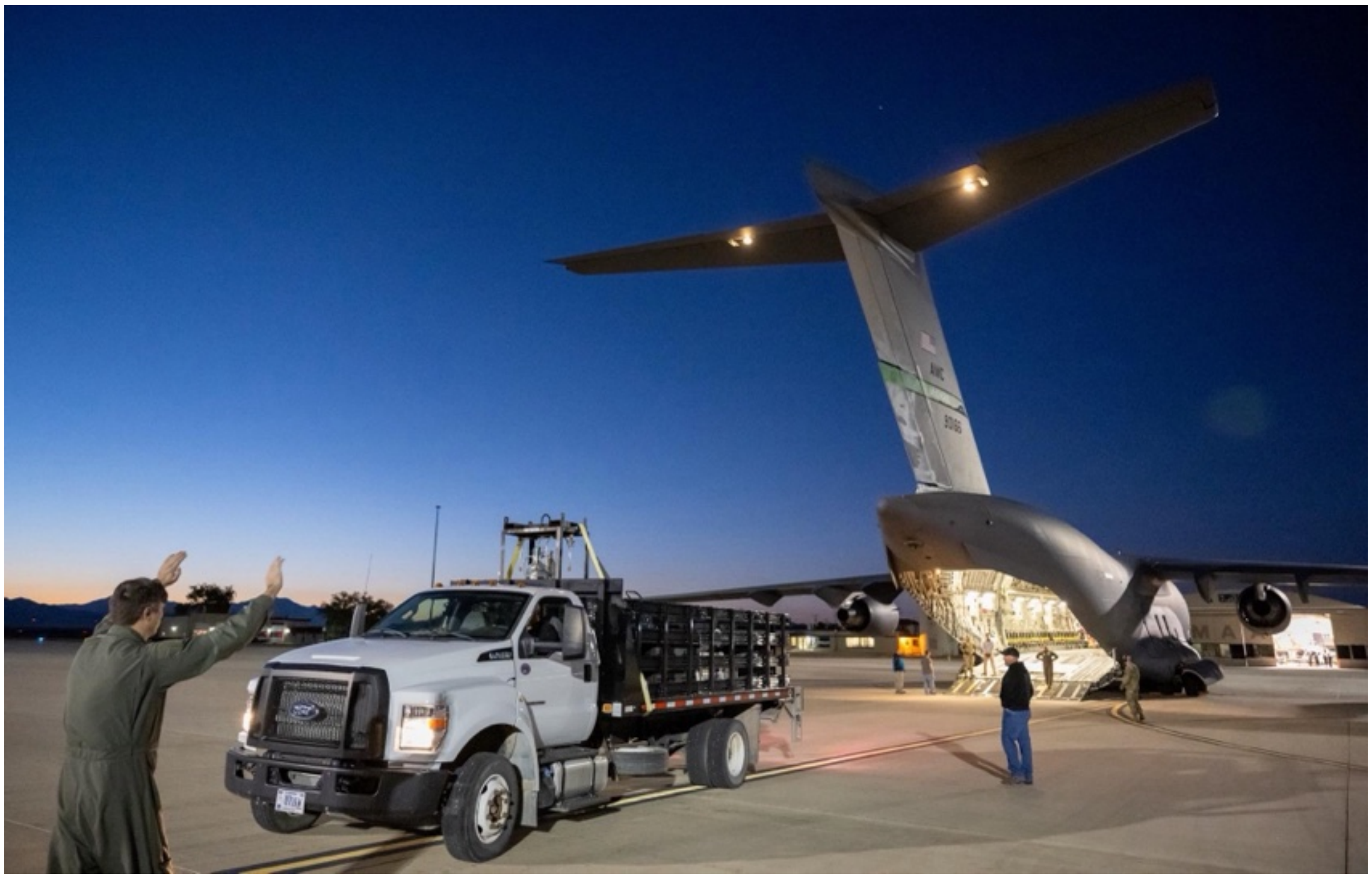

**Fig. S8-7** The sample canister and other hardware being loaded on a C-17 at the UTTR to fly to Houston, Texas. Image credit: NASA/Keegan Barber, https://www.flickr.com/photos/gsfc/albums/72177720310727975/

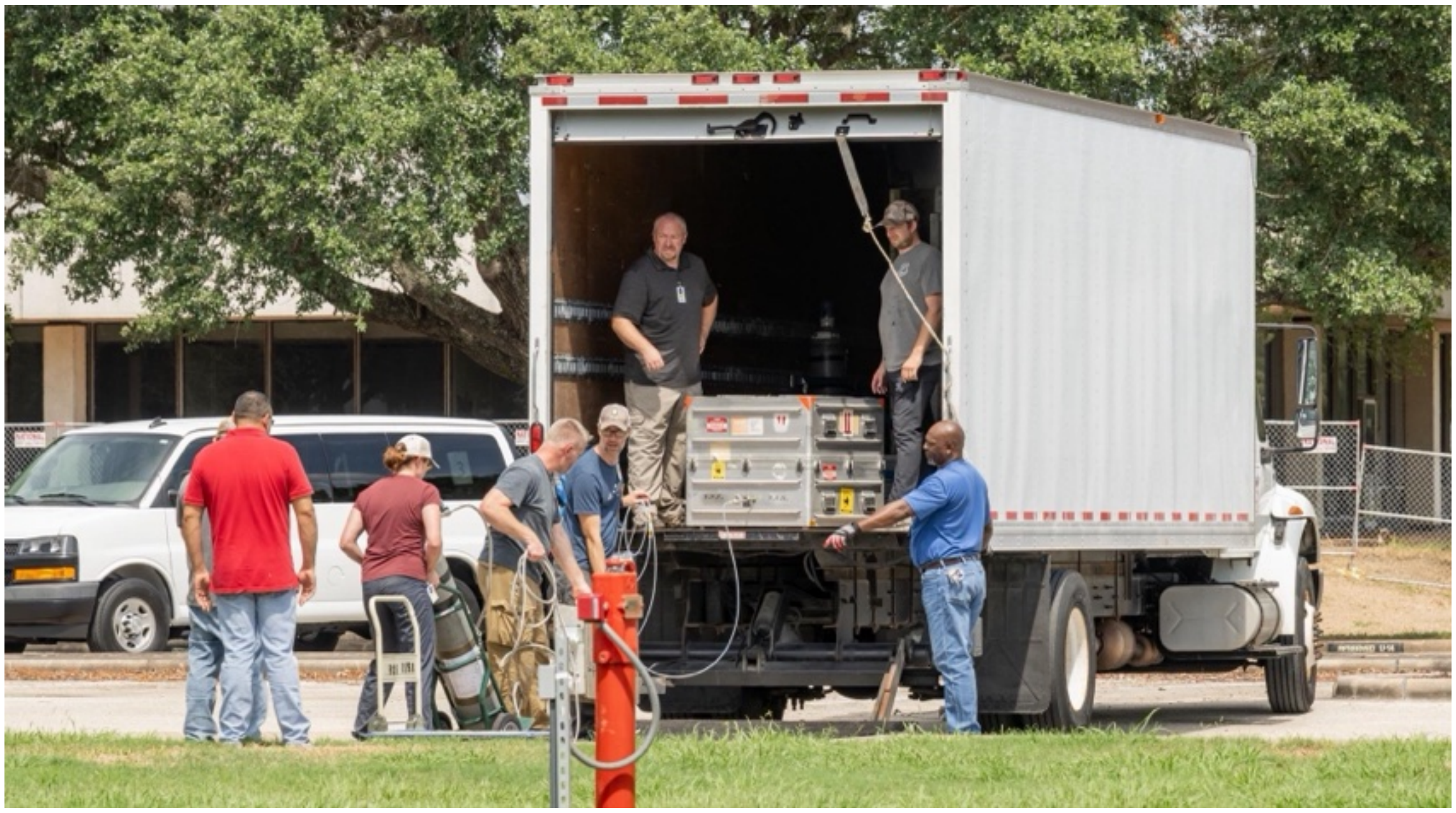

**Fig. S8-8** The sample still under purge being taken from the van to the cleanroom at JSC. Image credit: NASA/Keegan Barber, https://www.flickr.com/photos/gsfc/albums/72177720310727975/

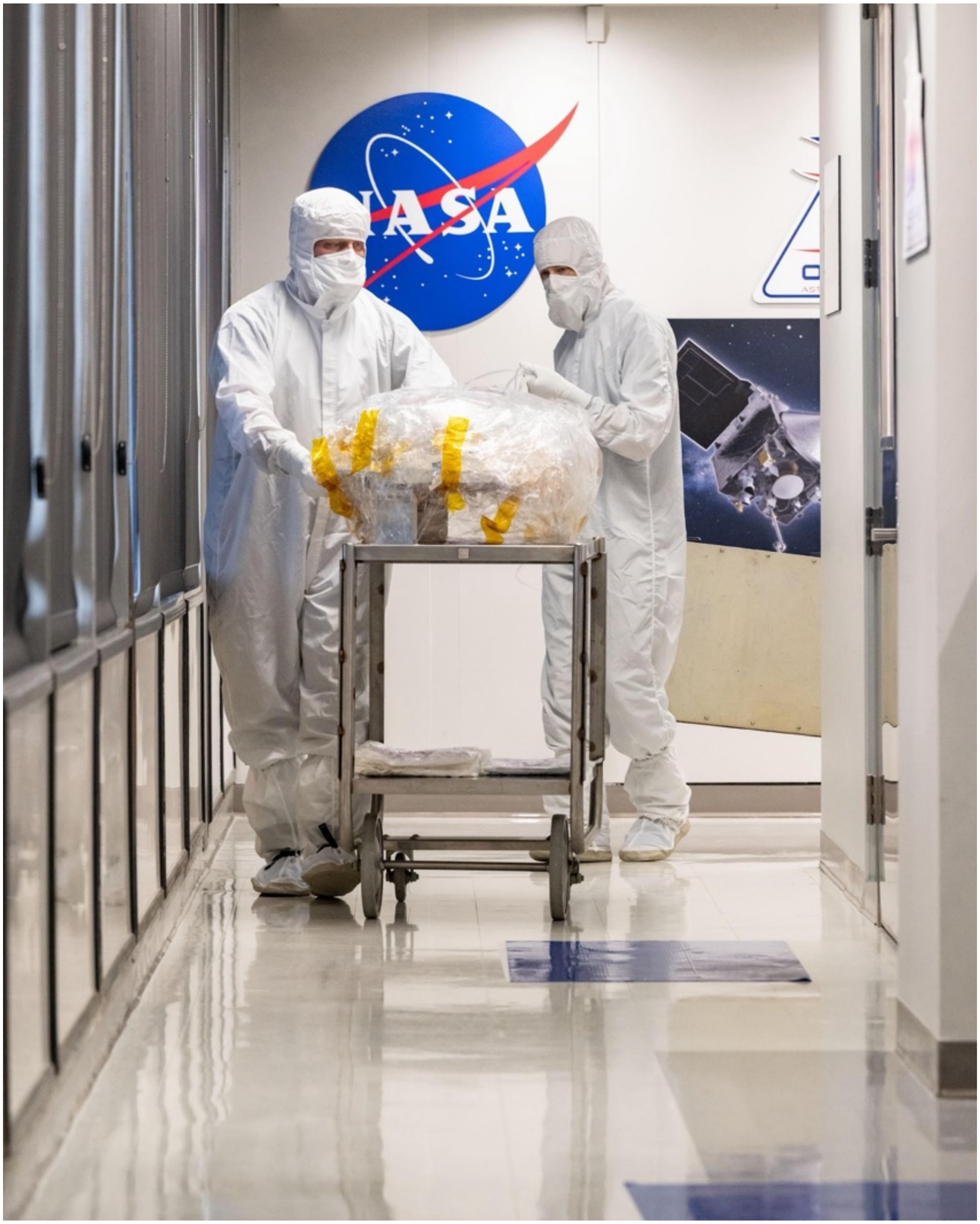

**Fig. S8-9** Since the latches had been removed at the UTTR, the wrapped sample canister was removed from the shipping container in the pedestrian corridor and taken directly to the permanent cleanroom at JSC without needing to break $GN_2$ purge. Image credit: NASA/Keegan Barber, https://www.flickr.com/photos/gsfc/albums/72177720310727975/

## **Supplement 9:** Methods for the analysis of Bennu samples and flight hardware

Most contamination testing was an outcome of the published analyses of Bennu samples (e.g. **Section 4**). However, there were a few sample analyses that were performed solely in support of contamination testing in addition to the analysis of spacecraft hardware from the materials archive conducted in preparation for sample return reported in **Supplement 6**.

**Flight hardware imaging**

The first identification of suspected contaminants came from visual inspection of the sample interior of the sample canister. Photography of the hardware and sample described in this work was performed at JSC from the Advanced Imaging and Visualization of Astromaterials (AIVA) system (JSC 2024; Lauretta & Connolly et al. 2024) Additional photography of the exterior of the sample canister and the witness plates was made with a Nikon Z7 II full-frame mirrorless camera.

**Suspected contaminant particle and sample mineral and elemental analysis**

Mineral and elemental distributions of Bennu particles and potential contaminants were analyzed by scanning electron microscopy with energy-dispersive X-ray spectroscopy (SEM/EDX) and low-angle backscatter electron (LABE) imaging.

A JEOL 7600 FESEM equipped with an Oxford Ultim Max 170 EDX spectrometer via the methods in Lauretta & Connolly et al. (2024) was used for the following analyses described in this work:

- OREX-220001-1 which contained particles exterior to the sample canister collected at UTTR.
- OREX-590073-0 which is an anomalous particle found in Bennu sample OREX-800150-6.
- OREX-800055-1 which is a polished section of a Bennu particle.

A Thermo Fisher FEI Quattro Analytical FE-SEM equipped with two opposing Thermo Fisher UltraDry 60 $mm^2$ EDX via the methods in McCoy & Russell (2025) at Smithsonian Institution National Museum of Natural History (NMNH) was used to investigate the origin of the fiber contamination in avionics deck aggregate samples described in this work:

- OREX-501030-0, which is an avionics aggregate sample containing fibers
- OR-MA-0019,72 which is a filtrete sample from the materials archive
- OR-MA-0168, which is SRC blanket sample from the materials archive
- OR-MA-0172, which is SRC blanket batting sample from the materials archive
- E50067, which is damaged SRC Kapton blanket from NASA's Genesis mission

**Alkanes, alkenes, aromatics, ketones, and alcohols**

Aggregate Bennu sample OREX-800128-104, was used for a direct comparison with the witness plate results obtained for alkanes, alkenes, aromatics, ketones, and alcohols by pyrolysis Gas Chromatography–Mass Spectrometry (GC-MS).

A Frontier Laboratories Mult-Shot Pyrolyzer (EGA/PY-3030D) feeding a Thermo Scientific TRACE 1610 gas chromatograph coupled to a Thermo Scientific 9610 triple-quadrupole mass spectrometer system at GSFC. A sub-mg fraction OREX-800128-104 (5.36 mg) was loaded into pyrolysis cups which was dropped into the heated pyrolysis furnace set to 600°C and held for 20 s via the method of Mojarro et al. (2025). The MS operated in simultaneous full scan (*m/z* 50–500) and multiple reaction monitoring (MRM) mode to both broadly search for compounds and identify target compounds via precursor ion → product ion transitions. MRM included transitions targeting described by Mojarro et al. (2025).

**Curation glovebox exhaust sampling**

Gas samples were collected from the JSC sample handling glovebox exhaust via trap consisting of passive adsorbent materials in a custom manifold downstream from the glovebox (**Supplement 6**). The trap was designed to minimize impact to curation operations and to exclude the possibility of contamination of the OSIRIS-REx sample at the expense of permitting a quantitative sample collection. As such, the trap was placed at a T-junction downstream from the glovebox (**Fig. S6-3**) in a position that was outside of the laminar flow of exhaust.

The trap was loaded with a range of Restek solid phase microextraction (SPME) fibers, Radiello cartridges, and adsorbent pellets designed to trap volatiles entrained in the $GN_2$ glovebox exhaust. Hydrocarbons were trapped with Radiello 145 to collect small aromatics, 20-45 mesh Supelco Carboxen 569 to collect $C_2$-$C_5$ volatiles. Polar species from 30-300 Da were trapped with five different SPME fibers made of polydimethylsiloxane (PDMS) and/or divinylbenzene (DVB) SPME fibers, carboxen, and/or polyacrylate. Radiello 165, Radiello 168, Radiello 170 were used to collect carbonyl-substituted hydrocarbons (aldehydes), $NH_3$, and $H_2S$, respectively. Traps were passively exposed to exhaust for several weeks or months of operation and periodically replaced with refreshed adsorbent materials. Data presented here is from the SPME fibers included in the first two collections: OREX-590040-0 from 10 October 2023 to 7 November 2023 and OREX-590054-0 from 6 December 2023 to 15 February 2024.

Samples were analyzed at GSFC via a Thermo Fisher Scientific Trace 1610 GC coupled to an Orbitrap Exploris 240 Mass Spectrometer. The instrument was equipped with an Agilent J&W PoraBOND Q PT GC Column 25 m column (ID 0.25 mm, film thickness 3 µm). Fibers were desorbed in a splitless (4 mm x 6.3 mm x 78.5 mm Topaz liner) injector inlet. The inlet temperature was specific to the fiber coating, ranging from 245-280 $^{o}$C. All scans were in positive mode; scan range was set to 30-300 m/z and resolution was 60,000. Transfer line temperature was 300 $^{o}$C, source temperature 325 $^{o}$C, He flow rate was 2 mL/min. The temperature program started at 50 with a one min hold, then a 5 min ramp to 150, followed by a 3 min hold, then a final 25 $^{o}$/min ramp to a 5 min hold at the maximum temperature specific to the fiber coating (ranging from 250-280 °C).

Given the extremely analyte-rich nature of these analyses that could not be replicated the data were treated using the GC Decon algorithm that compares co-eluted peaks and fragments to assign scores for the more likely parent molecule identities. Data were analyzed within the context of a “quiescent blank,” a collector that was installed on the JSC sample handling glovebox exhaust systems one month prior to Bennu arrival. For species that were present both in the asteroid sample and the quiescent blank a custom script was applied that measured the number of standard deviations from the mean of the quiescent blank and provided a score that indicated abundance increase for an analyte arising from the glovebox exhaust. Further, a set of SPME fibers from a fully assembled collector were analyzed and compared with the quiescent blank to understand analytes that may arise from the glovebox itself versus cross-talk between the different adsorbents in the trap.

Compound-specific isotope analyses of toluene were conducted using GC-MS/IRMS. dichloromethane (DCM) and methanol (MeOH) extracts and procedural blanks were analyzed without derivatization on a Thermo Trace gas chromatograph coupled to a Thermo DSQ II electron-impact quadrupole mass spectrometer for compound identification and a Thermo MAT 253 isotope ratio mass spectrometer for isotope measurements. Hydrogen isotope ratios (δD) were determined following the methodology of Sandford et al. (2026), whereas carbon isotope ratios ($\delta^{13}C$) were measured using identical chromatographic conditions. A sample of aggregate Bennu material (unsorted particles <5 mm in size) (OREX-800128-104, 6.425 g) had been powdered and homogenized to permit multiple parallel analyses (Ryan et al. 2026). Split OREX-800107-128 (198.6 mg) was used to measure the stable isotopic ratios of toluene via GC-MS/IRMS.

**Data table**

**Note:** Pyrolysis GC-MS data is from OREX-800128-104 is presented in **Supplement 11** with the witness plate and control data that was collected in parallel.

**Table S9-1** Selected compounds detected in the JSC sample handling glovebox exhaust with traps OREX-590040-0 (28 days, 10 October 2023 to 7 November 2023) and OREX-590054-0 (71 days, 6 December 2023 to 15 February 2024) compared with a quiescent blank of the glovebox collected during month before the sample canister arrived at JSC and a procedural blank of the assembled trap without any exposure but at this time the abundances cannot be quantitated, so only the presence of an analyte is noted. The relative abundances are given in arbitrary instrument counts in an integrated chromatographic peak. The complex fluid dynamics of the exhaust flow to the trap, different exposure times, and different absorptivity of the SPME fibers for different compounds contribute to the lack of quantitation. However, the ratio of each sample to the quiescent blank suggests that the Bennu sample was responsible for the toluene and perhaps isopropanol outgassing. Abundances (nmol/g) measured in Bennu aggregate OREX-800107-128 for hydrocarbons (Aponte et al. 2026) and acetone in Bennu aggregate OREX-800107-127 (Baczynski & Mcintosh et al. 2026) are shown for comparison.

| Analyte | Bennu Aggregates | Procedural Blank | Quiescent Blank | Trap OREX-590040-0 | Trap OREX-590054-0 | OREX-590040-0/ Blank | OREX-590054-0/ Blank |
|---|---|---|---|---|---|---|---|
| Acetone | 36.02 | Yes | $2.03 \times 10^8$ | $1.77 \times 10^8$ | $1.64 \times 10^8$ | 0.9 | 0.8 |
| Benzene | 13.5 | Yes | $2.18 \times 10^9$ | $1.13 \times 10^9$ | $1.16 \times 10^9$ | 0.5 | 0.5 |
| Isopropanol | | Yes | $1.81 \times 10^8$ | $3.27 \times 10^8$ | $7.69 \times 10^8$ | 2 | 4 |
| Toluene | 114.5 | Yes | $2.83 \times 10^8$ | $1.15 \times 10^{10}$ | $1.55 \times 10^9$ | 40 | 5 |
| Ethylbenzene | 1.2 [a] | Tentative [a] | $2.29 \times 10^7$ | $5.68 \times 10^7$ | [b] | 2 | |
| o-Xylene | [a] | Tentative [a] | $1.28 \times 10^7$ | $1.81 \times 10^8$ | $6.78 \times 10^7$ | 10 | |
| p-Xylene | [a] | Tentative [a] | $2.65 \times 10^7$ [b] | $1.16 \times 10^8$ [b] | [b] | 4 | |
| m-Xylene | [a] | Tentative [a] | $1.14 \times 10^{10}$ [b] | $2.61 \times 10^9$ [b] | [b] | 0.2 | |
| Naphthalene | 38.1 | Tentative | $1.07 \times 10^{10}$ | $1.23 \times 10^{10}$ | $7.30 \times 10^9$ | 1 | 0.7 |
| Methylnaphthalenes | 55.9 | Tentative | $1.68 \times 10^9$ | $1.38 \times 10^9$ | $4.19 \times 10^8$ | 0.8 | 0.2 |

[a] The four isomers of dialkylbenzene were not resolved so the combined value is shown.
[b] Coelutions made compound identification uncertain.

## **Supplement 10:** Flight witness plates, shims, and screens

**Table S-10.** Flight witness plates and associated hardware. The shaded *italicized* rows were analyzed in this study. The events which triggered the exposure or covering of a given witness plate are: 1) the initial installation in the Assembly Testing and Launch Operations (ATLO) cleanroom at Lockheed Martin Space (LMS), 2) after thermal-vacuum testing (TVAC) during ATLO, 3) the releasing of the TAGSAM arm latch that held the TAGSAM head inside the launch container during Bennu approach, 4) when the SRC was opened after the Touch and Go (TAG) sampling at Bennu, 5) when the TAGSAM head was severed from the TAGSAM arm, and 6) when the sample canister was disassembled at JSC. See also **Fig. 4**.

| Witness Description | Material | Length (mm) | Width (mm) | Thickness (mm) | Screen | Area ($cm^2$) | Curation ID Number | Exposure Start | Exposure End | Exposure Time (da) | Triggering Event |
|---|---|---|---|---|---|---|---|---|---|---|---|
| **TAGSAM (Circles)** | | | | | | | | | | | |
| Always Open | UHP aluminum | 30.16 | 30.16 | 1.4 | Yes | 7.1 | OREX-331001-0 | 18-Apr-16 | 02-Oct-23 | 2723 | Installation of TAGSAM Head @ LMS to Disassembly @ JSC |
| Always Open | Sapphire | 30.16 | 30.16 | 1.4 | Yes | 7.1 | OREX-341001-0 | 18-Apr-16 | 02-Oct-23 | 2723 | Installation of TAGSAM Head @ LMS to Disassembly @ JSC |
| Always Open | UHP aluminum | 30.16 | 30.16 | 1.4 | Yes | 7.1 | OREX-311001-0 | 18-Apr-16 | 02-Oct-23 | 2723 | Installation of TAGSAM Head @ LMS to Disassembly @ JSC |
| Always Open | Sapphire | 30.16 | 30.16 | 1.4 | Yes | 7.1 | OREX-321001-0 | 18-Apr-16 | 02-Oct-23 | 2723 | Installation of TAGSAM Head @ LMS to Disassembly @ JSC |
| Always Open Screen | 316L stainless steel | 30.16 | 30.16 | 0.023 | n/a | 7.1 | OREX-321001-0 | 18-Apr-16 | 02-Oct-23 | 2723 | Installation of TAGSAM Head @ LMS to Disassembly @ JSC |
| Always Open Screen | 316L stainless steel | 30.16 | 30.16 | 0.023 | n/a | 7.1 | OREX-316001-0 | 18-Apr-16 | 02-Oct-23 | 2723 | Installation of TAGSAM Head @ LMS to Disassembly @ JSC |
| Always Open Shims | 1145-0 aluminum | 30.16 | 30.16 | 0.13 | n/a | 9.1 | OREX-337001-0 | 18-Apr-16 | 02-Oct-23 | 2723 | Installation of TAGSAM Head @ LMS to Disassembly @ JSC |
| Always Open Shims | 1145-0 aluminum | 30.16 | 30.16 | 0.13 | n/a | 9.1 | OREX-337002-0 | 18-Apr-16 | 02-Oct-23 | 2723 | Installation of TAGSAM Head @ LMS to Disassembly @ JSC |
| Always Open Shims | 1145-0 aluminum | 30.16 | 30.16 | 0.13 | n/a | 9.1 | OREX-347001-0 | 18-Apr-16 | 02-Oct-23 | 2723 | Installation of TAGSAM Head @ LMS to Disassembly @ JSC |
| Always Open Shims | 1145-0 aluminum | 30.16 | 30.16 | 0.13 | n/a | 9.1 | OREX-347002-0 | 18-Apr-16 | 02-Oct-23 | 2723 | Installation of TAGSAM Head @ LMS to Disassembly @ JSC |
| Post-Stow | UHP aluminum | 25.4 | 25.4 | 0.51 | No | 5.1 | OREX-355001-0 | 28-Oct-20 | 03-Oct-23 | 1070 | Sever TAGSAM head @ Bennu to Disassembly @ JSC |
| Post-Stow | UHP aluminum | 25.4 | 25.4 | 0.51 | No | 5.1 | OREX-355002-0 | 28-Oct-20 | 03-Oct-23 | 1070 | Sever TAGSAM head @ Bennu to Disassembly @ JSC |
| Post-Stow | UHP aluminum | 25.4 | 25.4 | 0.51 | No | 5.1 | OREX-355003-0 | 28-Oct-20 | 03-Oct-23 | 1070 | Sever TAGSAM head @ Bennu to Disassembly @ JSC |
| Post-Stow | Sapphire | 25.4 | 25.4 | 0.51 | No | 5.1 | OREX-365001-0 | 28-Oct-20 | 03-Oct-23 | 1070 | Sever TAGSAM head @ Bennu to Disassembly @ JSC |
| Post-Stow | Sapphire | 25.4 | 25.4 | 0.51 | No | 5.1 | OREX-365002-0 | 28-Oct-20 | 03-Oct-23 | 1070 | Sever TAGSAM head @ Bennu to Disassembly @ JSC |
| Post-Stow | Sapphire | 25.4 | 25.4 | 0.51 | No | 5.1 | OREX-365003-0 | 28-Oct-20 | 03-Oct-23 | 1070 | Sever TAGSAM head @ Bennu to Disassembly @ JSC |
| Pre-Stow | UHP aluminum | 30.16 | 30.16 | 0.9 | No | 7.1 | OREX-334001-0 | 18-Apr-16 | 19-Oct-18 | 914 | Installation of TAGSAM Head @ LMS to Arm latch release @ Bennu |
| Pre-Stow | Sapphire | 30.16 | 30.16 | 0.9 | No | 7.1 | OREX-344001-0 | 18-Apr-16 | 19-Oct-18 | 914 | Installation of TAGSAM Head @ LMS to Arm latch release @ Bennu |
| Pre-Stow | UHP aluminum | 30.16 | 30.16 | 0.9 | No | 7.1 | OREX-314001-0 | 18-Apr-16 | 19-Oct-18 | 914 | Installation of TAGSAM Head @ LMS to Arm latch release @ Bennu |
| *Pre-Stow* | *Sapphire* | *30.16* | *30.16* | *0.9* | *No* | *7.1* | *OREX-324001-0* | *18-Apr-16* | *19-Oct-18* | *914* | *Installation of TAGSAM Head @ LMS to Arm latch release @ Bennu* |
| **Sample Canister (Rectangles)** | | | | | | | | | | | |
| Always Open | UHP aluminum | 30.16 | 20.06 | 0.76 | Yes | 6.1 | OREX-371001-0 | 16-Mar-16 | 04-Apr-24 | 2941 | Post-TVAC @ LMS to Disassembly @ JSC |
| Always Open | UHP aluminum | 30.16 | 20.06 | 0.76 | Yes | 6.1 | OREX-371002-0 | 16-Mar-16 | 04-Apr-24 | 2941 | Post-TVAC @ LMS to Disassembly @ JSC |
| Always Open | Sapphire | 30.16 | 20.06 | 0.76 | Yes | 6.1 | OREX-381001-0 | 16-Mar-16 | 04-Apr-24 | 2941 | Post-TVAC @ LMS to Disassembly @ JSC |
| Always Open | Sapphire | 30.16 | 20.06 | 0.76 | Yes | 6.1 | OREX-381002-0 | 16-Mar-16 | 04-Apr-24 | 2941 | Post-TVAC @ LMS to Disassembly @ JSC |
| Always Open Screen | 316L stainless steel | 30.16 | 20.06 | 0.023 | n/a | 6.1 | OREX-376003-0 | 16-Mar-16 | 04-Apr-24 | 2941 | Post-TVAC @ LMS to Disassembly @ JSC |
| Always Open Screen | 316L stainless steel | 30.16 | 20.06 | 0.023 | n/a | 6.1 | OREX-386003-0 | 16-Mar-16 | 04-Apr-24 | 2941 | Post-TVAC @ LMS to Disassembly @ JSC |
| Always Open Shims | 1145-0 aluminum | 30.16 | 20.06 | 0.13 | n/a | 6.1 | OREX-377001-0 | 16-Mar-16 | 04-Apr-24 | 2941 | Post-TVAC @ LMS to Disassembly @ JSC |
| Always Open Shims | 1145-0 aluminum | 30.16 | 20.06 | 0.13 | n/a | 6.1 | OREX-377003-0- | 16-Mar-16 | 04-Apr-24 | 2941 | Post-TVAC @ LMS to Disassembly @ JSC |
| Always Open Shims | 1145-0 aluminum | 30.16 | 20.06 | 0.13 | n/a | 6.1 | OREX-387001-0 | 16-Mar-16 | 04-Apr-24 | 2941 | Post-TVAC @ LMS to Disassembly @ JSC |
| Always Open Shims | 1145-0 aluminum | 30.16 | 20.06 | 0.13 | n/a | 6.1 | OREX-387003-0 | 16-Mar-16 | 04-Apr-24 | 2941 | Post-TVAC @ LMS to Disassembly @ JSC |
| Always Open Shims | 1145-0 aluminum | 30.16 | 20.06 | 0.13 | n/a | 6.1 | OREX-377002-0 | 16-Mar-16 | 04-Apr-24 | 2941 | Post-TVAC @ LMS to Disassembly @ JSC |
| Always Open Shims | 1145-0 aluminum | 30.16 | 20.06 | 0.13 | n/a | 6.1 | OREX-387002-0 | 16-Mar-16 | 04-Apr-24 | 2941 | Post-TVAC @ LMS to Disassembly @ JSC |
| Post-Stow | UHP aluminum | 30.16 | 20.06 | 1.14 | Yes | 6.1 | OREX-373001-0 | 27-Oct-20 | 04-Apr-24 | 1255 | Open SRC after TAG @ Bennu to Disassembly @ JSC |
| Post-Stow | UHP aluminum | 30.16 | 20.06 | 1.14 | Yes | 6.1 | OREX-373002-0 | 27-Oct-20 | 04-Apr-24 | 1255 | Open SRC after TAG @ Bennu to Disassembly @ JSC |
| Post-Stow | Sapphire | 30.16 | 20.06 | 1.14 | Yes | 6.1 | OREX-383001-0 | 27-Oct-20 | 04-Apr-24 | 1255 | Open SRC after TAG @ Bennu to Disassembly @ JSC |
| Post-Stow | Sapphire | 30.16 | 20.06 | 1.14 | Yes | 6.1 | OREX-383002-0 | 27-Oct-20 | 04-Apr-24 | 1255 | Open SRC after TAG @ Bennu to Disassembly @ JSC |
| Post-Stow Screen | 316L stainless steel | 30.16 | 20.06 | 0.023 | n/a | 6.1 | OREX-376001-0 | 27-Oct-20 | 04-Apr-24 | 1255 | Open SRC after TAG @ Bennu to Disassembly @ JSC |
| Post-Stow Screen | 316L stainless steel | 30.16 | 20.06 | 0.023 | n/a | 6.1 | OREX-386001-0 | 27-Oct-20 | 04-Apr-24 | 1255 | Open SRC after TAG @ Bennu to Disassembly @ JSC |
| Post-Stow Screen | 316L stainless steel | 30.16 | 20.06 | 0.023 | n/a | 6.1 | OREX-386002-0 | 27-Oct-20 | 04-Apr-24 | 1255 | Open SRC after TAG @ Bennu to Disassembly @ JSC |
| Post-Stow Screen | 316L stainless steel | 30.16 | 20.06 | 0.023 | n/a | 6.1 | OREX-376002-0 | 27-Oct-20 | 04-Apr-24 | 1255 | Open SRC after TAG @ Bennu to Disassembly @ JSC |
| *Pre Stow Shims* | *1145-0 aluminum* | *30.16* | *20.06* | *0.13* | *n/a* | *6.1* | *OREX-377025-0* | *16-Mar-16* | *27-Oct-20* | *1686* | *Post-TVAC @ LMS to Open SRC after TAG @ Bennu* |
| Pre Stow Shims | 1145-0 aluminum | 30.16 | 20.06 | 0.13 | n/a | 6.1 | OREX-387025-0 | 16-Mar-16 | 27-Oct-20 | 1686 | Post-TVAC @ LMS to Open SRC after TAG @ Bennu |
| Pre-Stow | UHP aluminum | 30.16 | 20.06 | 1.65 | No | 6.1 | OREX-372002-0 | 16-Mar-16 | 27-Oct-20 | 1686 | Post-TVAC @ LMS to Open SRC after TAG @ Bennu |
| Pre-Stow | Sapphire | 30.16 | 20.06 | 1.65 | No | 6.1 | OREX-382002-0 | 16-Mar-16 | 27-Oct-20 | 1686 | Post-TVAC @ LMS to Open SRC after TAG @ Bennu |
| Pre-Stow | UHP aluminum | 30.16 | 20.06 | 1.65 | No | 6.1 | OREX-372001-0 | 16-Mar-16 | 27-Oct-20 | 1686 | Post-TVAC @ LMS to Open SRC after TAG @ Bennu |
| Pre-Stow | Sapphire | 30.16 | 20.06 | 1.65 | No | 6.1 | OREX-382001-0 | 16-Mar-16 | 27-Oct-20 | 1686 | Post-TVAC @ LMS to Open SRC after TAG @ Bennu |

## **Supplement 11:** Witness plate analysis methods and data

## Witness sample preparation

Flight high-purity aluminum shim OREX-337025-0 (30.16 mm × 20.06 mm × 0.13 mm thick) was cut into three equal segments (OREX-337025-1, OREX-337025-2, and OREX-337025-3) at JSC (each ~10 × 20 mm) with stainless steel shears in a $GN_2$ glovebox and then each packed loosely with a double layer of ashed aluminum foil in separate sanitary stainless steel bottles from Eagle Stainless sealed with a Viton gasket under $N_2$. A similar 1100 pure aluminum piece (Design Interface, used as rolled) of the same thickness and size was used as a processing control. It was cut into three approximately equal pieces at NASA's Goddard Space Flight Center (GSFC), cleaned by sonication in isopropanol, and ashed overnight in air at 500 °C prior to use. The aluminum controls were processed in parallel with the flight shims. The surface area in square millimeters of a single side of each flight shim and control was calculated from the measured mass of each individual piece multiplied by the single-sided surface area of the uncut piece (30.16 mm × 20.06 mm = 605 $mm^2$) divided by its total mass (219.6 mg) to obtain single-sided surface areas of 227.84 $mm^2$, 176.05 $mm^2$, and 201.12 $mm^2$ for OREX-337025-1, OREX-337025-2, and OREX-337025-3, respectively, and 202.80 $mm^2$, 205.80 $mm^2$, and 196.36 $mm^2$ for the aluminum controls 1, 2, and 3, respectively.

Flight aluminum shim OREX-337025-1 and the corresponding aluminum shim control 1 were allocated for hot water extraction (100 °C for 24 h) and workup via the same methods used for the Bennu samples as discussed in Glavin & Dworkin et al. (2025) followed by gas chromatography–mass spectrometry/isotope ratio mass spectrometry (GC-MS/IRMS) and liquid chromatography–mass spectrometry (LC-MS) analyses (**Fig. S11-1**). The water extract was split first with the removal of 0.35% for ammonia analysis and 4% for amines and protein amino acid analysis. Of the remaining ~95%, ~5% of the water extract was kept in reserve and 90% was split into two equal portions with one of these two portions being dried under vacuum and acid hydrolyzed with 6 M HCl vapor at 150 °C for 3 h. Both the unhydrolyzed and acid hydrolyzed water extracts were desalted using cation-exchange resin (AG50W-X8, 100-200 mesh, hydrogen form) with the water washes from the unhydrolyzed and hydrolyzed fractions allocated for GC-MS analysis of carboxylic acids. The $NH_4OH$ desalting eluants from both unhydrolyzed and acid hydrolyzed water extracts were analyzed for free and total amino acid abundance and enantiomeric measurements using LC-MS. Flight shim OREX-337025-2 and the corresponding aluminum shim control 2 were allocated for non-volatile residue (NVR) analyses by thermal desorption (TD) GC-MS. Flight shim OREX-337025-3 and the corresponding aluminum shim control 3 were extracted in dichloromethane (DCM) and methanol (MeOH) (4:1 v/v) for two-dimensional high-resolution GC-MS (GC×GC-HRMS) analyses of aliphatic and aromatic hydrocarbons.

Flight sapphire witness plate OREX-324001-0 (30.16 mm diameter × 0.9 mm thick) was packed loosely with ashed aluminum foil in a large Eagle Stainless steel container under $GN_2$ at JSC. A control sapphire plate (UniversityWafer) of the same size and thickness as the flight plate, which was frosted on one side like the flight witnesses was processed in parallel. It was cleaned by sonication at GSFC in isopropanol and then ashed overnight at 500 °C. At GSFC, the control plate was scored with an ashed diamond etcher and broken into five pieces with an ashed steel chisel. After the control sapphire witness plate was split, the same procedure was to be conducted on the flight witness plate. However, the diamond etcher broke after the first score of the flight witness plate, creating an uneven split. The remaining cuts of the flight witness plate had to be made by applying force with the chisel. The single-sided surface area of each flight witness and control was calculated from the measured mass of each individual sapphire piece multiplied by the single-sided surface area of the circular sapphire plate (714.41 $mm^2$) divided by the total mass (2549.2 mg) of the plate to obtain the single-sided surface areas of 367.49 $mm^2$, 175.32 $mm^2$, 171.20 $mm^2$, and 0.25 $mm^2$ for OREX-324001-100, OREX-324001-101, OREX-324001-102, and OREX-324001-103, respectively, and 195.38 $mm^2$, 223.80 $mm^2$, 150.45 $mm^2$, and 16.14 $mm^2$ for the sapphire plate controls 0, 1, 2, and 3, respectively.

Sample OREX-324001-100 and control sapphire plate 0 were allocated for water extraction and workup via the methods of Glavin & Dworkin et al. (2025) followed by GC-MS and LC-MS analyses (**Fig. S11-2**). The water extract was split and analyzed as described above for the aluminum witness plate samples and controls. Sample OREX-324001-101 and control sapphire plate 1 were allocated for GC×GC-HRMS analysis. The sample was extracted in DCM and MeOH (4:1 v/v) for untargeted analysis according to the methods of Aponte et al. (2026). Sample OREX-324001-102 and control sapphire plate 2 were allocated for TD-GC-MS. Sample OREX-324001-103 and control sapphire plate 3 were allocated for pyrolysis-GC-MS analyses for alkanes, alkenes, aromatics, ketones, and alcohols. Control sapphire plate 4, which had a single-sided surface area of 127.13 $mm^2$, was held in reserve and not used because there was no analogous fourth fragment of OREX-324001-0 due to the difficulty in splitting.

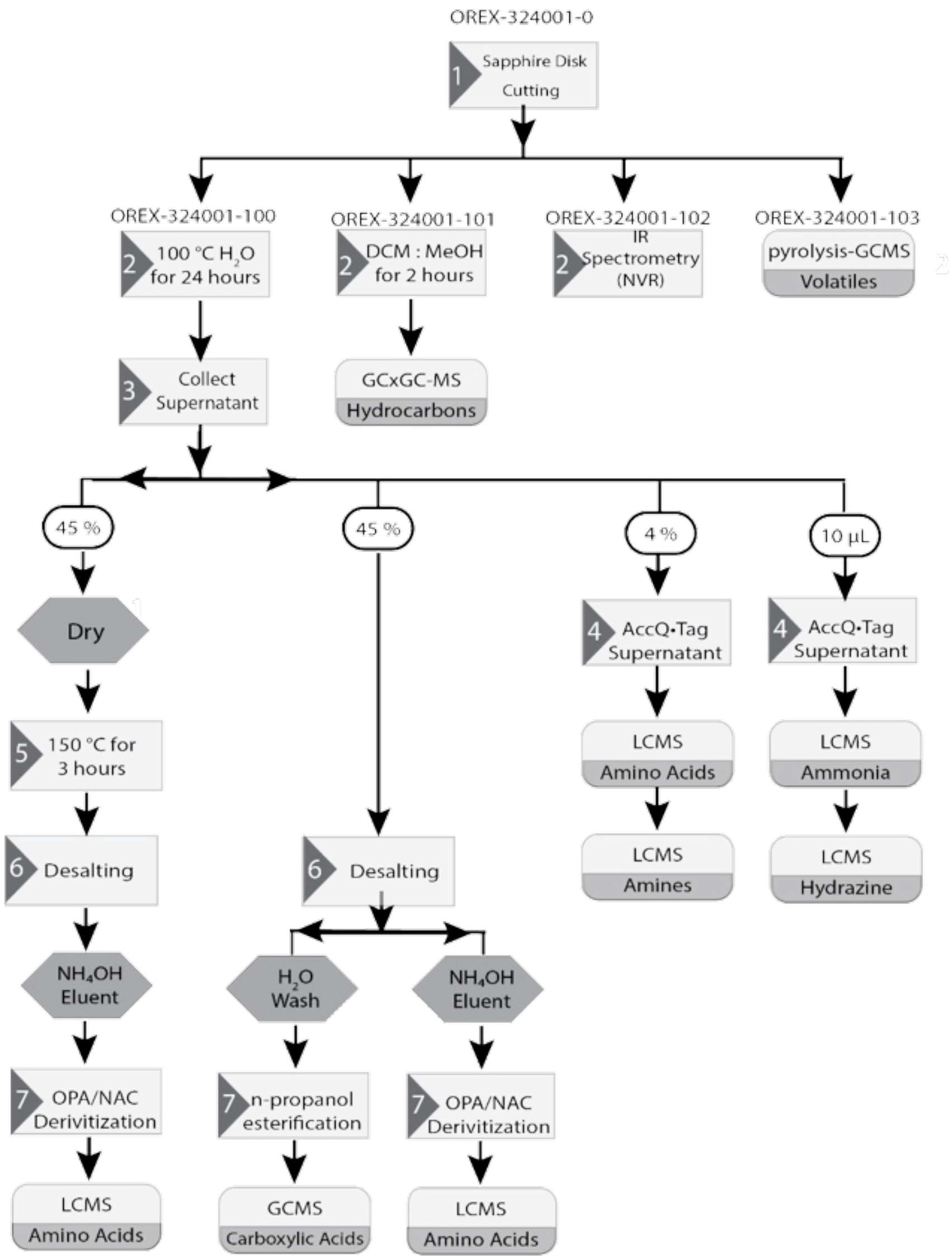

**Fig. S11-1** Analytical extraction and processing workflow for the sapphire witness plate. The protocol details sequential or parallel treatment steps including: (1) the cutting of the OSIRIS-REx sapphire witness plate (OREX-324001-0) via diamond-scribing; (2) parallel sample processing of allocated portions, including hot water extraction (100 °C for 24 h) for OREX-324001-100, DCM:MeOH extraction for 2 h for OREX-324001-101, direct non-volatile residue (NVR) IR spectrometry for a portion (OREX-324001-102), and pyrolysis-GCMS volatile profiling for OREX-324001-103 for a portion; (3) isolation of aqueous supernatant fractions; and (4) direct AccQ·Tag derivatization of 4 % of the supernatant for LC-MS quantification of amino acids, amines, and 10 µL for ammonia, and hydrazine. The primary aqueous supernatant is divided into equal 45% aliquots subjected to evaporation, acid hydrolysis (150 °C for 3 h), cation-exchange desalting chromatography, and subsequent functional-group-specific derivatization (OPA/NAC and BSTFA) to facilitate downstream LC-MS and GC-MS structural and enantiomeric analyses of amino acids and carboxylic acids.

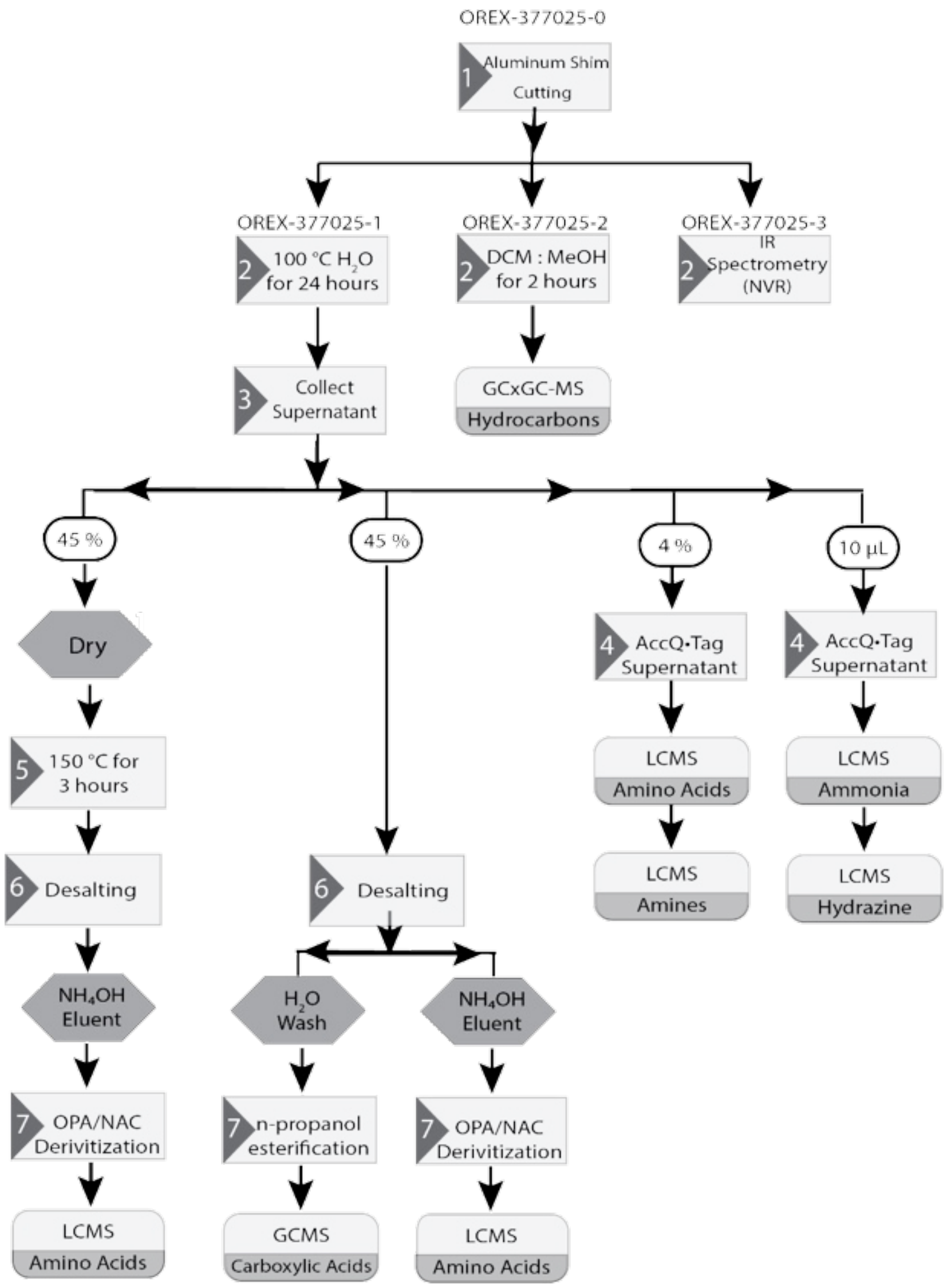

**Fig. S11-2** Analytical extraction and processing workflow for the aluminum shim. The protocol details sequential or parallel treatment steps including: (1) physical sample allocation via shim cutting for OREX-377025-0; (2) extraction at 100 °C for 24 h for OREX-377025-1, and organic solvents (DCM:MeOH for 2 h) for OREX-377025-2, alongside direct non-volatile residue (NVR) IR spectrometry for OREX-377025-3, (3) isolation of aqueous supernatant fractions; and (4) direct AccQ·Tag derivatization of 4 % of the supernatant for LC-MS quantification of amino acids, amines, and 10 µL for ammonia, and hydrazine. The primary aqueous supernatant is divided into equal 45% aliquots subjected to evaporation, acid hydrolysis (150 °C for 3 h), cation-exchange desalting chromatography, and subsequent functional-group-specific derivatization (OPA/NAC and BSTFA) to facilitate downstream LC-MS and GC-MS structural and enantiomeric analyses of amino acids and carboxylic acids.

## Analysis methods

Organic chemical analyses were performed at GSFC. All glass materials and sample handling tools were cleaned by rinsing in ultrapure water (Millipore SAS Milli-Q IQ 7005, 18.2 MΩ·cm, <3 parts per billion total organic carbon), and then ashed overnight at 500 °C in air in a muffle furnace to remove residual organics. Standards and reagents were purchased from Sigma-Aldrich and Thermo Scientific and used without further purification. All sample handling and subsampling occurred within an ISO 5 High-Efficiency Particulate Air (HEPA)–filtered laminar flow hood located in an ISO ~7 white room.

**Carboxylic acids** were analyzed by GC-MS/IRMS with a Thermo Trace 1610 GC coupled to a Thermo 7610 ISQ electron-impact triple-quadrupole mass spectrometer. Water extracts of the witness materials, controls, and procedural blank were alkalized with 2 M NaOH, dried under reduced pressure, and derivatized with an *n*-propanol esterification protocol via the methods of Buckner et al. (2026), modified from the 2-pentanol esterification used by Glavin & Dworkin et al. (2025) for GC-MS analysis of carboxylic acids in Bennu sample OREX-803001-0. Derivatized carboxylic acids were delivered to the GC-MS/IRMS as 3 µL injections made in splitless mode. The Thermo Trace 1610 GC was equipped with serial Rxi-5ms columns (Restek, 30 m, 0.25 mm ID, 0.25 µm) followed by a PoraBOND Q (Agilent, 25 m, 0.25 mm ID, 3 µm) connected with SilTite µ-union connectors (Restek). The GC was coupled to the MS via a transfer line (heated at 20 °C) and to the IRMS via a Thermo IsoLink II GC-combustion oxidation interface. The GC oven program was set at an initial temperature of 50 °C, held for 1 min, then ramped to 300 °C at 10 °C $min^{-1}$ and held for 9 min. The carrier gas was ultrahigh-purity helium (UHP; 5.0 grade) flowed at a constant rate of 1.4 mL $min^{-1}$, and the injector was heated at 250 °C. The MS used full scan mode with mass-to-charge ratio (*m/z*) ranging from 38 to 400 amu, the ionization source was heated at 250 °C at 70 eV in electron impact mode, and the MS filament was turned on 12 minutes after injection. Carboxylic acid abundances of all analytes were below the limit of detection with the IRMS (50 mV). GC-MS data was analyzed with Xcalibur software (Thermo Scientific), and carboxylic acids were identified and quantified by comparison to reference standards and three-point calibration curves (Aponte et al, 2019; Buckner et al. 2026)

**Aliphatic and aromatic hydrocarbons** were analyzed by GC×GC-HRMS using an Agilent 7890B gas chromatograph coupled to a LECO Pegasus HRT+ 4D time-of-flight mass spectrometer (ToF-MS) via the methods of Aponte et al. (2026).

**Alkanes, alkenes, aromatics, ketones, and alcohols** were measured by pyrolysis GC-MS via the methods described in **Supplement 9**.

**Non-volatile residue (NVR)** was measured by thermal desorption (TD) GC-MS on a Shimadzu QP 2010 Plus GC fitted with a Gerstel CIS4 TD unit and a Shimadzu QP 2010 Ultra MS. The TD oven was heated with a WATLOW THINBAND at 100 °C for 1 hour and purged with purified $GN_2$ at 250 standard $ft^3$/min prior to operation. The injection temperature was 330 °C and interface temperature of 350 °C. After a 1 min sampling time, analytes were focused onto a Restek RTX-5ms 5% phenyl 0.25 µm film thickness 30 m long GC column at 40–50 °C through a Tenax TA trap. The GC was configured with UHP (6.0 grade) He at 1 mL/min, 12 psi, and a 10:1 split between GC and TD. The MS was configured for electron ionization, with source temperature, and transfer line temperatures set to 180 °C and 250 °C, respectively. The quadrupole was calibrated with perfluorotributylamine (PFTBA) with a scan time of 0.3 s and an inter-scan delay of 0.1 s over the range of *m/z* 45–650. The oven program was held at 50 °C for 1 min, ramped at 50 °C/min to 230 °C, then ramped at 10 °C/min to a final temperature of 334 °C.

## Data tables

**Table S11-1** Carboxylic acids detected by GC-MS in the hot water extract of the sapphire witness plate subsample OREX-324001-100, aluminum shim OREX-377025-1, and Bennu aggregate OREX-803001-0 (Glavin & Dworkin et al. 2025).

| Compound | **OREX-324001-100** (nmol/cm$^2$) | **OREX-377025-1** (nmol/cm$^2$) | **OREX-803001-0** (nmol/g) |
|---|---|---|---|
| Formic acid | <1 | <1 | 4,106±91 |
| Acetic acid | <1 | <1 | 1,436±72 |
| Propanoic acid | <1 | <1 | 156±8 |
| Isobutyric acid | $1.7 \times 10^{-6} \pm 1.7 \times 10^{-6}$ | <1 | 42±2 |
| 2,2-Dimethylpropanoic acid | <1 | <1 | 40±3 |
| Butyric acid | <1 | <1 | 85±9 |
| 2-Methylbutyric acid | <1 | <1 | <0.1 |
| Isopentanoic acid | <1 | <1 | 95±7 |
| 2,2-Dimethylbutyric acid | <1 | <1 | <0.1 |
| 3,3-Dimethylbutyric acid | <1 | <1 | <0.1 |
| Pentanoic acid | <1 | <1 | 35±1 |
| 2-Ethylbutyric/2-methylpentanoic acid | <1 | <1 | <0.1 |
| 3-Methylpentanoic acid | <1 | <1 | <0.1 |
| 4-Methylpentanoic acid | <1 | <1 | <0.1 |
| Hexanoic acid | <1 | <1 | <0.1 |
| Benzoic acid | <1 | <1 | 346±10 |
| Oxalic acid | <1 | <1 | 844±44 |
| Malonic acid | <1 | <1 | <0.1 |
| Succinic acid | $1.5 \times 10^{-5} \pm 1.0 \times 10^{-6}$ | <1 | <0.1 |
| Fumaric/Maleic acid | <1 | <1 | <0.1 |
| Glutaric acid | <1 | <1 | 25±1 |
| **Sum carboxylic acids** | **$1.4 \times 10^{-5} \pm 2.0 \times 10^{-6}$** | **<1** | **7,210±125** |

**Table S11-2** Organic compounds in the DCM and MeOH (4:1 v/v) extracts of Bennu aggregate OREX-800107-128, sapphire flight witness plate subsample OREX-324001, and flight aluminum shim subsample OREX-377025-3. Shown are the retention times in accurate masses of diagnostic fragment ions (*m/z*) obtained by GC×GC-HRMS. Abundances are expressed in nmol/g of sample. Control samples were below the detection limit.

| # | Compounds | Retention Time (s) | Accurate Mass[a] | Samples [a] | | |
|---|---|---|---|---|---|---|
| | **Aliphatic hydrocarbons** | **(1st, 2nd)** | ***m/z*** | **OREX-800107-128[b]** | **OREX-324001-101** | **OREX-377025-3** |
| 1 | *n*-Hexane ($C_6$) | 480.05, 0.60 | 57.0697 | 0.6 | <0.01 | <0.01 |
| 2 | *n*-Nonane ($C_9$) | 2410.75, 0.65 | 43.0544 | 0.9 | <0.01 | <0.01 |
| 3 | *n*-Decane ($C_{10}$) | 3339.31, 0.62 | 57.0698 | 2.4 | <0.01 | <0.01 |
| 4 | *n*-Dodecane ($C_{12}$) | 4618.27, 0.67 | 57.0698 | 0.9 | <0.01 | <0.01 |
| 5 | *n*-Tetradecane ($C_{14}$) | 5602.90, 0.65 | 57.0698 | 1.4 | <0.01 | <0.01 |
| 6 | *n*-Hexadecane ($C_{16}$) | 6384.29, 0.62 | 57.0698 | 2.3 | <0.01 | <0.01 |
| 7 | *n*-Octadecane ($C_{18}$) | 7050.05, 0.69 | 57.0700 | 0.3 | <0.01 | <0.01 |
| 8 | *n*-Nonadecane ($C_{19}$) | 7354.90, 0.70 | 57.0700 | 0.5 | <0.01 | <0.01 |
| 9 | *n*-Eicosane ($C_{20}$) | 7645.73, 0.69 | 57.0700 | 0.1 | <0.01 | <0.01 |
| 10 | *n*-Docosane ($C_{22}$) | 8185.34, 0.73 | 57.0700 | 0.1 | <0.01 | <0.01 |
| 11 | *n*-Tetracosane ($C_{24}$) | 8679.41, 0.72 | 57.0700 | 0.1 | <0.01 | <0.01 |
| 12 | *n*-Hexacosane ($C_{26}$) | 9033.31, 0.78 | 57.0700 | 0.2 | <0.01 | <0.01 |
| 13 | *n*-Octacosane ($C_{28}$) | 9464.30, 0.78 | 57.0700 | 0.1 | <0.01 | <0.01 |
| 14 | *n*-Triacontane ($C_{30}$) | 10070.50, 0.81 | 57.0700 | <0.1 | <0.01 | <0.01 |
| | *Total aliphatic hydrocarbons (nmol/g)* | | | 9.8 | <0.01 | <0.01 |

| # | Compounds | Retention Time (s) | Accurate Mass[a] | Samples [a] | | |
|---|---|---|---|---|---|---|
| | **Aromatic Hydrocarbons [c]** | **(1st, 2nd)** | ***m/z*** | **OREX-800107-128[b]** | **OREX-324001-101** | **OREX-377025-3** |
| 15 | Benzene | 693.792, 0.696 | 78.0459 | 13.5 | <0.01 | <0.01 |
| 16 | Toluene | 1240.42, 0.888 | 91.0541 | 114.5 | <0.01 | <0.01 |
| 17 | $C_2$-Benzene isomers | 1969.25 - 2347.68, 0.791 | 91.0541 | 1.2 | <0.01 | <0.01 |
| 18 | Styrene | 2305.63, 1.356 | 104.062 | 0.3 | <0.01 | <0.01 |
| 19 | Indane | 3588.1, 0.928 | 117.0697 | 11.2 | <0.01 | <0.01 |
| 20 | Indene | 3647.66, 1.112 | 115.0539 | 6.6 | <0.01 | <0.01 |
| 21 | $C_3$-Benzene isomers | 2684.06 - 3787.82, 0.712 | 105.0695 | 2.1 | <0.01 | <0.01 |
| 22 | $C_4$-Benzene isomers | 4117.2, 0.928 | 119.0854 | 6.1 | <0.01 | <0.01 |
| 23 | Naphthalene | 4546.59, 1.152 | 128.0620 | 38.1 | <0.01 | <0.01 |
| 24 | 2-Methylnaphthalene | 5156.29, 1.040 | 142.0775 | 32.9 | <0.01 | <0.01 |
| 25 | 1-Methylnaphthalene | 5238.48, 1.232 | 142.0775 | 23.0 | <0.01 | <0.01 |
| 26 | Biphenyl | 5552.24, 1.048 | 154.0775 | 34.4 | <0.01 | <0.01 |
| 27 | $C_2$-alkylnaphthalenes | 5643.86 - 5766.66, 1.138 | 156.0932 | 47.2 | <0.01 | <0.01 |
| 28 | Diphenylmethane | 5778.1, 1.120 | 168.0930 | 48.0 | <0.01 | <0.01 |
| 29 | $C_3$-$C_4$-alkylnaphthalenes | 5859.42 - 6841.18, 1.055 | 155.0855 | 296.4 | <0.01 | <0.01 |
| 30 | Acenaphthylene | 5864.1, 1.200 | 152.0619 | 5.0 | <0.01 | <0.01 |
| 31 | Methylbiphenyl isomers | 5991.84 - 6110.98, 1.152 | 168.0928 | 15.6 | <0.01 | <0.01 |
| 32 | Acenaphthene | 6007.76, 1.048 | 153.0697 | 25.7 | <0.01 | <0.01 |
| 33 | $C_4$-$C_5$-alkylnaphthalenes | 6113.16 - 6981.23, 0.984 | 169.1012 | 271.0 | <0.01 | <0.01 |
| 34 | Fluorene | 6382.69, 1.128 | 165.0697 | 29.4 | <0.01 | <0.01 |
| 35 | Phenanthrene | 7069.47, 1.296 | 178.0775 | 57.1 | <0.01 | <0.01 |
| 36 | Anthracene | 7101.01, 1.296 | 178.0775 | 22.8 | <0.01 | <0.01 |
| 37 | $C_1$-Alkylphenanthrenes | 7386.72 - 7530.44, 1.504 | 192.0929 | 13.3 | <0.01 | <0.01 |
| 38 | $C_2$-Alkylphenanthrenes | 7808.05 - 7985.48, 1.488 | 206.1088 | 3.7 | <0.01 | <0.01 |
| 39 | 1-Phenylnaphthalene | 7873.49, 1.944 | 204.0567 | 0.4 | <0.01 | <0.01 |
| 40 | Fluoranthene | 7913.94, 1.472 | 202.0775 | 14.3 | <0.01 | <0.01 |
| 41 | $C_3$-Alkylphenanthrenes | 8004.45 - 8417.32, 1.445 | 220.124 | 3.2 | <0.01 | <0.01 |
| 42 | Pyrene | 8068.11, 1.520 | 202.0775 | 10.7 | <0.01 | <0.01 |
| 43 | Methylfluoranthenes and methylpyrenes | 8215 - 8486.82, 1.752 | 216.0929 | 9.5 | <0.01 | <0.01 |
| 44 | *p*-Terphenyl | 8231.36 - 1.040 | 230.1084 | 0.2 | <0.01 | <0.01 |
| 45 | $C_2$-Fluoranthenes,$C_2$-pyrenes, or Terphenyls | 8465.37 - 8917.41, 1.472 | 230.1091 | 10.2 | <0.01 | <0.01 |
| 46 | $C_3$-Fluoranthenes,$C_3$-pyrenes, or Methylterphenyls | 8605.82 - 9019.30, 1.512 | 244.1241 | 7.5 | <0.01 | <0.01 |
| 47 | Benz(a)anthracene | 8867.02, 1.480 | 228.0931 | 1.1 | <0.01 | <0.01 |
| 48 | Chrysene | 8888.05, 1.504 | 228.0931 | 0.8 | <0.01 | <0.01 |
| 49 | Benzo(b)fluoranthene | 9560.82, 2.616 | 252.0932 | 0.5 | <0.01 | <0.01 |
| 50 | Benzo(k)fluoranthene | 9578.34, 2.576 | 252.0932 | 0.5 | <0.01 | <0.01 |
| 51 | Benzo(a)pyrene | 9823.62, 3.224 | 252.0932 | 0.2 | <0.01 | <0.01 |
| 52 | Benzo(g,h,i)perylene | 11456.5, 0.624 | 276.0935 | 0.1 | <0.01 | <0.01 |
| | *Total aromatic hydrocarbons (nmol/g)* | | | 1,178.4 | <0.01 | <0.01 |

| # | Compounds | Retention Time (s) | Accurate Mass[a] | Samples [a] | | |
|---|---|---|---|---|---|---|
| | **Nitrogen [c]** | **(1st, 2nd)** | ***m/z*** | **OREX-800107-128[b]** | **OREX-324001-101** | **OREX-377025-3** |
| 53 | Pyridine | 1075.73, 0.992 | 52.031 | 289.6 | <0.01 | <0.01 |
| 54 | Pyridine, 3-methyl- | 1632.86, 1.224 | 93.0574 | 285.4 | <0.01 | <0.01 |
| 55 | Pyridine, 2-methyl- | 1997.28, 1.504 | 93.0573 | 247.8 | <0.01 | <0.01 |
| 56 | $C_2$-Alkylpyridines | 1840.76 -3394.32, 0.792 | 107.0728 | 491.8 | <0.01 | <0.01 |
| 57 | Aniline | 3150.10, 1.712 | 93.0572 | 1.2 | <0.01 | <0.01 |
| 58 | Benzonitrile | 3206.16, 2.264 | 103.0416 | 91.4 | <0.01 | <0.01 |
| 59 | Benzylamine | 3402.38, 1.000 | 106.065 | 67.1 | <0.01 | <0.01 |
| 60 | $C_{3\text{-}4}$ Alkylpyridines | 3493.98 - 4527.33, 0.952 | 107.0728 | 313.2 | <0.01 | <0.01 |
| 61 | Hexamethylenetetramine (HMT) | 4775.95, 0.640 | 140.1056 | 8.8 | <0.01 | <0.01 |
| 62 | Quinoline | 4881.54 - 1.336 | 129.0569 | 1.8 | <0.01 | <0.01 |
| 63 | Isoquinoline | 4965.17 - 1.356 | 129.0569 | 1.3 | <0.01 | <0.01 |
| 64 | Carbazole | 7230.66, 3.088 | 167.0728 | 9.4 | <0.01 | <0.01 |
| | *Total N-bearing (nmol/g)* | | | 1,808.7 | <0.01 | <0.01 |
| | **Oxygen** | | | | | |
| 65 | Benzaldehyde | 3001.33, 1.608 | 105.0334 | 12.4 | <0.01 | <0.01 |
| 66 | Acetophenone | 3772.21, 1.286 | 105.0334 | 115.3 | <0.01 | <0.01 |
| 67 | Methyl benzoate | 3949.01, 1.368 | 105.0334 | 12.2 | <0.01 | <0.01 |
| 68 | Propiophenone | 4415.04, 1.472 | 105.0336 | 43.0 | <0.01 | <0.01 |
| 69 | Dibenzofuran | 6126.9, 1.168 | 168.0568 | 462.4 | <0.01 | <0.01 |
| 70 | Benzophenone | 6534.96, 1.504 | 105.0335 | 9.0 | <0.01 | <0.01 |
| 71 | 9-Fluorenone | 6923.90 - 1768 | 180.0568 | 1.8 | <0.01 | <0.01 |
| 72 | Methyl dibenzofuran isomers | 7256.78 - 7361.9, 1.840 | 182.0724 | 183.5 | <0.01 | <0.01 |
| 73 | Anthraquinone | 7649.70 - 1.168 | 208.0518 | 0.1 | <0.01 | <0.01 |
| | *Total O-bearing (nmol/g)* | | | 839.7 | <0.01 | <0.01 |

| # | Compounds | Retention Time (s) | Accurate Mass[a] | Samples [a] | | |
|---|---|---|---|---|---|---|
| | **Sulfur [c]** | **(1st, 2nd)** | ***m/z*** | **OREX-800107-128[b]** | **OREX-324001-101** | **OREX-377025-3** |
| 74 | Thiophene | 721.824, 0.952 | 84.0029 | 3.6 | <0.01 | <0.01 |
| 75 | 2- or 3-Methylthiophene | 1275.46, 0.992 | 97.0108 | 6.2 | <0.01 | <0.01 |
| 76 | 3- or 2-Methylthiophene | 1324.51, 1.053 | 97.0108 | 5.1 | <0.01 | <0.01 |
| 77 | $C_2$-alkylthiophenes | 1993.78, 1.096 - 2466.82, 1.320 | 97.0108 | 2.3 | <0.01 | <0.01 |
| 78 | Benzothiophene | 4597.25, 1.55 | 134.0182 | 16.4 | <0.01 | <0.01 |
| 79 | Benzothiazole | 4782.96, 1.856 | 135.0137 | 59.8 | <0.01 | <0.01 |
| 80 | Methylbenzothiazoles | 4887.81 - 5073.1, 1.984 | 135.0137 | 8.1 | <0.01 | <0.01 |
| 81 | Dibenzothiophene | 6979.97, 1.520 | 184.0338 | 28.5 | <0.01 | <0.01 |
| 82 | Dimethyl sulfide | 364.416, 0.624 | 62.0184 | 4.9 | <0.01 | <0.01 |
| 83 | Dimethyl disulfide | 717.976, 870.552 | 63.9437 | 4.0 | <0.01 | <0.01 |
| 84 | Dimethyl trisulfide | 3087.02, 1.224 | 78.967 | 54.8 | <0.01 | <0.01 |
| 85 | Dimethyl tetrasulfide | 4744.42, 1.248 | 78.967 | 186.7 | <0.01 | <0.01 |
| 86 | Dimethyl sulfite | 1054.700, 1.368 | 78.9848 | 7,914.1 | <0.01 | <0.01 |
| 87 | $S_8$ and allotropes | 6517.97 - 8721.41, 1.14 | 63.9437 | [d] | <0.01 | <0.01 |
| | *Total S-bearing (nmol/g)* | | | 8,294.6 | <0.01 | <0.01 |

[a] HRMS *m/z* values were used to isolate compounds peaks within a range of ±10 ppm.

[b] Aponte et al. 2026

[c] $C_2$, $C_3$, and $C_4$ refers to the mixture of molecular isomers, where for example $C_2$ can be dimethyl or ethyl, $C_3$ can be trimethyl, methyl-ethyl, or propyl, etc.

[d] Concentrations reached saturation limits of mass spectrometer detector.

**Table S11-3** Compounds targeted by MRM by pyrolysis GC-MS of sapphire witness plate subsample OREX-324001-103, sapphire control 3, and Bennu aggregate OREX-800128-104.

| **Analyte** | **Scan window** (min) | **Masses** (*m/z*) | **Sapphire Control 3** | **OREX-324001-103 Sapphire Witness** | **OREX-800128-104 Bennu Sample** |
|---|---|---|---|---|---|
| Benzene | 4.3±1 | 78.1 | trace | trace | + |
| Thiophene | 4.4±1 | 84.1 | trace | trace | + |
| Dimethyl disulfide | 6.7 ±1 | 94 | - | - | + |
| Pyridine | 6.8±1.5 | 52.1, 79.1 | - | - | + |
| Toluene | 7.6±1 | 91.1, 92.1 | + | + | + |
| $C_1$-Alkylthiophenes | 8±1.5 | 97.1 | - | - | + |
| $C_2$-Alkylbenzenes | 12.5±3 | 91.1, 105.1, 106.1 | trace | + | + |
| $C_2$-Alkylthiophenes | 13±3 | 111.1, 112.1 | - | - | + |
| Styrene | 13.2±1 | 78.1, 104.1 | trace | + | + |
| $C_3$-Alkylbenzenes | 18.5±6 | 105.1, 120.2 | trace | trace | + |
| $C_3$-Alkylthiophenes | 19±8 | 125.1, 126.1 | - | - | + |
| Benzaldehyde | 16.5±1 | 77.1, 105.1, 106.1 | - | - | + |
| Dimethyl trisulfide | 17±1 | 79, 126 | - | - | + |
| Aniline | 17.4±1 | 93.1 | - | - | + |
| Phenol | 17.6±1 | 66.1, 94.1 | - | - | + |
| Benzonitrile | 17.7±1 | 76.1, 103.1 | - | - | + |
| Benzofuran | 18.3±1 | 90.1, 118.1 | - | - | + |
| $C_4$-Alkylbenzenes | 23±8 | 119.1, 134.2 | - | - | + |
| Naphthalene | 26.8±1 | 128.1 | - | + | + |
| Benzo[c]thiophene | 27.1±1 | 134.1 | - | - | + |
| Benzothiazole | 28.5±1 | 108, 135 | - | - | + |
| Thieno[n,n]thiophenes | 28.5±2 | 96.1, 140 | - | - | + |
| $C_1$-Alkylbenzo[b]thiophenes | 31.5±2 | 147.1, 148.1 | - | - | + |
| $C_1$-Alkylnaphthalenes | 31.6±1.5 | 141.1, 142.1 | trace | trace | + |
| Biphenyl | 34.5±1 | 76.1, 154.1 | - | trace | + |
| Bithiophene | 36±3 | 121, 166 | - | - | + |
| $C_2$-Alkylnaphthalenes | 36.1±4 | 141.1, 156.1 | - | - | + |
| Acenaphthylene | 37.2±1 | 76.1, 152.1 | - | - | + |
| Acenaphthene | 38.5±1 | 153.1, 154.1 | - | - | + |
| Dibenzofuran | 39.5±1 | 139.1, 168.1 | - | - | + |
| $C_3$-Alkylnaphthalenes | 40.5±5.5 | 155.2, 170.2 | - | - | + |
| Phenalene | 41.2±1.5 | 165.1, 166.1 | - | - | + |
| Fluorene | 41.8±1.5 | 165.1, 166.1 | - | - | + |
| Benzophenone | 43.4±1 | 105, 182.1 | - | - | + |
| $C_4$-Alkylnaphthalene | 44±6 | 169.1, 184.1 | - | - | + |
| Fluorenone | 46.8±1 | 152.1, 180.1 | - | - | + |
| Dibenzothiophene | 47.3 ±2 | 139.1, 184.1 | - | - | + |
| Phenanthrene + Anthracene | 48.3±2.5 | 178.1 | - | - | + |
| $C_1$-Alkylphenanthrenes | 52±4 | 191.1, 192.1 | - | - | + |
| Fluoranthene + Pyrene | 56.8±2.5 | 101.1, 202.1 | - | - | + |

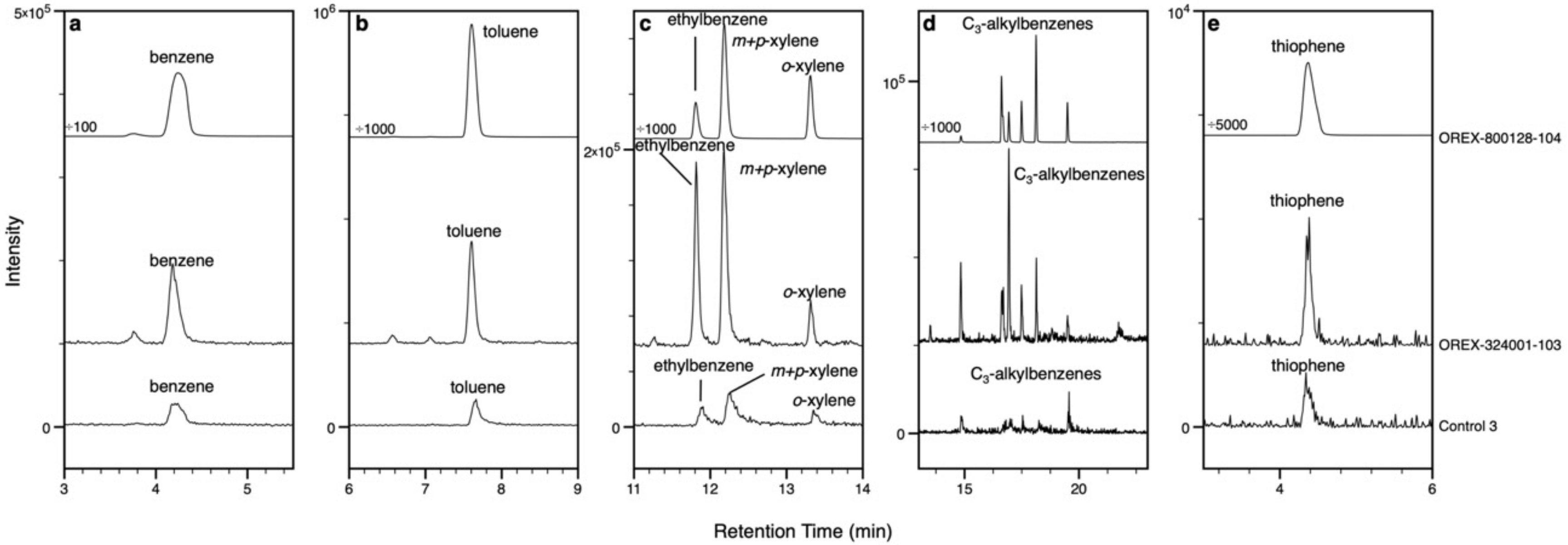

**Fig. S11-1** Pyrolysis GC-MS (**Supplement 9**) of compounds also detected in sapphire control 3 and seen in sapphire witness plate subsample OREX-324001-103 and Bennu aggerate sample OREX-800128-104. All peaks are identified based on their retention time compared with standards when available. **a.** Chromatogram of the MRM 78.1 → 77.1 transition characteristic of benzene. The Bennu aggregate chromatogram has been scaled down by a factor of 100 for clarity. **b.** Chromatogram of the MRM 92.1 → 91.1 transition characteristic of toluene. The Bennu aggregate chromatogram has been scaled down by a factor of 1,000 for clarity. **c.** Chromatogram of the MRM 106.1 → 91.1 transition characteristic of $C_2$-alkylbenzenes. The Bennu aggregate chromatogram has been scaled down by a factor of 1,000 for clarity. **d.** Chromatogram of the MRM 120.1 → 105.1 transition characteristic of various $C_3$-alkylbenzenes. The Bennu aggregate chromatogram has been scaled down by a factor of 1,000 for clarity. **e.** Chromatogram of the MRM 84.1 → 58.1 transition characteristic of thiophene. The Bennu aggregate chromatogram has been scaled down by a factor of 5,000 for clarity.

**Table S11-4** The samples for NVR analysis by TD-GC-MS showed low levels of NVR with low confidence as the procedural controls were higher than the samples. Each sample was spiked with 100 ng each of deuterated benzene ($d_6$), deuterated toluene ($d_7$), and deuterated naphthalene ($d_{10}$) for calibration. This permitted semi-quantitation by averaging the peak area of the deuterated benzene and deuterated toluene (the deuterated naphthalene peak area was significantly higher) of all the samples. An error of two standard deviations is applied to the total NVR. All of the samples, including the system blank, contained chloroform and potential diethylhexyl adipate (a handling contaminant) at ~12 min retention time as the major contaminant.

| Sample | Sample Area ($cm^2$) | Contaminant Mass (ng) | NVR ($ng/cm^2$) |
|---|---|---|---|
| Aluminum control 2 | 4.0 | 182 | 91±52 |
| Shim subsample OREX-337025-2 | 1.76 | 106 | 28±16 |
| Sapphire control 103 | 2.38 | 186 | 65±37 |
| Witness subsample OREX-324001-103 | 1.71 | 64 | 45±25 |

**Table S11-5** Blank subtracted free abundances of ammonia, hydrazine and amines measured by LC-MS, in the hot water extract of the sapphire witness subsample OREX-324001-100, aluminum shim subsample OREX-377025-1, and Bennu aggregate OREX-803001-0 (Glavin & Dworkin et al. 2025).

| **Compound** | **OREX-324001-100 Sapphire Witness (nmol/cm$^2$)** | **OREX-377025-1 Aluminum Shim (nmol/cm$^2$)** | **OREX-803001-0 Bennu Sample (nmol/g)** |
|---|---|---|---|
| ammonia | n.d. | n.d. | 13,613±357 |
| hydrazine | n.d. | n.d. | <0.1 |
| methylamine | 0.0019±0.00013 | n.d. | 914±88 |
| ethylamine | n.d. | n.d. | 121±7 |
| *iso*-propylamine | n.d. | n.d. | 6.8±0.3 |
| *n*-propylamine | n.d. | n.d. | 7.0±0.3 |
| *sec*-butylamine | n.d. | n.d. | 0.47±0.02 |
| *iso*-butylamine | n.d. | n.d. | 1.3±0.2 |
| *n*-butylamine | n.d. | n.d. | 1.0±0.3 |
| *tert*-butylamine | n.d. | n.d. | 3.4±0.4 |
| 3-aminopentane | n.d. | n.d. | 0.34±0.02 |
| 2-amino-3-methyl-*n*-butylamine | n.d. | n.d. | 0.44±0.02 |
| *sec*-pentylamine | n.d. | n.d. | 0.56±0.02 |
| 2-Methylbutylamine | n.d. | n.d. | 0.54±0.02 |
| *tert*-pentylamine | n.d. | n.d. | 1.3±0.1 |
| *iso*-pentylamine | n.d. | n.d. | 0.53±0.04 |
| *n*-pentylamine | n.d. | n.d. | 0.49±0.02 |
| *n*-hexylamine | n.d. | n.d. | 0.54±0.02 |
| **Sum Amines** | **0.0019±0.00013** | **n.d.** | **1,060±89** |

n.d. = not determined.

**Table S11-6** Blank subtracted free abundances of amino acids as measured by LC-MS, in the hot water extract of sapphire witness subsample OREX-324001-100, aluminum shim subsample OREX-377025-1, and Bennu aggregate OREX-803001-0 (Glavin & Dworkin et al. 2025).

| Amino Acid | **OREX-324001-100** **Sapphire Witness** ($nmol/cm^2$) | **OREX-377025-1** **Aluminum Shim** ($nmol/cm^2$) | **OREX-803001-0** **Bennu Sample** (nmol/g) |
|---|---|---|---|
| Glycine | n.d. | n.d. | 10.1±0.5 |
| D,L-Alanine | n.d. | n.d. | 1.60±0.03 |
| D,L-Valine | n.d. | n.d. | 0.081±0.004 |
| D,L-Leucine | n.d. | n.d. | 0.036±0.003 |
| D,L-Proline | n.d. | n.d. | 0.11±0.01 |
| D,L-Cysteine | n.d. | n.d. | <0.1 |
| D,L-Serine | n.d. | n.d. | tr. |
| D,L-Threonine | n.d. | n.d. | tr. |
| D,L-Asparagine | n.d. | n.d. | tr. |
| D,L-Glutamine | n.d. | n.d. | <0.1 |
| D,L-Aspartic acid | n.d. | n.d. | tr. |
| D,L-Glutamic acid | n.d. | n.d. | 0.01±0.01 |
| D,L-Arginine | n.d. | n.d. | <0.1 |
| D,L-Histidine | n.d. | n.d. | <0.1 |
| D,L-Lysine | n.d. | n.d. | <0.1 |
| D,L-Tyrosine | n.d. | n.d. | tr. |
| D,L-Methionine | n.d. | n.d. | <0.1 |
| **SUM PROTEIN AMINO ACIDS** | **n.d.** | **n.d.** | **11.9±0.5** |

n.d. = not determined. tr. = trace below the limit of quantitation.

**Table S11-7** Bank-corrected amino acid abundances, as measured by LC-MS, in the non-hydrolyzed (free) and 6 M HCl-vapor hydrolyzed (total) hot water extracts of sapphire witness plate subsample OREX-324001-100, aluminum shim subsample OREX-377025-1, and Bennu aggregate OREX-803001-0 (Glavin & Dworkin et al. 2025). Higher total amino acid concentration than free amino acids indicates that there were amino acid precursors or peptides that released monomers upon hydrolysis. The presence of lower total amino acid concentration than free amino acids may indicate that the small amount of destruction of amino acids during hydrolysis was significant due to the very low abundances such as was observed in some amino acids from Hayabusa2 (Parker et al. 2023).

| **Amino Acid** | **OREX-324001-100 Sapphire Witness** (nmol/cm$^2$) | | **OREX-377025-1 Aluminum Shim** (nmol/cm$^2$) | | **OREX-803001-0 Bennu Sample** (nmol/g) | |
|---|---|---|---|---|---|---|
| | ***Free*** | ***Total*** | ***Free*** | ***Total*** | ***Free*** | ***Total*** |
| Glycine | n.d. | n.d. | n.d. | 0.0039±0.00053 | 33±1 | 44±1 |
| D-Alanine | 0.00025±0.000034 | n.d. | n.d. | 0.00022±0.00001 | 2.8±0.4 | 4.0±0.6 |
| L-Alanine | 0.0054±0.00067 | n.d. | n.d. | 0.0011±0.00001 | 2.6±0.4 | 3.9±0.6 |
| β-Alanine | n.d. | n.d. | n.d. | 0.00034±0.00001 | 1.6±0.3 | 3.3±0.5 |
| D-Serine | 0.00094±0.00016 | 0.00011±0.000038 | n.d. | n.d. | 0.21±0.09 | 0.18±0.03 |
| L-Serine | 0.015±0.0011 | 0.0012±0.00028 | n.d. | n.d. | 0.24±0.01 | ≤0.7 |
| D-Aspartic Acid | n.d. | n.d. | n.d. | n.d. | 0.94±0.12 | 1.35±0.14 |
| L-Aspartic Acid | 0.0040±0.00035 | 0.0020±0.00017 | n.d. | n.d. | 0.79±0.16 | 1.22±0.12 |
| D-Threonine | n.d. | n.d. | n.d. | n.d. | <1 | <0.2 |
| L-Threonine | n.d. | 0.00082±0.00076 | n.d. | 0.00059±0.00001 | <1 | <0.2 |
| D,L-α-Amino-*n*-butyric acid | 0.0016±0.00082 | n.d. | n.d. | 0.00019±0.00001 | 0.55±0.03 | 0.80±0.02 |
| D-β-Amino-*n*-butyric acid | n.d. | n.d. | n.d. | n.d. | 0.34±0.05 | 0.55±0.08 |
| L-β-Amino-*n*-butyric acid | n.d. | n.d. | n.d. | n.d. | 0.33±0.05 | 0.53±0.08 |
| ɣ-Amino-*n*-butyric acid | n.d. | n.d. | 0.00087±0.0010 | n.d. | 0.37±0.04 | 3.03±0.13 |
| α-Aminoisobutyric acid | n.d. | n.d. | n.d. | n.d. | 0.21±0.03 | 0.61±0.05 |
| D-Glutamic Acid | n.d. | n.d. | n.d. | n.d. | 0.08±0.02 | 0.79±0.06 |
| L-Glutamic Acid | n.d. | n.d. | n.d. | n.d. | 0.07±0.01 | ≤0.8 |
| D-Valine | n.d. | n.d. | n.d. | n.d. | 0.08±0.01 | 0.16±0.01 |
| L-Valine | n.d. | n.d. | n.d. | n.d. | 0.09±0.01 | 0.32±0.06 |
| D-Isovaline | n.d. | n.d. | n.d. | n.d. | 0.015±0.003 | 0.080±0.007 |
| L-Isovaline | n.d. | n.d. | n.d. | n.d. | 0.016±0.002 | 0.074±0.006 |
| D-Norvaline | n.d. | n.d. | n.d. | n.d. | 0.024±0.004 | 0.052±0.005 |
| L-Norvaline | n.d. | n.d. | n.d. | n.d. | 0.029±0.007 | 0.051±0.004 |
| D-Leucine | n.d. | n.d. | n.d. | n.d. | tr. | 0.09±0.01 |
| L-Leucine | 0.0021±0.000053 | n.d. | n.d. | 0.00053±0.00001 | <0.3 | <0.3 |
| D-Isoleucine | 0.0021±0.000053 | n.d. | n.d. | n.d. | tr. | 0.069±0.005 |
| L-Isoleucine | n.d. | n.d. | n.d. | 0.00030±0.00001 | <0.1 | ≤0.4 |
| ε-Amino-*n*-caproic acid | n.d. | n.d. | n.d. | n.d. | <0.1 | 0.19±0.07 |
| **SUM $C_2$-$C_6$ AMINO ACIDS** | **0.030±0.002** | **0.0041±0.0008** | **0.00087±0.001** | **0.0068±0.0005** | **45±1** | **66±2** |

n.d. = not determined.

**Supplement 12:** List of the measurement data products from the Bennu and spacecraft samples analyzed in this study and corresponding DOIs available at https://astromat.org

### Images

| DOI | Product Name |
|---|---|
| **AIVA** | **Canister, TAGSAM** |
| 10.60707/11h6-hs71 | 20230926_AIVA_JSC-ARES_OREX-108001-0_2_AIVAImage_1886 |
| 10.60707/k0dp-cp12 | 20230926_AIVA_JSC-ARES_OREX-108001-0_1_AIVAImage_1 |
| 10.60707/s3zr-tk83 | 20230927_AIVA_JSC-ARES_OREX-108001-0_1_AIVAImage_1885 |
| **Curation Images** | **Canister, TAGSAM** |
| 10.60707/tc82-ev79 | 20230924_CPD_JSC-ARES_OREX-108001-0_1_CPDImage_1 |
| 10.60707/txsm-0a43 | 20230926_CPD_JSC-ARES_OREX-108001-0_1_CPDImage_1 |
| 10.60707/jham-pv51 | 20230926_CPD_JSC-ARES_OREX-108001-0_2_CPDImage_2 |
| 10.60707/f5jn-ba74 | 20231002_CPD_JSC-ARES_OREX-310000-0_1_CPDImage_1 |
| 10.60707/wfe1-3a78 | 20240328_CPD_JSC-ARES_OREX-372001-0_1_CPDImage_1 |
| 10.60707/vmwz-1e35 | 20240328_CPD_JSC-ARES_OREX-372001-0_2_CPDImage_2 |
| 10.60707/sg6j-4y95 | 20240328_CPD_JSC-ARES_OREX-373001-0_1_CPDImage_1 |
| 10.60707/s5kw-7a20 | 20240401_CPD_JSC-ARES_OREX-377025-0_1_CPDImage_1 |
| 10.60707/f114-py90 | 20240328_CPD_JSC-ARES_OREX-382001-0_1_CPDImage_1 |
| 10.60707/a38h-7t39 | 20231016_CPD_JSC-ARES_OREX-580001-0_1_CPDImage_1 |

### SEM/EDX

| DOI | Product Name |
|---|---|
| **SEM/EDX** | **OREX-220001-1, OREX-501030-0, OREX-590073-0, OREX-800055-1** |
| 10.60707/4x7v-mk29 | 20250730_SEM_JSC-ARES_OREX-220001-1_1_SEMImageCollection_2096 |
| 10.60707/NPYE-7N94 | 20231025_SEM_NMNH_OREX-501030-0_1_SEMEDSElementalMaps_415 |
| 10.60707/P16K-K193 | 20231025_SEM_NMNH_OREX-501030-0_1_SEMEDSPointData_223 |
| 10.60707/YQSP-RX12 | 20231025_SEM_NMNH_OREX-501030-0_1_SEMImageCollection_414 |
| 10.60707/a76e-8g85 | 20250331_SEM_JSC-ARES_OREX-590073-0_1_SEMImageCollection_2097 |
| 10.60707/fk0w-8x81 | 20240924_SEM_JSC-ARES_OREX-800055-1_1_SEMImageCollection_2098 |

### GC-MS

| DOI | Product Name |
|---|---|
| **GC-MS/IRMS** | **OREX-324001-100, OREX-377025-1, OREX-800107-128** |
| 10.60707/tw8z-7211 | 20260420_GC-MS_GSFC_OREX-324001-100_1 |
| 10.60707/z908-pt31 | 20260421_GC-MS_GSFC_OREX-377025-1_1 |
| 10.60707/paf0-x016 | 20240911_GC-MS_GSFC_OREX-800107-128_1_GCMSCollection_2046 |
| **GC-MS** | **OREX-590040-0, OREX-590054-0** |
| 10.60707/4b8s-q597 | 20231222_590040_qb_grey |
| 10.60707/w4zf-m622 | 20231222_590054_lc1_grey |
| **TD-GC-MS** | **OREX-324001-102, OREX-377025-2** |
| 10.60707/w8qs-x350 | 20260424_GC-MS_TU_OREX-324001-102_1 |
| 10.60707/m11b-5w42 | 20260424_GC-MS_TU_OREX-377025-2_1 |
| **GC×GC-HRMS** | **OREX-324001-101, OREX-377025-3** |
| 10.60707/t32t-4f32 | 20260430_GC-MS_GSFC_OREX-324001-101_1_GCMSCollection_1771 |
| 10.60707/dxb6-t563 | 20260430_GC-MS_GSFC_OREX-377025-3_1_GCMSCollection_1786 |
| **Pyrolysis GC-MS** | **OREX-324001-103, OREX-800128-104** |
| 10.60707/zmf1-7773 | 20260224_GC-MS_GSFC_OREX-324001-103_1 |
| 10.60707/my67-xz65 | 20260304_GC-MS_GSFC_OREX-800128-104_1 |

## LC-MS

| DOI | Product Name |
| --- | --- |
| **LC-MS** | **OREX-324001-100, OREX-337025-1** |
| 10.60707/bcj5-9y29 | 20260324_LC-MS_GSFC_OREX-324001-100_1_LCMSCollection_1 |
| 10.60707/7daz-3j29 | 20260421_LC-MS_GSFC_OREX-324001-100_1_LCMSCollection_1 |
| 10.60707/qy40-eq51 | 20260422_LC-MS_GSFC_OREX-324001-100_1_LCMSCollection_1 |
| 10.60707/sph9-f103 | 20260422_LC-MS_GSFC_OREX-324001-100_2_LCMSCollection_1 |
| 10.60707/y52w-dy48 | 20260428_LC-MS_GSFC_OREX-324001-100_1_LCMSCollection_1 |
| 10.60707/hsak-y928 | 20260428_LC-MS_GSFC_OREX-324001-100_2_LCMSCollection_1 |
| 10.60707/42k4-t296 | 20260429_LC-MS_GSFC_OREX-377025-1_1_LCMSCollection_1 |
| 10.60707/f2gn-d537 | 20260325_LC-MS_GSFC_OREX-377025-1_1_LCMSCollection_1 |
| 10.60707/ncrr-7236 | 20260421_LC-MS_GSFC_OREX-377025-1_1_LCMSCollection_1 |
| 10.60707/s94c-5497 | 20260422_LC-MS_GSFC_OREX-377025-1_1_LCMSCollection_1 |
| 10.60707/mb50-1c61 | 20260422_LC-MS_GSFC_OREX-377025-1_2_LCMSCollection_1 |
| 10.60707/hmyn-xn61 | 20260428_LC-MS_GSFC_OREX-377025-1_1_LCMSCollection_1 |